\newif\ifphysrep
\physrepfalse

\ifphysrep
\documentclass[onecolumn,float,floatfix,preprintnumbers,aps,rmp,nofootinbib,amsmath,amssymb,amsfonts,superscriptaddress,tightenlines]{revtex4}
\else
\documentclass[twocolumn,float,floatfix,preprintnumbers,aps,rmp,nofootinbib,amsmath,amssymb,amsfonts,superscriptaddress,tightenlines]{revtex4}
\fi

\usepackage[pass,letterpaper]{geometry}

\usepackage[utf8]{inputenc} 
\usepackage[T1]{fontenc}
\usepackage{feynmp-auto}
\usepackage[caption=false]{subfig}
\usepackage{siunitx}
\usepackage{dcolumn}
\newcolumntype{d}[1]{D{.}{.}{#1}}

\usepackage{hyperref}
\usepackage{slashbox}
\usepackage{amsmath}
\usepackage{amssymb}
\usepackage{amsfonts}

\usepackage{booktabs}
\usepackage{tabularx}
\newcolumntype{A}{>{\centering\arraybackslash}X}

\usepackage{graphicx}
\usepackage[usenames]{color}
\usepackage{dsfont}
\usepackage{bm}

\usepackage{subeqnarray}
\DeclareMathOperator\arccosh{arccosh}

\newcommand{\mycolumnwidth}{\ifphysrep 0.65 \textwidth \else \columnwidth \fi}

\newcommand{\myk}{K}

\newcommand{\be}{\begin{equation}}
\newcommand{\ee}{\end{equation}}

\newcommand{\bea}{\begin{eqnarray}}
\newcommand{\eea}{\end{eqnarray}}

\newcommand{\fourp}[1]{\underline{#1}}

\newcommand{\beas}{\begin{subeqnarray}}
\newcommand{\eeas}{\end{subeqnarray}}

\newcommand{\gr}[1]{\mathbf{#1}}

\newcommand{\ii}{\mathrm{i}}
\newcommand{\myxi}{\xi}

\newcommand{\dd}{\mathrm{d}}

 \newcommand{\Gap}{\Delta}
 \newcommand{\Fermi}{{\cal F}}
\newcommand{\YP}{Y_{\rm P}}
\newcommand{\YPfour}{Y_{\rm P}^{(4)}}

 \newcommand{\coeffa}{\alpha}

 \newcommand{\ma}{u}

\newcommand{\RC}{{\rm RC}}
\newcommand{\FD}[1]{g^+\left(#1\right)}
\newcommand{\BE}[1]{g^-\left(#1\right)}

\newcommand{\deu}{${\rm D}$}
\newcommand{\tro}{$^3{\rm He}$}
\newcommand{\qua}{$^4{\rm He}$}
\newcommand{\six}{$^{6}{\rm Li}$}
\newcommand{\sep}{$^{7}{\rm Li}$}
\newcommand{\bery}{$^{7}$Be}

\newcommand{\neu}{$^{9}$Be}
\newcommand{\dix}{$^{10}$B}
\newcommand{\onz}{$^{11}$B}

\newcommand{\zdag}{D$(\alpha,\gamma)^6$Li}
\newcommand{\npg}{{$^1$H(n,$\gamma)^2$H}}
\newcommand{\ddn}{{D(d,n)$^3$He}}
\newcommand{\ddp}{{D(d,p)$^3$H}}
\newcommand{\dpg}{{D(p,$\gamma)^3$He}}
\newcommand{\hag}{{$^3$He($\alpha,\gamma)^7$Be}}
\newcommand{\hli}{{$^4$He, D, $^3$He and $^{7}$Li}}
\newcommand{\sfac}{$S$--factor}
\newcommand\beq{\begin{equation}}
\newcommand\eeq{\end{equation}}
\makeatletter
\def\l@subsubsection#1#2{}
\makeatother

\begin{document}
\unitlength = 1mm
\title{Precision big bang nucleosynthesis with improved Helium-4 predictions}

\author{Cyril Pitrou}
\email{pitrou@iap.fr}
\affiliation{Institut d'Astrophysique de Paris,
CNRS UMR 7095, 98 bis Bd Arago, 75014 Paris,
France\\
Sorbonne Universit\'e, Institut Lagrange de Paris, 98 bis Bd Arago, 75014 Paris, France}

\author{Alain Coc}
\email{coc@csnsm.in2p3.fr}
\affiliation{Centre de Sciences Nucl\'eaires et de Sciences de la
  Mati\`ere (CSNSM), CNRS IN2P3,\\ 
  Univ. Paris-Sud,  Universit\'e
  Paris-Saclay, B\^atiment 104, F-91405 Orsay  Campus France}

\author{Jean-Philippe Uzan}
\email{uzan@iap.fr}
\affiliation{Institut d'Astrophysique de Paris,
CNRS UMR 7095, 98 bis Bd Arago, 75014 Paris,
France\\
Sorbonne Universit\'e, Institut Lagrange de Paris, 98 bis Bd Arago, 75014 Paris, France}

\author{Elisabeth Vangioni}
\email{vangioni@iap.fr}
\affiliation{Institut d'Astrophysique de Paris,
CNRS UMR 7095, 98 bis Bd Arago, 75014 Paris,
France\\
Sorbonne Universit\'e, Institut Lagrange de Paris, 98 bis Bd Arago, 75014 Paris, France}

\date{\today}

%\pacs{95.30.Jx, 98.70.Vc, 98.80.-k, 98.80.Es}

%\pacs{Y'en a pas dans Phys. ReP.!}

%%%%%%%%%%%%%%%%%%%%%%%%%%%%%%%%%%%%%%%%%%%%%%%%%%%%%%%%%%%%%%%%%%%%%%%
\begin{abstract}
Primordial nucleosynthesis is one of the three historical evidences for the big bang model, together with the expansion of the universe
and the cosmic microwave background. There is a good global agreement
between the computed primordial abundances of helium-4, deuterium, helium-3 and their values deduced from observations. Now that the number of neutrino families
and the baryonic densities have been fixed by laboratory measurements
or CMB observations, the model has no free parameter and its predictions are rigid.
Since this is the earliest cosmic process for which we  {\it a priori} know all the physics involved,  departure from 
its predictions could provide hints or constraints on new physics or
astrophysics in the early universe. Precision on primordial abundances deduced from 
observations has recently been drastically improved and reach the percent level for both deuterium and helium-4.  
Accordingly, the BBN predictions should reach the same level of
precision. For most isotopes, the dominant sources of uncertainty
come from those on the laboratory thermonuclear reactions. This
article focuses on helium-4 whose predicted primordial abundance depends
essentially on weak interactions which control the neutron-proton
ratio. The rates of the various weak interaction processes depend on
the experimentally measured neutron lifetime, but also includes
numerous corrections that we thoroughly investigate here. They are the radiative, zero--temperature,  corrections, finite nucleon mass  corrections, finite temperature radiative corrections,
weak-magnetism, and QED plasma effects, which are for the first time all included and calculated in a self consistent way, allowing
to take into account the correlations between them, and verifying that all satisfy detailed balance. 
Finally, we include the incomplete neutrino decoupling and claim to reach a $10^{-4}$
accuracy on the helium-4 predicted mass fraction of $0.24709\pm0.00017$ (when including the uncertainty on the neutron lifetime). 
In addition, we provide a {\it Mathematica} primordial nucleosynthesis code that incorporates,
not only these corrections but also a full network of reactions, using
the best available thermonuclear reaction rates, allowing the predictions of primordial abundances of helium-4, deuterium, helium-3
and lithium-7 but also of heavier isotopes up to the CNO region.  
\end{abstract}
\maketitle

%%%%%%%%%%%%%%%%%%%%%%%%%%%%%%%%%%%%%%%%%%%%%%%%%%%%%%%%%%%%%%%%%%%%%%%
%\newpage

\tableofcontents
%\newpage

\section{Introduction}

Besides the universal spatial expansion and the cosmic microwave
background (CMB) radiation, the third historical evidence for 
the hot big bang model comes from primordial, or big bang nucleosynthesis (BBN). During the first $\approx$20 minutes of the
Universe, when it was dense and hot enough for nuclear reactions to
take place, BBN describes the production of the so called  ``light
elements'',  \hli, together with  only minute traces of heavier
nuclei (see e.g. \citet{Oli10,Ste07,Cyb16,PDG17,CV17} for recent reviews). 
The number of free parameters that entered in standard BBN has
now been reduced to zero. Indeed, the number
of light neutrino families is now known from the measurement of the $Z^0$
width by LEP experiments at CERN: $N_\nu$ = 2.984$\pm$0.008 \cite{PDG17}.
The lifetime of the neutron entering in weak reaction rate calculations and many nuclear reaction rates have been measured in nuclear physics laboratories \cite{Coc15,Cyb16,Serpico:2004gx,Des04}.
The last parameter to have been independently determined is the baryonic density
of the Universe which is now deduced from the analysis of the anisotropies
of the CMB radiation from the {\it Planck} satellite data \cite{Planck2016}.
Hence, there is no more free parameter in standard BBN  and the calculated 
primordial abundances are in principle only affected by the moderate uncertainties in
nuclear cross sections. 
Keeping in mind that abundances span a range of nine orders of magnitude, the agreement between primordial abundances,
either deduced from observations, or from primordial
nucleosynthesis calculations, is an outstanding support to the hot big bang model \cite{Cyb16}. 
Hence, BBN is an invaluable tool for probing the physics of the early
Universe, and it has been very efficient to constrain physics beyond
the standard model. When we look back in time,
it is the ultimate process for which we {\it a priori} know all the
physics involved, given that it is very well tested in laboratories, so that departures from its predictions provide hints for new physics or astrophysics  \citep{IoccoReport,Pos10,MKK17,Nakamura2017}.

Great progresses have been made in the precision of both observations and laboratory measurements. 
The precision on deuterium observations have now reached the percent level \cite{Coo18}, a precision hardly 
reached in nuclear physics measurements.  The determination of the primordial abundance 
of \qua\ has been reduced to less than 2\% by the inclusion of an additional atomic infrared line \cite{Ave15}. However, there is still a significant discrepancy on lithium, for which 
predictions are a factor of  $\approx$3 higher than observations.
The previous paper \citet{Coc15} investigated the uncertainties on D/H predictions which
directly reflects the experimental uncertainties on few reaction rates that have all been re-evaluated \cite{Bayes16,Bayes17},
leading to a small but significant decrease of the predicted deuterium abundance with reduced uncertainty.
This review aims at reducing the uncertainties, this time on the \qua\ abundance prediction which are
dominated by theoretical uncertainties on the weak reactions that interconvert neutrons and protons.

To that goal, we implemented the BBN equations into a {\it
Mathematica} code\footnote{{\tt PRIMAT} : PRImordial MATter, freely available at~\url{http://www2.iap.fr/users/pitrou/primat.htm} }, in addition to the Fortran code, that has been
used recently, e.g. in \citet{Coc09,CV10,Coc15}. This Fortran code originates from the
model created at IAP by Elisabeth Vangioni~\citep{Van00} and further
developed by Alain Coc \citep{Coc02}. Hence, it was possible to cross--check the implementations and after tuning
the parameters to reach similar precisions, verify that the results were virtually identical, and next to focus on
nuclear reaction rate uncertainties.  
Leaving aside, the ``lithium problem'' that has not yet found a fully satisfactory solution \cite{Fie11}, it has
become essential to improve the precision on deuterium and \qua\ primordial abundances (\tro\ is not pertinent here). 
As nuclear uncertainties affecting deuterium BBN have already been investigated recently \cite{Bayes16,Bayes17,Coc15},
we will just summarize the situation. On the contrary the uncertainties impacting the \qua\ abundance predictions
are limited by the experimental value of the neutron lifetime \cite{PDG17,Wie11} but also by the numerous corrections 
that have to be introduced in the theoretical weak--interaction reaction rates.  
The main aim of this article is to calculate in details all these corrections. In our previous works, either only a fraction of them was
taken into account \cite{Coc09,CV10} or were also supplemented by a final correction to the \qua\ abundance \citep{Coc14}, based on
other works \cite{Dicus1982,LopezTurner1998}. 

With the now precise theoretical and observational determinations of both the deuterium and \qua\ primordial abundances,
it is now possible to better constrain the standard models of particle physics and cosmology and eventually provide
hints of physics beyond the standard model.

We define $n_i$ as {\em volume} density of isotope $i$, and
$n_{\rm b}$, the baryon or nucleon {\em volume} density. The (pseudo-)mass fraction of isotope $i$ is defined as
\be\label{DefYX}
X_i \equiv A_i \frac{n_i}{n_{\rm b}}\,,
\ee
where $A_i$ is the {\em dimensionless} mass integer {\em number} of nuclear physics, or baryon number of particle physics,  (i.e. not the atomic mass). Baryon number
conservation requires that $\sum_iA_in_i=n_{\rm b}$ or $\sum_i
X_i=1$. For ${}^4 {\rm He}$ it is customary to define
\ifphysrep
\be
\YP \equiv X_{{}^4 {\rm He}}\,,\quad \mathrm{and~to~introduce~for~convenience}\qquad \YPfour
\equiv 10^4 \YP\,.
\ee
\else
\be
\YP \equiv X_{{}^4 {\rm He}} \,,
\ee
and to introduce for convenience
\be
\YPfour
\equiv 10^4 \YP\,. \nonumber
\ee
\fi
For other elements, it is customary to use the density ratio with ${}^1{\rm H}$, that is $n_i/n_{{}^1{\rm H}}$, abbreviated as $i/{\rm H}$.

\subsection{Observed abundances}
\label{s:obs}

During the galactic evolution, massive stars are the main source of enrichment of the interstellar medium, when they explode as supernovae,
out of which next generations of stars are born. In this process, they release matter, enriched in heavy elements that they
have synthesized during the various phases of their evolution.   
Accordingly, the abundance of {\em metals} (elements heavier than
helium) in star forming gas increases with time. The observed {\em metallicity} is therefore an indication of age: the older, the lower the {\em metallicity}.
Hence, primordial abundances are extracted from observations of objects with low metallicity, but depending on their 
galactic chemical evolution, pristine abundances have to be derived from different classes of objects as detailed below. 
 
\subsubsection{${}^4{\rm He}$ observations}\label{SecHe4Obs}

After BBN, stars produce also \qua\, and its primordial abundance can
be measured thanks to observations in H\textsc{ii} (ionized hydrogen)
regions inside compact blue galaxies. It is thought that galaxies are formed by the merging of such dwarf galaxies, in a hierarchical 
structure formation paradigm, hence these are considered to be more primitive.
First, for each H\textsc{ii} region, the  \qua\ abundance has to be determined within a model aiming at reproducing
the observed \qua\ emission lines, that also depends on parameters like e.g. the electron density.     
Then, to account for stellar production, the \qua\  deduced abundances should be extrapolated to zero metallicity.
Recently, the observation \cite{Izo14} of an additional atomic infrared line ($\lambda$10830) in 45 low--metallicity H{\sc ii}  regions have
allowed to better constrain the thermodynamic conditions that prevail in the emission regions. After selecting 28 object, \citet{Izo14}
obtained $\YP=0.2551\pm0.0022$ for the \qua\ mass fraction. 
However, \citet{Ave15} (see also \citet{Cyb16}), starting from the same observational data made a stricter selection, based, 
in particular on goodness of fit for the emission model of each object. For the selected 16 objects, the uncertainties were 
so reduced that a slope in the $Y_p$ versus metallicity data could be considered. After extrapolation to zero metallicity, 
  \citet{Ave15} obtained, 
\be\label{YPAbundance}
\YP=0.2449\pm 0.0040\,,
\ee 
that we use here.

\subsubsection{Deuterium observations}

\begin{figure}[!htb]
\begin{center}
% paw/BBN/obs/deuobs2017.kumac
\includegraphics[width=\mycolumnwidth]{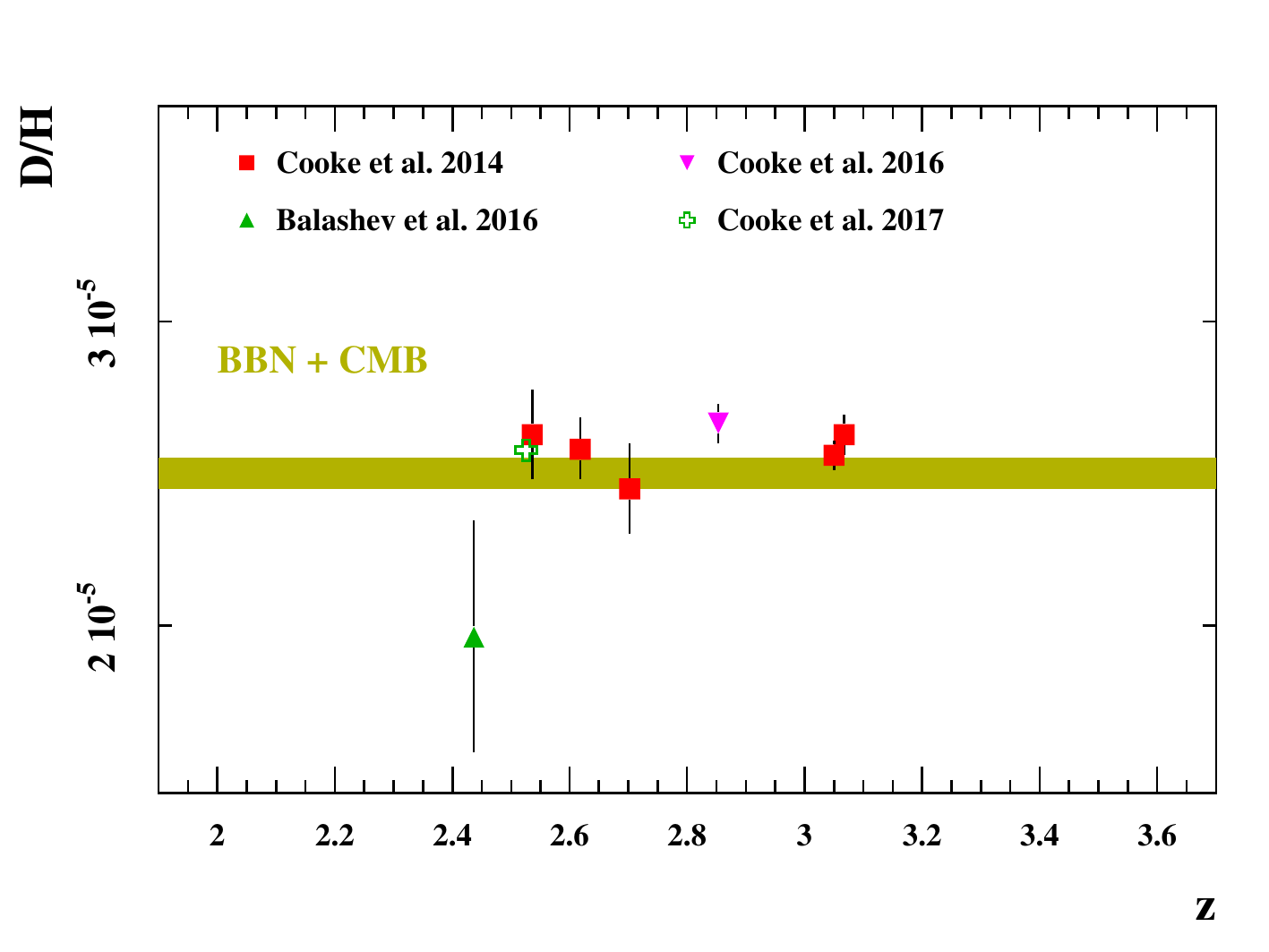}
\end{center}
\caption{D/H observations, as a function of the redshift of the absorber.
These most recent observations \cite{Coo14,Coo16,Coo18} have very small error bars and show 
very few dispersion compared to previous determinations (except \citet{Bal16}), and are in fair agreement with our BBN calculations when
using \textit{Planck} \citep{Planck2016} baryonic density.}
\label{f:deuobs}
\end{figure}

Deuterium is a very fragile isotope. It can only be destroyed after
BBN thanks to stellar evolution. The deuterium abundance closest to primordial abundance is determined from the observation of a few
cosmological clouds (absorbers) at high redshift on the line of sight of distant quasars (emitters). 
Figure~\ref{f:deuobs} shows the observed D/H values as a function of the redshift of the absorber.
Until recently, the distribution of D/H observations showed a significant scatter (see e.g. \citet{Pet12}).
It allowed 
\citet{Oli12} to adopt a weighted mean of D/H = $(3.02 \pm 0.23) \times 10^{-5}$ without excluding
an upper limit of  D/H = $4 \times 10^{-5}$. This is not possible anymore thanks to the recent new
observations or reanalyses of existing data \cite{Coo14,Bal16,Coo16,Rie17,Coo18} 
that now display a plateau as a function of redshift (and metallicity)  with a very small scatter (Fig.~\ref{f:deuobs}).
Moreover, in \citet{Dvo16}, the comparison with the available measurements is consistent with the cosmic merger 
tree model of structure formation. It is shown that at redshift higher than 2 (see their Fig. 2), the dispersion in the 
cosmic deuterium abundance is very tiny leading to think that at these redshifts the observation of the D 
abundance is probably primordial.
Hence, we adopt the new recommended value provided by  \citet{Coo18}:  
\be\label{DeuteriumAbundance}
{\rm D}/{\rm H} = (2.527 \pm 0.030) \times 10^{-5}\,,
\ee
lower and with smaller uncertainties than in previous determinations \footnote{A recent reanalysis and new  D/H observations
towards the Q1009+2956 quasar \citep{Zav17,Zav18}
provided a new value of D/H = $(2.48^{+0.41}_{-0.35}) \times 10^{-5}$
with a limited precision (17\%). If included, it shifts the abundance
downward to D/H= $(2.545 \pm 0.025) \times 10^{-5}$
}. 
If such a precision of $\sim$1\% in observations is confirmed, great care should be paid to 
nuclear cross sections affecting deuterium nucleosynthesis.

%%%%%%%%%%%%%%%%%%%%%%%%%%%%%%%%FIGURE%%%%%%%%%%%%%%%%%%%%%%%%%%%%%%%%%%%%%%%%
%%%%%%%%%%%%%%%%%%%%%%%%%%%%%%%%FIGURE%%%%%%%%%%%%%%%%%%%%%%%%%%%%%%%%%%%%%%%%

\subsubsection{${}^3{\rm He}$ observations}
\label{s:obsh}

Contrary to the case of \qua, stars can both produce and destroy ${}^3{\rm He}$, so that
the evolution of its abundance in time is not known precisely \citep{Van03}.
Since observing helium is difficult, and given the small 
$^3$He/$^4$He ratio, \tro\ was only observed in our Galaxy and
the bounds obtained are ${}^3{\rm He}/{\rm H}= (0.9-1.3)  \times10^{-5}$,
keeping in mind that this is an upper limit extracted from a single object \cite{Ban02}.   
However, the next generation of 30+ m 
telescope facilities may allow to extract the $^3$He/$^4$He ratio from observations of extra-galactic metal poor HII regions \citep{Coo15}.

\subsubsection{${}^7{\rm Li}$ observations}

${}^7{\rm Li}$ is peculiar because it has three distinct sources: BBN but also spallative nuclear reactions between galactic 
cosmic rays and the interstellar medium, and a stellar source (Asymptotic giant branch stars and novae) \cite{Fu18}. 
For instance, recent observations \cite{Taj16,Izz15} have confirmed Li production by novae, 
at a level even higher than model predictions \cite{HJCI96}.
Hence, after BBN,  \sep\ can be produced but can also easily be destroyed in the interior of stars by proton 
capture at temperatures as low as 2.5~MK.

%%%%%%%%%%%%%%%%%%%%%%%%%%%%%%%%FIGURE%%%%%%%%%%%%%%%%%%%%%%%%%%%%%%%%%%%%%%%%
\begin{figure}[!htb]
\begin{center}
\ifphysrep
\includegraphics[width=\textwidth]{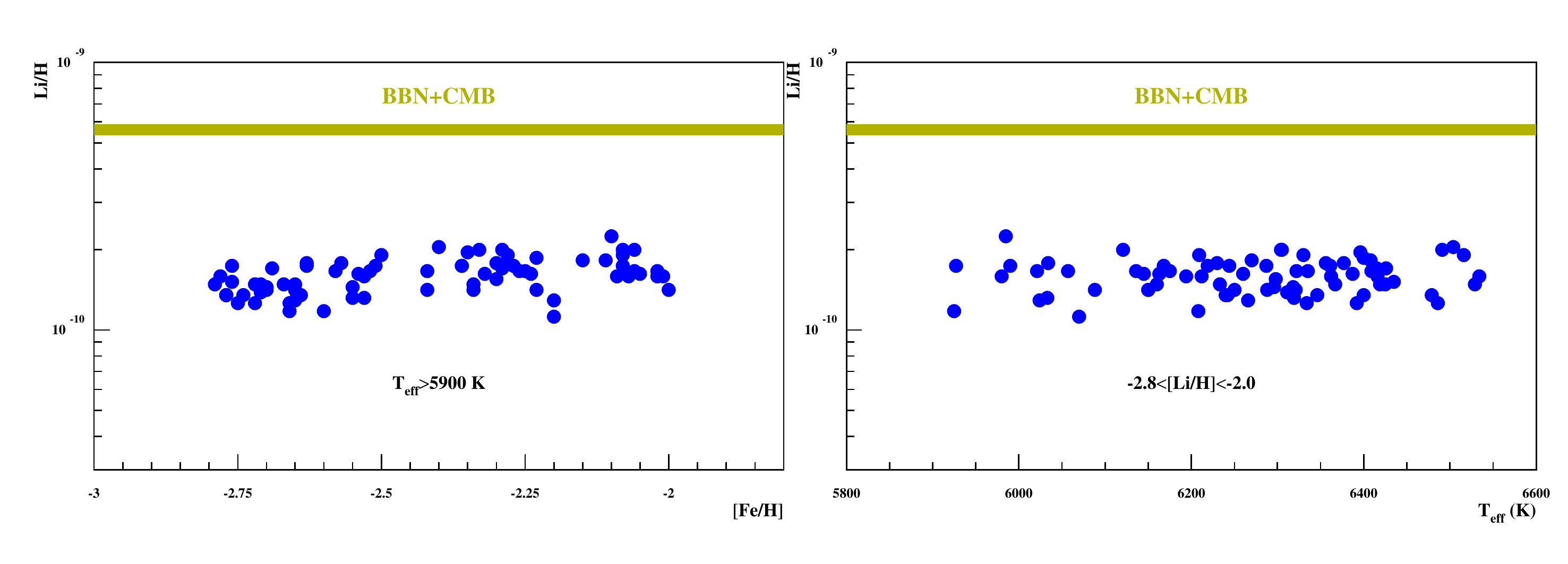}
\else
\includegraphics[width=\mycolumnwidth]{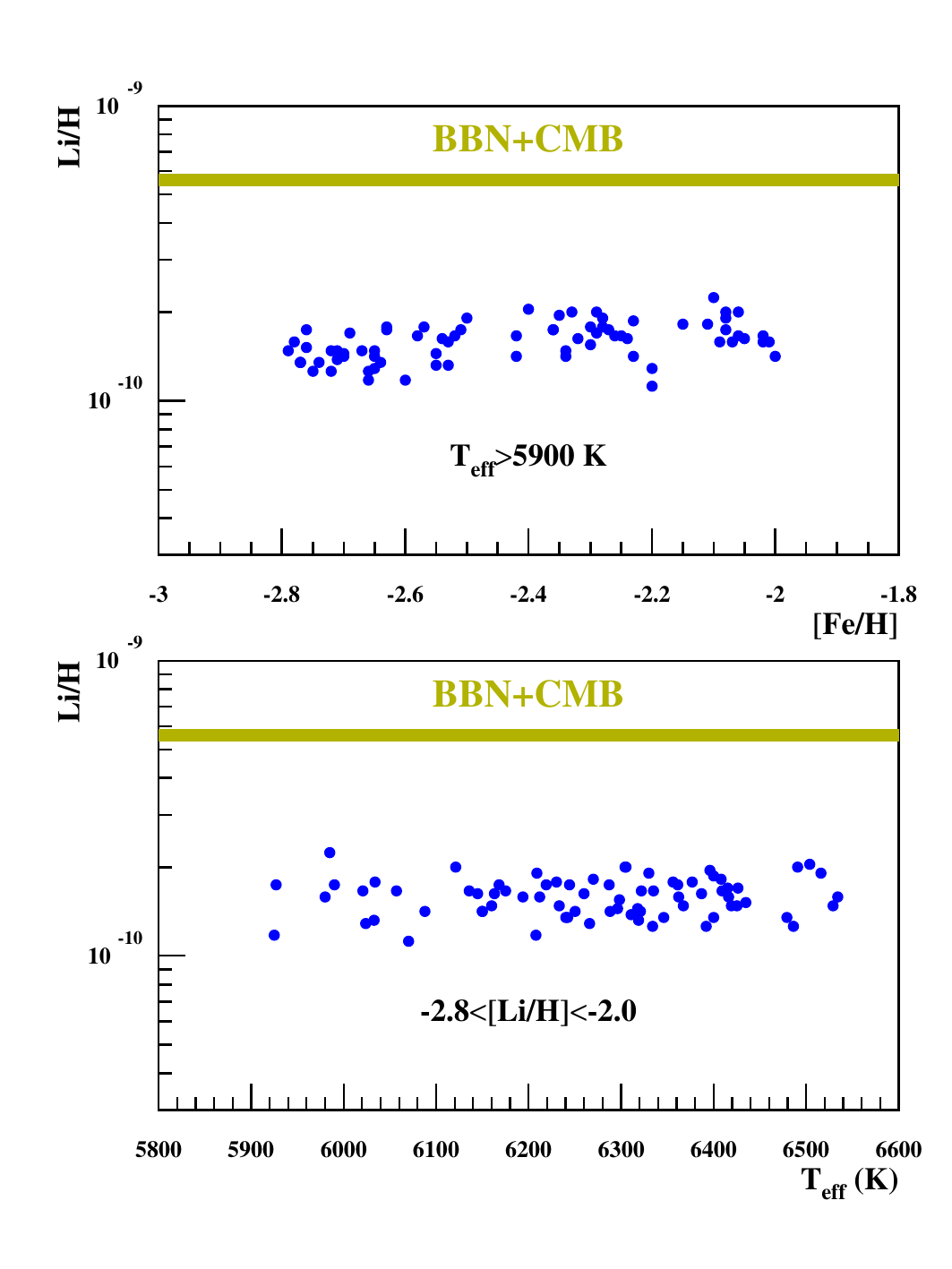}
\fi
\end{center}
\caption{Li/H observations, as a function of metallicity\textsuperscript{\ref{foot1}} and effective temperature. Data \citep{Aok09,Bon07,Cha05,Hos09,Mel10,Sbo10,Sch12,Asp06,Sch15} come from a compilation by \citet{Spi12,Spi15}.}
\label{f:liobs}
\end{figure}
%%%%%%%%%%%%%%%%%%%%%%%%%%%%%%%%FIGURE%%%%%%%%%%%%%%%%%%%%%%%%%%%%%%%%%%%%%%%%

The life expectancy of stars with masses lower than our Sun is larger than the age of the 
Universe so that very old such stars can still be observed in the halo of our Galaxy.
In this context, lithium can be observed at the surface of these stars and its abundance was found to be
independent of metallicity and effective temperature in the ranges 
-2.4$\leq$[Fe/H]$\leq$-1.4\footnote{Logarithm of the ratio, relative to its solar value: [$X/Y]\equiv\log\left((X/Y)/(X_\odot/Y_\odot)\right)$.\label{foot1}} 
(i.e. between $\approx4\times10^{-3}$  and $4\times10^{-1}$ of the solar metallicity)  and  5700 $\leq{T_\mathrm{eff}}
\leq$6800~K \citep{Fu18}. 
This {\em plateau} was discovered by Fran\c{c}ois and Monique Spite \cite{Spi82} and this
constant Li abundance is interpreted as corresponding to the BBN \sep\ production.
The thinness of the ``Spite plateau'' has generally been considered as an indication that surface Li depletion may not have been 
very effective so that it should reflect the primordial value.
However, as shown in Fig.~\ref{f:liobs}, there is a discrepancy of a factor $\approx$3 between the BBN predicted value and 
the lithium abundance derived in metal-poor main-sequence (MS) stars. 
This discrepancy could be alleviated if the stars observed today had undergone photospheric depletion of lithium.
Different observations in globular clusters (see \citet{Gru16}) and stellar scenarios concerning
the formation of the early metal poor stars have been developed, \citep{Fu15} trying to follow lithium evolution in metal-poor stars, 
from pre-main sequence to the Spite plateau. 
Li evolution could affect its abundance by introducing
the effects of convective overshooting, atomic diffusion and mixing \citep{Ric05}. In this context, \sep\
could be depleted and it could be possible to move closer to the
observed Li Spite plateau. 
Indeed,  several globular clusters (NGC 6397, 6752, M30) show that the discrepancy  could be reduced by atomic 
diffusion leading to a factor of 1.6 instead 3 \citep{Mic84}. Finally the pre-MS depletion can be efficient and this is presently an open question.
Note that recent lithium 
observations \cite{How12} have been done in the Small Magellanic Cloud which has a quarter of the sun's 
metallicity and a Li abundance nearly equal to the BBN predictions.

Recently, \citet{Reg17} have constrained cosmic scatter in the
galactic halo using a differential analysis of metal poor for several
elements. Regarding lithium, they find a very low scatter (0.04 dex)
and a mean value $1.86\times10^{-10}$ which is compatible with other
studies. We adopt here the analysis of \citet{Sbo10}, namely~\footnote{All uncertainties are given with one standard deviation.} 
\be\label{LiAbundance}
{}^7{\rm Li}/{\rm H}=(1.58\pm0.3)\times 10^{-10}\,.
\ee

\subsection{Outlook on weak-rates corrections}

Given the observational precision on ${}^4{\rm He}$, we aim at
predictions with a precision better than $0.1 \%$. This amounts to considering all effects which affect $\YPfour$ by units,
and to treat carefully all effects which modify the abundance by an
order $10^{-4}$. $\YP$ is almost exclusively controlled by weak reactions, because nearly all neutrons end up in ${}^4{\rm He}$. By varying artificially the weak rates ($\Gamma$) 
we find the relation\footnote{Using Eq.~(\ref{EqDYPOverYP}) and $\Gamma \propto \tau_{\rm n}^{-1}$ from Eq.~(\ref{Getmyk}).}
\be\label{RuleOfThumb}
\frac{\delta \YP}{\YP} \simeq -0.73 \frac{\delta \Gamma}{\Gamma}\,.
\ee 
Hence we  need to focus on all corrections which affect the weak rates by $10^{-4}$ or more, on top of the experimental uncertainty on the neutron lifetime.

As for ${}^2{\rm H}$ and ${}^3{\rm He}$, we aim at predictions of order
$10^{-3}$ if we leave aside the uncertainty in nuclear rates, and of
order of a few $10^{-2}$ when including uncertainties in nuclear rates. In all
our computations, we must also make sure to maintain numerical errors much below $10^{-3}$. 

There are many effects to take into account for the weak rates and we
must also consider their possible couplings since they cannot always
be summed linearly. These effects have several origins, namely
\begin{enumerate}
\item radiative corrections,
\item finite nucleon mass corrections, 
\item finite temperature radiative corrections,
\item weak-magnetism,
\item QED plasma effects,
\item incomplete neutrino decoupling.
\end{enumerate}

{\it Radiative corrections} correspond to the contribution of virtual
photons in weak reactions, together with the emission of photons in
the final state (bremsstrahlung) from the electron line. They are
typically order $10^{-2}$ effects (see \S~\ref{SecResults})
  because of the value of the fine-structure constant $\alpha_{\rm  FS} \simeq 1/137$. They are well established in the context of neutron
  beta decay~\citep{Czarnecki2004}, and even with some resummed
  effects which are higher orders in $\alpha_{\rm FS}$ for increased
  precision~\citep{Ivanov:2012qe}. In the context of BBN, these
  effects were originally estimated by \citet{Dicus1982}.\\

{\it Finite nucleon mass} correction correspond to the effect of nucleon
  recoil and nucleon thermal distribution of velocity in the
  theoretical computation of the weak rates. These corrections are
  also of order $10^{-2}$ (see \S~\ref{SecResults}) even if
  smaller than radiative corrections. \citet{Wilkinson1982} provided a
  comprehensive list of these effects in the context of neutron beta decay, and \citet{Seckel1993} reviewed
  these corrections in the context of BBN. These were later estimated
  numerically by \citet{Lopez1997} using a Monte-Carlo estimation of
  multidimensional integrals.  We introduce a new efficient method  which
  relies on a Fokker-Planck expansion in the energy transfer, and
  which relies only on one-dimensional integrals.\\

{\it Finite temperature radiative corrections} correspond to the
interactions with the bath of electrons and positrons during weak
reactions. They lead to a long subject of controversy in the literature. They were initially computed in
  \citet{Dicus1982} and \citet{Cambier1982}. Discrepancies between these approaches
  were analyzed by \citet{Kernan} and a numerical estimation was
  provided by \citet{LopezTurner1998}. However, \citet{BrownSawyer}
  pointed incoherences and the lack of detailed balance, and they proposed
  to set the finite temperature radiative correction on firm ground by
  providing a comprehensive theoretical computation from
  finite-temperature quantum
  field theory. We find, and this is new, that to be complete and satisfy
  detailed balance, one must also consider corrections to the
  bremsstrahlung effects and add them to finite temperature
  corrections. Overall, when finite-temperature corrections and
  bremsstrahlung corrections are added, they almost perfectly cancel,
  leaving $\YPfour$ nearly unchanged (see \S~\ref{SecResults}).\\

{\it Weak-magnetism} which arises from the internal structure of
  nucleons is part of finite mass corrections and is order $10^{-3}$ (see \S~\ref{SecResults}).\\

{\it Quantum Electrodynamics} (QED) is responsible for two
  effects. First the interaction with the plasma modifies the electron
  mass. However this effect is strictly speaking part of finite
  temperature radiative corrections in our computations. Second, QED
  effects modify the thermodynamics of the plasma, that is it modifies
  the pressure and energy density. We find that this effect, taken
  alone, results in a negligible modification of $\YPfour$ which is of
  order $10^{-5}$ (see \S~\ref{SecResults}). However it affects the number of effective neutrinos
  for subsequent cosmology such as the physics of cosmic microwave
  background (CMB).\\

{\it Incomplete neutrino decoupling} corrections occur when electrons
and positrons annihilate, because neutrinos are not fully  decoupled
from the plasma and some annihilations end up heating the neutrino
bath. This effect increases the neutrino temperature and produces
distortions in their spectrum. In order to track the neutrinos
spectral distortions, it is necessary to use the full machinery of
coupled Boltzmann equations, especially when considering neutrino
oscillations. Since this effect also modifies the effective number of
neutrinos for the subsequent cosmology, several authors among which
\citet{Dolgov1997,Mangano2005,Grohs:2015tfy,deSalas:2016ztq} focused
%:
%:
on it. We shall use a fit taken from \citet{Parthenope} for the heating rate of
  neutrinos. Eventually we find that this correction is of order $10^{-3}$ (see \S~\ref{SecResults}).\\%This amounts to ignoring the spectral  distortions and
             %this affects $\YPfour$by approximately one unit. 

All these effects are detailed in \S~\ref{SecWeak}.

\subsection{Main eras of BBN}

\begin{figure}[!htb]
\includegraphics[width=\mycolumnwidth]{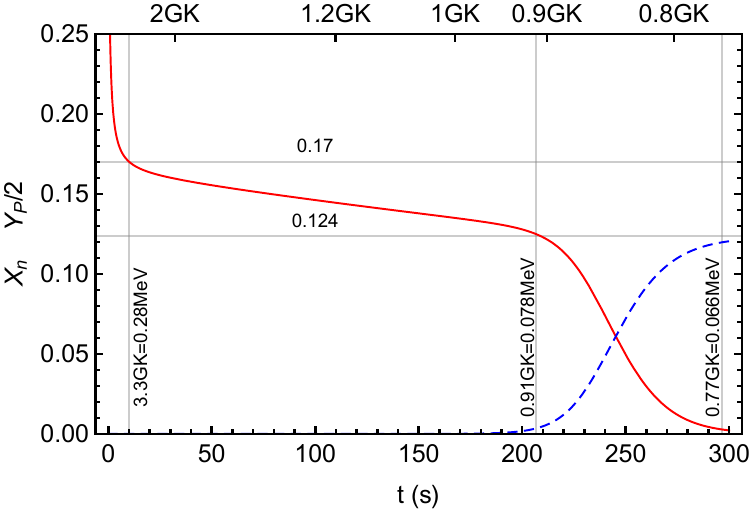}
     \caption{Evolution of $X_n$ (red continuous line with nuclear
       reactions and red dotted line without nuclear reactions) and
       $\YP/2$ (blue dashed line).}
\label{FigYnCoc}
\end{figure}

During BBN, one can distinguish different eras depending on the
dominant physical processes.
\begin{enumerate}
\item For plasma temperatures $T$ in the range $2\, {\rm MeV }\lesssim T
  \lesssim  {\rm GeV}$, nucleons are formed and their density is only
  affected by expansion. Neutrinos are still in thermal equilibrium with the plasma (electron-positrons and protons), that is $T_\gamma=T_\nu$  and the neutron to
  proton ratio $X_n/X_p$ is enforced to be the thermodynamical equilibrium value since weak interactions interconvert efficiently neutrons in protons
  and vice-versa, through the reactions~(\ref{AllWeakReactions}).
\item In the range, $0.8\, {\rm MeV }\lesssim T \lesssim  2\,{\rm
    MeV}$ neutrinos have essentially decoupled but weak interactions
  maintain neutrons and proton in thermodynamical equilibrium. However
  at weak-interactions freeze-out temperature $T_F\simeq 0.8 {\rm
    MeV}$, defined by the equality between weak interaction rates and
  the cosmological expansion rate, the abundance of neutrons are mostly affected by neutron beta
  decay. In practice, freeze-out is not instantaneous and the neutron
  abundance is subject only to neutron beta decay around $T_F\simeq
  0.28\,{\rm MeV}$. The abundance of neutrons is then about $X_n \simeq
  0.17$ and it is eventually reduced to $X^{\rm Nuc}_n \simeq 0.125$ by beta
  decay when nucleosynthesis starts (see Fig. \ref{FigYnCoc}).
\item Around $0.5 {\rm MeV}$, electrons and positrons annihilate and
  heat up the photon bath, resulting in a different temperature
  between photons and neutrinos ($T_\gamma > T_\nu$). This affects
  directly the expansion history of the Universe since the energy
  content of massive particles (electrons and positrons) is replaced
  by massless particles (photons), but it also affects weak-interaction rates since this is concomitant with the freeze-out period.
\item As long as $T \gtrsim 0.078 {\rm MeV}$, deuterium dissociation is
  too efficient to allow for deuterium synthesis. Even though the binding energy of deuterium is
  about $2.2 {\rm MeV}$, deuterium is efficiently destroyed by the high-energy
  tail of the Bose-Einstein distribution of photons, because the ratio between baryon number density and
  photon number density $\eta$ is smaller than $10^{-9}$.
\item Below $T\lesssim T_{\rm Nuc} = 0.078 {\rm MeV}$, deuterium can be
  formed. Then since the binding energy per nucleon of ${}^4{\rm He}$ is much
  larger than for deuterium, a network of reactions ends up in
  producing nearly only ${}^4{\rm He}$ and very tiny amounts of other
  light elements. Since ${}^4{\rm He}$ is made of two neutrons and two
  protons, the (pseudo-)mass abundance satisfies $\YP\simeq 2 X^{\rm Nuc}_n$, hence leading
  to a final value $\YP \simeq 0.25$. Nucleosynthesis is completely
  over for all elements when $T \lesssim 0.01{\rm MeV}$ or $T \lesssim
  10^8 {\rm K}$ but for prediction with $10^{-3}$ precision on
  deuterium, we found that we must wait until $T\lesssim 6\times 10^7 {\rm K}$.
\end{enumerate}

\subsection{Resolution strategy and outline}

This outline of nucleosynthesis implies that we do not need to solve
jointly the abundance of all species at all times. Indeed, given that the matter-radiation equivalence occurs
around $T \simeq 3000 {\rm K}$, this means that baryonic matter accounts for
less than $10^{-4}$ of the total energy density at the end of BBN and
even less during freeze-out. In principle, one should account for the
fact that neutron abundance evolves even though the nucleonic density
(protons and baryons) is only affected by cosmic expansion. Since the mass difference between
neutrons and protons is of order $10^{-3}$ of their rest mass energy,
we can ignore this effect when evaluating the total energy
density of the Universe. Similarly, given that the binding
energy in nuclei is around or less than $10^{-2}$ of the rest mass
energy, we can also consider to a very good approximation that nuclear reactions do not affect the
baryon energy density. Hence it is possible to compute the evolution of the cosmological background,
without having to compute in details the abundance of neutrons nor
the details of nuclear reactions. In \S~\ref{SecThermo} we study the
thermodynamics of all species and the dynamics of the cosmic expansion.
Then in \S~\ref{SecWeak} we focus on weak interactions and
detail all corrections. Since the weak interaction rates depend on the plasma
of electrons, positrons and neutrinos only, they can be studied ignoring
nuclear reactions. Finally in \S~\ref{SecNucleo} we focus on nuclear reactions and we present the
results obtained when coupling the background dynamics, the weak
interactions and the nuclear reactions. Most technical details can be found
in the appendices for ease of reading. We use natural units
in all expressions, that is we work in units in which 
\be
k_B=\hbar=c=1\,,
\ee
except when we judge instructive to write them explicitly.

\section{Background thermodynamics}\label{SecThermo}

We consider a homogeneous and isotropic cosmology, more precisely a
flat Friedmann-Lema\^itre (FL) spacetime characterized by the scale factor
$a(t)$ where $t$ is the cosmic time. The Hubble expansion rate is
$H = \dot a/a$, where a dot indicates a derivative with respect to
$t$. 
%We adopt the convention that today the scale factor is unity, that is
%$a_0 = 1$, hence whenever $a$ appears alone we
%should read $a/a_0$ if using
%a different convention. 
Since cosmological perturbations are of order $10^{-5}$, it is
fully justified to ignore their effect and consider an homogeneous cosmology given
our precision goal.

\subsection{Thermodynamics in a FL spacetime}
 
\subsubsection{Distribution function and Boltzmann equation}
 
Relativistic species, which encompass neutrinos, photons and
electrons with positrons during the relevant BBN era, are best
described with a distribution function $f(t,p)$ for each
species, and where the dependence is only on the magnitude of momenta
$p = \sqrt{E^2-m^2}$ given the symmetries of the FL spacetime. This description is valid even out of thermodynamical
equilibrium. The distribution function satisfies the general Boltzmann
equation in a FL spacetime
\be\label{BoltzmannFL}
L[f] \equiv \frac{\partial f}{\partial t} +\dot p \frac{\partial
  f}{\partial p} = C[f]\,,\qquad \dot p =-Hp\,.
\ee
We have used that $p \propto 1/a$ because of cosmic expansion and $C[f]$ is the
collision term of the species under scrutiny. 
In order to relate this description to the fluid description, thermodynamical quantities such as number density $n$,
energy density $\rho$ and pressure $P$, can be formed from the distribution function as summarized in App. \ref{AppThermo}.

\subsubsection{Number density evolution}

From the definition of thermodynamic quantities (Eqs.~\ref{ThermoFromf}), considering $\int L[f] 4\pi p^2\dd
p/(2\pi)^3$ with the Boltzmann equation~(\ref{BoltzmannFL}) leads
after integration by parts to the number conservation equation
\be\label{Eqdotn}
\dot n + 3 H n = {\cal J}\,,\qquad {\cal J} \equiv \int C[f]\frac{4\pi p^2 \dd p}{(2\pi)^3}\,.
\ee
${\cal J}$ is the net creation rate of particles per unit of
physical volume. When the collision term vanishes or when it conserves
the number of particles because it describes elastic scattering, 
\be\label{Eqdotn2}
\dot n + 3 H n = 0 \quad \Rightarrow \quad \frac{\dd (n
  a^3)}{\dd t}=0\,.
\ee
In that case, for a comoving volume $a^3$, the total number of particles $N \equiv n
a^3$ remains constant.
%Eqs.~(\ref{Eqdotn}) and (\ref{Eqdotn2}) are general and do not rely on the distribution function taking a particular form.

\subsubsection{Energy density evolution}

Similarly, starting from the definition of thermodynamic quantities
(Eqs.~\ref{ThermoFromf}), and considering $\int L[f] E 4\pi p^2\dd
p/(2\pi)^3$ with the Boltzmann equation~(\ref{BoltzmannFL}), leads after integration by parts
(and using $E \dd E = p \dd p$) to the energy conservation equation
\be\label{Eqdotrho}
\dot \rho + 3 H (\rho + P) = \dot q\,,\qquad \dot q \equiv \int C[f]
\frac{4\pi E p^2 \dd p}{(2\pi)^3}\,.
\ee
$\dot q$ is the volume heating rate. For massless particles, such as photons or neutrinos (which can be considered massless during the BBN), $E=p$ and $P=\rho/3$, implying that when the collision term is vanishing
\be\label{Eqdotrho2}
\dot \rho + 4 H \rho = 0 \quad \Rightarrow \quad \frac{\dd (\rho
  a^4)}{\dd t}=0\,.
\ee
Eqs.~(\ref{Eqdotrho}) and (\ref{Eqdotrho2}) are general and do not
rely on a specific distribution function.

\subsubsection{Entropy evolution}

Volume entropy is defined (see e.g.~\citet{Grohs:2015tfy}) from the
distribution function by
\be
s = - \int S_{\rm B} (f) \frac{4\pi p^2 \dd p}{(2\pi)^3}\,,
\ee
where the Boltzmann entropy is defined as 
%(TODO check that this entropy is called Boltzmann entropy
\be
S_{\rm B}(f) \equiv \left[f \ln f\pm (1\mp f)\ln (1\mp f)\right]
\ee
with upper (lower) sign for fermions (bosons). Using the identity
\be
\partial_p[S_{\rm B} (f)] = \ln\left(\frac{f}{1\mp f}\right) \partial_p f 
\ee
and multiplying Eq.~(\ref{BoltzmannFL}) by $S_{\rm B} (f)$ we get after
integration by parts
\be\label{sdotneutrinos}
\dot s + 3 H s = -\int C[f] \ln\left(\frac{f}{1\mp f}\right)\frac{4\pi p^2 \dd p}{(2\pi)^3}\,,
\ee
which dictates the evolution of volume entropy. If there are no
collisions, it is only affected by dilution as $s \propto 1/a^3$ such
that the total entropy in a given comoving volume, $S\equiv s a^3$, is
conserved ($\dot S=0$).

\subsubsection{Local thermodynamical equilibrium}

In case of local thermodynamical equilibrium (LTE), fermions (bosons) follow a
Fermi-Dirac (Bose-Einstein) distribution~(\ref{FDBE}) and
\be
\ln\left(\frac{f}{1\mp f}\right) = \frac{\mu-E}{T}\,,
\ee 
where $\mu$ is the chemical potential. Hence the evolution of volume
entropy for a given species satisfies, using Eqs. (\ref{Eqdotn}) and (\ref{sdotneutrinos}),
\be\label{Eqdots}
\dot s + 3 H s = \frac{\dot q}{T} -\frac{\mu}{T}(\dot n + 3H
n)\,,
\ee
with $\dot q$ defined in Eq.~(\ref{Eqdotrho}).
From Eq.~(\ref{Eqdots}) multiplied by $a^3$, we recover the usual thermodynamical identity 
\be\label{DotS}
T \dot S = \dot Q -\mu \dot N
\ee 
with $\dot Q \equiv \dot q a^3$. Hence, the entropy for a given species is
conserved if there is no heat exchange, that is no interactions, and
either a vanishing chemical potential (in practice $\mu\ll T$) or
conservation of the number of particles (i.e. $\dot N=0$). From the conservation of the total stress-energy tensor, we must have
$\sum_i \dot q_i =0$, that is there can be no global production of
heat. Hence, in case of local thermodynamical equilibrium, the total entropy is conserved whenever for all species
either the chemical potential is negligible $\mu_i \ll T$ or the
number of particles is conserved.

If there is LTE, entropy for a given species $i$ can then be linked to other thermodynamical quantities
thanks to
\be\label{sexpression}
s_i = \frac{P_i+\rho_i-\mu_i n_i}{T_i}\,.
\ee
The total entropy is obtained from the extensivity of $\rho$ and $n$,
given that pressure and temperature of all species are equal  at equilibrium
\be
s = \sum_i \frac{P_i + \rho_i }{T} + \sum_i \frac{\mu_i n_i}{T}=\frac{\rho+P}{T}+ \sum_i \frac{\mu_i n_i}{T}\,.
\ee
During nucleosynthesis, given the very low value of the
baryon-to-photon ratio $\eta$ (see \S~\ref{SecBaryonDensity} below),
we can totally ignore the entropy of protons and neutrons. This amounts to neglecting the electrons, positrons and
neutrinos created and destroyed in weak interactions, and any energy exchange with
these particles and the photons.  Hence we can focus only on the thermodynamics of electrons,
positrons, photons and neutrinos. During nucleosynthesis, neutrinos
are nearly fully decoupled from other species and their entropy is separately conserved
if we can assume that they are totally decoupled. See~\S \ref{SecIncompleteDecoupling} for the small modifications
induced by incomplete decoupling (ID). 

Conversely, electrons, positrons and photons are tightly coupled by electromagnetic interactions and
they behave collectively as a plasma whose common temperature is
defined as $T$. Photons decouple from electrons only around
recombination much later when the CMB is formed and $T\lesssim {\rm eV}$. Due to electron-positron
annihilations, the number of particles in the plasma is not conserved,
and it is crucial that we can neglect the chemical potential of
plasma particles to claim that the plasma entropy is conserved. The
chemical potential of photons is always vanishing due to processes
which do not conserve the number of photons so we need only to make
sure that the chemical potential of electrons and positrons can be
neglected. In App.~\ref{AppChemicalPotential}, we evaluate these
chemical potentials and show they are completely negligible, implying that
we can use entropy conservation.

For convenience we define dimensionless reduced energy density, pressure, number
density and entropy as
\be\label{DefReducedThermo}
\bar n_i \equiv \frac{n_i}{T_i^3}\quad \bar P_i \equiv \frac{P_i}{T_i^4}\quad \bar \rho_i \equiv \frac{\rho_i}{T_i^4}\quad \bar s_i \equiv \frac{s_i}{T_i^3}\,
\ee
such that the relation~(\ref{sexpression}) reads simply in case of
vanishing chemical potential
\be\label{sreducedexpression}
\bar s_i = \bar \rho_i+\bar P_i\,.
\ee
For photons the reduced thermodynamic variables do not
depend on temperature and are constants when there is LTE. We find
\beas\label{ReducedValues}
\bar n_\gamma &=&
\frac{2}{2\pi^2}I_-^{(1,1)} =\frac{2 \zeta(3)}{\pi^2}\,\\
\bar \rho_\gamma &=&
\frac{2}{2\pi^2}I_-^{(2,1)} =\frac{\pi^2}{15}\,\\
\bar P_\gamma &=&
\frac{2}{6\pi^2}I_-^{(0,3)} =\frac{\pi^2}{45}\,\\
\bar s_\gamma &=&\bar \rho_\gamma+\bar P_\gamma =\frac{4 \pi^2}{45}\,.
\eeas
where the definitions for the $I_\pm^{(p,q)}$ are given in
App.~\ref{AppThermoQuantities}. Similarly for neutrinos, and assuming that they have a
negligible chemical potential, the reduced variables defined in case
of LTE, take the constant values
\beas\label{ReducedValuesnu}
\bar n_\nu &=&
\frac{2}{2\pi^2}I_+^{(1,1)} =\frac{3}{4} \bar n_\gamma\,\\
\bar \rho_\nu &=&
\frac{2}{2\pi^2}I_+^{(2,1)} =\frac{7}{8}\bar \rho_\gamma\,\\
\bar P_\nu &=&
\frac{2}{6\pi^2}I_+^{(0,3)} =\frac{7}{8}\bar P_\gamma\,\\
\bar s_\nu &=&\bar \rho_\nu +\bar P_\nu =\frac{7}{8}\bar s_\gamma\,,
\eeas
where we used $g_\nu=1$ since only left-handed neutrinos contribute to the
relativistic species, but we have multiplied by a
factor $2$ since conventionally we add together the contributions of
neutrinos and antineutrinos\footnote{If neutrinos were Majorana
  particles, neutrinos would be their own antiparticle but they would
  possess both helicities and one would take $g=2$, resulting in
  the same final result}. In particular we deduce from Eq.~(\ref{Eqdotrho2}) that for photons
or (massless) neutrinos in thermodynamical equilibrium
\be\label{TaConstant}
\frac{\dd(a T)}{\dd t}=\frac{\dd(a T_\nu)}{\dd t} =0\,,
\ee
if they are completely decoupled from other species. During BBN, $a
T$ varies because of electron-positron annihilations. It is customary to define~\cite{Mangano2005,deSalas:2016ztq}
\be\label{Defz}
z\equiv a(T) T \qquad  z_\nu\equiv a(T_\nu) T_\nu
\ee
to characterize this total variation of $aT$ and $aT_\nu$, with the
convention that long before decoupling when all species were coupled
together and at the same temperature, that is when $T=T_\nu\gg m_e$,
we had $a(T) T = a(T_\nu)T_\nu=1$, that is $z=z_\nu=1$ at early times.

Additionally, in a first approximation, one can also assume that during BBN neutrinos are
fully decoupled, implying that $a T_\nu$ is constant and $z_\nu=1$ remains always true. The tiny effect of incomplete decoupling induces in fact a small variation of $z_\nu$ that we shall take into
account in \S~\ref{SecIncompleteDecoupling}. Ignoring it, the
neutrino temperature scales simply as
\be\label{ScalingTnu}
T_\nu = \frac{a_0 T_0}{a\,z_0}\qquad z_0 \equiv z(T
\ll m_e) \,,
\ee
where $T_0$ is the photons temperature today and $z_0$ is the value of
$z$ long after BBN is finished. 

\subsection{Plasma temperature}\label{Secat}

\begin{figure}[!htb]
     \includegraphics[width=\mycolumnwidth]{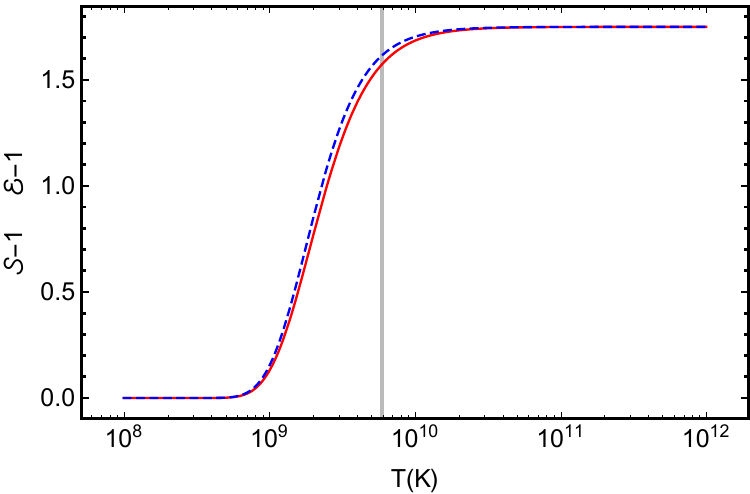}
     \caption{{\it Red continuous line} : ${\cal S}(T)-1$. {\it Blue dashed
         line} :   ${\cal E}(T)-1$. The vertical bar corresponds to
       $T=m_e\simeq 0.511\, {\rm MeV}$.}
\label{FigCalSCalE}
\end{figure}

Since for the plasma $S= s a^3$ is conserved, we can obtain the
relation between temperature and scale factor. 
From Eq.~(\ref{sreducedexpression}) and expressions in App.~\ref{AppThermo}, we find that the ratio
between plasma entropy at a given time of BBN and photons entropy today is given by
\beas
\frac{s}{s_0} &=& \frac{T^3}{T_0^3}{\cal S}(T)\\
{\cal S}(T) &\equiv&\frac{\bar s_{\rm pl}}{\bar s_\gamma}=\frac{\bar s_\gamma+\bar s_{e^+}(T)+\bar
  s_{e^-}(T)}{\bar s_\gamma}\slabel{DefCalS}\\
&=& 1 + \frac{2}{\pi^2 \bar s_\gamma}\left[\frac{1}{3}I_+^{(0,3)}(x)+I_+^{(2,1)}(x)\right]\nonumber
\eeas
where $x\equiv m_e/T$. The function ${\cal S}$ is plotted in Fig.~\ref{FigCalSCalE}. For
$T \gg m_e$, ${\cal S}\to 11/4$ whereas for $T\ll m_e$, ${\cal S} \to
1$. Entropy conservation ($s a^3 = s_0 a_0^3$) allows then to relate
the scale factor to the plasma temperature as
\be\label{EqaT}
\frac{z^{\rm stand}_0}{z^{\rm stand}}=\frac{a_0T_0}{aT}={\cal S}^{1/3}\quad \Rightarrow\quad a(T)=\frac{a_0 T_0 }{T {\cal S}(T)^{1/3}}\,.
\ee
Here $z^{\rm stand}$ denotes $z$ in the case where we can ignored
incomplete neutrino decoupling and QED plasma effects which we analyze
in \S~\ref{SecQEDplasma} and \ref{SecIncompleteDecoupling} respectively. 
The inverse relation $T(a)$ is obtained by numerical
inversion. $z^{\rm stand}_0$ takes the value
\be
z_0^{\rm stand} = S^{1/3}(T \gg m_e)= \left(\frac{11}{4}\right)^{1/3}\simeq 1.40102\,.
\ee
From Eqs.~(\ref{EqaT}) and (\ref{ScalingTnu}) the ratio between neutrino and plasma temperatures is given by
\be\label{RatioTnuT}
\frac{T_\nu}{T} = \frac{{\cal
  S}^{1/3}(T)}{z_0^{\rm stand}}\,,
\ee
and tends to $1/z_0^{\rm stand} =(4/11)^{1/3}$ at low temperatures,
which is the celebrated result of instantaneous decoupling.

\subsection{Baryon density}\label{SecBaryonDensity}

In order to obtain the energy density of baryons, it is sufficient to use that for non-relativistic and cold species $\rho
\propto 1/a^3$. Hence
\be\label{EqBaryonsScaling}
\rho_{\rm b}= \left(\frac{a_0}{a}\right)^3\rho_0^{\rm crit}
\Omega_{\rm b}=\left(\frac{a_0}{a}\right)^3\rho_{100}^{\rm crit} \times \left(\Omega_{\rm b} h^2\right)
\ee
where we defined the critical densities $\rho_0^{\rm crit} \equiv 3
H_0^2/(8\pi G)$ and $\rho_{100}^{\rm crit} \equiv 3 H_{100}^2/(8\pi
G)$, with the reduced Hubble rate $h\equiv H_0/H_{100} $ (with the
standard definition $H_{100}\equiv 100\, {\rm
  km}/{\rm s}/{\rm Mpc}$). The numerical value of $\rho_{100}^{\rm
  crit}$ can be found in appendix~\ref{ParticleValues}.
The baryon energy density can be converted into a number density by estimating
the average mass of nucleons 
\be\label{Defnb}
n_{\rm b} = \frac{\rho_{\rm b}}{m_{\rm b}}\,.
\ee
More details about the definition of $m_{\rm b}$ are given in
App. \ref{AppReactionConventions}. We define the ratio between baryons and photons by
\be\label{DefEta}
\eta \equiv \frac{n_{\rm b}}{n_\gamma}\,.
\ee
Since $n_{\rm b} \propto 1/a^3$ and $n_\gamma \propto T^3$ it can be
rewritten in the form
\be
\eta = \eta_0 \left(\frac{z_0}{z}\right)^3 \,.
\ee
Using Eq. (\ref{Eqxivalue}), the value today is approximately found to be
%\bea
%\eta_0 &\simeq& 6.0913\times10^{-10}\left(\frac{\Omega_{\rm b}h^2}{0.02225}\right)
%\left(\frac{2.7255}{T_0}\right)^3\,.\nonumber\\
%&&\times \left(\frac{1.00605 \ma}{m_{\rm b}}\right)
%\eea
\ifphysrep
\be
\eta_0 \simeq 6.0913\times10^{-10}\left(\frac{\Omega_{\rm b}h^2}{0.02225}\right)
\left(\frac{2.7255}{T_0}\right)^3  
\left(\frac{1-1.759\times 10^{-3}}{1-1.759\times 10^{-3}\frac{\YP}{0.24709}}\right)\,.
\ee
\else
\bea
\eta_0 &\simeq& 6.0913\times10^{-10}\left(\frac{\Omega_{\rm b}h^2}{0.02225}\right)
\left(\frac{2.7255}{T_0}\right)^3  \nonumber\\
&&\times \left(\frac{1-1.759\times 10^{-3}}{1-1.759\times 10^{-3}\frac{\YP}{0.24709}}\right)\,.
\eea
\fi
Finally, note that we can safely ignore the thermal energy of baryons since its ratio with the
energy density of photons is of order
\be
\frac{n_{\rm b} T}{\rho_\gamma} = \frac{\bar n_{\rm b}}{\bar
  \rho_\gamma}\propto \eta \ll 1\,,
\ee
thus justifying the use of the scaling (\ref{EqBaryonsScaling}) which
is the one of cold particles.

\subsection{Cosmology and scale factor}

The evolution of the scale factor is dictated by the Friedmann equation
\be\label{Friedmann}
H^2 = \frac{8\pi G}{3} \rho\,.
\ee
It determines the evolution of the scale factor $a(t)$, and to solve
it we need the expressions of the total energy density as a function of
$a$. Since we already have found the relations $a(T)$, it is
sufficient to express the energy density in terms of temperature.
This is fortunate, because the energy density of the plasma depends
only on temperature as it is always in local thermodynamical equilibrium. From
App.~\ref{AppThermo}, we find that for the plasma
\beas\label{EqDefEPlasms}
\rho_{\rm pl} &=&  {\cal E}(T)\bar \rho_\gamma T^4\,,\\
{\cal E}(T)&\equiv&\frac{\bar \rho_\gamma+\bar \rho_{e^+}+\bar \rho_{e^-}}{\bar \rho_\gamma} = 1+\frac{30}{\pi^4} I_+^{(2,1)}(x)\,,
\eeas
where we use the definitions (\ref{DefIpq1})-(\ref{DefIpq3}). The
function ${\cal E}$ is plotted in Fig. \ref{FigCalSCalE} and in the
limit $T \gg m_e$ it also tends to $11/4$. The energy density of the $N_\nu$ generations of (massless) neutrinos is 
\be\label{EqEnergyNeutrinos}
\rho_\nu =N_\nu \bar \rho_\nu T_\nu^4\,.
\ee
As for the energy density of baryons, we use the simple dilution
relation (\ref{EqBaryonsScaling}) which amounts to neglecting their
thermal energy. Hence a similar scaling can be used for cold dark matter.

Using that\footnote{The effect of the cosmological constant is also
  totally negligible as it accounts for less than $10^{-30}$ of the energy
content during BBN even if it dominates today.}
\be\label{TotalRho}
\rho = \rho_\nu+\rho_{\rm pl}+\rho_{\rm b} + \rho_{\rm cdm},
\ee
we obtain $\rho(T)$ that we combine with the relation $T(a)$ obtained
in section~\ref{Secat} to get $\rho(a)$, such that the Friedmann equation~(\ref{Friedmann}) is in the
form of an ordinary differential equation for the function $a(t)$ that we can solve
numerically. The relation $t(a)$ is then deduced by numerical inversion.

\subsection{QED corrections for the plasma thermodynamics}\label{SecQEDplasma}

\begin{figure}[!htb]
%\centering  
\subfloat[Vacuum]{
\begin{fmffile}{MS1}
\begin{fmfgraph*}(30,10)
\fmfstraight
\fmfbottom{ei,v1,v2,eo}

\fmf{fermion}{ei,v1,v2,eo}
\fmf{photon,tension=0,left}{v1,v2}

\fmflabel{$e^\pm$}{ei}
\fmflabel{$e^\pm$}{eo}
  \end{fmfgraph*}
\end{fmffile}
\label{FDMS1}}

\subfloat[Interaction with photons]{
\begin{fmffile}{MSS2}
\begin{fmfgraph*}(30,10)
\fmfstraight
\fmfbottom{ei,v1,v2,eo}
\fmfleft{ei,pi}
\fmfright{eo,po}

\fmf{fermion}{ei,v1,v2,eo}
\fmf{photon,tension=0}{v1,pi}
\fmf{photon,tension=0}{v2,po}

\fmflabel{$e^\pm$}{ei}
\fmflabel{$e^\pm$}{eo}
 \end{fmfgraph*}
\end{fmffile}
\label{FDMS2}}\qquad\ifphysrep\qquad\qquad\qquad\else\quad\fi
\subfloat[Interaction with antiparticles]{
\begin{fmffile}{MS3}
\begin{fmfgraph*}(30,10)
\fmfstraight
\fmfbottom{ei,v1,v2,eo}
\fmfleft{ei,pi}
\fmfright{eo,po}

\fmf{fermion}{ei,v1}
\fmf{photon}{v1,v2}
\fmf{fermion}{v2,eo}
\fmf{fermion,tension=0}{v1,pi}
\fmf{fermion,tension=0}{v2,po}

\fmflabel{$e^\pm$}{ei}
\fmflabel{$e^\pm$}{eo}
  \end{fmfgraph*}
\end{fmffile}
\label{FDMS3}}

\caption{{\it Top :} electron/positron self-energy. {\it Bottom :} electron/positron mass
  shift from interaction with plasma.}

\label{FigFD4}
\end{figure}
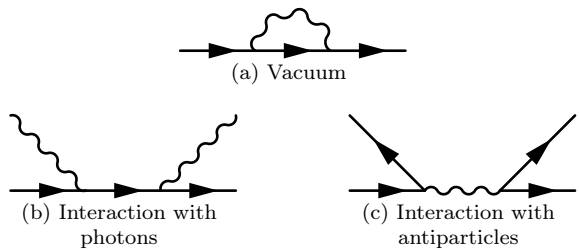

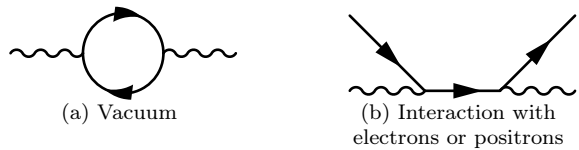
\begin{figure}[!htb]
%\centering  
\subfloat[Vacuum]{
\begin{fmffile}{MS4}
\begin{fmfgraph*}(30,10)
\fmfleft{pi}
\fmfright{pf}
\fmf{photon}{pi,v1}
\fmf{photon}{v2,pf}
\fmf{fermion,left,tension=0.45}{v1,v2,v1}

\fmflabel{$\gamma$}{pi}
\fmflabel{$\gamma$}{pf}
  \end{fmfgraph*}
\end{fmffile}
\label{FDMS4}}\qquad\ifphysrep\qquad\qquad\qquad\else\quad\fi
\subfloat[Interaction with electrons or positrons]{
\begin{fmffile}{MS5}
\begin{fmfgraph*}(30,10)
\fmfstraight
\fmfbottom{ei,v1,v2,eo}
\fmfleft{ei,pi}
\fmfright{eo,po}

\fmf{photon}{ei,v1}
\fmf{fermion}{v1,v2}
\fmf{photon}{v2,eo}
\fmf{fermion,tension=0}{pi,v1}
\fmf{fermion,tension=0}{v2,po}

\fmflabel{$\gamma$}{ei}
\fmflabel{$\gamma$}{eo}
  \end{fmfgraph*}
\end{fmffile}
\label{FDMS5}}

\caption{{\it Left :} photon self-energy. {\it Right :} photon mass shift from interaction with electron/positron plasma.}

\label{FigFD5}
\end{figure}

The thermal bath of electrons and positrons has electromagnetic
interactions with the bath of photons. Using QED, this leads to a
modification of the plasma thermodynamics which depend on its temperature.
First this leads to an effective modification of the electron and
photon masses, which come from the diagrams~\ref{FigFD4} and
\ref{FigFD5}.

The electron mass is shifted by (see \citet[Eq. 12]{Mangano2001}, \citet[Eq. 35]{LopezTurner1998} or
\citet{Heckler1994,Fornengo:1997wa} for further details on its derivation)
\ifphysrep
\be\label{deltame2}
\frac{\delta m_e^2(p,T)}{T^2} = \frac{4 \alpha}{\pi} \left[I_-^{(0,1)}+I_+^{(0,1)}(x)\right]-\frac{2 \bar m_e^2 \alpha}{\pi \bar p}\int_0^\infty \dd \bar q
\frac{\bar q}{{\cal E}_{\bar q}} \ln
\left|\frac{\bar p+\bar q}{\bar p-\bar q} \right|\FD{{\cal E}_{\bar q}}
\ee
\else
\bea\label{deltame2}
\frac{\delta m_e^2(p,T)}{T^2} &=& \frac{4 \alpha}{\pi} \left[I_-^{(0,1)}+I_+^{(0,1)}(x)\right]\\
&-&\frac{2 \bar m_e^2 \alpha}{\pi \bar p}\int_0^\infty \dd \bar q
\frac{\bar q}{{\cal E}_{\bar q}} \ln
\left|\frac{\bar p+\bar q}{\bar p-\bar q} \right|\FD{{\cal E}_{\bar q}}\nonumber
\eea
\fi
where we used the definitions  (\ref{DefIpq1})-(\ref{DefIpq3}). In
this expression, ${\cal E}_{\bar q} \equiv \sqrt{\bar q^2+x^2}$, $\bar
m_e=x=m_e/T$ 
%TODO see if this definition is really needed,
 $\bar p \equiv p/T
$ and $g^\pm({\cal E}_{\bar q}) \equiv 1/({\rm e}^{{\cal E}_{\bar
    q}}\pm1)$.
%The second contribution (second line), which depends
The second contribution, which depends
explicitly on the $e^\pm$ momentum, accounts for less than $10\%$ to
the total mass shift (see ~\citet{Mangano2001,LopezTurner1998}). The mass shift (\ref{deltame2}) can be further simplified using
\be\label{Indkdk}
I_-^{(0,1)} \equiv \int_0^\infty \BE{\bar k} \bar k \dd \bar k  = \frac{\pi^2}{6}\,.
\ee

\begin{figure}[!htb]
     \includegraphics[width=\mycolumnwidth]{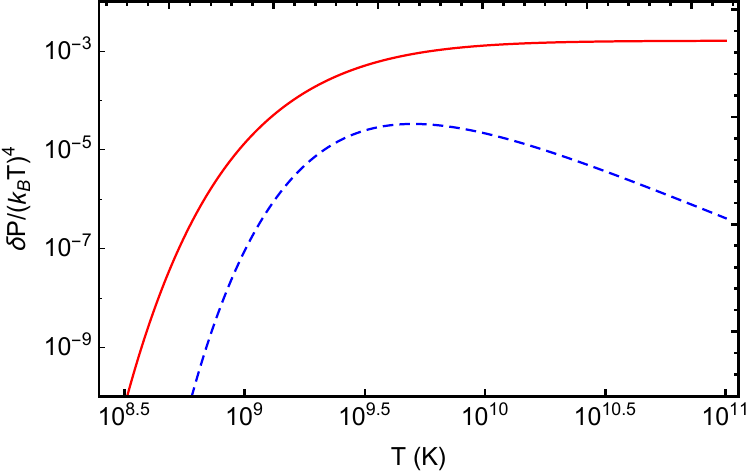}
     \caption{{\it Red continuous line} : $\overline{\delta P}_{\rm
         d}$. {\it Blue dashed line} : $\overline{\delta P}_{\rm
         s}$. Definitions are given in Eq.~(\ref{EqdPab}).}
\label{dPadPb}
\end{figure}

The photon mass shift is instead given by
\be\label{deltamgamma2}
\frac{\delta m_\gamma^2}{T^2} = \frac{8 \alpha}{\pi}  I_+^{(0,1)}(x) \,.
\ee
These mass shifts induce a change in the pressure of the plasma whose
expression is given by \citet[Eq. 13]{Heckler1994} or \citet[Eq. 16]{Mangano2001}. The pressure shift reads
\ifphysrep
\be\label{PressureShift}
\overline{\delta P}\equiv\frac{\delta P}{T^4} = -\int_0^\infty
\frac{\dd \bar p}{2
  \pi^2 }\frac{\bar p^2}{{\cal E}_{\bar p}}\frac{\delta m_e^2(p,T)}{T^2}
  \FD{{\cal E}_{\bar p}}\,\,-\int_0^\infty \frac{\dd \bar k}{2
  \pi^2}\frac{\bar k}{2}\frac{\delta m_\gamma^2(T)}{T^2}\BE{\bar k}\,.
\ee
\else
\bea\label{PressureShift}
\overline{\delta P}\equiv\frac{\delta P}{T^4} &=& -\int_0^\infty
\frac{\dd \bar p}{2
  \pi^2 }\frac{\bar p^2}{{\cal E}_{\bar p}}\frac{\delta m_e^2(p,T)}{T^2}
  \FD{{\cal E}_{\bar p}}\nonumber\\
&&-\int_0^\infty \frac{\dd \bar k}{2
  \pi^2}\frac{\bar k}{2}\frac{\delta m_\gamma^2(T)}{T^2}\BE{\bar k}\,.
\eea
\fi
Hence from the expressions of the mass shifts we get
\be
\overline{\delta P} =\overline{\delta P}_{\rm d} + \overline{\delta
  P}_{\rm s}\,,
\ee
where the dominant and subdominant terms are respectively 
\bea\label{EqdPab}
\overline{\delta P}_{\rm d}&\equiv& \frac{\alpha_{\rm FS}}{\pi}\left[-\frac{2}{3}
  I_+^{(0,1)}(x) -\frac{2}{\pi^2}\left( I_+^{(0,1)}(x)
  \right)^2\right]\\
\overline{\delta P}_{\rm s}&\equiv&\frac{\alpha_{\rm FS}}{\pi^3}\frac{\bar m_e^2}{\bar p \bar q}
\int\frac{\bar p^2 \dd \bar p}{{\cal E}_{\bar p}} \frac{\bar q^2 \dd
  \bar q}{{\cal E}_{\bar q}} \ln\left|\frac{\bar p+\bar q}{\bar p
    -\bar q}\right|\FD{{\cal E}_{\bar p}}\FD{{\cal E}_{\bar q}}\,.\nonumber
\eea
These pressure modifications are plotted in Fig. \ref{dPadPb}. We can check that the
subdominant contribution is indeed negligible since QED plasma corrections are already very small (see \S~\ref{SecResults}), but we
include it anyway for completeness.

\begin{figure}[!htb]
     \includegraphics[width=\mycolumnwidth]{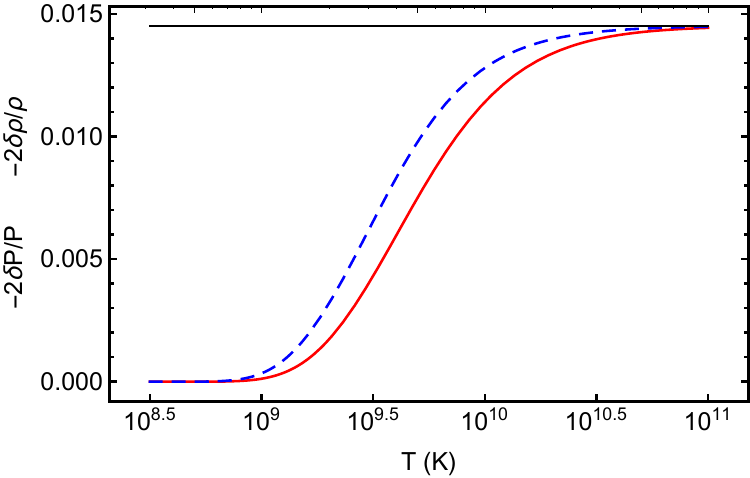}
     \caption{Effective degrees of freedom in the plasma from QED
       corrections. {\it Red continuous line} : $-\delta g_P=-2\delta P /P$.
{\it Blue dashed line} : $-\delta g_\rho=-2\delta \rho /\rho$. {\it
  Black thin line} : High temperature asymptotic value $25\alpha/(4\pi)$.}
\label{FigQEDPlasma}
\end{figure}

The modification of the plasma energy density is then obtained from the thermodynamic identity\footnote{Choosing $T$ and $V$ to describe
  the state of a system, then $P$ and $\rho$ are functions of $T$ only
since they are intensive quantities. It follows from $T \dd S = \dd
(\rho V) + P \dd V$ that $\dd S = (P+\rho)/T \dd V+V/T (\dd \rho/\dd
T) \dd T$. From the integrability conditions we then get
$\partial_T[(\rho+P)/T]=\partial_V(V/T \dd \rho/\dd T)$ and the
identity~(\ref{SortOfClapeyron}) follows.}
\be\label{SortOfClapeyron}
\rho=-P + T \frac{\dd P}{\dd T}\,,
\ee
so that we get immediately
\be
\overline{\delta \rho}\equiv \frac{\delta \rho}{T^4} =
3\overline{\delta P} + \frac{\partial \overline{\delta P}}{\partial \ln
T}\,.
\ee
\citet{LopezTurner1998} defined corrected relativistic degrees of freedom by
\be
\delta g_\rho \equiv 2\frac{\overline{\delta
    \rho}}{\bar\rho_{\gamma}}=\frac{30}{\pi^2}\overline{\delta
    \rho}\,,\quad \delta g_P \equiv
  2\frac{\overline{\delta P}}{\bar P_{\gamma}}=\frac{90}{\pi^2}\overline{\delta
    P}\,.
\ee
At high temperatures, that is for $T \gg m_e$, we obtain using
$I_+^{(0,1)}(x=0)=\pi^2/12$ (see table \ref{FigImn} in appendix~\ref{AppThermoQuantities}) the limits
\be
\overline{\delta \rho} \simeq -\frac{5}{96}(4\pi\alpha)\,,\qquad \overline{\delta P} \simeq \frac{1}{3} \overline{\delta \rho} \,,
\ee
which expressed in terms of corrected relativistic degrees of freedom
read
\be
\delta g_\rho \simeq \delta g_P \simeq -\frac{25 \alpha}{4\pi}\simeq 0.01452\,.
\ee
These corrections are plotted in Fig.~\ref{FigQEDPlasma} together with
the high temperature limit. 

The QED plasma corrections enter in four places.
\begin{itemize}
\item  First they imply that the entropy of the
plasma as given in Eq.~(\ref{DefCalS}) must be modified and we must use 
\be\label{SQED}
{\cal S}^{\rm QED} = {\cal S}+\frac{\bar \rho_\gamma}{\bar s_\gamma}\frac{\delta g_\rho}{2}+\frac{\bar P_\gamma}{\bar s_\gamma}\frac{\delta g_P}{2}= {\cal S}+\frac{\delta g_P+3 \delta g_\rho}{8}
\ee
where the last equality follows from Eqs.~(\ref{ReducedValues}). ${\cal S}^{\rm QED}$ must be used instead
of ${\cal S}$ in Eq.~(\ref{EqaT}) in order to obtain $a(T)$.

\item Second, this entropy modification also affects the evolution of
  $z$ and we find
\be
(z_0^{\rm QED})^3 = S ^{\rm QED} (T \gg m_e)= \frac{11}{4}\left(1-\frac{4}{11}\frac{25\alpha}{8\pi}\right)\,,
\ee 
that is $z_0^{\rm QED} \simeq 1.39979$. This affects the neutrino
temperature scaling which then becomes
\be\label{TnuQED}
T^{\rm QED}_\nu=\frac{a_0 T_0}{a\,z_0^{\rm QED}}\,.
\ee 
This neutrino temperature must be used in Eq.~(\ref{EqEnergyNeutrinos})
when computing the energy density of neutrinos for the Friedmann equation~(\ref{Friedmann}).

\item Third, the plasma QED correction must also be incorporated in the energy
density of the plasma when solving the Friedmann equation~(\ref{Friedmann}), meaning that we must modify
Eqs.~(\ref{EqDefEPlasms}) and use instead
\be\label{CalEQED}
{\cal E}^{\rm QED} ={\cal E} + \frac{\delta g_\rho}{2}\,,
\ee
when computing $\rho_{\rm pl}$ for the Friedmann equation~(\ref{Friedmann}).
\item Finally, the electron mass shifts modifies the statistics of
  electrons and positrons in the weak-interactions. 
\end{itemize} 
An important comment is in order here on the last effect. Since the mass shift (\ref{deltame2}) is exactly Eq.~(5.12) of
  \cite{BrownSawyer} (that is $2E \Delta E = \delta m_e^2$), the
  effect of the electron mass shift is one of the several effects
  involved in finite-temperature radiative corrections for the weak
  rates. Hence, following \citet{BrownSawyer}, we consider that the
  modification of the statistics through the mass shift is part of the
  finite-temperature corrections and we do not consider that it is
  part of the QED plasma corrections. This point of view is similar to
  \citet{Esposito1999,Serpico:2004gx} but different from
  \citet{LopezTurner1998}. Hence comparisons on the magnitude of
  the correction must take into account the point of view on the effect
  of mass shifts, since either they are reported as QED effects or
  finite-temperature radiative corrections.
 
It turns out that the QED plasma corrections (that is without the
effect of the mass shift on the weak-interaction rates), coming from the modifications (\ref{SQED}),(\ref{TnuQED}) and (\ref{CalEQED}), are very small since they modify the Helium production by approximately $\Delta
  \YPfour=-0.1$ (see \S~\ref{SecResults}). 

\subsection{Incomplete neutrino decoupling}\label{SecIncompleteDecoupling}

The effect of incomplete neutrino decoupling has been studied in
details in \citet{1992PhRvD..46.3372D,Fields:1992zb,Hannestad1995,Gnedin1997,Dolgov1997,Dolgov1998,Esposito:2000hi,Mangano2001,Hannestad2001,Mangano2005,Mangano2006,Birrell:2014uka,Grohs:2015tfy,deSalas:2016ztq}. Electron-positron
annihilations lead to a small reheating of the neutrino bath which
must be studied in the context of coupled Boltzmann equations since it
also leads to spectral distortions in the neutrino
spectrum. Furthermore the complete study, that we do not reproduce here, requires to consider neutrino
flavor oscillations.

\begin{figure}[!htb]
     \includegraphics[width=\mycolumnwidth]{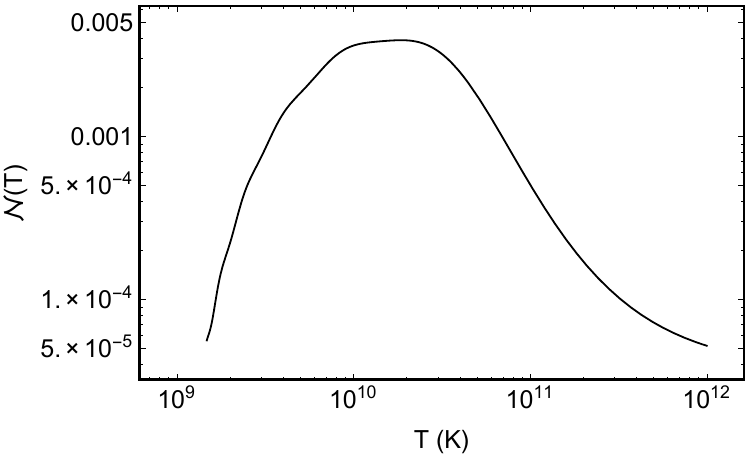}
     \caption{${\cal N}(T)$ as defined in Eq.~(\ref{EqDefN}).}
\label{FigCalN}
\end{figure}

Let us first study the effect of incomplete neutrino decoupling (ID)
on the plasma. If there is heat exchange, that is energy exchange between the plasma
and the neutrinos, then the evolution of the plasma entropy is
dictated by Eq.~(\ref{DotS}) with the chemical potential
neglected, that is
\be\label{dotSplasma}
\dot S_{\rm pl} = \frac{\dot Q_{\rm pl}}{T}\,.
\ee 
This is always true for the plasma because it is always at local
thermodynamic equilibrium, but not necessarily true for the neutrinos
for which we should rely on Eq.~(\ref{sdotneutrinos}) if we were to
compute the evolution of their entropy. The numerical
studies of the effect of incomplete neutrino decoupling gives a fit to
the heating rate of the plasma in \citet{Parthenope}. More
precisely, we define the dimensionless function ${\cal N}(T)$ related to the heating rate as
\be\label{EqDefN}
\frac{\dot q_{\rm pl}}{H T} =- T^3 {\cal N}(T)\,,
\ee
where we remind $\dot Q = a^3 \dot q$. The function $T^4 {\cal N}$ can be viewed as the volume heating rate in units of the
Hubble rate $H$. The fit given in \citet[Eqs. A24-A25]{Parthenope} is a $13$-order polynomial in $x=m_e/T$ valid for
$x<4$, and we plot it in Fig.~\ref{FigCalN}.

From Eqs.~(\ref{dotSplasma}) and (\ref{EqDefN}), the evolution of the
plasma reduced entropy is dictated by
\be
\frac{1}{(aT)^3} \frac{\dd [\bar s_{\rm pl }(aT)^3]}{\dd \ln a}=-{\cal
  N}\,.
\ee
From the definition (\ref{DefCalS}) we obtain
\be\label{dlnaTdlnT}
\frac{\dd \ln (aT)}{\dd \ln T} = \frac{{\cal N}-\bar s_\gamma\frac{\dd {\cal
      S}}{\dd \ln T}}{{\cal N}+3 \bar s_\gamma {\cal S}}
\ee
where ${\cal S}$ must be replaced by ${\cal S}^{\rm QED}$ when
including plasma QED effects. We recover when ${\cal N}=0$, that is
for complete neutrino decoupling, that $z=aT \propto{\cal S}^{-1/3}$.
This equation can be integrated numerically to obtain $a(T)$  in the case of incomplete neutrino
decoupling, and a numerical inversion allows to obtain subsequently
$T(a)$. $z$ is modified because of the non-conservation
of plasma entropy, and we get $z^{\rm ID}_0 \simeq 1.39911$ if QED plasma
effects are not included and $z^{\rm ID,QED}_0 \simeq 1.39788$ if they are also
included, in agreement (except for the last digit) with table 1 of
\citet{deSalas:2016ztq}. Our results are summarized in table \ref{Tablez}.

Let us now focus on the effect of incomplete neutrino decoupling on
neutrinos. Since the energy taken from the plasma is gained by the neutrinos, we
have necessarily $q_{\rm pl}=-q_{\nu}$. Hence we can solve for the
evolution of the neutrino energy density from Eq.~(\ref{Eqdotrho}) which is
\be\label{drhonu}
\frac{1}{a^4}\frac{\dd (a^4 \rho_\nu)}{\dd \ln a}= -\frac{\dot q_{\rm pl}}{H}=T^4 {\cal N}\,.
\ee
Since $T^4 {\cal N}$ is a function of $T$ it can be considered as a
function of $a$ using the $T(a)$ previously obtained. Solving
numerically this differential equation gives $\rho_\nu(a)$. Together
with the previously solved $T(a)$ [which allows to get $\rho_{\rm
  pl}(a)$ from Eq.~(\ref{EqDefEPlasms})] and the QED correction
(\ref{CalEQED}), we obtain $\rho(a)$ and thus the relation $a(t)$ and its inverse
$t(a)$ from numerically solving the Friedmann equation
(\ref{Friedmann}). 

This approach to compute the variation of the neutrino energy density from Eq.~(\ref{drhonu}) is correct even if
incomplete neutrino decoupling and neutrino reheating by
electron-positron annihilations create spectral distortions in the
neutrino spectra. That is, the gravitational effect of incomplete
neutrino decoupling is taken into account correctly with our
approach. However it cannot fully take into account the effect on the
weak rates which we discuss further in \S~\ref{SecWeakRatesID} and \ref{SecEffectID}.

\subsection{Effective description of neutrinos}\label{SecNeff}

Neutrinos are not in local thermodynamical equilibrium and their
temperature is not well defined. Indeed, the full characterization of
the neutrinos requires to solve the coupled Boltzmann equations dictating the evolution of neutrino distribution functions,
and we must in principle build a temperature definition from the energy spectrum (see e.g. \citet{PitrouStebbins}).
In the BBN context it is convenient to define a brightness
temperature, being the temperature of the Fermi-Dirac distribution
with no chemical potential that would have the same energy
density. Hence we define $T_\nu$ as in Eq.~(\ref{EqEnergyNeutrinos}) by
\be\label{DefTnueff}
\rho_\nu =  N_\nu \bar \rho_\nu T_\nu^4\quad \Rightarrow \quad \rho_\nu a^4= N_\nu \bar \rho_\nu  z^4_\nu\,,
\ee
where $\rho_\nu$ results from the integration of
Eq.~(\ref{drhonu}). $z_\nu$\footnote{Note that we still define $z_\nu$
  as in the LTE case [Eq.~(\ref{Defz})].} is deduced from the value of $\rho_\nu a^4$ obtained
numerically, and we find $z_{\nu,0}=1.00144$. It corresponds to a
neutrino energy density increase of a factor $z^4_{\nu,0}= 1.00576$ compared to the complete decoupling
scenario.

\begin{figure}[!htb]
     \includegraphics[width=\mycolumnwidth]{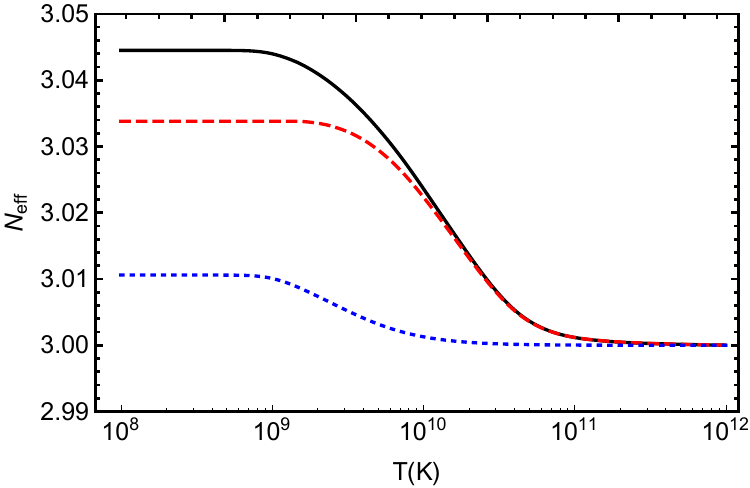}
     \caption{$N_{\rm eff}(T)$. {\it Black continuous line} : both QED plasma effects and incomplete
     neutrino decoupling. {\it Red long dashes}: Incomplete
     decoupling but no QED corrections. {\it Blue small dashes}: QED
     correction with complete decoupling}
\label{FigNeff}
\end{figure}

The ratio between neutrino and photon temperature is simply
\be\label{Eqznufromrhonu}
\frac{T_\nu}{T}=\frac{z_\nu}{z}\,.
\ee
Using this ratio, Eq.~(\ref{DefTnueff}) can be rewritten to define an effective number
of neutrinos as
\be\label{DefNeff}
\rho_\nu = N_{\rm eff}\bar \rho_\nu \left(\frac{T}{z^{\rm
      stand}}\right)^4\ifphysrep\qquad\qquad\else\quad\fi N_{\rm eff} \equiv N_\nu\left(\frac{z_\nu
    z^{\rm stand}}{z}\right)^4\,.
\ee
This is the number of neutrinos that would be required to have the
same energy density if the ratio $T_\nu/T$ was taken from the complete decoupling result
without QED correction, that is $1/z^{\rm stand}$. $N_{\rm eff}$ today is completely determined by the
values of $z_0$ and $z_{\nu, 0}$. Its evolution is plotted in Fig. \ref{FigNeff}.

We find $N^{\rm ID}_{\rm eff} = 3.0337$ today if QED plasma effects are
ignored [$z$ obtained from Eq.~(\ref{dlnaTdlnT}) and $z_\nu$ from
Eqs.~(\ref{drhonu}) and (\ref{Eqznufromrhonu})] but non-instantaneous
decoupling is taken into account. We find $N^{\rm QED}_{\rm
  eff}=3.0106$ for the reverse situation ($z_\nu=1$ but $z=z^{\rm
  QED}$). If both effects are taken into account  [$z$ obtained from
Eq.~(\ref{dlnaTdlnT}) with ${\cal S}^{\rm QED}$ instead of ${\cal S}$, and $z_\nu$ from
Eqs.~(\ref{drhonu}) and (\ref{Eqznufromrhonu})], we find $N^{\rm ID,QED}_{\rm eff} = 3.0444$, in very close agreement with
the last results of \citet{deSalas:2016ztq}. 

After BBN, e.g. for the evolution of initial perturbations through
baryon acoustic oscillations, only the energy density of neutrinos is important
through its gravitational effect. Distortions in the neutrino spectra
  can have in principle an effect on CMB and structure formation when
  considering its joint effect with neutrino masses at late times. However this
effect is expected to be extremely small and it is enough to assume
that after BBN neutrinos follow a Fermi-Dirac distribution.  Hence $N_{\rm
eff}$ is used in CMB codes such as CAMB~\cite{CAMB,Lewis:1999bs} and
CLASS \cite{CLASS,CLASSPaper} to take into account the effect of
incomplete decoupling during BBN. From the definition~(\ref{DefNeff}) the total energy
density of radiation after BBN is
\be
\rho_{\rm R} =\rho_\nu+\rho_\gamma= \rho_\gamma\left(1+\frac{7}{8}N_{\rm eff}\left(\frac{4}{11}\right)^{4/3}\right)\,
\ee 
meaning that we can forget about the QED plasma corrections and the
incomplete neutrino decoupling during BBN if we use $N_{\rm eff}$ instead of $N_\nu=3$. 

The values of $z_0$, $z_{\nu,0}$ and $N_{\rm eff}$ for the various
cases are reported in table \ref{Tablez}. Let us comment on the value
$N_{\rm eff}\simeq 3.02$ found in \citet{Grohs:2015tfy} for the
effect of QED alone. It is important to realize that we must be careful not to replace the mass
shifts (\ref{deltame2}) and (\ref{deltamgamma2}) directly in the distribution function inside the general
expression~(\ref{DefPressure2}) for pressure to obtain the pressure
shift (\ref{PressureShift}), since this would overestimate it by a
factor $2$. This subtlety is detailed in \citet{Heckler1994} and
is correctly taken into account in \citet{Mangano2001}. Since this is what
has been done in \citet{Grohs:2015tfy}, it explains why the effect
of QED on the plasma as measured by $N_{\rm eff}$ found in this
reference is around $3.02$ instead of $3.01$.
\begin{table}
\centering
\caption{$z_0$, $z_{\nu,0}$ and $N_{\rm eff}$ depending on the effects. \label{Tablez}}
%\resizebox{\columnwidth}{!}{
\begin{tabularx}{\mycolumnwidth}{AAAAA}
\toprule
 QED & Decoupling & $z_0=aT$ & $z_{\nu,0}=a T_\nu$ & $N_{\rm eff}$ \\\midrule
No & Yes & $1.40102$ & $1.00000$ & $3.0000$ \\
Yes & Yes & $1.39979$ & $1.00000$ & $3.0106$ \\
No & No & $1.39911$ & $1.00144$ & $3.0338$ \\
Yes & No &$1.39788$ & $1.00144$ & $3.0445$ \\\bottomrule
\end{tabularx}
%}
\end{table}

\section{Weak Interactions}\label{SecWeak}

Once the dynamics of the background has been solved, it is possible to
study the evolution of the abundance of neutrons. This determines the
final amount of chemical species since atomic nuclei form from the fusion of neutrons
and protons. Throughout we assume that all particles are in local
thermodynamical equilibrium. If this is certainly true for the plasma of
photons strongly coupled with electrons and positrons, this is not exactly true for
neutrinos. Indeed, as we have seen in \S~\ref{SecIncompleteDecoupling}, neutrinos are not fully decoupled
when BBN takes place and there is a residual heating of neutrinos
which cannot be fully described as an increased neutrino temperature, since
it also leads to a  distorted neutrino spectrum. 
%Hence, a complete resolution
%would require to solve the coupled system of Boltzmann equations, taking
%into account neutrino oscillations, to keep track of the neutrino
%spectrum, as performed in details in \citet{Mangano2005,Grohs:2015tfy,deSalas:2016ztq}.

\subsection{General formulation}\label{WeakGeneral}

The weak interaction reactions correspond to a set of reactions
which are all related by crossing symmetry. These are the six reactions
\beas\label{AllWeakReactions}
n+\nu &\leftrightarrow& p + e^-\slabel{EqWeak1}\\
n&\leftrightarrow& p + e^-+\bar \nu \slabel{EqWeak2}\\
n+e^+ &\leftrightarrow& p + \bar \nu \slabel{EqWeak3}
\eeas
where $n$ ($p$) stands for neutrons (protons). The two reactions $n+e^++\nu
\leftrightarrow p$ are not possible energetically since $m_n> m_p$, as
can be seen considering the reaction in the proton rest mass
frame. However it can exist if there is a photon in the final state of the
forward reaction, see appendix~\S~\ref{AppBS}.

It follows that the number density of neutrons and protons
is not just diluted by expansion, but varies according to the reaction
rates. The general form for the neutron-proton density evolution is
\beas
\dot n_n + 3 H n_n &=& -n_n \Gamma_{n\to p} + n_p\Gamma_{p\to n}\\
\dot n_p + 3 H n_p &=& -n_p\Gamma_{p\to n} +n_n \Gamma_{n\to p}
\eeas
where the reaction rates $\Gamma$ are associated to the corresponding collision terms in the Boltzmann
equation. Here we have gathered the rates according to
\beas
\Gamma_{n \to p} &=& \Gamma_{n+\nu \to p + e} + \Gamma_{n\to p +
  e+\bar \nu}+\Gamma_{n+e \to p + \bar \nu}\\
\Gamma_{p \to n} &=& \Gamma_{ p + e\to n+\nu } + \Gamma_{p +
  e+\bar \nu \to n}+\Gamma_{p + \bar \nu \to n+e}.
\eeas

The general expression for these reaction rates is detailed in
App.~\ref{Apprates}. For all rates with the nucleon $a$ in the
initial state and nucleon $b$ in the final state, they take the form
\begin{widetext}
\be\label{GeneralGammaMainText}
n_a \Gamma_{a \to b} = \int \frac{ \dd^3 \gr{p}_a \dd^3 \gr{p}_e \dd^3
  \gr{p}_\nu }{2^4 (2\pi)^8}\delta\left({E}_a-E_b + \coeffa_e E_e +
  \coeffa_\nu E_\nu\right) \frac{\left| M\right|^2_{a \to b}}{{E}_n {E}_p {E}_e
  {E}_\nu}  f_a(E_a)
f_\nu(\coeffa_\nu {E}_\nu) f_e(\coeffa_e {E}_e) 
\ee
\end{widetext}
where $\coeffa_\nu=1$ if the neutrino is in the initial state and
$\coeffa_\nu=-1$ if it is in the final state, and a similar definition
for $\coeffa_e$ according to the ${\rm e}^\pm$ position. These coefficients
appear obviously in the Dirac delta function ensuring energy
conservation in reactions, and they are also used to express the final
nucleon momentum as $p_b = p_a+\coeffa_\nu p_\nu + \coeffa_e p_e$.
They also appear in the distribution functions of the electrons and neutrinos, because either the particle
is in the initial state and the corresponding distribution function
appears in the expression, or it is in the final state and it becomes a
Pauli-blocking factor thanks to the relation
\be\label{PauliBlockingMagic}
f(-E) = 1-f(E)\,,
\ee
valid for a Fermi-Dirac distribution with vanishing chemical potential.
%TODO maybe better notation than $f(E)$ here
For a given reaction, $\left| M\right|^2_{a \to b}$ is the corresponding matrix-element of the weak interaction summed over all initial and
final states. For weak interactions in the Fermi theory, it is of the
form \citep{Fidler:2017pkg}
\be\label{DefMLLMRRMLR}
\frac{|M|^2}{2^7 G_F^2}= c_{LL }{\cal M}_{LL} +
c_{RR} {\cal M}_{RR} +c_{LR} {\cal M}_{LR}\,,
\ee
with the coupling factors $c_{..}$ given in Eqs.~(\ref{DefgLLgRRgLR}). The expressions for ${\cal M}_{LL}, {\cal M}_{RR}, {\cal M}_{LR}$ are
reported in Eqs.~(\ref{papbpcpd}). The $LL$ term (resp. $RR$) corresponds to purely
left-chiral (resp. right-chiral) couplings, and the $LR$ term is an
interference term.

\subsection{Infinite nucleon mass approximation}\label{SecBornMainText}

Let us define the energy gap
\be
\Gap\equiv m_n  - m_p\simeq 1.29333\,\,{\rm MeV}\,.
\ee
Throughout we use $g(E)$ for the Fermi-Dirac distribution at
temperature of electrons $T$ and $g_\nu(E)$ for the Fermi-Dirac
distribution at the neutrino temperature $T_\nu$. 
\be\label{DefgnuANDg}
g(E) \equiv \frac{1}{\left({\rm e}^{\frac{E}{T}}+1\right)}\qquad g_\nu(E) \equiv  \frac{1}{\left({\rm e}^{\frac{E}{T_\nu}}+1\right)}\,.
\ee
In \S~\ref{SecThermo}, details are given on how these temperatures can
be computed in function of the scale factor $a$ and the time
$t$.  
%\cite{BrownSawyer} assume it is the same and that is bad.

In the infinite nucleon mass limit (but keeping $\Gap$ constant), also
called the Born approximation, the reaction rates take simple forms (see App. \ref{Apprates}). First, the
factors entering the matrix element (\ref{DefMLLMRRMLR}) are in that
limit simply
\be
\frac{{\cal M}_{LL}}{E_n E_p E_\nu E_e}=\frac{{\cal M}_{RR}}{E_n E_p E_\nu E_e}=\frac{{\cal M}_{LR}}{E_n E_p E_\nu E_e}=1\,.
\ee
The last equality is correct only if it is understood that an angular
average either on electrons momentum or neutrino momentum is
performed\footnote{The LR coupling is at the origin of the asymmetry (\ref{MLRBorn}) in the decay product of
  neutron beta decay~\cite{Ivanov:2012qe} from which the value of
  $g_A$ is inferred.} (see the detailed explanation at the end of
\S~\ref{SecFokker}).
Hence from Eq.~(\ref{GeneralGammaMainText}), we find the Born rates~\cite{BrownSawyer,LopezTurner1998,Weinberg1972,Bernstein1989}
\bea\label{GammanBorn}
\overline{\Gamma}_{n \to p}&=&\overline{\Gamma}_{n \to
  p+e}+\overline{\Gamma}_{n +e \to p}\\
&=&\myk \int_0^\infty p^2 \dd p [\chi_+(E) + \chi_+(-E)]\,,
\eea
with $E=\sqrt{p^2+m_e^2}$ and
\bea\label{Defchipm}
\chi_\pm(E) &\equiv& (E_\nu^\mp)^2 g_\nu(E^\mp_\nu) g(-E)\,,\\
E^\mp_\nu &\equiv& E\mp\Gap\,,\\
\myk &\equiv& \frac{4 G_W^2(1+3 g_A^2)}{(2\pi)^3} \label{Defk}\,.
\eea
The first contribution in Eq.~(\ref{GammanBorn}) corresponds
to the $n\to p$ processes (\ref{EqWeak1}) and (\ref{EqWeak2}) added, that is
for all processes where the electron is in the final state. It can be
checked indeed that the electron distribution is evaluated as $g(-E)=1-g(E)$. Furthermore, if the neutrino is in
the initial state (when $E>\Gap$) its energy is $E_\nu = E-\Gap$ and its
distribution function appears as $g_\nu(E_\nu)$, but if it is in the final state (when $E<\Gap$) its energy is $E_\nu =
\Gap-E$ and the neutrino distribution function is evaluated as
$g_\nu(E-\Gap)=1-g_\nu(\Gap-E)$. 

The second term of Eq.~(\ref{GammanBorn})  corresponds to the reaction
(\ref{EqWeak3}), that is to the process where the positron is in the
initial state. The energy of the positron is $E$ and
its distribution function appears as an initial state [$g(E)$], whereas the neutrinos in the final state have energy $E_\nu = \Gap+E$ and
their distribution function appear thus as Pauli-blocking factor
$g_\nu(-E-\Gap)=1-g_\nu(E+\Gap)$.

The reaction rate for protons, that is $\Gamma_{p \to n}$, is obtained
by the simple replacement $\Gap \to -\Gap$, which amounts to $\chi_+
\to \chi_-$. We give it for completeness
\bea\label{GammanBornp}
\overline{\Gamma}_{p \to n} &=& \overline{\Gamma}_{p \to n+e} +
\overline{\Gamma}_{p +e \to n}  \\
&=&\myk \int_0^\infty
p^2 \dd p [\chi_-(E) + \chi_-(-E)]\,.
\eea
Similarly the second term corresponds to the reverse processes
(\ref{EqWeak1}) and (\ref{EqWeak2}) added since the electron
distribution function is always in an initial state [$g(E)$], and the
neutrino is in the initial or final state depending on the sign of
$E_\nu=-E+\Gap$. The first term corresponds to the reverse process
(\ref{EqWeak3}) with the positron always in the final state
[$g(-E)=1-g(E)$] and the neutrino always in the initial state [$g_\nu(E+\Gap)$].

Finally, note that using
\be\label{BasicLawDetailedBalance}
g(-E) = 1-g(E) = {\rm e}^{E/T} g(E) \,,
\ee
we get in the case of thermal equilibrium between neutrinos and the plasma (that is when $T_\nu = T$)
\be\label{BasicLawDetailedBalance2}
\chi_+(E) ={\rm  e}^{\Gap/T} \chi_-(-E)\,.
\ee
This implies that if neutrinos have the same temperature as the
plasma, the reaction rates satisfy the Born approximation detailed
balance relation
\be\label{MagicDetailedBalanceBorn}
\overline{\Gamma}_{p \to n} = {\rm e}^{-\Gap/T} \overline{\Gamma}_{n \to p}\,.
\ee

It is important that detailed balance be satisfied in our estimation
of the weak rates since it is at the origin of the enforcement of the
thermodynamical equilibrium between neutrons and protons. When adding corrections to the Born approximation weak
rates in the form $\Gamma \equiv \delta \Gamma + \overline{\Gamma}$,
then the relative corrections of the forward and reverse rates must be
equal, that is
\be
\frac{\delta \Gamma_{p \to n}}{\overline{\Gamma}_{p \to n}} = \frac{\delta \Gamma_{n \to p}}{\overline{\Gamma}_{n \to p}}\,,
\ee
in order that the corrected rates also satisfy the detailed balance
property (\ref{MagicDetailedBalanceBorn}). When estimating the corrections to the Born rates, we
systematically discuss how this detailed balance is kept valid with
the corrections and we highlight our differences on that crucial property with
previous literature.

\subsection{Calibration from free neutron decay rate}

The interaction rates are proportional to the factor $\myk$ defined in
Eq.~(\ref{Defk}). It is proportional to 
\be
G_W^2 \equiv  G_F^2 \cos^2 (\theta_{\rm C})
\ee 
where $G_F$ is the Fermi constant and $\theta_{\rm
  C}$ is a CKM angle, and also to $1+3 g_A^2$ where
$g_A$ is the axial current constant for the nucleons (see
App. \ref{ParticleValues} for numerical values). 
Given the uncertainty in these parameters, it is more precise to
obtain $\myk$ from the free neutron decay\footnote{The relative precision on $\cos (\theta_{\rm C})$ and $g_A$ are
  respectively $2.0 \times 10^{-4}$ and $1.8\times 10^{-3}$. If
  $\tau_n$ was to be obtained theoretically from these constants, then
its uncertainty $\delta \tau_n/\tau_n \simeq 2 \delta \cos
(\theta_{\rm C})/\cos (\theta_{\rm C}) + 1.66 \delta g_A/g_A$ would be
of order $3 \times 10^{-3}$ which is larger than the experimental
uncertainty which is of order $1.0 \times 10^{-3}$. However
improvements in the measurements of $g_A$ by a factor 3 would make
this direct estimation of $\tau_n$ competitive.}.
Indeed, at low temperatures, $\Gamma_{n \to p}$ should be equal to the free neutron decay rate
$1/\tau_n$.  The low temperature corresponds to (\ref{GammanBorn}) restricted to $E_\nu
<0$, that is
\be\label{BornNeutronDecay}
\frac{1}{\tau_n} = \Gamma_{n \to p}(T=0)=\myk \int_0^{\sqrt{\Gap^2-m_e^2}}
p^2 E_\nu^2 \dd p\,,
\ee
with $E_\nu = \sqrt{p^2+m_e^2}-\Gap$. We define
\be\label{Deflambdazero}
\lambda_0 \equiv  m_e^{-5 }\myk^{-1 }\Gamma_{n \to p}(T=0)
\ee
so $\myk$ is obtained from the free neutron decay rate as
\be\label{Getmyk}
\myk=1/(\tau_n \lambda_0 m_e^5)
\ee 
instead of Eq.~(\ref{Defk}). At the Born
approximation level, that is using $\overline{\Gamma}_{n \to p}$ in
Eq.~(\ref{Deflambdazero}), $\lambda_0$ takes the value~\cite{Bernstein1989}
\bea
\bar \lambda_0 &\equiv& m_e^{-5}\int_0^{\sqrt{\Gap^2-m_e^2}}
p^2 E_\nu^2 \dd p \nonumber\\
&=&\sqrt{\bar \Gap^2-1}\left(\frac{-8-9\bar \Gap^2+2\bar
    \Gap^4}{60}\right)+\frac{\bar \Gap}{4}\arccosh \bar \Gap\nonumber\\
&\simeq& 1.63609, 
\eea
where $\bar \Gap \equiv \Gap/m_e$.

\subsection{Neutron abundance and freeze-out}

\begin{figure}[h]
\centering
    \includegraphics[width=\mycolumnwidth]{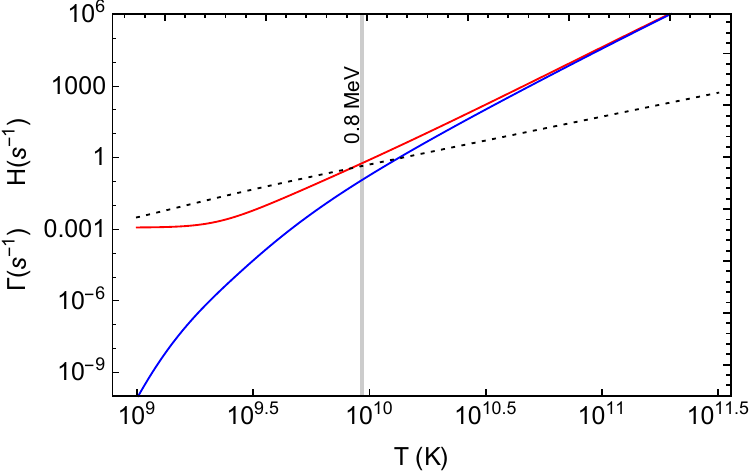}
    \caption{$\Gamma$. {\it Continuous line} : Upper red curve is $n
       \to p$ rate and lower blue curve is $p \to n$ rate . {\it Dashed  line} :
       Hubble rate.}
\label{FigRates}
\end{figure}

It is enough to consider the weak rates in the Born approximation to estimate the
freeze-out temperature. The Born rates are plotted in Fig. \ref{FigRates} together with the
Hubble rate.
A first estimation of the freeze-out temperature consists in noting that expansion overcomes both rates for $T_F\simeq
8\times 10^{9}{\rm K}$ which we can take as the freeze-out
temperature. 
\begin{figure}[!htb]
     \includegraphics[width=\mycolumnwidth]{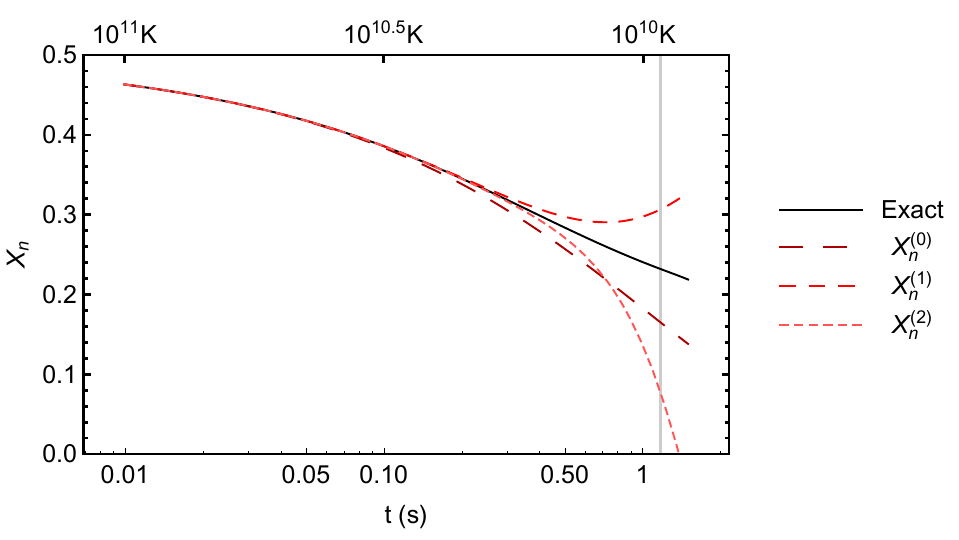}
     \caption{{\it Black continuous line} : numerical solution. {\it
         red dashed   lines} : solutions (\ref{Ynp}) for $p=0,1,2$. The vertical bar
       corresponds to $T_F=0.8 \,{\rm MeV}$}
\label{FigTC}
\end{figure}

We can justify this estimation by finding an approximate solution for
the neutron abundance. Its evolution, in absence of nuclear reactions,
is dictated by
\be
\dot X_n = X_p \Gamma_{p \to n}-X_n \Gamma_{n \to p}\,.
\ee
Using $X_p=1-X_n$, we get
\be
X_n = \frac{\Gamma_{p \to n}}{\Lambda} -\frac{1}{\Lambda} \dot
X_n\,,\qquad \Lambda \equiv \Gamma_{p \to n} + \Gamma_{n \to p}\,.
\ee
We obtain tight-coupling solutions to order $m$ by recursive
replacement of $X_n$ and we get
 \be\label{Ynp}
X_n^{(m)}=\sum_{k=0}^m \left(\frac{-1}{\Lambda}\frac{\dd}{\dd
    t}\right)^k \frac{\Gamma_{p\to n}}{\Lambda}\,.
\ee
$m=0$ corresponds to the pure thermodynamical equilibrium and higher order
corrections are due to the fact that the neutron abundance has less and
less time to approach this equilibrium value. This expansion is
equivalent to \citet[Eq. 2.8]{Bernstein1989}. We can estimate the
freeze-out temperature, that is the temperature at which weak
interactions fail to maintain the thermodynamical equilibrium value,
by computing the temperature for which the first correction is of
order of the equilibrium value, that is when $|X_n^{(1)}-X_n^{(0)}| =
X_n^{(0)}$. Assuming that the neutrino temperature does not differ
significantly from the plasma temperature before freeze-out, we can
use Eq.~(\ref{MagicDetailedBalanceBorn}) to estimate this condition
which reads
\be
H\frac{Q}{T}\frac{{\rm e}^{Q/T}}{1+{\rm e}^{Q/T}} = \Lambda\,.
\ee
We find $T_F \simeq 8.9 \times 10^9{\rm K}\simeq 0.77\,{\rm
  MeV}$ and replacing in the thermodynamical equilibrium abundance we get
\be\label{DefFOYn}
X_n^F\equiv X_n^{(0)}(T_F) \simeq 0.156\,.
\ee
Finally, another estimation consists in determining visually when the
neutron abundance is only affected by beta decay. In
Fig.~\ref{FigYnCoc}, this leads to $T_F \simeq 3.3\times 10^9\,{\rm
  K}\simeq 0.28\,{\rm MeV}$ and $X_n^F \simeq 0.17$.

After the freeze-out, the neutron abundance is only affected by the neutron beta
decay. The nucleosynthesis starts approximately when $T< T_{\rm
  Nuc}=0.078\,{\rm MeV}$, or 0.9~GK (see \S~\ref{SecTnuc}) corresponding to roughly $t_{\rm  Nuc}=200\,{\rm s}$. The neutron abundance when the nucleosynthesis starts is thus approximately
given by
\be\label{FreezeOutYnModel}
X_n(T_{\rm Nuc}) = X^F_n {\rm exp}\left(-\frac{t_{\rm nuc}-t_F}{\tau_n}\right)\approx0.13\,.
\ee

\subsection{Radiative corrections at $T=0$}\label{SecCCRTnull}

\subsubsection{Standard computation}

\begin{figure}[h]
%\centering
\subfloat[Born]{      
\begin{fmffile}{BornVP}
 \begin{fmfgraph*}(30,15)
   \fmfleft{n,nu}
   \fmfright{p,e}
   \fmf{fermion}{n,v1,p}
   \fmf{fermion}{nu,v2,e}
  \fmf{boson,label=$W$,tension=0}{v2,v1}

        \fmflabel{$n$}{n}
        \fmflabel{$p^+$}{p}
        \fmflabel{$e^-$}{e}
        \fmflabel{$\nu$}{nu}
\end{fmfgraph*}
\end{fmffile}
\label{FDBorn}}
\ifphysrep\qquad\qquad\else\fi\qquad\quad\subfloat[]{
     \begin{fmffile}{VPa}
\begin{fmfgraph*}(30,15)
\fmfstraight
   \fmfbottom{n,vb,vb1,vb2,p}
   \fmftop{nu,vt,vt1,vt2,e}
  \fmffreeze
   \fmf{fermion}{n,vb,p}

   \fmf{fermion}{nu,vt}
 \fmf{plain}{vt,v1}
\fmf{plain}{v1,v2}
\fmf{fermion}{v2,e}

\fmf{photon,label=$\gamma$,right,tension=0}{v1,v2}
\fmf{boson,label=$W$,tension=0}{vt,vb}

\fmflabel{$n$}{n}
        \fmflabel{$p^+$}{p}
        \fmflabel{$e^-$}{e}
       \fmflabel{$\nu$}{nu}
\end{fmfgraph*}
\end{fmffile}
\label{FDVPb}}

\vspace{0.5cm}

\subfloat[]{
     \begin{fmffile}{VPb}
\begin{fmfgraph*}(30,15)
\fmfstraight
   \fmfbottom{n,vb,vb1,vb2,p}
   \fmftop{nu,vt,vt1,vt2,e}
  \fmffreeze
   \fmf{fermion}{n,vb}
\fmf{plain}{vb,vb1}
\fmf{plain}{vb1,vb2}
\fmf{fermion}{vb2,p}

   \fmf{fermion}{nu,vt}
 \fmf{plain}{vt,v1}
\fmf{plain}{v1,v2}
\fmf{fermion}{v2,e}

\fmf{photon,label=$\gamma$,tension=0}{vb2,v2}
\fmf{boson,label=$W$,tension=0}{vt,vb}

\fmflabel{$n$}{n}
        \fmflabel{$p^+$}{p}
        \fmflabel{$e^-$}{e}
       \fmflabel{$\nu$}{nu}
\end{fmfgraph*}
\end{fmffile}
\label{FDVPc}}
\ifphysrep\qquad\qquad\else\fi\qquad\quad
\subfloat[]{
     \begin{fmffile}{VPd}
\begin{fmfgraph*}(30,15)
\fmfstraight
   \fmfbottom{n,vb,vb1,vb2,p}
   \fmftop{nu,vt,vt1,vt2,e}
  \fmffreeze
   \fmf{fermion}{n,vb}
\fmf{plain}{vb,vb1}
\fmf{plain}{vb1,vb2}
\fmf{fermion}{vb2,p}

   \fmf{fermion}{nu,vt}
 \fmf{plain}{vt,v1}
\fmf{plain}{v1,v2}
\fmf{fermion}{v2,e}

\fmf{photon,label=$\gamma$,left,tension=0}{vb1,vb2}
\fmf{boson,label=$W$,tension=0}{vt,vb}

\fmflabel{$n$}{n}
        \fmflabel{$p^+$}{p}
        \fmflabel{$e^-$}{e}
       \fmflabel{$\nu$}{nu}
\end{fmfgraph*}
\end{fmffile}
\label{FDVPd}}
\caption{Born diagram and virtual photon radiative corrections.}
    \label{FigFD1}
%  \end{center}
\end{figure}
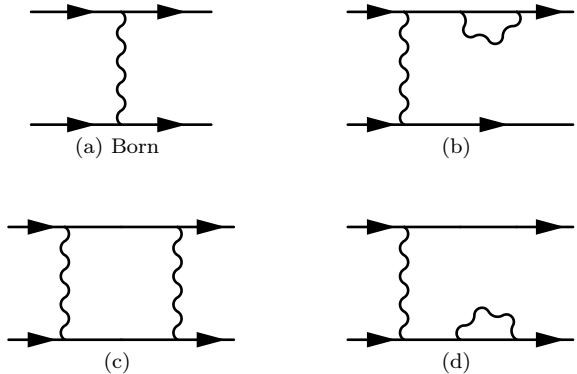

Radiative corrections at $T=0$ correspond to two types of
corrections~\citep{Sirlin1967,Dicus1968,Dicus1982,Kernan}. First the
radiative corrections per se, that is for which a virtual photon is
emitted and absorbed inside the interaction and which interfere
with the lowest order or Born diagram. Hereafter we call
these {\it pure radiative corrections} and the relevant diagrams in the
infinite nucleon mass limit are depicted in Fig.~\ref{FigFD1} (see
also \citet[App. C]{Ivanov:2012qe}). As shown in the seminal article
by \citet{Sirlin1967}, the other diagrams involving virtual
photons\footnote{These extra diagrams correspond first to a virtual photon
  exchange between the electron and the gauge boson, and between the
  electron and the neutron \citep[Fig. 1]{Sirlin1967}, and second to
  a virtual photon exchange between the proton and the
gauge boson and between the proton and the neutron
\citep[Fig. 3]{Sirlin1967}.} can be reabsorbed in the redefinitions of
$G_F$ and $g_A$. 

\begin{figure}[h]
%\centering
\subfloat[]{
\begin{fmffile}{BS1}
\begin{fmfgraph*}(30,20)
\fmfstraight
 \fmfleft{n,b1,b2,nu,a1}
\fmfright{p,b3,b4,e,a5}
\fmffreeze
\fmfbottom{n,vb,vb1,p}
 \fmftop{a1,a2,a3,a5}
%  \fmffreeze
   \fmf{fermion}{n,vb,p}
   \fmf{fermion}{nu,vt}
 \fmf{plain}{vt,v1}
\fmf{fermion}{v1,e}

\fmf{photon,tension=0}{a5,v1}

\fmf{boson,label=$W$,tension=0}{vt,vb}
\fmflabel{$n$}{n}
        \fmflabel{$p^+$}{p}
        \fmflabel{$e^-$}{e}
        \fmflabel{$\nu$}{nu}
  \end{fmfgraph*}
\end{fmffile}
\label{FDBS1}}
\ifphysrep\qquad\qquad\else\fi\qquad\quad
\subfloat[]{
\begin{fmffile}{BS2}
\begin{fmfgraph*}(30,20)
\fmfstraight
 \fmfleft{n,b1,b2,nu,a1}
\fmfright{p,b3,b4,e,a5}
\fmffreeze
\fmfbottom{n,vb,vb1,p}
 \fmftop{a1,a2,a3,a5}
%  \fmffreeze
   \fmf{fermion}{n,vb,p}
   \fmf{fermion}{nu,vt}
 \fmf{plain}{vt,v1}
\fmf{fermion}{v1,e}

\fmf{photon,tension=0}{v1,a2}

\fmf{boson,label=$W$,tension=0}{vt,vb}
\fmflabel{$n$}{n}
        \fmflabel{$p^+$}{p}
        \fmflabel{$e^-$}{e}
        \fmflabel{$\nu$}{nu}
  \end{fmfgraph*}
\end{fmffile}
\label{FDBS2}}

\caption{Bremsstrahlung (left) and absorption (right).}
    \label{FigFD2}
%  \end{center}
\end{figure}
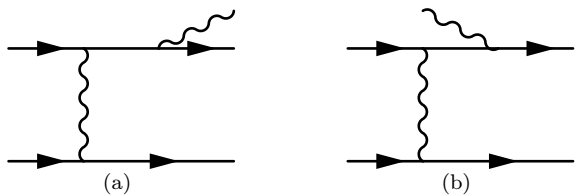

Second it also contains the corrections due to the emission of real photon \citep[Fig 2e,2f]{Dicus1968}.
These processes are generically called bremsstrahlung (BS), and in the
infinite nucleon mass approximation only the emission from the
electron line (Fig.~\ref{FDBS1}) contributes.

It is necessary to handle these two types of corrections simultaneously, because
each of these contributions contains an infrared divergence in the photon momentum. But as is
well known, when both are considered, these divergences exactly cancel
provided they are correctly regularized. The usual procedure consists
in letting the photon have a mass for the computation of each type of
correction, and then noticing that the log-divergences which appear are exactly
opposite, leading to a finite result in the limit $m_\gamma \to 0$ \citep{Dicus1968}; \citep[App. B]{Ivanov:2012qe}. These radiative corrections are very well understood for the
neutron beta decay~\cite{Sirlin1967,Dicus1968,Czarnecki2004,Marciano:2005ec}. The main result
is that radiative corrections can be taken into
account by a multiplicative factor whose expression involves Sirlin's
universal function \citep[Eq. 20b]{Sirlin1967}. This was
subsequently refined mainly to improve the
expression of the high energy cut-off which was  related to the mass of
the vector boson once weak interactions were properly understood, and
we use \citet{Czarnecki2004} as the most accurate account on this issue.

A careful analysis however indicates that only pure radiative
corrections take a universal form which can be used for all the reactions
(\ref{AllWeakReactions}). They are given by \citet[Eq. 6.2]{Dicus1968}
and depend only on the electron velocity $\beta = p/E$~\footnote{Even
  though they have been derived for neutron beta decay, they remain
  formally identical for other reactions because they are related by
  crossing symmetry under which $\beta$ is left invariant.}. The total
corrections for the neutron beta decay include also the bremsstrahlung
\citep[Eq. 6.6]{Dicus1968} and these are computed specifically for the neutron beta decay, with a
maximum photon energy being $\Gap-E$. 

Hence, the radiative corrections for the reactions
(\ref{AllWeakReactions}) should in principle be recomputed
to take into account the correct bremsstrahlung corrections for each
reaction. However it proves much more convenient to temporarily forget
about this issue and assume that the radiative corrections computed
for the neutron beta decay can be used without modification for all
weak reactions.  We postpone the correct treatment of bremsstrahlung
to the next section. 

Note that even for the neutron beta decay this approximation amounts to ignoring the
Pauli-blocking effects of the neutrinos which is modified by the
energy subtracted from the photon. In this approximation, the maximum energy of emitted
photons is chosen to be the maximum energy of the neutrino when in the
final state. For the neutron beta decay this is $\Gap-E$, but for
reaction (\ref{EqWeak3}) this is $\Gap+E$.

Under this assumption, our method
to compute the radiative correction at $T=0$ follows exactly the
approach by \citet{Dicus1982} and subsequent literature based on it, for which the same assumption was implicitly made.
The radiative corrections can be separated between a Coulomb correction,
which accounts for the motion of the electron in the Coulomb field of
the proton, and the other corrections. 
It is customary to factorize the Coulomb corrections as they can be taken into account multiplicatively
by the Fermi function\footnote{This amounts to including some
  radiative corrections which are higher order in $\alpha_{\rm
    FS}$.}. The diagram \ref{FDVPc} corresponds to the interaction
between the proton and the electron, and since this is accounted for by
the Fermi function, its Coulomb part is subtracted so as to avoid double counting.

\begin{figure}[!htb]
     \includegraphics[width=\mycolumnwidth]{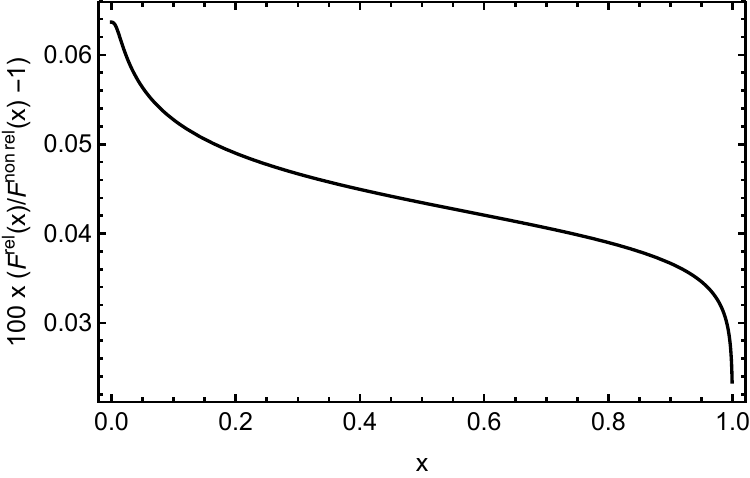}
     \caption{Relative difference between the relativistic and the
       non-relativistic Fermi function.}
\label{FigFermi}
\end{figure}

The non-relativistic Fermi function 
%(TODO une bonne citation historique) 
is
\be
F(E) \equiv \frac{y}{1-{\rm e}^{-y}}\,,\qquad y \equiv
\frac{2\pi \alpha_{\rm FS}}{\beta}\quad \beta = \frac{p}{E}\,.
\ee
The relativistic Fermi function is given by \citep[Eq. 5]{Ivanov:2012qe}
%TODO look at \cite{SmithFuller} which uses a different Fermi function
%and see if it makes a difference.
\ifphysrep
\be
F(E) \equiv \frac{4 \left(1+\tfrac{\gamma}{2}\right) (2 r_p m_e
  \beta)^{2\gamma}}{\Gamma^2(3+2\gamma)}\frac{{\rm e}^{\pi \alpha_{\rm
    FS}/\beta}}{(1-\beta^2)^\gamma}\left|\Gamma\left(1+\gamma +\ii \frac{\alpha_{\rm FS}}{\beta}\right)\right|^2
\ee
\else
\bea
F(E) &\equiv& \frac{4 \left(1+\tfrac{\gamma}{2}\right) (2 r_p m_e
  \beta)^{2\gamma}}{\Gamma^2(3+2\gamma)}\frac{{\rm e}^{\pi \alpha_{\rm
    FS}/\beta}}{(1-\beta^2)^\gamma}\nonumber\\
&&\qquad\times\left|\Gamma\left(1+\gamma +\ii \frac{\alpha_{\rm FS}}{\beta}\right)\right|^2
\eea
\fi
where $\gamma \equiv \sqrt{1-\alpha_{\rm FS}^2}-1$ and $r_p$ is the proton radius.
In practice the non-relativistic function is enough as long as we do
not focus on a precision on $\YPfour$ better than unity. The modification on light elements yields is
evaluated in \S~\ref{SecResults} and we find $\Delta \YPfour \simeq
0.3$ when using the relativistic Fermi function (see also
\citet{SmithFuller} which reports $\Delta \YPfour \simeq 0.4$). In this work we use the relativistic
Fermi function whenever it is not specified differently. The relative difference between the two Fermi functions is plotted in
Fig. \ref{FigFermi} and it remains smaller than $0.06 \%$.

Considering the contributions of Coulomb and other radiative
corrections, the Born rates are modified as
\ifphysrep
\be\label{CCRn}
{\Gamma}^{{\RC} 0}_{n \to p}=\myk \int_0^\infty p^2 \dd p[\Fermi_+(E)R(E,|\Gap-E|)\chi_+(E) + \Fermi_+(-E)R(E,\Gap+E)\chi_+(-E)]\,.
\ee
\else
\bea\label{CCRn}
{\Gamma}^{{\RC} 0}_{n \to p}&=&\myk \int_0^\infty p^2 \dd p[\Fermi_+(E)R(E,|\Gap-E|)\chi_+(E) \nonumber\\
&&\qquad+ \Fermi_+(-E)R(E,\Gap+E)\chi_+(-E)]\,.
\eea
\fi
The Coulomb corrections occur only if both the electron and the
proton are either in the initial or the final state. We have thus defined the compact notation
%\be
%\Fermi_s(E) = 1+[F(|E|)-1] \theta(sE)
%\ee
\be
\Fermi_\pm(x)=\begin{cases} \label{Fermi}
      F(|x|) & {\rm if} \quad \pm x >0 \\
      1 & {\rm if}  \quad \pm x \le 0\,, \\
   \end{cases}
\ee
which ensures that this is the case. 

The factor $R(.,.)$ takes into account pure radiative corrections and
bremsstrahlung (but only in the form of neutron beta decay in vacuum
as evoked above) and is of the form
\be\label{DefRC}
R(E,k_{\rm max}) \equiv1+\frac{\alpha_{\rm FS}}{2\pi}{\cal
  C}\left(E,k_{\rm max}\right)\,.
\ee
The function ${\cal C}$ is given in Eq.~(\ref{CSirlin}). However we
find it better to use the most recent form
(\ref{RadiativeMarcianoSirlin}) for the radiative correction
$R(E,k_{\rm kmax})$
which amounts to resumming higher order corrections, since it is more
accurate \cite{Esposito1998, Czarnecki2004}. We show in
\S~\ref{SecResults} that it modifies the ${}^4{\rm He}$ production by
$\Delta \YPfour \simeq 0.2$.

\begin{figure}[!htb]
     \includegraphics[width=\mycolumnwidth]{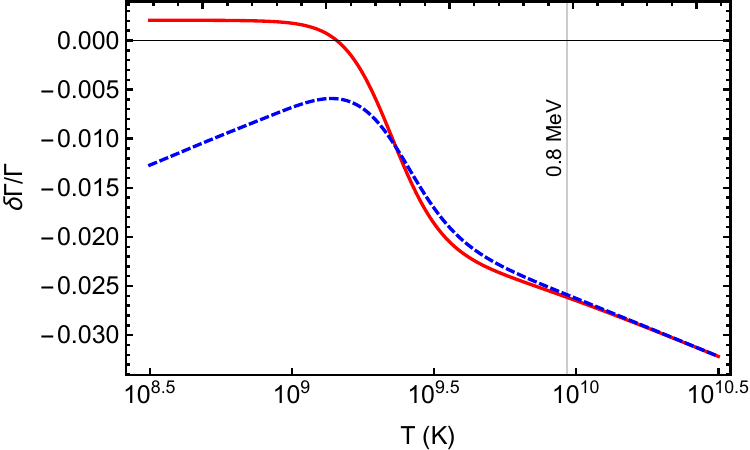}
     \caption{$\delta \Gamma/\Gamma$ due to Coulomb and
       null-temperature radiative corrections.{\it Red continuous line} : $n
       \to p$. {\it Blue dashed   line} :   $p \to n$. The relative
       variations are equal at high temperature, when $T_\nu\simeq T$,
       and the detailed balance is still satisfied when including
       these corrections. This plot reproduces \citet[Fig. 8]{LopezTurner1998}.}
\label{FigCorrectionCCRRates}
\end{figure}

The rate for protons is obtained by the replacement $\Gap\to -\Gap$ in
(\ref{CCRn}), that is from $\chi_+ \to \chi_-$, together with the
replacement $\Fermi_+ \to \Fermi_-$ to ensure that the Fermi function
appears when the proton and the electron are on the same
side. Hence it is given by
\ifphysrep
\be\label{CCRp}
{\Gamma}^{{\RC} 0}_{p \to n}=\myk \int_0^\infty p^2 \dd p[\Fermi_-(E)R(E,E+\Gap)\chi_-(E) + \Fermi_-(-E)R(E,|E-\Gap|)\chi_-(-E)]\,.
\ee
\else
\bea\label{CCRp}
{\Gamma}^{{\RC} 0}_{p \to n}&=&\myk \int_0^\infty p^2 \dd p[\Fermi_-(E)R(E,E+\Gap)\chi_-(E) \nonumber\\
&&\quad+ \Fermi_-(-E)R(E,|E-\Gap|)\chi_-(-E)]\,.
\eea
\fi
Given our choice to take into account the bremsstrahlung corrections
of neutron beta decay for all processes and our choice for the maximum
energy of emitted photons $k_{\rm max}$, then by construction the detailed balance
property (\ref{MagicDetailedBalanceBorn}) is preserved using again the
property (\ref{BasicLawDetailedBalance2}), that is we get by construction
\be
{\Gamma}^{\rm RC0}_{p \to n} = {\rm e}^{-\Gap/T}
{\Gamma}^{\rm RC0}_{n \to p}\,.
\ee 
The relative variations of the weak rates from radiative corrections are plotted in Fig.~\ref{FigCorrectionCCRRates}.

The radiative corrections also affect the free decay rate of neutrons and thus the value of $\lambda_0$ which is increased to
\bea\label{lambdazeroCCR}
\lambda^{{\RC} 0}_0 &\equiv& m_e^{-5}\int_0^{\sqrt{\Gap^2-m_e^2}}F(E)R(E,E_\nu) p^2
E_\nu^2 \dd p \nonumber\\
&\simeq& 1.75838\,,
\eea
with $E_\nu = \Gap-E$.

\subsubsection{Bremsstrahlung corrections}\label{SecBS}

It was argued in \citet{BrownSawyer} that the detailed balance
relation (\ref{MagicDetailedBalanceBorn}) should hold even including radiative corrections at
order $\alpha_{\rm FS}$, as long as we work in the infinite nucleon
mass approximation. Hence the derivation of radiative corrections
due to interactions with the surrounding bath of particles, which are
usually called finite-temperature corrections, must satisfy this
detailed balance relation. These corrections correspond to real photon
processes which are the absorption diagram~\ref{FDBS2} together with the stimulated emission part of
\ref{FDBS1}, and to mass shift effects illustrated by the diagrams \ref{FigFD3}.

The full computation of the finite temperature radiative corrections
was carefully carried out in \citet{BrownSawyer}, following the
guideline of detailed balance and with a detailed discussion on
misconceptions in earlier efforts~\citep{Dicus1982,Cambier1982, Sawyer1996, Chapman1997,Esposito1999} concerning the so called
wave-function renormalization. However the authors of
\citet{BrownSawyer} reported that when combining their results
with the null-temperature radiative corrections discussed in the
previous section, the rates failed to satisfy the detailed balance
relation. We argue that this is because bremsstrahlung
was inconsistently taken into account. As long as we take into account
only null-temperature radiative corrections, it is reasonable to
consider that the bremsstrahlung effects are those of the neutron beta
decay for all reactions, so as to maintain the detailed balance
relation (\ref{MagicDetailedBalanceBorn}). This is what has been done
in the previous section.

Strictly speaking it is a mistake because part of the
null-temperature corrections are left out. However it is the
most reasonable procedure if one ignores the full details of the
finite-temperature radiative corrections. Here since we take into account
the interactions with the bath of photons and electrons in the finite
temperature radiative corrections, we must also correct for this
ad-hoc and incorrect treatment of bremsstrahlung in the previous section. The
details are reported in appendix~\ref{AppBS} and the result is that we need to add the
corrections
\ifphysrep
\beas\label{BSnpFormal}
\delta \Gamma^{\rm BS}_{n \to p} &=&\frac{\alpha_{\rm FS}
  \myk}{2\pi}\int_{m_e}^\infty \dd E \left[g(-E) \Fermi_+(E) \gamma^{\rm BS}_{n \to p+e}+g(E) \Fermi_+(-E) \gamma^{\rm BS}_{n+e\to p}\right]\\
\delta \Gamma^{\rm BS}_{p \to n}  &=&\frac{\alpha_{\rm FS}
  \myk}{2\pi}\int_{m_e}^\infty \dd E \left[g(-E) \Fermi_-(E) \gamma^{\rm BS}_{p \to n+e}+g(E) \Fermi_-(-E) \gamma^{\rm BS}_{p+e\to n}\right]\,,
\eeas
\else
\beas\label{BSnpFormal}
\delta \Gamma^{\rm BS}_{n \to p} &=&\frac{\alpha_{\rm FS}
  \myk}{2\pi}\int_{m_e}^\infty \dd E \left[g(-E) \Fermi_+(E) \gamma^{\rm BS}_{n \to p+e}\nonumber\right.\\
&&\qquad \qquad \left.+g(E) \Fermi_+(-E) \gamma^{\rm BS}_{n+e\to p}\right]\\
\delta \Gamma^{\rm BS}_{p \to n}  &=&\frac{\alpha_{\rm FS}
  \myk}{2\pi}\int_{m_e}^\infty \dd E \left[g(-E) \Fermi_-(E) \gamma^{\rm BS}_{p \to n+e}\nonumber\right.\\
&&\qquad \qquad\left.+g(E) \Fermi_-(-E) \gamma^{\rm BS}_{p+e\to n}\right]\,,
\eeas
\fi
with the definitions~(\ref{BSCorrection}-\ref{BSCorrection2}). Note that we have added the
Fermi factor contribution for consistency with the
rest of the radiative corrections. Their temperature dependence is plotted in
Fig.~\ref{FigCorrectionCCRTRates}. Even though these BS corrections
correspond to null temperature radiative corrections, unless specified
we include them as part of the finite temperature radiative corrections discussed
in the next section, since they must be added only to maintain consistently the detailed balance relation when finite temperature
radiative corrections are included. We report in \S~\ref{SecResults}
that these BS corrections are responsible for a non-negligible modification of
${}^4{\rm He}$ production which is $\Delta \YPfour \simeq -3.1$.

\subsection{Finite temperature radiative corrections}\label{SecRadTfinite}

The finite-temperature corrections can be separated in three
parts. Using a notation similar to \citet{BrownSawyer},
these three contributions are noted
\be\label{DefGammaT}
\Gamma^{T}_{n \to p}  \equiv \Gamma^{\gamma, T}_{n \to p}
+\Gamma^{\Delta E, T}_{n \to p} +\Gamma^{ep+ee, T}_{n \to p}\,, 
\ee
and similar notation for the $p \to n$ processes.

\begin{figure}[h]
%\begin{center}
  
\subfloat[]{
\begin{fmffile}{TRC1}
\begin{fmfgraph*}(30,20)
\fmfstraight
 \fmfleft{n,b1,b2,nu,a1}
\fmfright{p,b3,b4,e,a5}
\fmffreeze
\fmfbottom{n,vb,vb1,vb2,p}
 \fmftop{a1,a2,a3,a4,a5}
%  \fmffreeze
   \fmf{fermion}{n,vb,p}
   \fmf{fermion}{nu,vt}
 \fmf{plain}{vt,v1}
\fmf{plain}{v1,v2}
\fmf{fermion}{v2,e}

\fmf{photon,tension=0}{a2,v1}
\fmf{photon,tension=0}{a5,v2}

\fmf{boson,label=$W$,tension=0}{vt,vb}
\fmflabel{$n$}{n}
        \fmflabel{$p^+$}{p}
        \fmflabel{$e^-$}{e}
        \fmflabel{$\nu$}{nu}
  \end{fmfgraph*}
\end{fmffile}
\label{FDTRC1}}
\ifphysrep\qquad\qquad\else\fi\qquad\quad
\subfloat[]{
\begin{fmffile}{TRC2}
\begin{fmfgraph*}(30,20)
\fmfstraight
 \fmfleft{n,b1,b2,nu,a1}
\fmfright{p,b3,b4,e,a5}
\fmffreeze
\fmfbottom{n,vb,vb1,vb2,p}
 \fmftop{a1,a2,a3,a4,a5}
%  \fmffreeze
   \fmf{fermion}{n,vb,p}
   \fmf{fermion}{nu,vt}
 \fmf{fermion,label=$e^-$}{vt,v1}
\fmf{photon}{v1,v2}
\fmf{fermion}{v2,e}

\fmf{fermion,label=$e^+$,tension=0}{v1,a2}
\fmf{fermion,label=$e^+$,tension=0}{v2,a5}

\fmf{boson,label=$W$,tension=0}{vt,vb}
\fmflabel{$n$}{n}
        \fmflabel{$p^+$}{p}
        \fmflabel{$e^-$}{e}
        \fmflabel{$\nu$}{nu}
  \end{fmfgraph*}
\end{fmffile}
\label{FDTRC2}}
\caption{Feynman diagram interpretation of finite temperature radiative
  corrections, that is of the effect of the interactions with the
  surrounding plasma.}

\label{FigFD3}
%  \end{center}
\end{figure}
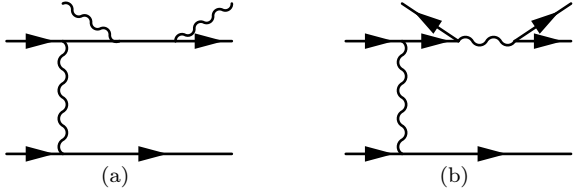

The first part is the stimulated emission and absorption of real
photons \citep[Eq. 5.9]{BrownSawyer}, see Fig.~\ref{FigFD2}. When
combined with a contribution coming from the diagram~\ref{FDTRC1} it gives
\ifphysrep
\beas\label{RCT1}
&&\Gamma^{\gamma, T}_{n \to p} = \frac{\alpha_{\rm FS}\myk}{2
  \pi}\int_0^\infty \frac{\dd k}{k}\int_{m_e}^\infty \dd E\times\\
&&\qquad\qquad\left\{A(E,k)\left[g(-E) \widetilde{\chi}_+^A(E,k)+g(E)
    \widetilde{\chi}_+^A(-E,k)\right]\right.-\left.k B(E,k) \left[g(-E) \widetilde{\chi}_+^B(E,k)+g(E)
    \widetilde{\chi}_+^B(-E,k)\right]\right\}\nonumber\\
&&\Gamma^{\gamma, T}_{p \to n} = \left.\Gamma^{\gamma, T}_{n \to
  p}\right|_{\widetilde{\chi}_+^{A/B} \to \widetilde{\chi}_-^{A/B}}
\eeas
\else
\beas\label{RCT1}
&&\Gamma^{\gamma, T}_{n \to p} = \frac{\alpha_{\rm FS}\myk}{2
  \pi}\int_0^\infty \frac{\dd k}{k}\int_{m_e}^\infty \dd E\times\\
&&\quad\left\{A(E,k)\left[g(-E) \widetilde{\chi}_+^A(E,k)+g(E)
    \widetilde{\chi}_+^A(-E,k)\right]\right.\nonumber\\
&&\qquad-\left.k B(E,k) \left[g(-E) \widetilde{\chi}_+^B(E,k)+g(E)
    \widetilde{\chi}_+^B(-E,k)\right]\right\}\nonumber\\
&&\Gamma^{\gamma, T}_{p \to n} = \left.\Gamma^{\gamma, T}_{n \to
  p}\right|_{\widetilde{\chi}_+^{A/B} \to \widetilde{\chi}_-^{A/B}}
\eeas
\fi
with
\bea
\widetilde{\chi}_\pm^A(E,k) &\equiv& \widetilde{\chi}_\pm(E-k)+\widetilde{\chi}_\pm(E+k)-2 \widetilde{\chi}_\pm(E)\\
\widetilde{\chi}_\pm^B(E,k) &\equiv& \pm\left[\widetilde{\chi}_\pm(E-k)-\widetilde{\chi}_\pm(E+k)\right]\,.
\eea
It does not satisfy the detailed balance relation. However when the
bremsstrahlung corrections (\ref{BSnpFormal}) are added the sum does satisfy
it. This crucial point is detailed in App. \ref{SecCorrectionBS}.

The second part of finite temperature radiative corrections is made of
the electron energy shift \citep[Eq. 5.13]{BrownSawyer}, and it
originates partly from the diagrams \ref{FigFD3}. It reads
\ifphysrep
\beas\label{RCT2}
\Gamma^{\Delta E, T}_{n \to p} &=&-\frac{2 \alpha_{\rm FS}
  \myk}{\pi}\int \dd p\left[\chi_+(E)+\chi_+(-E)\right]\left[\frac{\pi^2 T^2}{6}+\int_{m_e}^\infty \dd E' p' g(E')\left(1-\frac{m_e^2}{{E'}^2-E^2}\right)\right]\\
\Gamma^{\Delta E, T}_{p \to n} &=&\left.\Gamma^{\Delta E, T}_{n \to
    p}\right|_{\chi_+ \to \chi_-}\,,
\eeas
\else
\beas\label{RCT2}
\Gamma^{\Delta E, T}_{n \to p} &=&-\frac{2 \alpha_{\rm FS}
  \myk}{\pi}\int \dd p\left[\chi_+(E)+\chi_+(-E)\right]\\
&\times& \left[\frac{\pi^2 T^2}{6}+\int_{m_e}^\infty \dd E' p' g(E')\left(1-\frac{m_e^2}{{E'}^2-E^2}\right)\right]\nonumber\\
\Gamma^{\Delta E, T}_{p \to n} &=&\left.\Gamma^{\Delta E, T}_{n \to
    p}\right|_{\chi_+ \to \chi_-}\,,
\eeas
\fi 
where $E' \equiv \sqrt{{p'}^2+m_e^2}$.

The third and last part is made of proton-electron interactions, which
is a finite temperature correction to the diagram \ref{FDVPc}, and electron self-energy and wave-function
renormalization \citep[Eq. 5.10]{BrownSawyer}, which come partly from
the diagram~\ref{FDTRC2}. It reads
\ifphysrep
\beas\label{RCT3}
\Gamma^{ep+ee,T}_{n \to p} &=&\frac{\alpha_{\rm
    FS}\myk}{2\pi}\int_{m_e}^\infty \dd
E\left[\chi_+(E)+\chi_+(-E)\right]\int_{m_e}^\infty \dd E' g(E')\frac{E}{{E'}^2-E^2}\\
&\times& \left\{8p
  p'-({E'}^2+E^2)\ln\left(\frac{p+p'}{p-p'}\right)^2\right.-\left.\frac{E'}{E}({E'}^2+E^2)\ln\left[\frac{(E E' + p p')^2-m_e^4}{(E E' - p p')^2-m_e^4}\right]\right\}\nonumber\\
\Gamma^{ep+ee,T}_{p \to n} &=&\left. \Gamma^{ep+ee,T}_{n \to
    p}\right|_{\chi_+ \to \chi_-}\,.
\eeas
\else
\beas\label{RCT3}
\Gamma^{ep+ee,T}_{n \to p} &=&\frac{\alpha_{\rm
    FS}\myk}{2\pi}\int_{m_e}^\infty \dd
E\left[\chi_+(E)+\chi_+(-E)\right]\nonumber\\
&\times& \int_{m_e}^\infty \dd E' g(E')\frac{E}{{E'}^2-E^2}\nonumber\\
&\times& \left\{8p
  p'-({E'}^2+E^2)\ln\left(\frac{p+p'}{p-p'}\right)^2\right.\\
&-&\left.\frac{E'}{E}({E'}^2+E^2)\ln\left[\frac{(E E' + p p')^2-m_e^4}{(E E' - p p')^2-m_e^4}\right]\right\}\nonumber\\
\Gamma^{ep+ee,T}_{p \to n} &=&\left. \Gamma^{ep+ee,T}_{n \to
    p}\right|_{\chi_+ \to \chi_-}\,.
\eeas
\fi
The second and third contributions [Eqs.~(\ref{RCT2}) and (\ref{RCT3})]
satisfy manifestly the detailed balance relation thanks to the
property (\ref{BasicLawDetailedBalance2}). Furthermore, they can be recast in
a form which does not involve principal parts of integrals and are thus better suited for numerical integration \citep[Eq 5.15]{BrownSawyer}]. We report it in App. \ref{AppFullRCT}. 

\begin{figure}[!htb]
     \includegraphics[width=\mycolumnwidth]{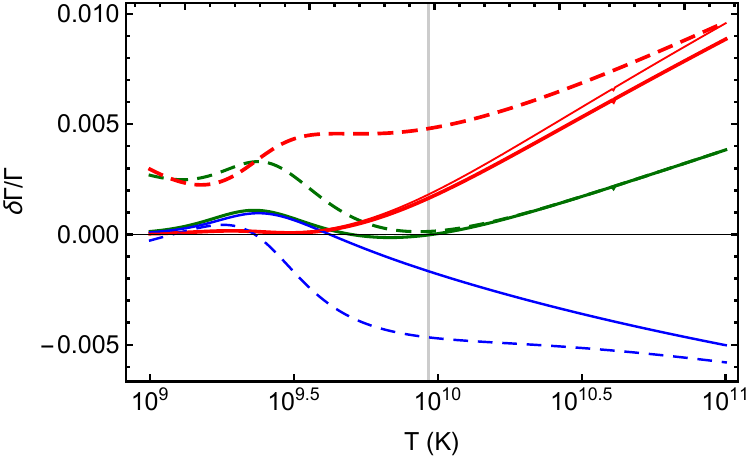}
     \caption{$\delta \Gamma/\Gamma$ from finite-temperature radiative
       corrections. {\it Continuous line} : $n
       \to p$. {\it Dashed   line} :   $p \to n$. The red lines are the corrections of
       \citet{BrownSawyer} in very good agreement with their
       Fig.~2. In the thin continuous red line, the $n+\bar \nu
       +e^+ \to p+\gamma$ has been added to the $n\to p$ corrections,
       as advocated in \citet{BrownSawyer} and we check that it
       results in a very small modification, as also found in their
       Fig.~3, and most notably it is insufficient to satisfy detailed balance. The blue lines are the Bremsstrahlung
       corrections (\ref{BSnpFormal}). Finally the green lines
       are the total corrections, that is the corrections of
       \cite{BrownSawyer} with all bremsstrahlung corrections added. We
       check  that above $T>10^{10} {\rm K}$, that is when $T_\nu=T$,
       the  relative variations of the total corrections are equal, implying that the detailed
       balance is satisfied.}
\label{FigCorrectionCCRTRates}
\end{figure}

The modification of the rates from these finite temperature corrections is plotted
in Fig. \ref{FigCorrectionCCRTRates}. We also show the BS corrections
and the sum of the two. By construction for the sum, it can be checked that
whenever the neutrinos have the same temperature, that is for
$T>10^{10}\,{\rm K}$, the relative variation of the rates are equal
and the detailed balance property (\ref{MagicDetailedBalanceBorn}) is preserved.

\subsection{Finite nucleon mass corrections}\label{SecFM}

It is not fully correct to consider that nucleons have an infinite mass. Indeed, the
typical energy transfer in weak-interactions to electrons and
neutrinos is of the order of the mass gap $\Gap\simeq 1.29 \,{\rm MeV}$,
which is $1.4 \times 10^{-3}$ smaller than the
nucleon mass. It corresponds to a temperature of $1.5 \times 10^{10}
{\rm K}$ which is not much larger than the freeze-out temperature. In the infinite
nucleon mass approximation, we have thus neglected factors of the type
$E_\nu/m_N$, $E_e/m_N$ or $\Gap/m_N$ (where $m_N$ is the average
nucleon mass $m_N\equiv (m_p+m_n)/2$) which represent order $10^{-3}$
corrections with respect to the leading one around $10^{10}{\rm K}$ and
even larger corrections at higher temperature. Our method consists in expanding
the full reaction rate in power of a small parameter $\epsilon \equiv
T/m_N$. Terms of the type $E_\nu/m_N$ or $E_e/m_N$ are obviously of
order $\epsilon$ and terms of the type $\Gap/m_N$ are also treated as
being of order $\epsilon$. Our implementation of the finite mass corrections consists
in including all the terms up to order $\epsilon$, but neglecting
terms of order $\epsilon^2$. This means that we neglect terms whose
importance is of order $10^{-6}$ which is justified given our goal in precision.

\begin{figure}[!htb]
     \includegraphics[width=\mycolumnwidth]{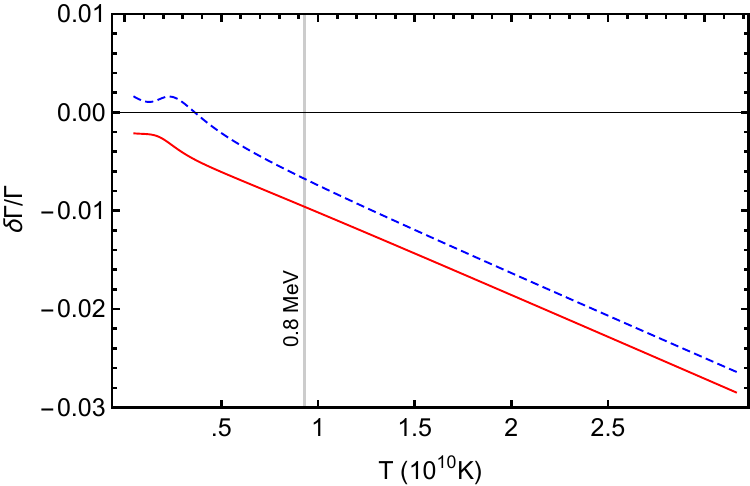}
     \caption{$|\delta \Gamma/\bar \Gamma|$ from finite nucleon mass
       effects. {\it Red continuous line} : $n
       \to p$. {\it Blue dashed   line} :   $p \to n$. This plot is to be
       compared with \citet[Fig. 9]{LopezTurner1998}.}
\label{FigFM}
\end{figure}

\begin{figure}[!htb]
     \includegraphics[width=\mycolumnwidth]{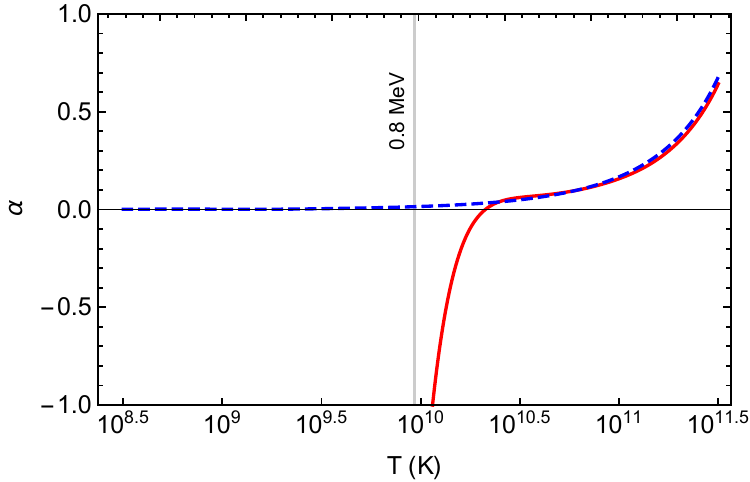}
     \caption{$\alpha$ defined in Eq.~(\ref{DefAlpha}). {\it Red continuous line} : With $T_\nu$ differing
       from $T$ when electron-positron annihilations occur. {\it
         Blue dashed   line} :   With $T_\nu=T$ artificially enforced. This
       plot is to be compared with \citet[Fig. 4]{Lopez1997}. }
\label{FigDetailedBalance}
\end{figure}

If we ignore radiative corrections at null temperature, these corrections take the form
\ifphysrep
\beas\label{GammaFMnoRC}
\delta\Gamma^{\rm FM}_{n \to p} &=& \myk \int_0^\infty p^2 \dd p \,[\chi^{\rm FM}_+(E,g_A) +\chi^{\rm FM}_+(-E,g_A)]\\
\delta\Gamma^{\rm FM}_{p \to n} &=& \myk \int_0^\infty p^2 \dd p \, [\chi^{\rm FM}_-(E,-g_A)  +\chi^{\rm FM}_-(-E,-g_A)]\,,
\eeas
\else
\beas\label{GammaFMnoRC}
\delta\Gamma^{\rm FM}_{n \to p} &=& \myk \int_0^\infty p^2 \dd p\\
&\times&[\chi^{\rm FM}_+(E,g_A) +\chi^{\rm FM}_+(-E,g_A)]\nonumber\\
\delta\Gamma^{\rm FM}_{p \to n} &=& \myk \int_0^\infty p^2 \dd p \\
&\times&[\chi^{\rm FM}_-(E,-g_A)  +\chi^{\rm FM}_-(-E,-g_A)]\,,\nonumber
\eeas
\fi
and the functions $\chi^{\rm FM}_\pm$ are reported in App. \ref{AppFM}.
However since finite mass effects and radiative corrections can be
rather important, they cannot be added linearly and one should also
include radiative corrections inside finite nucleon mass
corrections\footnote{In principle one should rederive all radiative corrections
  including consistently all finite nucleon mass effects. This is
  absent from the literature, and is certainly a daunting task. Our procedure consists in postulating that the multiplicative factor
  introduced by radiative corrections also multiplies the finite
  nucleon mass correction as in \citet{2010PhRvC..81c5503C,Czarnecki2004}}.
Hence the corrections to the rates are
\ifphysrep
\beas\label{GammaFMRC}
\delta\Gamma^{\rm RC+FM}_{n \to p} &=& \myk \int_0^\infty p^2 \dd p [\Fermi_+(E)R(E,|E-\Gap|)\chi^{\rm FM}_+(E,g_A) +\Fermi_+(-E)R(E,E+\Gap)\chi^{\rm
  FM}_+(-E,g_A)] \\
\delta\Gamma^{\rm RC+FM}_{p \to n} &=& \myk \int_0^\infty p^2 \dd p [\Fermi_-(E)R(E,E+\Gap)\chi^{\rm FM}_-(E,-g_A)  +\Fermi_-(-E)R(E,|E-\Gap|)\chi^{\rm FM}_-(-E,-g_A)]\,.
\eeas
\else
\beas\label{GammaFMRC}
\delta\Gamma^{\rm RC+FM}_{n \to p} &=& \myk \int_0^\infty p^2 \dd p\\
&\times& [\Fermi_+(E)R(E,|E-\Gap|)\chi^{\rm FM}_+(E,g_A) \nonumber\\
&&+\Fermi_+(-E)R(E,E+\Gap)\chi^{\rm
  FM}_+(-E,g_A)] \nonumber\\
\delta\Gamma^{\rm RC+FM}_{p \to n} &=& \myk \int_0^\infty p^2 \dd p\\
&\times&[\Fermi_-(E)R(E,E+\Gap)\chi^{\rm FM}_-(E,-g_A)  \nonumber\\
&&+\Fermi_-(-E)R(E,|E-\Gap|)\chi^{\rm FM}_-(-E,-g_A)]\,.\nonumber
\eeas
\fi
The expression for $\chi^{\rm FM}_\pm(E,g_A) $ is given in App. \ref{AppFM}. The relative modifications to the rates are depicted
in Fig.~\ref{FigFM}. It is not obvious that the finite nucleon mass
corrections preserve the detailed balance relation. In fact when
including these corrections the detailed balance ratio between
neutrons and protons is given by
Eq.~(\ref{DetailedBalanceCorrections}). Since this must also be the
ratio $\Gamma_{p \to n}/\Gamma_{n \to p}$ we define $\alpha$ following
\citet{Lopez1997} as
\be\label{DefAlpha}
\frac{\overline{\Gamma}_{p \to n} +\delta\Gamma^{\rm FM}_{p \to n}
}{\overline{\Gamma}_{n \to p} +\delta\Gamma^{\rm FM}_{n \to p} }
  \equiv {\rm  e}^{-\frac{\Gap}{T}} \left[1+(1+\alpha)\frac{3\Gap}{2 m_N}\right]\,.
\ee
$\alpha$ characterizes the deviation from  the detailed balance
equality and must vanish if detailed balance with finite mass
corrections is satisfied. It is plotted in Fig.~\ref{FigDetailedBalance}. We observe that for $T <
m_e$ deviations from detailed balance occur because the neutrino temperature is not equal to the plasma temperature. Hence we have also plotted $\alpha$ with
$T_\nu=T$ artificially enforced. At low temperature, that is below
$T=m_e$, detailed balance is very well satisfied because the parameter
$\epsilon = T/m_N$ is small. At higher temperatures we see deviations,
and this comes from the fact that we considered only corrections which
are first order in $\epsilon$. 

Finally, the total neutron rate at null temperature is
\be\label{GammanpRCFMTnull}
\Gamma_{n\to p}(T=0) = \Gamma_{n \to p}^{{\RC 0}}(T=0)+\delta\Gamma^{\rm FM}_{n \to p}(T=0)\,, 
\ee
that is there is a correction to $\lambda_0$ due to finite
nucleon mass effects. If Eq.~(\ref{GammaFMnoRC}) is used, this is
\be\label{GammaFMTneutrondecay}
\delta\lambda_0^{\rm FM} = m_e^{-5} \int_0^{\sqrt{\Gap^2-m_e^2}} p^2 \dd p \,\left.\chi^{\rm FM}_+(E,g_A)\right|_{T=0}
\ee
combined with Eq. (\ref{chiFMneutrondecay}), and we obtain  $\delta \lambda^{\rm FM}_0 \simeq -3.3828\times 10^{-3}$, that is
$\delta \lambda^{\rm FM}_0/\bar \lambda_0 \simeq -0.002068$. This is in agreement with
the result $\delta \lambda^{\rm FM}_0/\bar \lambda_0 \simeq  -0.00206$ found in \citet{Lopez1997} with an exact method to compute the
finite nucleon mass effects. We recomputed this ratio with the method
of \citet[Eq. 20]{Lopez1997}  and found it is more precisely given
by $-0.0020637$. The tiny difference with our value $\delta \lambda^{\rm
  FM}_0/\bar \lambda_0$ is only due to the fact that we kept only first order finite nucleon mass corrections, that is we kept
terms of order $\epsilon$, and ignored terms of higher order
$\epsilon^2$. Furthermore, if we also include radiative corrections, that is if we use $\delta\Gamma^{\rm RC+FM}_{n \to p} $ instead of
$\delta\Gamma^{\rm FM}_{n \to p} $ in Eq.~(\ref{GammanpRCFMTnull}),
hence corresponding to an extra factor $F(E)R(E,\Gap-E)$ in the integrand of Eq.~(\ref{GammaFMTneutrondecay}), we
get $\delta \lambda^{\rm FM}_0 \simeq -3.6201\times
10^{-3}$. These values are summarized in table \ref{TablelambdazeroFM}.

\subsection{Weak magnetism}\label{SecMainWM}

We show in \ref{AppWM} that the effect of weak magnetism is exactly
similar to the finite mass correction which arises from the coupling between the
axial current and the vector current. This has been noticed earlier by
\citet{Seckel1993}. That is, it brings corrections which are
exactly of the same type as those proportional to $g_A$ in Eq.~(\ref{DefMLLMRRMLR}).
The weak magnetism corrections amount to a simple redefinition of the
constant factors (\ref{DefgLLgRRgLR}) as 
\beas\label{DefgLLgRRgLRwithWM}
c_{LL} &\equiv& \frac{(1+g_A)^2}{4}+ f_{\rm wm} g_A\\
c_{RR} &\equiv& \frac{(1-g_A)^2}{4}- f_{\rm wm} g_A\\
c_{LR} &\equiv &\frac{g_A^2-1}{4}
\eeas
when computing the finite nucleon mass corrections. The
associated couplings (\ref{gtildeLLRRLR}) must be replaced accordingly.

The weak magnetism, if considered independently of radiative corrections, induces no modification of $\lambda_0$, at least up to the
order $\epsilon$ of our finite nucleon mass expansion, as explained in
appendix~\S~\ref{AppWM}. However when this is combined with the radiative
corrections, this brings a residual increase to $\delta \lambda^{\rm FM}_0 \simeq -3.6333\times
10^{-3}$.
Summing this value of the finite nucleon mass corrections (which
includes weak-magnetism and which is coupled to radiative
corrections), to the radiative corrections themselves (\ref{lambdazeroCCR}) leads to
\be\label{lambda0good}
\lambda_0^{{\RC}+{\rm FM}}=\lambda_0^{{\RC}}+ \delta \lambda^{\rm FM}_0 \simeq 1.75474\,.
\ee
This is to be compared with \citet{2010PhRvC..81c5503C} where it is reported
$\lambda_0 \simeq 1.03887 \times 1.6887 \simeq 1.75434$, a modest
$0.023\%$ smaller than the value (\ref{lambda0good}). This close agreement can also be seen by noting that
  $(\hbar 2 \pi^3)/(m_e^5 c^{10} G_F^2 \lambda_0^{\rm RC+FM} ) \simeq
  4907.4 \,{\rm s}$, which is very close to the value $4908\,{\rm s}$ given by
  \citet[Eq. 17]{Czarnecki2004} or the refined value $4908.7 {\rm s}$
  of \citet[Eq. 18]{Marciano:2005ec}.
%
%Note that if we use the value (\ref{lambda0good}) and the values for
%$g_A$ and $\theta_C$ given in appendix~\ref{ParticleValues}, then we find theoretically
%\be
%m_e^5 \myk \simeq 6.45390\times 10^{-4}{\rm s}^{-1}\quad \Rightarrow\quad \tau_n \simeq
%880.1\,{\rm s}\,,
%\ee
%which is very close to the most recent measured value of the neutron
%decay rate $\tau_n \simeq 880.2\pm 1.1\,{\rm s}$.

\begin{table}
\centering
\caption{Value of $\delta \lambda_0^{\rm FM}$ depending on the effects considered. \label{TablelambdazeroFM}}
\begin{tabularx}{\mycolumnwidth}{cAAA}\toprule
 RC & FM & WM & FM+WM\\\midrule
No &$ -3.3828\times 10^{-3}$ & $0$ & $-3.3828\times 10^{-3}$\\
Yes &  $-3.6201\times 10^{-3}$& $-0.0132 \times 10^{-3}$ & $-3.6333\times 10^{-3}$ \\\bottomrule
\end{tabularx}
\end{table}

\subsection{Effect of incomplete neutrino decoupling}\label{SecWeakRatesID}

Neutrino heating also induces modification of the weak rates. In our
thermal approximation of the incomplete neutrino decoupling
(see~\S~\ref{SecIncompleteDecoupling}), this is taken into account by
putting the effective neutrino temperature defined in
Eq.~(\ref{DefTnueff}) in all expressions for the weak rates. This
amounts to assuming that all neutrinos receive the same share of the
heating and ignoring the spectral distortions. We postpone a more detailed discussion
on the effect of incomplete neutrino decoupling on the weak rates and thus the ${}^4{\rm He}$
production in \S~\ref{SecEffectID}. Briefly, the most notable effect of
incomplete neutrino decoupling is to affect the time-temperature
relation, and this is also the case from QED corrections in the plasma
but the effect is much smaller. Indeed, since for a given plasma temperature, neutrino heating induces an increase of the total energy density, the Hubble rate is
increased and so is the rate of variation $\dd T/\dd t$. In
practice this means that the Universe is younger when nucleosynthesis
starts around $T_{\rm Nuc} = 0.078 {\rm MeV}$ and the neutron beta
decay results in a lower loss of neutrons, and thus a higher production
of ${}^4 {}\rm He$ at the end of the BBN. This is the {\it clock effect}
explained in \citet{Fields:1992zb,1992PhRvD..46.3372D}.

\subsection{Total correction to the weak rates}

The total weak reaction rates are given by summing the various
effects [Eqs.~(\ref{CCRn}), (\ref{DefGammaT}), (\ref{BSnpFormal}) and (\ref{GammaFMRC})], that is
\be\label{SumAllCorrections}
\Gamma_{n \to p} = \Gamma^{\rm RC0}_{n \to p}+\Gamma^{T}_{n \to
  p}+\delta\Gamma^{\rm BS}_{n \to p}+\delta\Gamma^{\rm FM}_{n \to p}\,,
\ee
and a similar sum for $p \to n$ processes. The constant $\myk$
involved in all contributions is then calibrated on neutron beta decay
from $\lambda_0$ given by Eq.~(\ref{lambda0good}) replaced in
Eq.~(\ref{Getmyk}). We recall that when including weak-magnetism in finite nucleon mass
effects, that is in Eqs.~(\ref{GammaFMRC}), one must use the constants~(\ref{DefgLLgRRgLRwithWM}) in
(\ref{gtildeLLRRLR}), so as to evaluate the $\chi^{\rm FM}_\pm$ given by Eqs.~(\ref{chiFMFull}).

\begin{figure}[h]
\centering
       \includegraphics[width=\mycolumnwidth]{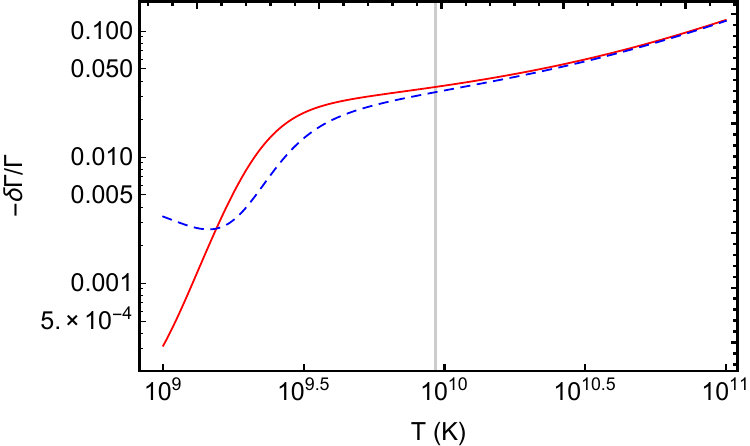}
\caption{Total relative rate corrections, including all the effects discussed, $-\delta
  \Gamma/\overline{\Gamma}$. $n \to p$ in red continuous
  line and $p \to n$ in blue dashed line. The vertical line corresponds to
  $T_F=0.8{\rm MeV}$.}
\label{FigRates2}
\end{figure}

The size of all corrections relative to the Born approximation is plotted in Fig.~\ref{FigRates2}. Around
$T\simeq 3.3 \times 10^{9}{\rm K}$ for which the neutron are only subject to beta decay, we find $\delta \Gamma_{n \to p}/\bar \Gamma_{n
\to p}
\simeq -0.023$. From the rule (\ref{RuleOfThumb}) this should lead to $\Delta \YPfour \simeq 42$, quite in nice agreement with the total correction $\Delta \YPfour \simeq 44.7$ that we report in \S~\ref{SecResults}.

\section{Nucleosynthesis}\label{SecNucleo}

Once the weak interaction rates are determined and computed with great
precision, it is possible to solve for the nucleosynthesis by adding
the effect of nuclear reactions which form nuclei. Hence we need to
build the differential system which rules the evolution of all
(neutrons, protons and main isotopes) abundances. We describe how this
is performed in this section and turn to report and discuss the
numerical results obtained for final abundances in \S~\ref{SecNumericalResults}.

We remind that, for a general reaction of the type
\be
A+b \rightarrow C+d
\ee
the conventional notation in experimental nuclear physics is $A(b,d)C$ to keep in mind that
$A$ is the target nucleus at rest, $b$ is the projectile from the beam, $d$ is the outgoing,
detected, particle and $C$ is left generally undetected, possibly not escaping the target.
Hence, usually, but not always, $A$ and $C$ are the heaviest nuclei. The same notation
is used in theoretical nuclear physics, regardless of the experimental details.

\subsection{Thermonuclear reaction rates}
\label{s:rates}

We summarize here a few results, to be used in this review, and
refer to \cite{chbook,NACRE,Eval1,Clayton} for a detailed treatment.
It is assumed that the medium is in local thermodynamical equilibrium so that
the distribution of ion velocities/energies follows a
Maxwell--Boltzmann distribution (see \S~\ref{SecMaxwellian}),
\begin{equation}
\phi_\mathrm{MB}(v)v \dd v=
\sqrt{8\over{\pi m}}{{1}\over{(k_B T)^{3/2}}}e^{-\tfrac{E}{k_B T}}E \dd E.
\label{q:mb}
\end{equation}
It is understood that for the distribution of relative velocities between
two reacting nuclei, $m$ is their reduced mass. 
In such conditions, one defines the thermonuclear reaction rate by
\begin{equation}
N_A\langle \sigma v \rangle  =
N_A\int_0^\infty\sigma(v)\phi_\mathrm{MB}(v)vdv
\label{q:taux}
\end{equation}
in cm$^3$s$^{-1}$mole$^{-1}$ units where  $N_A$ is Avogadro's number 
(mole$^{-1}$). 
For reactions involving charged particles, since the kinetic energies are below the Coulomb barrier, the 
energy dependence of the cross section is dominated by the tuneling effect
through the barrier. The  Coulomb plus centrifugal barrier penetration probability is given by 
\begin{equation}
P_\ell(E)={kR\over{F^2_\ell(\eta,kR})+G^2_\ell(\eta,kR)}
\label{q:pen}
\end{equation}
where $F_\ell$ and $G_\ell$ are the Coulomb functions \citep{Fro55}, $k= \sqrt{2 m E}/\hbar$ is the wave number,
$\ell$ the orbital angular momentum and 
\begin{equation}
\eta\equiv{{\textstyle Z_{1}Z_{2}e^2}\over{\textstyle \hbar v}}
\end{equation}
the Sommerfeld parameter. 
To account for this strong energy dependency of the cross section, it is convenient to introduce
the astrophysical \sfac:
\begin{equation}
\sigma(E)\equiv {\textstyle{S(E)}\over{E}}
\exp\left(-2\pi\eta\right)
%\eqno(2-4)
\label{q:se}
\end{equation}
which, in the absence of resonances removes most of the energy dependence. 
A resonance, associated to a nuclear level in the compound nucleus formed by the fusion 
of the projectile and target nuclei, induces a strong but localized variation of the \sfac. 
Hence, the presence of a resonance can increase by several orders of magnitude a reaction rate.

\subsection{General form}

The evolution of abundances, defined in Eq.~(\ref{DefYX}), is deduced from the evolution of
number densities found from Eq.~(\ref{Eqdotn}). Since both nuclear and
weak reactions preserve the number of baryons, we obtain
\begin{equation} 
\frac{\dd n_{\rm b}}{\dd t}+3 H n_{\rm b}=0,
\end{equation} 
that is baryon volume density is only affected by dilution and $n_{\rm
b}
\propto 1/a^3$. 

For a given isotope $i$, the evolution of the volume density depends
on the reaction rates in which it is involved. It is of the form
\be
\frac{\dd n_i}{\dd t}+3 H n_i= {\cal J}_i\,,
\ee
where ${\cal J}_i$ is the net rate of evolution of number density
due to all nuclear reactions. A decay of species $i$ ($i\to \dots$)
and a decay reaction in which species $i$ is the end product ($j
\to i + \dots$)  contribute as
\be\label{Decayreaction}
{\cal J}_i \supset -n_i \Gamma_{i\ \to \dots} + n_j \Gamma_{j \to i +\dots}
\ee
where the $\Gamma$ are the decay rates (usually given in ${\rm s}^{-1}$). 
A two-body reaction of the type $i+j \leftrightarrow k+l$ contributes instead as
\be\label{Ni2body}
{\cal J}_i \supset n_k n_l \gamma_{kl \to ij}  - n_i n_j \gamma_{ij \to kl} \,.
\ee
with
\be
\gamma_{ij \to kl} \equiv \langle \sigma v\rangle_{ij \to kl}\,.
\ee

Since both the individual $n_i$ and the total $n_{\rm b}$ are affected
similarly by expansion, it proves much simpler to study directly the
evolution of abundances defined as~\footnote{For neutrons and protons
  $Y_n=X_n$ and $Y_p=X_p$.}
\be
Y_i \equiv \frac{n_i}{n_{\rm b}}\,,
\ee
since one finds
\be
\dot Y_i = C[Y_i]\,,\qquad C[Y_i] \equiv \frac{{\cal
    J}_i}{n_{\rm b}}\,.
\ee 
Obviously the decay reactions considered in Eq.~(\ref{Decayreaction}) contribute as
\be
C[Y_i]  \supset -Y_i \Gamma_{i\ \to \dots} + Y_j \Gamma_{j \to i +\dots}\,.
\ee
However two-body reactions contribute as
\be\label{CYi}
C[Y_i]  \supset Y_k Y_l \Gamma_{kl \to ij} - Y_i Y_j \Gamma_{ij \to kl} \,,
\ee
where we related the rates for abundances to those of number densities through
\be\label{Gammatogamma2}
\Gamma_{ij \to kl} \equiv n_{\rm b} \gamma_{ij \to kl} \,.
\ee

Eq.~(\ref{Ni2body}) is straightforwardly generalized to reactions with
more bodies. In full generality, and without restricting to decay reactions or two-body
reactions, the evolution of abundances takes the form~\citep{Fow67,Wag69}
\begin{widetext}
\bea\label{GeneralYidot}
\dot Y_{i_1} = \sum_{i_2\dots i_p,j_1\dots j_q}
N_{i_1}\left(\Gamma_{j_1\dots j_q\to
    i_1\dots i_p}\frac{Y_{j_1}^{N_{j_1}}\dots
    Y_{j_q}^{N_{j_q}}}{N_{j_1}! \dots N_{j_q}!} -\Gamma_{i_1 \dots
    i_p\to j_1 \dots j_q}\frac{Y_{i_1}^{N_{i_1}}\dots Y_{i_p}^{N_{i_p}}}{N_{i_1}!\dots N_{i_p}!}\right)\,,
\eea
\end{widetext}
where $N_i$ is the stoichiometric coefficient of species $i$ in the
reaction and with the relation between abundance rates and number
density rates given by
\be\label{GammaTogamma}
\Gamma_{i_1 \dots  i_p\to j_1 \dots j_q} = n_{\rm b}^{(N_{i_1}+\dots +N_{i_p})-1}\gamma_{i_1 \dots  i_p\to j_1 \dots j_q} \,.
\ee
Note that for a decay reaction $\Gamma_{i \to \dots}=\gamma_{i \to \dots}$.

In practice the reaction rates for many-body reactions are given as
tables for the quantities $N_A^{(N_{i_1}+\dots +N_{i_p})-1} \gamma_{i_1 \dots  i_p\to j_1 \dots
  j_q}$, as detailed in App.~\ref{AppReactionConventions}
on the steps required to deduce them from the nuclear
physics tables. Furthermore the rates are only given for forward reactions. Indeed, since
when there is nuclear statistical equilibrium (NSE) reverse reactions
should balance with forward reactions, we can always relate the reverse
reactions to the forward reactions from the detailed balance relation
\be\label{ReverseNSE1}
\frac{\gamma_{i_1 \dots i_p\to j_1 \dots j_q}}{ \gamma_{j_1\dots j_q\to
  i_1\dots i_p}} = \frac{N_{i_1}!\dots N_{i_p}!}{N_{j_1}! \dots N_{j_q}!} \left[\frac{n_{j_1}^{N_{j_1}}\dots
  n_{j_q}^{N_{j_q}}}{n_{i_1}^{N_{i_1}}\dots
  n_{i_p}^{N_{i_p}}}\right]^{\rm NSE}   
\ee
where the NSE densities are given in Eq.~(\ref{niThermal}).
Since nuclear reactions in many-body reactions do not change the
nature of nucleons\footnote{This is not the case for decay reactions
  but for these we neglect the reverse reactions.}, that is they conserve the number of protons and neutrons, then using the
relation~(\ref{GammaTogamma}) the reverse reactions are
related to the forward reactions by 
\ifphysrep
\be\label{ReverseNSE2}
\frac{\gamma_{j_1\dots j_q\to
  i_1\dots i_p}}{\gamma_{i_1 \dots i_p\to j_1 \dots j_q}}= \frac{\Pi_{i=i_1 \dots i_p}
\frac{1}{N_i!}\left[g_i\left(\frac{m_i
      T}{2\pi}\right)^{3/2}\right]^{N_i}}{\Pi_{j=j_1 \dots j_q}
\frac{1}{N_j!}\left[g_j\left(\frac{m_j
      T}{2\pi}\right)^{3/2}\right]^{N_j}}\,{\rm exp}\left(\frac{\sum_{j=1}^q m_j-\sum_{i=1}^p m_i}{T}\right).
\ee
\else
\bea\label{ReverseNSE2}
\frac{\gamma_{j_1\dots j_q\to
  i_1\dots i_p}}{\gamma_{i_1 \dots i_p\to j_1 \dots j_q}}&=& \frac{\Pi_{i=i_1 \dots i_p}
\frac{1}{N_i!}\left[g_i\left(\frac{m_i
      T}{2\pi}\right)^{3/2}\right]^{N_i}}{\Pi_{j=j_1 \dots j_q}
\frac{1}{N_j!}\left[g_j\left(\frac{m_j
      T}{2\pi}\right)^{3/2}\right]^{N_j}}\nonumber\\
&\times &{\rm exp}\left(\frac{\sum_{j=1}^q m_j-\sum_{i=1}^p m_i}{T}\right).
\eea
\fi
We check easily that we recover the particular cases of~\citet[Eqs. 15, 18, 25]{Fow67}.
Hence the relation between forward and reverse reactions is of the form
\be
\frac{ \gamma_{j_1\dots j_q\to
  i_1\dots i_p}}{\gamma_{i_1 \dots i_p\to j_1 \dots j_q}} = \alpha
\left(\frac{T}{10^9{\rm K}}\right)^\beta {\rm exp}\left(\frac{\gamma
    \times 10^9{\rm K}}{T}\right)\,,
\ee
which depends only on 3 constants. In practice, for a given reaction, the forward reaction rate is
tabulated for various values of $T$ or is approximated by an analytic
fit, and the constants $(\alpha,\beta,\gamma)$ needed to obtain the reverse reaction
rate are also computed once for all from Eq.~(\ref{ReverseNSE2}) using the tabulated
masses and spins of isotopes [we use the table {\tt nubase2016.asc} described in \citet{Aud17}].

\subsection{Nuclear network and reaction rates uncertainties}
\label{s:nucl}

The nuclear reaction network used here has been fully described in \citet{Coc12a}: 
it includes 59 nuclides from neutron to $^{23}$Na (see table \ref{TableNuclides}), 
linked by 391 reactions involving neutrons, protons, deuterium,
tritium (${}^3{\rm H}$), \tro\ and 
$\alpha$--particles induced reactions, and 33 $\beta$-decay processes. 
The complete list of reactions  can be found in Table~4 of  \citet{Coc12a}, together with the references 
for the values of the reaction rates. Each of these reactions is systematically supplemented ($ij{\to}kl$) 
by its reverse ($kl{\to}ij$) whose rate is obtained as described
above, except for decay reactions whose rates are from \citet{Aud17}.

\begin{figure}[htb!]
\begin{center}
\includegraphics[width=\mycolumnwidth]{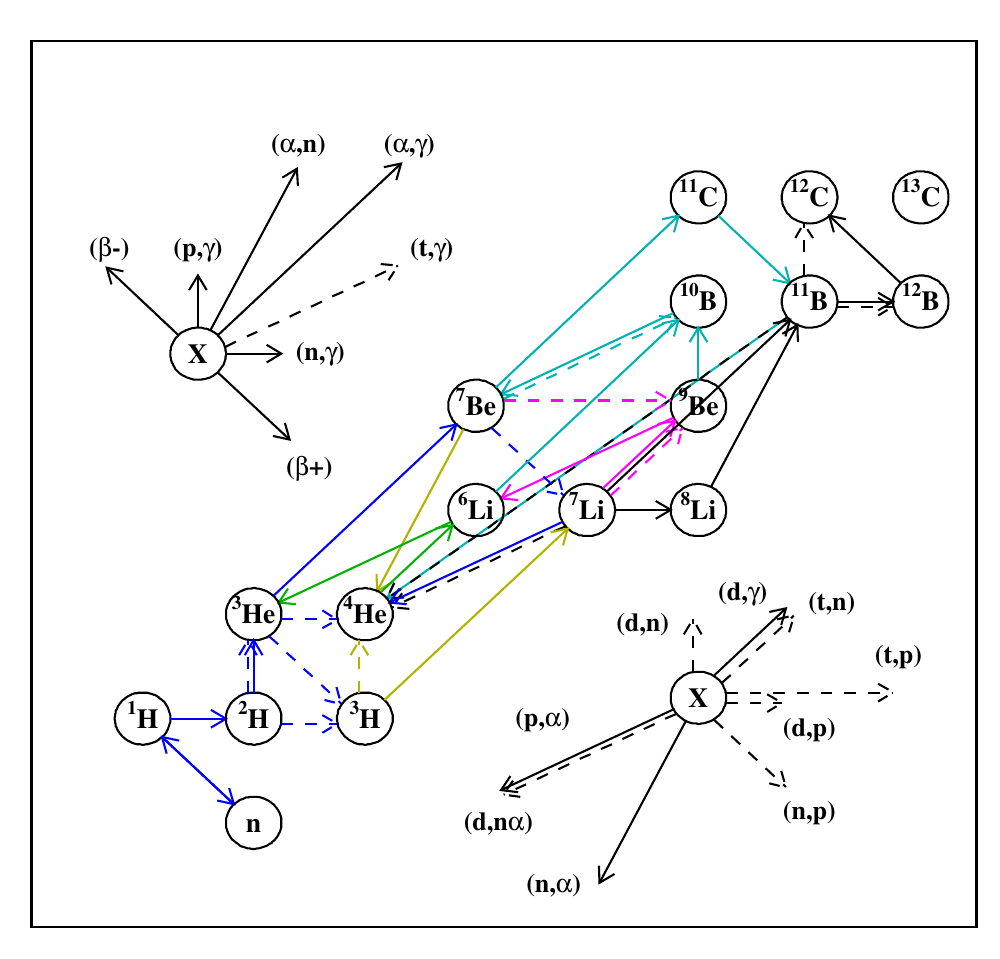}
\caption{Nuclear network  of the most important reactions in BBN (out of the 424) up to\sep\ (blue), 
including \six\ (green), $^{10,11}$B (light blue), \neu\ (pink) and up to CNO (black).
The yellow arrows indicate the reactions that are now considered as unimportant.\label{f:netw}}
\end{center}
\end{figure}

This network is adapted to the prediction of the primordial abundances of the light elements, but also to the calculation 
of the abundances of the \six, \neu, \dix, \onz\ and CNO isotopes. 
As listed in Table~4 of  \citet{Coc12a}, 
reaction rates and their associated uncertainties were taken primarily from \citet{NACRE,Des04,Eval2,NACRE2} evaluations 
when available. For many reactions, in the absence of sufficient experimental data, the rates come from theory (TALYS code) 
\cite{TALYS}. An extensive sensitivity study, performed by \citet{Coc12a}, identified ten reactions that needed
further analyses, and it was followed by their re--evaluations (shown in boldface in  Table~4 of  \citet{Coc12a}). 
Finally, out of these hundreds of reactions, the most important ones are displayed in Fig.~\ref{f:netw}. They were identified in sensitivity studies, i.e. 
 \cite{Nol00,Cyb04,CV10} for the light elements (\hli) and \citet{Coc12a,Coc14} for others and correspond to the main nuclear flow.

\begin{table}[!htb]
%\centering
\begin{center}
\caption{Nuclides considered in the nuclear network.\label{TableNuclides}}
\end{center}
\ifphysrep
\else
\resizebox{\columnwidth}{!}{
\fi
\begin{tabular}{|c|c|c|c|c|c|c|c|c|c|c|c|c|c|c|}
 \hline\hline
\backslashbox{$Z$}{$N$} & 0 & 1 & 2 & 3 & 4 & 5 & 6 & 7 & 8 &9&10&11&12&13\\
\hline 0 & &  n & & & & & & & &&&&&\\
\hline
1 & $\,\,{\rm H}\,\,$ & $^2$H & $^3$H & & & & & & &&&&&\\
\hline
2 & & $^3$He & $^4$He & $^5$He& $^6$He& & & & &&&&&\\
\hline
3 & & & &$^6$Li & $^7$Li& $^8$Li&$^9$Li & & &&&&&\\
\hline
4 & & & &$^7$Be &  $^8$Be& $^9$Be& $^{10}$Be &$^{11}$Be & $^{12}$Be&&&&&\\
\hline
5 & & & & $^8$B & $^{9}$B & $^{10}$B& $^{11}$B& $^{12}$B & $^{13}$B & $^{14}$B& $^{15}$B&&&\\
\hline
6 & & & &$^{9}$C &$^{10}$C &$^{11}$C&$^{12}$C &$^{13}$C&$^{14}$C&$^{15}$C&$^{16}$C&&&\\
\hline
7 & & & & & & $^{12}$N& $^{13}$N &$^{14}$N& $^{15}$N&$^{16}$N&$^{17}$N&&&\\
\hline
8 & & & & & &$^{13}$O & $^{14}$O & $^{15}$O&$^{16}$O&$^{17}$O&$^{18}$O&$^{19}$O&$^{20}$O&\\
\hline
9 & & & & & & & & &$^{17}$F&$^{18}$F&$^{19}$F&$^{20}$F&&\\\hline
10 & & & & & & & &&$^{18}$Ne &$^{19}$Ne&$^{20}$Ne&$^{21}$Ne&$^{22}$Ne&$^{23}$Ne\\\hline
11 & & & & & & & &&&$^{20}$Na&$^{21}$Na&$^{22}$Na&$^{23}$Na&\\
 \hline\hline
\end{tabular}
\ifphysrep
\else
}
\fi
\end{table}
%%%%%%%%%%%%%%%%%%%%%%%%%%%%%%%%%%%%%%%%%%%%%%%%%%%%%%%%%%%%%%%%%%%%%%%

%%%%%%%%%%%%%%%%%%%%%%%%%%%%%%%%%%%%%%%%%%%%%%%%%%%%%%%%%%%%%%%%%%%%%%%

Experimental uncertainties on $\YP$ are due to the \npg, \ddn\ and \ddp\ reaction rates and to the neutron lifetime.
For these three reactions, one finds \citep[e.g.][]{Cyb04,CHK}
%%%%%%%%%%%%%%%%%%%%%%%%%%%%%%%%%%%%%%%%%%%%%%%%%%%%%%%%%%%%%%%%%%%%%%%
\be\label{q:dsigv}
\Delta{\YP}\approx(1.5\times10^{-3})\frac{\Delta\langle\sigma{v}\rangle}
{\langle\sigma{v}\rangle}\,,
\ee
%%%%%%%%%%%%%%%%%%%%%%%%%%%%%%%%%%%%%%%%%%%%%%%%%%%%%%%%%%%%%%%%%%%%%%%
and $\Delta{\YP}\approx0.18\Delta\tau_n/\tau_n$ from Eq.~(\ref{RuleOfThumb}). Since the experimental uncertainties  
are of the order of $10^{-2}$ for the rates and  $10^{-3}$ for the lifetime, they may all contribute to the error budget. 

There are basically two methods for the experimental determination of the neutron lifetime: detecting dying neutrons
i.e. their decay rate (``beam experiments") or surviving neutrons after being left in a magnetic ``bottle" for a certain time.
Both methods produce slightly different results \cite[see e.g. Fig.~8 in][]{You14}. This is most probably due to
different systematic uncertainties but it might also be real and be explained by an undetected decay into a dark 
sector \cite[see e.g.][]{For18}. Note that in that case it would be the ``bottle" results, corresponding to the surviving neutrons
counting, that would be important for BBN.   
Hence, the value of the neutron lifetime has been revised several
times by the Particle Data Group (PDG) from  
885.7$\pm$0.8~s~\citep{PDG08}, used, e.g.  in \citet{CV10}, to
879.4$\pm$0.6~s resulting from an average of ``bottle" experiments \citep{Cza18}, slightly below the current PDG
average~\citep{PDG17} which is $880.2\pm1.0\,{\rm s}$.  We use the
average over post-2000 experiments, $879.5\pm0.8\,{\rm s}$
\citep{Ser17}, which is extremely close to the average on bottle
experiments, but with more conservative errors.

The cross section of the \npg\ reaction is obtained from calculations in the framework of Effective Field Theory 
whose results are estimated to be reliable to within 1\% error \cite{AndoEtAl2006}. 
Indeed,  the few experimental information available for this cross section at BBN energies
are in very good agreement with theory (see Fig.~1 in \citet{Zakopane}).  

The  \dpg, \ddn\ and \ddp\ reactions are the main source of nuclear uncertainty for deuterium nucleosynthesis 
while the two last one may affect the error budget of $\YP$.
 The relative variations of D/H are related to the variation of these
rates \cite[see e.g.][]{CV10} by
%%%%%%%%%%%%%%%%%%%%%%%%%%%%%%%%%%%%%%%%%%%%%%%%%%%%%%%%%%%%%%%%%%%%%%%
\bea
\frac{\Delta{\rm (D/H)}}{\rm D/H}&=&-0.32\frac{\Delta\langle\sigma{v}\rangle_{\mathrm{d(p,}\gamma)^3\mathrm{He}}}
{\langle\sigma{v}\rangle_{\mathrm{d(p,}\gamma)^3\mathrm{He}}}\nonumber\\
\frac{\Delta{\rm (D/H)}}{\rm D/H}&=&-0.54\frac{\Delta\langle\sigma{v}\rangle_{\mathrm{d(d,n)}^3\mathrm{He}}}{\langle\sigma{v}\rangle_{\mathrm{d(d,n)}^3\mathrm{He}}}
                                                   -0.46\frac{\Delta\langle\sigma{v}\rangle_{\mathrm{d(d,p)}^3\mathrm{H}}}{\langle\sigma{v}\rangle_{\mathrm{d(d,p)}^3\mathrm{H}}}\nonumber
\eea
%%%%%%%%%%%%%%%%%%%%%%%%%%%%%%%%%%%%%%%%%%%%%%%%%%%%%%%%%%%%%%%%%%%%%%%
so that to achieve the $\sim$1\% precision required by observations, one needs a similar precision on reaction rates.  
None of them are affected by resonances, so that the only questions are to model the slowly varying energy dependence
of the \sfac{s} and precisely determine their absolute scale. There are basically two options: either empirically
fit both the energy dependence and scale so as to follow closely the data, or use theoretical energy dependences
from nuclear physics models and only determine the absolute normalization. We adopted the rates from the
new evaluations of \citet{Bayes16} and \citet{Bayes17} that use both the second option, together with Bayesian methods.    
The theoretical, {\it ab initio} energy dependences were taken from \citet{Mar05} for  \dpg\ and from \citet{Ara11} for \ddn\ and \ddp.
The main difficulty to determine the absolute scale of the \sfac{s} is that  one needs to combine results from 
different experiments. 

\begin{figure}[tbh]
\begin{center}
\includegraphics[width=\mycolumnwidth]{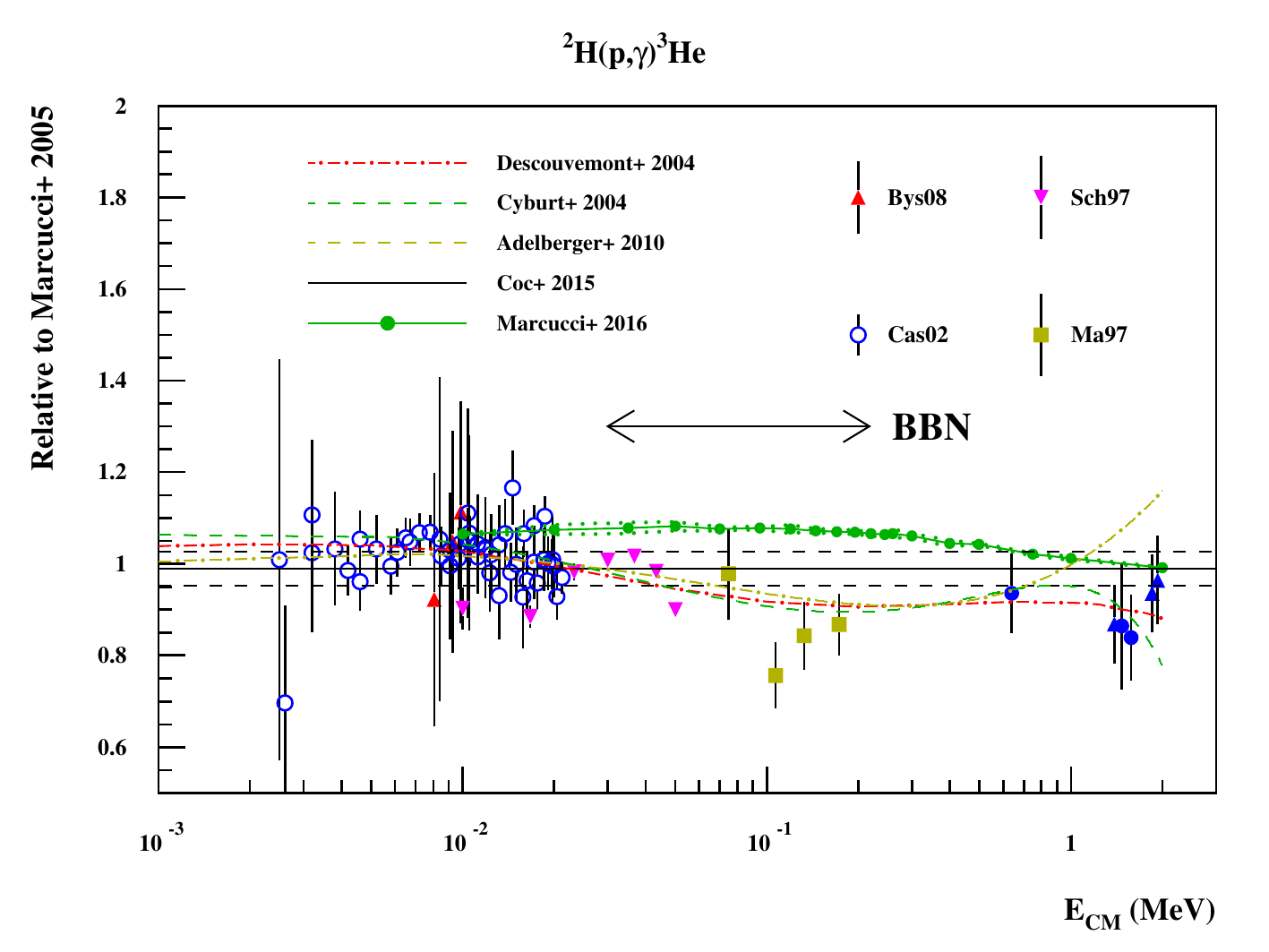}
\caption{Ratio of experimental \cite{Bys08b,Cas02,Sch97,Ma97}, fitted \cite{Des04,Cyb04,Ade11,Coc15} and new theoretical \cite{Mar16}
$S$--factors to the theoretical one \cite{Mar05}; 
the horizontal lines correspond to the theoretical $S$--factor scaled according to \citet{Coc15}. (Systematic uncertainties, i.e. global
normalization errors,
in the range 4.5--9\% are shown in keys).}
\label{f:dpg}
\end{center}
\end{figure}

\citet{Coc15}, using traditional statistics found a normalization factor of 0.9900$\pm$0.0368 for the \dpg\ theoretical \sfac\ of \citet{Mar05},
while the Bayesian analysis gives 1.000$^{+0.038}_{-0.036}$ \cite{Bayes16}. This shows that, starting from the 
same experimental data and theoretical model, different statistical analyses can lead to a, significant,  1\% difference.
Figure~\ref{f:dpg} displays the \dpg\ experimental \sfac\ normalised to the theoretical model of \citet{Mar05}. 
The solid horizontal line corresponds to the scaling of the theoretical \sfac\ adopted by \citet{Coc15}. It is obvious that 
experimental data are scarce at BBN energies and slightly below the scaled \sfac\ 
(an overall 9\% systematic uncertainty is not shown however), while the empirical fit by \citet{Ade11} or \cite{Des04}
follows closely, by construction, the experimental data. Note also that \citet{Mar16} have included higher order terms
in their {\it ab initio} model resulting in a $\approx$10\% increase with respect with their previous result \cite{Mar05},
this time well above the experimental data (see Fig.~\ref{f:dpg}). Using this new theoretical \sfac, one would obtain an 
additional reduction of  $\Delta$(D/H) = -0.072$\times10^{-5}$ that nevertheless would vanish 
if we rescale it (by 0.915) to fit experimental data. This rate is thus a major source of uncertainty for D/H prediction
that should be resolved when the new experimental data \cite{Gus17}  from LUNA at the Gran Sasso underground facility will be released,
supplementing the low energy ones \cite{Cas02}. 

In a similar way, the \ddn\ and \ddp\ rates have been evaluated 
by \cite{Coc15} and later by \cite{Bayes17}, using the  {\it ab initio} \sfac\ from \citet{Ara11} scaled according to
experimental data. They found negligible differences in scaling factors: 0.959$\pm$0.010 and 0.955$\pm$0.010 for 
the traditional analysis to be compared with  0.961$\pm$0.010 and 0.956$\pm$0.010 for the Bayesian one.
However, the theoretical work of  \citet{Ara11} was focused on low energies and does not correctly reproduce the \ddn\ and \ddp\  
experimental data above $\approx$600~keV. It is highly desirable that these calculations be extended up to $\approx$2~MeV,
to cover the range of experimental data that encompass BBN energies.

%%%%%%%%%%%%%%%%%%%%%%%%%%%%%%%%%%%%%%%%%%%%%%%%%%%%%%%%%%%%%%%%%%%%%%%
%%%%%%%%%%%%%%%%%%%%%%%%%%%%%%%%%%%%%%%%%%%%%%%%%%%%%%%%%%%%%%%%%%%%%%%

At the CMB deduced density, \sep\ is produced through the formation of $^7$Be via the $^3$He($\alpha,\gamma)^7$Be reaction as $^7$Be
will decay much later to \sep. The destruction of $^7$Be occurs through the $^7$Be(n,p)$^7$Li(p,$\alpha)^4$He channel which is limited by
the scarcity of late time neutron abundance.  
The most influential reaction rates on \sep\ nucleosynthesis are  \citep[e.g. Table~1 in ][]{CV10} \npg\ (indirectly by affecting the neutron abundance) and \hag, 
but large deviations from their nominal cross sections are strongly constrained by experiments.  
Even though, there has not been new experimental data, since it is the major source of uncertainty on the \sep\ 
production, the \hag\ reaction rate has also been recently re-evaluated using Bayesian methods \cite{Bayes17} to
scale the theoretical \sfac\ of \citet{Nef11}.
It was known that the $^7$Be(n,$\alpha)^4$He reaction could not help solve the lithium problem, but its rate was highly uncertain
and affected the \sep\ production at the few percent level.
Until recently, the only published rate came from an evaluation by  \citet{Wag69} based on very scarce data.
We used either this rate or the one obtained by TALYS  \cite{TALYS}  in previous publications \cite{Coc12a,Coc14}.
A new re-evaluation \cite{Hou15} and experiments \cite{Bar16,Kaw17} confirmed that the  $^7$Be(n,$\alpha)^4$He
rate is approximately, one order of magnitude below the Wagoner one, rendering negligible the effect this reaction. 
Hence, we now use the rate provided by the n\_TOF collaboration \cite{Bar16} that now has no impact on $^7$Be.

Lithium--6 nucleosynthesis is quite simple given that it is only produced by the \zdag\ reaction and destroyed
by  $^6$Li(p,$\alpha)^4$He. While the rate of the latter has been precisely known for a long time, the rate of
the former suffered from large uncertainties \citep{NACRE}. This has now been solved, thanks to experiments 
\citep{Ham10,And14}, and theory \cite{Muk16}.

Elements with atomic number above 7 are not expected to be  significantly produced in BBN, unless some
of the reaction rates involved in their production differ strongly from their current estimates \cite{Ioc07,Coc14}. Indeed, some of 
them rely on theoretical models not well adapted to low mass nuclei.   
Figure \ref{f:netw} displays the main reactions producing or destructing the beryllium, boron and C, N and O stable
isotopes. 

Table~\ref{t:rates} displays the few reaction rates that have been updated in \citet{Coc14,Coc15}, and now in this work,
with respect to the Table~4 of \citet{Coc12a}.

\begin{table}[h]
\caption{Updated reaction rates with respect to \citet{Coc12a}} 
\label{t:rates}
\ifphysrep
\else
\resizebox{\columnwidth}{!}{
\fi
\begin{tabular}{lll} 
\toprule
Reaction & Previous  & Present \\
\midrule
 \dpg & \citet{Des04}$^{a,b}$ &  \citet{Bayes16} \\
  & \citet{Coc15}$^c$ &  \\
\midrule
 \ddp\ and & \citet{Des04}$^{a,b}$ &  \citet{Bayes17} \\
 \ddn & \citet{Coc15}$^c$ &  \\
\midrule
\hag\ & \citet{Cyb08a}$^{a,b}$ & \citet{Bayes16} \\
  & \citet{deB14}$^c$ &  \\
\midrule
$^8$Li(p,n)2$\alpha$ & \citet{Bec92}$^a$ & \citet{Men12}$^c$ \\
\midrule
 $^7$Be(n,$\alpha)^4$He & \citet{TALYS}$^a$ &  \citet{Bar16} \\
  & \citet{Wag69}$^b$ &  \\
  & \citet{Hou15}$^c$ &  \\
\midrule
 $^{14}$C($\alpha,\gamma)^{18}$O and & Error in tabulated  rates$^{a,b,c}$ &  \citet{TALYS} \\
$^{14}$C(p,$\gamma)^{15}$N &  &  \\
\bottomrule
\end{tabular}
\ifphysrep\else
}
\fi
\\
Used in: $^a$\citet{Coc12a}, $^b$\citet{Coc14}, or $^c$\citet{Coc15}
\end{table}

\section{Numerical results}\label{SecNumericalResults}

\subsection{Overview of {\tt PRIMAT}}

To our knowledge, apart from the Kawano code \citep{Kawano:1992ua}
which is based on the historical code of Wagoner~\citep{Wagoner1967,Wag69,Wagoner1973}, there exist only two
public BBN codes, which are {\tt PArthENoPE}~\cite{Parthenope,Parthenope2} and
{\tt AlterBBN}~\cite{AlterBBN}. The method followed in {\tt PRIMAT} differs
slightly from these two recent implementations, in that we integrate
directly differential equations in time, instead of integrating differential
equations in the plasma temperature which are obtained by the replacement of $\dd T /\dd t$ derived from the Friedmann equation. The
code is abundantly commented and refers to equations of the previous
sections of this article. Let us summarize the main steps of the code.

\begin{itemize}
\item First we solve for the thermodynamics of the plasma following
  the details provided in \S~\ref{SecThermo}. This allows to obtain $a(T)$ either using
entropy conservation~(\ref{EqaT}) if the effect of incomplete neutrino decoupling
is neglected, or using the variation of entropy from the heat transfer
between the plasma and the neutrinos~(\ref{dlnaTdlnT}). When QED
plasma effects are included we use the corresponding (\ref{SQED}). The
relation $T(a)$ is obtained by a numerical inversion. The temperature
of neutrinos is either deduced from (\ref{ScalingTnu}), or
(\ref{TnuQED}) when QED plasma effects are included, if we assume they are
fully decoupled. If we take into account the incomplete neutrino
decoupling, then their effective temperature is obtained from
Eqs.~(\ref{drhonu}) and (\ref{DefTnueff}). The evolution of baryon
energy density follows the simple scaling (\ref{EqBaryonsScaling}),
and similarly for cold dark matter. Then the cosmological expansion is solved using the total energy density
(\ref{TotalRho}) [with QED effects included for the plasma energy
density (Eq. \ref{CalEQED}) if the choice is made], inside the
Friedmann equation (\ref{Friedmann}), so to obtain $a(t)$ by numerical
resolution of the differential equation. The relation $t(a)$ is
obtained by numerical inversion. Eventually we obtain $T(t)$ as
$T(a(t))$ that we can use later in the reaction rates since they
depend on temperature. 
%TODO Hence the integration of equations gives $T(a)$, $T_\nu(a)$, $\rho_b(a)$, $n_b(a)$, H(a),

\item Once the thermodynamics and the cosmological expansion are known, we
compute the weak rates for a grid of plasma temperatures so as to
interpolate them. The total rates are obtained by considering all relevant
corrections discussed in \S~\ref{SecWeak}, which are added as in
Eq.~(\ref{SumAllCorrections}). The constant $\myk$ involved in all
rates is obtained from Eq.~(\ref{Getmyk}) with $\lambda_0$ given by
the value~(\ref{lambda0good}).  Since the weak rates computation can
be very long, especially when including finite temperature radiative
corrections which involve two-dimensional integrals, we store them on
hard disk for
later use. An option allows to recompute them when desired and various
options allow to switch on or off the various corrections.

\item Finally we build the system of equations for the nuclear
network, that is the system of Eqs.~(\ref{GeneralYidot}) including the weak rates. The nuclear reaction rates are read from tabulated values in
function of temperature, or from analytical fits and the reverse rates are obtained from
(\ref{ReverseNSE2}). The evolution of the nuclides abundances is numerically solved in three periods.

\begin{enumerate}
\item For $10^{11}{\rm K} \ge T \ge 10^{10}{\rm K}$ we solve only for
  the abundance of neutrons and protons and totally ignore the nuclear
  reactions. Photodissociation reactions are too strong for this period and
  only the abundance of neutrons and protons is relevant at this stage. Nuclear reactions are anyway not tabulated above $10^{10}{\rm K}$.

\item Then for $\SI{e10}{\kelvin} \ge T \ge \SI{1.25e9}{\kelvin}$ we
  solve for a small nuclear network, made of light elements only
  (protons, neutron, ${}^2{\rm H}$, ${}^3{\rm H}$, ${}^3{\rm He}$,
  ${}^4{\rm He}$, ${}^6{\rm Li}$, ${}^7{\rm Li}$ and ${}^7{\rm Be}$) and
  starting from the NSE abundance, apart for neutrons and protons
  which are taken from the previously solved period. For this period
  the system is very stiff and we use a first order BDF scheme
  (backward differentiation formula), which is equivalent to a
  backward Euler method.

\item Finally for $1.25\times10^{9}{\rm
    K}  \ge T \ge 6\times10^7{\rm K} $ we solve numerically for all nuclides, using the full network
  of reactions. The system is less stiff and it is possible to
  fasten the numerical integration by using a second order BDF scheme.
\end{enumerate}
\end{itemize}

\subsection{Temperature of nucleosynthesis}\label{SecTnuc}

\begin{figure}[!htb]
\includegraphics[width=\mycolumnwidth]{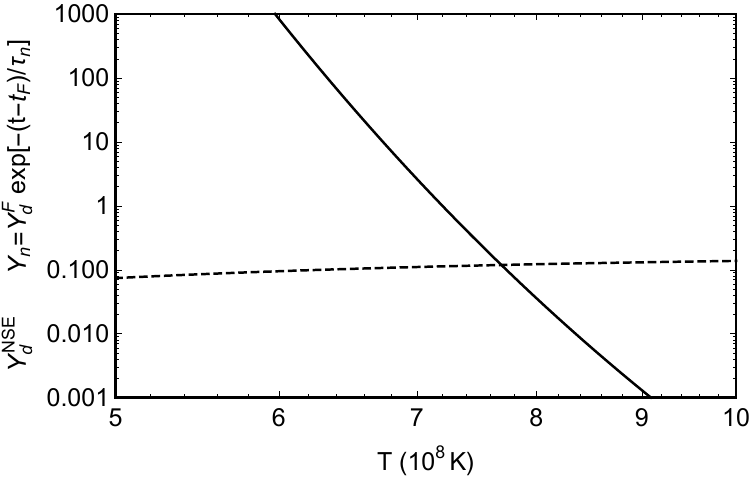}
     \caption{Evolution of NSE deuterium abundance (continuous
       line) and $Y_n^F$ found from a simple freeze-out and beta decay
       model (dashed line).}
\label{FigTNuc}
\end{figure}

\begin{figure*}[!htb]
\centering
    \includegraphics[width=0.47\linewidth]{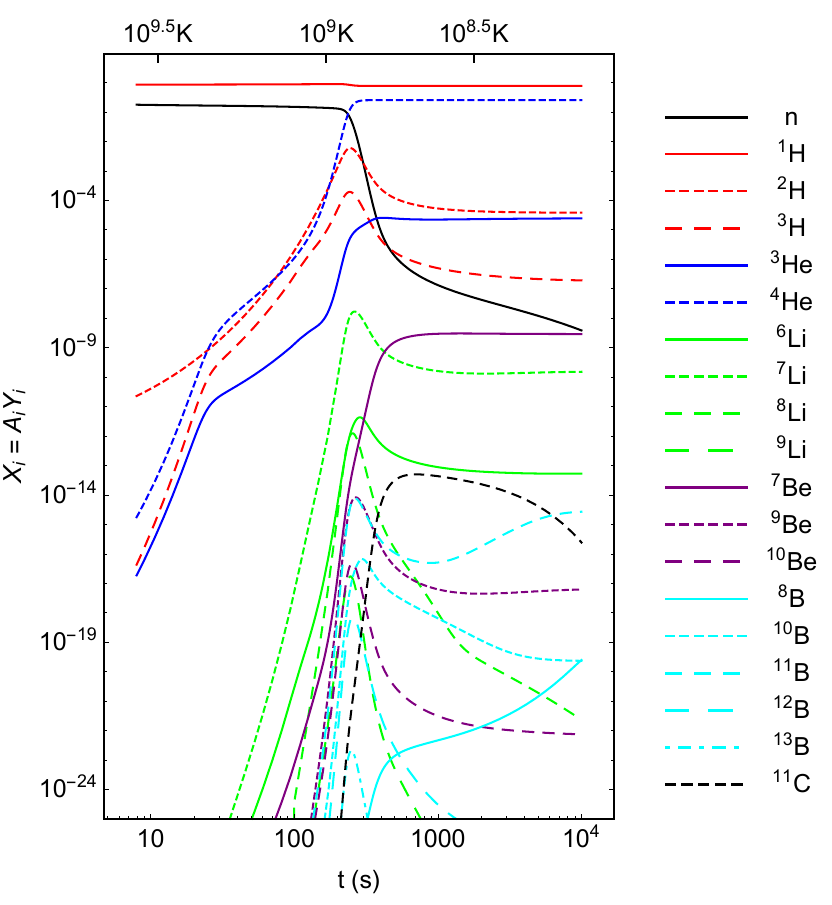}\hfil
    \includegraphics[width=0.47\linewidth]{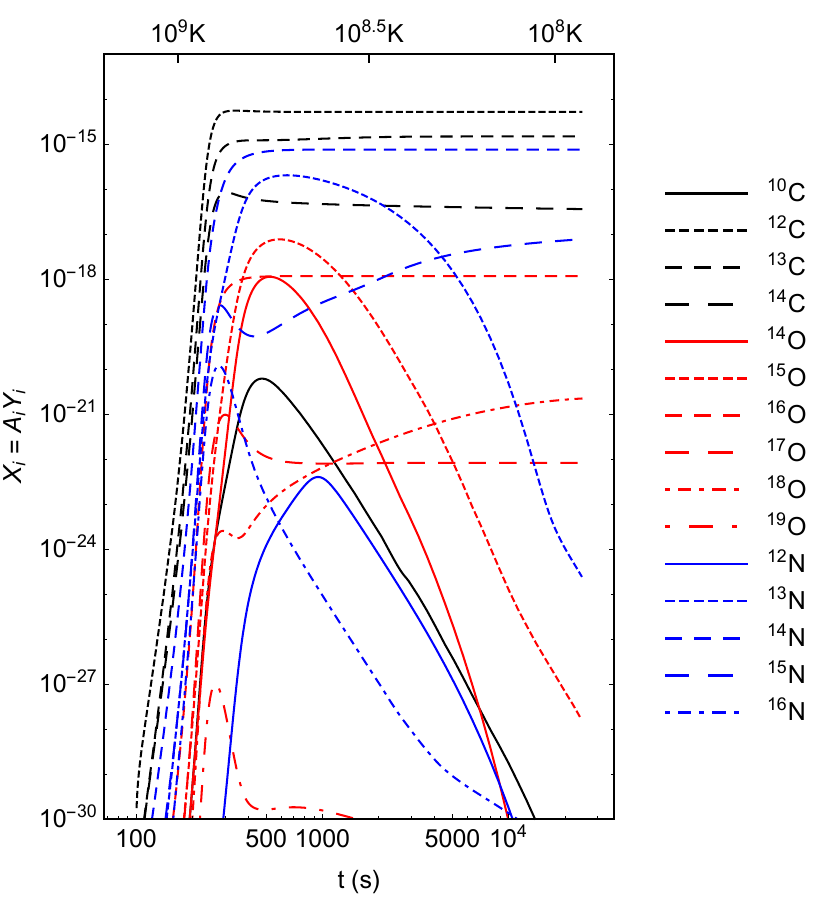}\par\medskip
\caption{{\it Left} : Evolution of light elements abundances. {\it Right} : Evolution of heavier elements abundances\textsuperscript{\ref{foot2}}.}
\label{FigYs}
\end{figure*}

One can estimate the temperature of nucleosynthesis as the temperature for
which the NSE value of deuterium is equal to the abundance of neutrons
found from the simple freeze-out plus beta decay model
(\ref{FreezeOutYnModel}). At this temperature deuterium would have
gobbled up all free neutrons and this definition corresponds
essentially to the end of nucleosynthesis. Hence $T_{\rm Nuc}$ is defined by
\be
Y_n(T_{\rm Nuc}) \equiv Y_d^{\rm NSE}(T_{\rm Nuc})\,.
\ee
These two abundances are plotted in Fig.~\ref{FigTNuc} and we find numerically $T_{\rm Nuc} \simeq 7.7 \times 10^8{\rm
  K}\simeq 0.066\, {\rm MeV}$ and a corresponding $t_{\rm Nuc}
\simeq 300\,{\rm s}$. On Fig.~\ref{FigYnCoc}, we see that nearly all
neutrons are hidden in ${}^4{\rm He}$ which is energetically more
favored than deuterium. Overall the deuterium is only a catalytic
agent, necessary to convert free neutrons in ${}^4{\rm He}$, and we
find only traces of it at the end of BBN. The nucleosynthesis of all
rare light elements is over when the temperature is around $6\times
10^7{\rm K}$, corresponding to $t_{\rm end} \simeq 5\times 10^4\,{\rm s}$. In
our code, the decay reactions of elements whose half-life is much longer than
$t_{\rm end}$ are not included. However when reporting the final
abundances of elements, it is customary to consider that these
elements have been fully converted into stable elements. For instance
the half-life of ${}^7{\rm Be}$ is $53.22$ days, as it decays into
${}^7{\rm Li}$. Hence when reporting the latter abundances, we add the
former ones. Similarly, tritium (${}^3{\rm H}$) decays in ${}^3{\rm
  He}$ in $12.32$ years and the former is added to the latter in the
final abundances reported. The evolution of the isotopes as a function
of time is depicted in Fig. \ref{FigYs}\footnote{The time evolution of $^{14}$C in Fig.~\ref{FigYs}, left panel, strongly differs from the one in Fig.~13
of \citet{Coc12a}. This was due to an error in the  $^{14}$C($\alpha,\gamma)^{18}$O and  $^{14}$C(p,$\gamma)^{15}$N
reaction rates that are now fixed. It has, however no consequences on the total CNO production.\label{foot2}}.

\subsection{Effect of corrections on abundances}\label{SecResults}

\begin{table*}
\caption{Final abundances depending on the corrections included. ID
  is incomplete decoupling of neutrinos. FM is finite nucleon mass
  effect without weak-magnetism, WM is weak-magnetism, and FM+WM are
  both effects. RC are radiative corrections. ThRC are finite
  temperature radiative corrections without bremsstrahlung
  corrections, and BS are bremsstrahlung corrections. QED-MS is the
  QED electron mass shift effect considered alone when replaced
  directly in distribution functions (see discussion in \S \ref{SecEffectRCT}), and QED-Pl are the
  QED effects on the plasma thermodynamics (\S \ref{SecQEDplasma}).\label{TableCorrections}}
\centering
\resizebox{\textwidth}{!}{
\begin{tabular}{|l|ccccccc|}
  \hline \hline
 Corrections & $\YP$  & $\delta \YP\times 10^4$ & $\delta \YP/\YP (\%)$ & ${\rm  D}/{\rm H}  \times 10^5$ &$\Delta \left({\rm  D}/{\rm H}\right) \,(\%)$&  ${}^3{\rm He}/{\rm  H} \times 10^5$ &  ${}^7{\rm Li}/{\rm  H} \times 10^{10}$ \\
  \hline
  Born & $0.24262$ & $0$ & $0$ & $2.423$ &$0$& $1.069$  & $5.635$ \\ \hline
  Born+ID& $0.24274$ & $1.2$& $0.05$ & $2.432$ &$0.37$& $1.070$  & $5.613$ \\\hline
  Born+FM& $0.24374$ & $11.2$& $0.46$ & $2.430$ &$0.25$& $1.070$  & $5.651$\\
  Born+FM+WM & $0.24390$ & $12.8$& $0.53$ & $2.430$&$0.29$ & $1.070$  & $5.654$ \\ \hline
  RCa [Eq.~(\ref{CSirlin}), Non. Rel. Fermi] & $0.24572$ & $31.0$ & $1.27$ & $2.440$&$0.70$ & $1.071$  & $5.681$ \\ 
  RCb [Eq.~(\ref{RadiativeMarcianoSirlin}), Non. Rel. Fermi] & $0.24575$ & $31.3$ & $1.29$ & $2.440$ &$0.70$& $1.071$  & $5.682$ \\ 
  RC  [Eq.~(\ref{RadiativeMarcianoSirlin}), Rel. Fermi]& $0.24577$ &  $31.5$ & $1.30$ & $2.440$ &$0.70$& $1.071$  & $5.682$ \\ \hline
  RC+QED-MS  & $0.24591$ & $32.9$ & $1.36$ & $2.441$ &$0.74$& $1.071$  & $5.684$\\ 
  RC+QED-Pl  & $0.24577$ & $31.5$ & $1.30$ & $2.443$&$0.82$ & $1.072$  & $5.674$\\ 
  RC+ID  & $0.24588$ & $32.6$ & $1.34$ & $2.449$ &$1.07$& $1.073$  &  $5.660$ \\  
  RC+ID+QED-Pl & $0.24588$ & $32.6$ & $1.34$ & $2.452$&$1.19$ & $1.073$  &  $5.652$ \\ \hline
  RC+FM+WM  & $0.24705$ & $44.3$ & $1.82$ & $2.447$ &$0.99$& $1.072$  &$5.701$  \\ \hline 
  RC+FM+WM+QED-MS  & $0.24718$ & $45.6$ & $1.87$ & $2.448$&$1.03$ & $1.073$  & $5.701$ \\ 
  RC+FM+WM+QED-Pl  & $0.24704$ & $44.2$ & $1.81$ & $2.450$ &$1.11$& $1.073$  & $5.693$ \\ 
  RC+FM+WM+ID & $0.24710$ & $44.8$ & $1.84$ & $2.456$ &$1.36$& $1.074$  & $5.678$ \\ 
  RC+FM+WM+ThRC (No BS)  & $0.24736$ & $47.4$ & $1.95$ & $2.449$ &$1.07$& $1.073$  &  $5.706$ \\ 
  RC+FM+WM+ThRC+BS  & $0.24705$ & $44.3$ & $1.82$ & $2.447$ &$0.99$& $1.072$ & $5.701$ \\ \hline
  RC+FM+WM+ThRC+BS+ID+QED-Pl& $0.24709$ & $44.7$ & $1.84$ & $2.459$&$1.49$ & $1.074$  & $5.670$ \\ \hline\hline
\end{tabular}
}
\end{table*}

The effects of the various corrections in the weak rates have been
estimated numerically and are reported in table \ref{TableCorrections}. We have assessed the
importance of individual corrections as well as the interplay of some
corrections when they do not add simply linearly. Let us now comment in
details the effect of these corrections.

\subsubsection{Radiative corrections}\label{SecEffectRC}

Using the non-relativistic Fermi function and Sirlin's function
(\ref{CSirlin}) without resummation of higher order corrections, we
find $\Delta \YPfour = 31.0$, exactly as in \citet[Table V, line
2]{LopezTurner1998}. With the resummed radiative corrections
(\ref{RadiativeMarcianoSirlin}) and using the relativistic Fermi
function brings an extra $\Delta \YPfour = 0.5$, worth being
considered for precise predictions.

\subsubsection{Finite nucleon mass corrections}\label{SecEffectFM}

We find that the cumulated effect of finite nucleon mass corrections and
weak-magnetism brings $\Delta \YPfour = 12.8$ corresponding to
$+0.53\%$ This is slightly more than found by \citet[Table V, line
3]{LopezTurner1998} which is $\Delta \YPfour = 12$ corresponding to
$+0.50\%$, also evoked in \citet{Lopez1997}. Given the smallness of
the difference we consider that our results are in agreement with
these references. Note that weak-magnetism accounts for around $12\%$ of the finite
nucleon mass corrections.

It is worth commenting that when finite nucleon mass effects are
considered together with incomplete neutrino decoupling, the total
effect is less than the sum of the effects taken individually and
there is a reduction $\Delta \YPfour \simeq -0.6$. We found that this is because finite nucleon mass corrections are very
sensitive to the neutrino temperature (see
Fig. \ref{FigDetailedBalance}). Hence we advocate that a correct treatment of neutrino
decoupling needs not only to be performed with the full machinery of
the Boltzmann equation, with the inclusion of neutrino oscillations,
but it should also be performed jointly with the inclusion of finite
nucleon mass corrections. So far this joint treatment is lacking in
the literature.

\subsubsection{Finite temperature radiative corrections}\label{SecEffectRCT}

In table~\ref{TableCorrections}, we report that the finite radiative corrections
bring $\Delta \YPfour \simeq 3.1$, reduced to $\Delta \YPfour \simeq 0$ when bremsstrahlung corrections are added to obtain a consistent detailed
balance of weak rates. Indeed in Fig.~(\ref{FigCorrectionCCRTRates}) we check that around
freeze-out, finite radiative corrections and bremsstrahlung
corrections are almost opposite. In both cases this is very different from the
modification computed in \citet{Esposito1999} where $\Delta \YPfour \simeq -4$ is reported.

It is important to realize that the electron mass shift~(\ref{deltame2}) is part of the
finite-temperature corrections when it comes to considering the
corrections to the weak rates. Hence we adopt the point of view of \citet{BrownSawyer} which is different from \citet{LopezTurner1998}. In order to
allow a comparison with that reference, we evaluated independently the
effect of the electron mass shift following their method, that is replacing the electron mass shift directly in the distribution
functions, and we found a modification $\Delta \YPfour \simeq +1.4$. Hence from the estimation  of
\citet{LopezTurner1998} for finite-temperature radiative
corrections $\Delta \YPfour \simeq 3$ one should instead use $\Delta
\YPfour \simeq 4.4$ to compare with our results. This is slightly larger than
our value  $\Delta \YPfour \simeq 3.1$ without BS corrections but
certainly larger than our value  $\Delta \YPfour \simeq 0$ with BS corrections.

In general, an enhancement of the rates induces a decrease of
$\YP$ thanks to Eq.~(\ref{RuleOfThumb}). In \citet{LopezTurner1998} the rates are increased by the finite
temperature radiative corrections, and one would expect a decrease
$\YP$. However since the corrected rates no longer satisfy the
detailed balance relation (\ref{MagicDetailedBalanceBorn}) as they
should, Eq.~(\ref{RuleOfThumb}) cannot be used and the effect is not
opposite.  In our case this can
be seen when using the radiative corrections, without the BS
corrections added. The weak rates are increased (see
Fig. \ref{FigCorrectionCCRTRates}) and still it leads to $\Delta
\YPfour \simeq 3.1$. This highlights the extreme importance of
constructing corrections which respect the detailed balance, in order
to obtain meaningful results. Any corrections added which does not
satisfy the correct detailed balance relation Eq.~(\ref{RuleOfThumb})
is somehow equivalent to a modification of the value of the mass gap
$\Gap$, modifying artificially the thermal equilibrium value. Stated more directly, it is obvious that an overestimated enhancement of
$\Gamma_{p \to n}$ as in \citet{LopezTurner1998} leads to an
artificial increase of $Y_n$ and thus of $\YP$. When a correction
does not satisfy the detailed balance relation, the primary effect is
no more a delayed or advanced freeze-out, that is a lower or larger
freeze-out temperature, but an artificially different freeze-out
abundance for the same freeze-out temperature since one rate is
overestimated and the thermal equilibrium is artificially displaced.

\subsubsection{QED effects on plasma thermodynamics}\label{SecEffectQED}

The effect of QED corrections on $\YP$ is negligible. In
\citet{LopezTurner1998}, it is estimated to be around $\Delta \YPfour
\simeq 1$ only because it is also cumulated with the electron mass shift effects which
are part of finite temperature corrections of the weak rates in our
terminology. We find $\Delta \YPfour \simeq 1.4$ when computing it
with the same method.
 
\subsubsection{Incomplete neutrino decoupling}\label{SecEffectID}

The pure clock effect evoked at the end of \S~\ref{SecWeakRatesID} is not the only effect, because one must also consider the effect of
incomplete neutrino decoupling on the weak rates. In our approximate
description, we have implicitly assumed in~\S~\ref{SecNeff} that all
neutrino flavors share the same ratio of heating, that is we assume that $\delta
\rho_{\nu_e}/\rho_{\nu_e}=\delta \rho_{\nu_\mu}/\rho_{\nu_\mu}=\delta
\rho_{\nu_\tau}/\rho_{\nu_\tau}$ so that it is meaningful to define a
common effective temperature. It is certainly not correct since there
is more energy gained by electronic neutrino than other types of
neutrino. Indeed electronic neutrinos couple to electrons and positron with charged and
neutral currents whereas the other flavors of neutrinos couple only through neutral
currents. However, first this is less the case when considering neutrino
flavor mixing~\cite{Mangano2005}, and second in the early stage of
neutrino heating by electron-positron annihilations, the heating is
more efficiently redistributed among the three species. Hence it is
not a too bad approximation to assume that the heating is distributed in
the same ratio among flavors when considering the effect on weak
rates. This means that we assume that all neutrino flavors share
always $1/N_\nu$ of the total energy density. Even though we know that
they have distorted spectra, we still defined an effective neutrino
temperature from their energy density. That is we still use
Eq.~(\ref{EqEnergyNeutrinos}) to define the neutrino temperature
$T_\nu$, and by construction it is the temperature for thermalized
neutrino distributions that would have the same energy density. 

Apart from the clock effect mentioned
earlier, there are two competing effects in the weak rates which
nearly fully cancel~\cite{Fields:1992zb,1992PhRvD..46.3372D,Mangano2005}. First the
higher energy density in neutrinos results in an increase of the weak
rates, inducing a freeze-out which happens later, with less neutrons
and thus producing less ${}^4 {\rm He}$. However the energy gained by
neutrinos is lost from the plasma, and the reduction in
electrons-positrons energy density results in lower weak rates,
inducing an earlier freeze-out which results in more neutrons and then
more ${}^4{\rm He}$ production. It has been shown in
\citet{Fields:1992zb} that both effects level off when the freeze
out is complete, that is around $T=3.3 \times 10^9 {\rm K}$. Indeed we
find that in our simple thermal approximation, the relative variation in the neutron fraction at that temperature is a modest relative increase of
$2 \times 10^{-5}$. Without the clock effect this would result in a
negligible $\Delta \YPfour \simeq 0.05$. It is only the clock effect
(see \S~\ref{SecResults}), that is the fact that neutrons have
slightly less time to beta decay, which results in $\Delta \YPfour
\simeq 1.2$. When taking more carefully into account the fact that
neutrinos do not get the same share of the heating, and that
furthermore there are spectral distortions which affect the neutrino 
distribution functions entering the weak-rates, it is found a slightly
larger increase in $\YPfour$~\cite{Mangano2005}. 

However note that our variations for $\YP$, ${\rm D}/{\rm H}$, ${}^3{\rm He}/{\rm H}$ and
${}^7{\rm Li}/{\rm H}$ are in very close agreement with those found in
\citet{Grohs:2015tfy} (Table IV, second line), where these spectral
distortions effects (but not the flavor oscillations) have been fully
taken into account. It is puzzling that with our thermal
approximation based on a heating rate found from~\citet{Parthenope}, that
is from the result of \citet{Mangano2005}, we do not recover the
results of table 3 in \citet{Mangano2005} but we recover with very good
agreement the results of \citet{Grohs:2015tfy}. We found that the argument presented in \citet{Grohs:2015tfy} for the variations
of  ${\rm D}/{\rm H}$, ${}^3{\rm He}/{\rm H}$ and ${}^7{\rm Li}/{\rm
  H}$ are very convincing and we recover them in our numerics. Indeed the clock effect results in less
time to destroy deuterium through \ddn, \dpg\ and \ddp, ending up in
more deuterium being left out at the end of BBN. Since two of
these deuterium destroying reactions are producing ${}^3{\rm He}$
[namely \ddn\ and \dpg], this results in more ${}^3{\rm He}$. As for ${}^7 {\rm Li}$ it is
reduced only because the production of ${}^7 {\rm Be}$ has less time
to proceed, but we found that the ${}^7 {\rm Li}$ (without ${}^7 {\rm
  Be}$ added) is increased since it has also less time to be destroyed
down from its peak value during BBN. Hence the signs of variations for
${\rm D}/{\rm H}$, ${}^3{\rm He}/{\rm H}$ and ${}^7{\rm Li}/{\rm
  H}$ have a clear physical origin, when incomplete neutrino
decoupling is taken into account, and these signs are opposite to
those reported in table 3 of \citet{Mangano2005}.

\subsection{Dependence on main parameters}

The variations of the yields for small variations of the main
parameters can be given with expansions of the type
\be\label{TaylorAbundances}
\frac{\Delta \YP}{\overline{\YP}} =\sum_{pqr} C_{pqr}\left(\frac{\Delta \Omega_{\rm b}h^2}{\overline{\Omega_{\rm b} h^2}}\right)^p
\left(\frac{\Delta N_\nu}{\overline{N_\nu}}\right)^q \left(\frac{\Delta \tau_n}{\overline{\tau_n}}\right)^r
\ee
with similar expansions for the other abundances (D/H,${}^3{\rm
  He}$/H,${}^7{\rm Li}$/H). The reference abundances are given by the
last line of table~\ref{TableCorrections} and the reference parameters
are $\overline{\Omega_{\rm b} h^2}=0.02225$, $\overline{\tau_n}=879.5\,{\rm s}$ and $\overline{N_\nu}=3$. The
meaning of varying the number of neutrinos is further discussed in
\S~\ref{SecNumberNeutrinos}. The first coefficients of these expansions are given in table~\ref{TableTaylor}.

\begin{table}
\caption{First coefficients of Eq.~(\ref{TaylorAbundances}). These
  provide abundances with precision better than $0.01\%$ for $\YP$
  and $0.03\%$ for other abundances, in the range of $10\%$
  variations in $\Omega_{\rm b}h^2$, $2\%$ variation in $\tau_\nu$ and
  $2\le N_\nu \le 4$. However these are still subject to reaction
  rates uncertainties which are estimated below in table~\ref{TableUncertainty}.\label{TableTaylor}}
\ifphysrep
\else
\resizebox{\columnwidth}{!}{
\fi
%\begin{center}
\addtolength{\tabcolsep}{-1pt} 
\begin{tabular}{c *{4}{S[table-format=-1.6,table-space-text-post=***]}}\toprule
  & $\YP$ & {${\rm D}/{\rm H}\quad$} & {${}^3{\rm He}/{\rm H}\quad$} & {${}^7{\rm Li}/{\rm H}\quad$}\\\midrule
$C_{100}\quad$  & 0.039039 & -1.64550 &-0.56699  & 2.07605 \\
$C_{010}\quad$  & 0.163552 & 0.40901 & 0.13587 &  -0.27675\\\midrule
$C_{110}\quad$  &-0.000044  &-0.61229 & -0.12157 & -0.29277\\
$C_{200}\quad$  & -0.029351 & 2.04137 & 0.53303 & 0.58639\\
$C_{020}\quad$  & -0.036124 & -0.00599 & -0.01265 & 0.03888\\\midrule
$C_{300}\quad$  & 0.017891 & -2.40817 & -0.51855 & -0.88243\\
$C_{210}\quad$  & -0.001037 & 0.80150 & 0.12083 & 0.51082\\
$C_{120}\quad$  & -0.000354 & -0.00477 & 0.00928 & -0.10335\\
$C_{030}\quad$  & 0.009938 & 0.00224 & 0.00270 & 0.00884\\
\midrule\midrule
$C_{001}\quad$  & 0.731614 & 0.42220 & 0.14052 & 0.43829\\
$C_{101}\quad$  &-0.009741  & -0.66030 & -0.12390 & 1.22619\\
$C_{011}\quad$  & 0.018321 & 0.19366 & 0.04040 & -0.33069\\
$C_{111}\quad$  & -0.003423 & -0.33158 & -0.03879 & -0.70923\\
$C_{201}\quad$  & 0.004189 & 0.90617 & 0.12121 & 0.92463 \\
$C_{021}\quad$  & -0.011981 & 0.00498 & -0.00328 & 0.13530\\
\bottomrule
\end{tabular}
%\end{center}
\ifphysrep\else
}
\fi
\end{table}

Leaving aside the non-standard BBN physics which corresponds to a
variation of the number of neutrinos, we can restrict this expansion
to the linear behaviour to estimate rapidly the variations of the
abundances in function of changes in the baryon abundance or the
neutron lifetime. Hence if these parameters are slightly modified in
the future, our results can be transposed easily. We find
\beas
\frac{\Delta \YP}{\overline{\YP}} &=&0.0390\frac{\Delta \Omega_{\rm b}
  h^2}{\overline{\Omega_{\rm b} h^2}} +0.732\frac{\Delta \tau_n}{\overline{\tau_n}}\slabel{EqDYPOverYP}\\
\frac{\Delta {\rm D}/{\rm H}}{\overline{{\rm D}/{\rm H}}} &=&-1.65\frac{\Delta
  \Omega_{\rm b} h^2}{\overline{\Omega_{\rm b} h^2}}+0.422 \frac{\Delta \tau_n}{\overline{\tau_n}} \\
\frac{\Delta {}^3{\rm He}/{\rm H}}{\overline{{}^3{\rm He}/{\rm H}}}&=&-0.567\frac{\Delta \Omega_{\rm b} h^2}{\overline{\Omega_{\rm b} h^2}}+0.141\frac{\Delta \tau_n}{\overline{\tau_n}} \\
\frac{\Delta {}^7{\rm Li}/{\rm H}}{\overline{{}^7{\rm Li}/{\rm H}}} &=&2.08\frac{\Delta \Omega_{\rm b} h^2}{\overline{\Omega_{\rm b} h^2}} +0.438\frac{\Delta \tau_n}{\overline{\tau_n}}\,.
\eeas 
As expected, $\YP$ is very sensitive to the weak rates and thus to
$\tau_n$ but not to the baryon abundance. Hence even though
uncertainty is larger in baryon abundance, the theoretical uncertainty
of $\YP$ is dominated by the uncertainty in the determination of the
neutron lifetime. From the measured value of $\tau_n$ reported in appendix~\ref{ParticleValues}, this corresponds
to a $0.068\, \%$ uncertainty in $\YP$ or $\sigma(\YP)\simeq 0.00017$
which is lower than the value $\sigma(\YP) \simeq 0.0003$ used in
\citet[p47]{Planck2016}, implying that we find errors approximately
$40\%$ smaller in $\YP$. The predicted and observed abundances as functions of $\eta$ are plotted in Fig.~\ref{FigEta}.

\subsection{Distribution of abundance predictions}\label{SecStatistics}

We use the method described in \citet{Coc14} to estimate the
uncertainty in light elements productions during BBN due to
uncertainty in nuclear rates and weak-rates (that is the uncertainty
on $\tau_n$ that we assume to follow a normal distribution). 
It has been found \cite{Eval1}, that probability density functions of reaction rates can be well approximated  by lognormal distributions
\begin{equation}
f(x) = \frac{1}{\sigma \sqrt{2\pi}} \frac{1}{x} e^{-(\ln x - \mu)^2/(2\sigma^2)}  \label{lognormalpdf}
\end{equation}
(with $x\equiv{N_A}\langle\sigma{v}\rangle$ for short is the rate).
This is equivalent to the assumption that  $\ln(x)$ is Gaussian distributed with expectation value $\mu$ and variance $\sigma^2$ (both functions of temperature).
The lognormal distribution allows to cope with large uncertainty factors (${\equiv}e^\sigma$) 
together with ensuring that the rates remain positive.
If these parameters are  tabulated as a function of the temperature, they can be used to perform subsequent
Monte-Carlo nucleosynthesis calculations.
The quantile of the distributions of the abundances obtained by such a
method are reported in table \ref{TableUncertainty} for which we used
$20\,000$ Monte-Carlo points. 

\begin{table}[h]
\centering
\caption{Monte-Carlo estimation of light elements uncertainties due to
nuclear rates, and $\tau_n$ (aka weak rates) uncertainties. \label{TableUncertainty}}
\ifphysrep
\else
\resizebox{\columnwidth}{!}{
\fi
\begin{tabular}{llllllll}\toprule
 {\rm Quantile} & $2.275\%$
  &$15.865\%$&$50\%$&$84.135\%$&$97.725\%$&${\rm mean}$&$\tfrac{\sigma}{{\rm mean}}$\\\midrule
$\YP$ &$0.24676$&$0.24693$ &$0.24709$&$0.24726$&$0.24742$ &$0.24709$&$0.068\%$\\
${\rm D}/{\rm H} \times 10^5$ &$2.386$ &$2.423$ &$2.460$&$2.496$&$2.532$ &$2.459$&$1.49\%$\\
${}^3{\rm He}/{\rm H} \times 10^5$ & $1.023$& $1.048$ &$1.074$&$1.100$&$1.127$&$1.074$&$2.43\%$ \\
${}^7{\rm Li}/{\rm H} \times 10^{10}$ & $5.123$& $5.392$ &$5.627$&$5.858$&$6.105$ &$5.623$&$4.39\%$\\
${}^6{\rm Li}/{\rm H} \times 10^{14}$&$0.61$&$0.85$&$1.20$&$1.68$&$2.35$&$1.27$&$35\%$ \\
${\rm CNO}/{\rm H} \times 10^{15}$ &$0.14$&$0.52$  &$1.02$&$3.07$&$65.6$&$15.3$&$13.4$\\\bottomrule
\end{tabular}
\ifphysrep\else
}
\fi
\end{table}

\begin{figure}[!htb]
\centering
    \includegraphics[width=\mycolumnwidth]{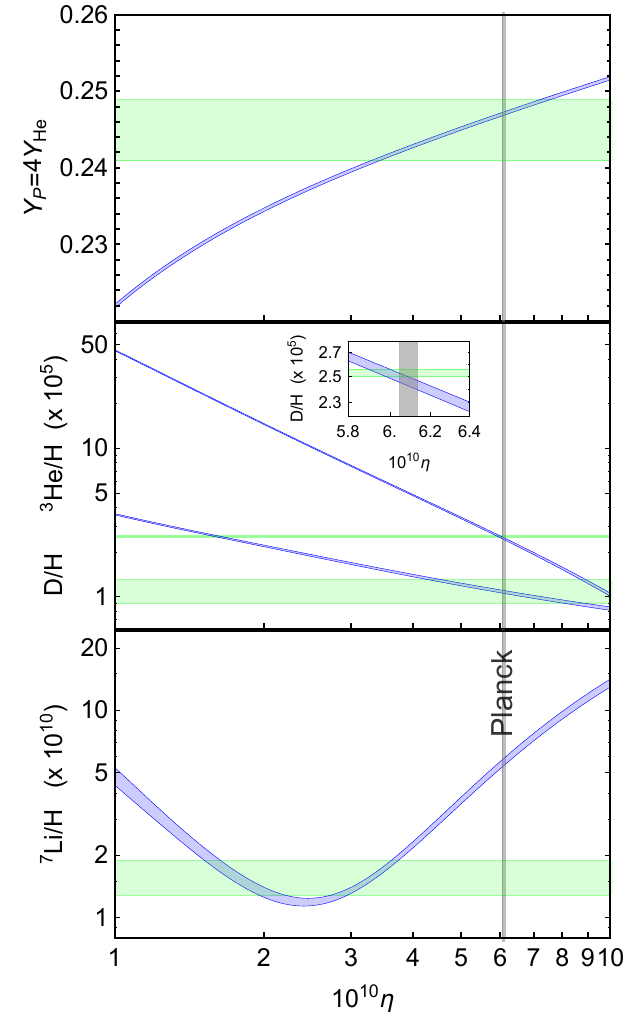}
\caption{{\it Top } : Dependence of $\YP=4 Y_{{}^4 {\rm He}}$ in
  $\eta$ and observational constraints. {\it Middle} : Dependence of deuterium (to curve) and
  ${}^3{\rm He}$ (bottom curve) in
  $\eta$ with observational constraints. The ${}^3 {\rm H}$ has been added since it decays
  radioactively in ${}^3{\rm He}$. {\it Bottom} : Dependence of ${}^7{\rm Li}$ in
  $\eta$ with observational constraints. The ${}^7 {\rm Be}$ has been added since it decays
  radioactively in ${}^7{\rm Li}$. In all these plots, the width of
  the curves represents the $\pm\sigma$ uncertainty from nuclear rates and neutron lifetime.
\label{FigEta}}
\end{figure}

The variations of $\YP$ are nearly entirely
due to the uncertainty on $\tau_n$ because it is almost completely
controlled by weak rates which set the abundance of free neutrons
before nucleosynthesis starts. This Monte-Carlo method allows to construct the
probability $P(Y_i | \omega_{\rm b})$ for each species, where $\omega_{\rm b} \equiv
\Omega_{\rm b} h^2$. This probability reflects the uncertainty in all
parameters affecting the reaction rates, independently of cosmology
and it is used to plot the $\pm\sigma$ width of curved in
Fig.~\ref{FigEta} by computing the quantiles $\{0.15865,0.84135\}$.

For elements for which the uncertainty is small, it proves useful to
approximate these probabilities by a normal distribution as
\be\label{ApproxNormal1}
P(Y_i|\omega_{\rm b}) \simeq {\cal N}\left[\bar{Y}_i \left(\omega_{\rm b}\right),\sigma^{\rm th}_i\left(\omega_{\rm b}\right)\right](Y_i)
\ee
where the normal distribution is noted
\be
{\cal N}[\mu,\sigma](x) \equiv \frac{1}{\sqrt{2 \pi}\sigma}{\rm
  exp}\left[-\frac{(x-\mu)^2}{2 \sigma^2}\right]\,.
\ee
If we use the CMB to obtain a prior distribution on $\omega_{\rm b}$ that we assume
to follow a normal distribution (the mean and standard deviation are
reported in appendix~\ref{ParticleValues}), then we can build the joint probability
\be
P(Y_i , \omega_{\rm b}|{\rm CMB}) =P(Y_i | \omega_{\rm b}) P(\omega_{\rm b}|{\rm CMB})\,,
\ee
and marginalizing over $\omega_{\rm b}$ we get probabilities in
predicted abundances
\be
P(Y_i|{\rm CMB}) = \int P(Y_i , \omega_{\rm b}|{\rm CMB}) \dd \omega_{\rm b}\,.
\ee
In practice we use also a Monte-Carlo method to obtain directly
$P(Y_i|{\rm CMB})$. We only need to vary $\omega_{\rm b}$, according to
its normal distribution, in addition to varying the reaction
rates. With this method we can predict the underlying abundances that
are reported in table~\ref{t:heli}. The predicted abundance for
deuterium is noticeably lower than the one in
\citet[Eq. 74]{Planck2016} which uses {\tt PArthENoPE} \citep{Parthenope}, and this has
consequences when inferring the chemical potential of neutrinos (see \S~\ref{SecNeutrinoPhi}).

%%%%%%%%%%%%%%%%%%%%%%%%%%%%%%%%%%%%%%%%%%%%%%%%%%%%%%%%
\begin{table*}[htbp!] 
\caption{\label{t:hlix} Primordial abundances compared to observations.}
\begin{center}
\begin{tabularx}{\textwidth}{AAAAAAAAA}
\toprule
& Observations & $\quad {\rm (a)}\quad$ $\tau_n = 880.3(1.1)\, {\rm s}$& {\bf This work (h)} $\tau_n = 879.5(8)\, {\rm
                       s}$ & $\quad {\rm (f)}\quad$ $\tau_n = 880.3(1.1)\, {\rm s}$& Planck 2015 (g) $\tau_n = 880.3(1.1)\, {\rm s}$\\
\midrule
$\YP$    & 0.2449$\pm$0.0040(b)  & 0.2484$\pm$0.0002 &{\bf  0.24709}$\pm${\bf   0.00018}& 0.24709$\pm$0.00025 &  0.24667$\pm$0.00062 \\
\deu/H   ($ \times10^{-5})$ & 2.527$\pm$0.030 (c)& 2.45$\pm$0.05&{\bf  2.460}$\pm${\bf     0.046}   &2.58$\pm$0.13& 2.614$\pm$0.13  \\
${}^3{\rm He}/{\rm H}$    ($ \times10^{-5}$)  & <1.1$\pm$0.2 (d)&   1.07$\pm$0.03&{\bf     1.074}$\pm${\bf  0.026}  & 1.0039$\pm$0.0090& \\
\sep/H ($\times10^{-10}$)  &   1.58$^{+0.35}_{-0.28}$ (e)
               &5.61$\pm$0.26& {\bf 5.627}$\pm${\bf 0.259}& 4.68$\pm$0.67& \\
\bottomrule
\end{tabularx}\\%}\\
(a) \citet{Coc15}, (b) \citet{Ave15}, (c) \citet{Coo18},
(d)\citet{Ban02}, (e) \citet{Sbo10}, (f)  \citet{Cyb16},
(g)\citet[TT+TE+EE+LowP, 95$\%$CL]{Planck2016}, (h) we get
$\YP=0.24726$ when using $\tau_n = 880.3\,{\rm s}$
\end{center}
\label{t:heli}
\end{table*}

%%%%%%%%%%%%%%%%%%%%%%%%%%%%%%%%%%%%%%%%%%%%%%%%%%%%%%%%

\subsection{Comparison with observations}

Except for \qua, and the weak rates, the predicted abundances, displayed in Table~\ref{t:heli}, do not differ significantly from those of \citet{Coc15}. 
As can be seen in Table~\ref{t:rates}, very few reaction rates have been updated, and if so, either the rate changes are tiny or
the reactions have, in any case, a negligible effect. On the contrary, there are significative differences with the results of \citet{Cyb16} which
are mostly, and probably entirely, due to the different choice of reaction rates. Apparently,  \citet{Cyb16} use for 
some of the most important reactions, \dpg, \ddn, \ddp, and \hag, the rates from NACRE~\textsc{ii} \citep{NACRE2}, now superseded 
for these reactions by \citet{Coc15,Bayes16,Bayes17}. This is clearly the origin of the differences concerning D and \sep. 
There is an even greater difference with the D/H value reported by
\textit{Planck} \cite{Planck2016,DiV14} using the {\tt PArthENoPE} code \citep{Parthenope}.
This is again, probably due to a different choice of reaction rates for deuterium destruction.

Since the results for deuterium and lithium  are not significantly different from our earlier works \citep{Coc14,Coc15} the 
comparison with observations is very similar. As explained in
\S~\ref{s:obsh} we do not consider \tro\ as a constraint.

\subsubsection{Helium}

Thanks to the re--evaluation of the corrections to the weak rates, we
claim a precision of a few $10^{-4}$ on $\YP$ i.e. smaller than one
unit on $\YPfour$ before taking into account the experimental uncertainty on the neutron lifetime and a few reaction rates (\S~\ref{s:nucl}).
We finally obtained $\YP=0.24705\pm$0.00019, fully consistent with the value $\YP=0.2449\pm 0.0040$ deduced from \citet{Izo14} observations
by \citet{Ave15} (see Fig.~\ref{FigEta}), without the need for extra relativistic degrees of freedom as in \cite{Izo14}.

Since this work is focused on \qua\ and is an improved continuation of previous works it is worthwhile tracking the 
evolution of our calculation of $\YP$. In our earliest works (e.g. \citet{Coc04}), since the observational uncertainties on $\YP$ 
were large, we neglected some of the corrections discussed here because we were unable to calculate them (e.g. including six--fold integrals).
The corrections that we took into account corresponds to line RCa line in Table~\ref{TableCorrections} and our (Fortran) calculated 
corrections did amount to $\delta\YPfour=31.6$ \citep{Coc14}. When the observational uncertainties were reduced, it
become important to include the neglected corrections, by artificially increasing $\YPfour$ by 18 at the the very end of the calculation.
It corresponded to  the finite-nucleon mass correction ($\delta\YPfour=12$ \citep{LopezTurner1998}), 
finite-temperature radiative correction ($\delta\YPfour=3$) \citep{LopezTurner1998}), QED plasma ($\delta\YPfour=1$ 
\citep{LopezTurner1998}) and neutrino decoupling ($\delta\YPfour=2$ \citep{Mangano2005}), for a total of $\delta\YPfour=18$,
that we could not easily directly re-calculate. Hence, before this final correction, the $\YP$ value from \citet{Coc15}  was 0.2466
(i.e. 0.2484 minus 0.0018) to be compared to $0.24572$ (RCa in
Table~\ref{TableCorrections}). This difference of $\delta\YPfour=9$, which is in fact only $\delta\YPfour=7$ if we account for the
different $\tau_n$ used, was reduced to $\delta\YPfour \simeq 0.5$ by improving the Fortran code (time steps, temperature grid,...) at the expense of execution time\footnote{Details to appear elsewhere.}.

\subsubsection{Deuterium}

Our present result differs by less than 3\%, and agrees within error bars with the latest value inferred from \citet{Coo18} observations.
Since the theoretical value lies on the lower bound of the observational one, this leaves only little room for models of lithium (i.e. \bery)
destruction that byproduct extra deuterium (see below). 
Tiny differences with previous results \cite{Coc15} come in part from the corrections to the weak rates (see Table \ref{TableCorrections}),
to the re-evaluation of the three nuclear reaction rates \citep{Bayes16,Bayes17} that govern deuterium destruction and improvements
in the numerical method. With the high precision reached by D/H observations, these reaction rates [\dpg, \ddn\ and \ddp] need to 
be known at the percent level! 
This demands accurate measurement at BBN energies where data are scarce and theoretical improvement to better constrain
the energy dependence of the \sfac{s}.

\subsubsection{Lithium}

There remains a factor of 3.6 between the predicted and observed Li/H values. This discrepancy has not yet found a fully 
satisfactory solution. On the contrary, the problem had worsened because of an updated reaction rate \citep{Cyb08a,Cyb08b} and
because many solutions have been ruled out by the improved precision on D/H observations \citep{Coo18}. 
We present below the kinds of solutions that have been considered so far. 

\begin{itemize}
\item {\em There is no nuclear solution to the lithium problem.} Extensive sensitivity studies \cite{Coc04,Coc12a} have not
identified reactions, beyond those already known, that could have a strong impact on lithium nucleosynthesis.
The most promising was $^7$Be(d,p)$2\alpha$ \cite{Coc04,Cyb12}: an increase of its rate by a factor of $\approx$100 would have 
solved the problem.  However, measurements of its average cross section \cite{Ang05} or properties of candidate 
resonances \cite{OMa11,Sch11,Kir11} ruled out this possibility.
More generally, other  destruction channels have recently been proposed \cite{Cha11} : $^7$Be+n, p, d, t, \tro\ and \qua.
In particular, the existence of a relatively narrow state around 15 MeV in the compound nucleus $^{10}$C formed by 
$^7$Be+$^3$He or the existence of a state close to 8 MeV in the compound 
nucleus $^{11}$C formed by $^7$Be+$^4$He could help reduce the  $^7$Be production. 
However, a recent search \cite{Ham13} for missing levels in the relevant excitation energy regions of $^{10}$C and $^{11}$C,  
via the reactions $^{10}$B($^3$He,t)$^{10}$C and $^{11}$B($^3$He,t)$^{11}$C, respectively, did not find any new level,
whose corresponding resonances, in any case, would have too low strengths \cite{Bro12} because of the Coulomb barrier.      
It seems now that all extra $^7$Be destructing reactions have been considered and found inefficient.

\item {\em The effect of electron screening or modification of decay lifetime is negligible.} For reactions of interest to BBN, screening 
affects the laboratory cross sections at too low energies
[e.g. $\lesssim$ 20~keV for \ddp\ \cite{Gre95} or  $^3$He(d,p)$^4$He \citep{Ali01}] to affect
measurement at BBN energies [$\approx$100~keV], on the one hand.  On
the other hand, the effect of screening during BBN is 
completely negligible \cite{Wan11,Fam16}. 
It is well known that the lifetime of \bery\  that decays by electron capture depends on the probability of presence of an electron (from an atomic
$s$ orbital or from a plasma) inside the nucleus; for instance it is increased to $\sim100$~days at the center of the Sun \cite{Ade11}.
To have an impact on lithium prediction, the \bery\   lifetime must be reduced to a value of $\sim10^3$~s. 
However, because of  the Boltzmann suppression factor, at $T<$0.5 GK, when \bery\ is present, the electron density becomes smaller than
in the Sun so that one can expect an even longer lifetime. This can be
confirmed if one extrapolates the results of \citet[Fig. 1]{Sim13},
to  $T\lesssim500\times10^6$K and $\rho\lesssim10^{-5}$~g/cm$^3$.

\item Many exotic solutions to the lithium problem have been investigated (e.g. \citet{Yam14}), but most rely on extra neutron sources 
to boost $^7$Be destruction  through the $^7$Be(n,p)$^7$Li(p,$\alpha)^4$He channel. 
However, these extra neutrons, inevitably, also boost the D and $^3$H production through the  
$^1$H(n,$\gamma)^2$H and $^3$He(n,p)$^3$H channels, respectively \cite{Kus14,Coc15}.
This is shown in Figs.~1 in \citet{Oli12}, 4 in \citet{Kus14}, and 14 in \citet{Coc15} that display results of various
types of models that succeed in solving the lithium problem, but at the expense of deuterium overproduction
to levels now excluded by observations \cite{Coo14,Coo16,Coo18}.
Very few solutions, even beyond the Standard Model, that do not suffer from this drawback
are left, e.g. \citet{Gou16}.

\item {\em Stellar physics solutions require a uniform reduction of surface lithium over a wide range of effective temperature and metallicity.}
Some amount of surface lithium destruction is unavoidable, because of atomic diffusion that transport lithium down to deeper and hotter layers
where it is  destroyed by a factor of 1.5 to 2 \citep{Mic84}. The difficulty comes from the small thickness of the lithium plateau over a wide
range of metallicity and temperature and the absence of stars between the plateau and the BBN prediction (Fig.~\ref{f:liobs}). 
This could possibly be circumvented if an additional mixing process is included in the outer layers of these stars \cite{Ric05}.
This is supported by the comparison between lithium observations in the metal--poor globular cluster NGC 6397 with stellar depletion 
models  \citep{Kor06}.
Another recent proposition relies on full lithium destruction but followed by a self--regulated re-enrichment of lithium by late time  accretion 
from the interstellar gas \citep{Fu15}.  
In addition, the Spite plateau does not exist anymore at the metallicity below [Fe/H]=-3, and is replaced by an increased spread of abundances, 
below the plateau value \citep{Sbo10}. This ``meltdown'' of the Spite plateau is not understood yet. All this suggests that lithium observations
cannot be used anymore to constrain BBN models.

\item {\em There is no \six\ problem anymore.} 
A few years ago, observations \cite{Asp06}  of \six\ in a few metal poor stars had suggested the presence of a plateau, at 
typically  \six/H $\approx10^{-11}$, orders of magnitude higher than the BBN predictions of 
 \six/H $\approx1.3\times10^{-14}$ \cite{Ham10}.
However,
later, the observational \six\  plateau has been questioned  
due to line asymmetries which were neglected in previous abundance analyses.
Hence, there is no remaining evidence for a plateau at very low metallicity \cite{Lin13} that can be used to derive a primordial \six\  abundance.
\end{itemize}

\subsubsection{Other elements}

Leaving aside the \neu, \dix\ and \onz\ isotopes for which no primordial abundance can be inferred from observations~\citep{Coc14}, 
it is worth mentioning the CNO abundance. Here, we call CNO all isotopes with masses larger than 12 ($^{11}$C mostly decays to \onz).
Even though, there are no primordial CNO abundance either,  it is of peculiar interest since it may affect the evolution of the first stars 
(Population III) within the first structures of the Universe. Hydrogen burning in the first generation of stars proceeds through the slow 
pp--chains until enough carbon is produced (through the triple-alpha reaction) to activate the CNO cycle. The minimum value of the 
initial CNO mass fraction that would affect Population III stellar evolution is estimated  to be 10$^{-11}$ \cite{Cas93}  or
even as low as  10$^{-13}$ (in number of atoms relative to hydrogen, CNO/H) for the less massive stars \cite{Eks08}. 
Table~\ref{TableUncertainty} shows that the median abundance (0.5 quantile) is in agreement  with previous works \cite{Ioc07,Coc12a},
if taking into account that the distribution is not Gaussian \cite{Coc14,NIC2014}. As a result, the 0.97725 quantile corresponds to
CNO/H$\approx0.7\times10^{-13}$, close to the limit to have an impact on some first stars.     
As this is explained in \citet{Coc14}, this value, much larger than the median comes from the simultaneous variations, during the 
Monte Carlo,  of a few reaction rates around $^{10}$Be. Since very few or no experimental data are available for these reactions,
the actual rates may differ by a few orders of magnitudes, from the TALYS \cite{TALYS} theoretical predictions. Experimental
efforts are needed to confirm or infirm this possibility.

%%%%%%%%%%%%%%%%%%%%%%%%%%%%%%%%%%%%%%%%%%%%%%%%%%%%%%%%%

\section{Cosmology with BBN}

\subsection{Cosmological perturbations}

Cosmological fluctuations are of order $10^{-5}$, and whatever their
effect, we have neglected them given our precision goal. However it is easy
to estimate the effect of adiabatic perturbations seeded by inflation. Indeed, since the Hubble
radius is extremely small during BBN, most modes can be considered as
super-Hubble modes, that is longer than the Hubble radius. In that case their
effect can be described by a simple coordinate transformation applied
on a homogeneous cosmology~\citep{Weinberg:2003sw,Creminelli:2011sq,Mirbabayi:2014hda}.

Let us consider a perturbed metric (expressed in conformal time
defined by $\dd t = a \dd \eta$)
\be
\dd s^2 = a^2(\eta)\left(-(1+2 \Phi)\dd \eta^2 + (1-2\Psi)\delta_{ij} \dd x^i \dd
x^j\right)\,.
\ee
It can be put in the
homogeneous form using a coordinate transformation. Or conversely the
dynamics of a spacetime perturbed by long modes can be deduced from
the homogeneous dynamics thanks to a coordinate transformation. Not
all such transformations lead to a physical long mode dynamics, and one
must ensure some conditions. In the case where the Universe dynamics
is dominated by radiation, as is the case for BBN, the relevant
coordinate transformation is \citep{Creminelli:2011sq}
\be\label{DefChangeCoordinates}
\tilde \eta = \eta\left(1-\frac{\zeta}{3}\right) \,,\qquad \tilde x^i = x^i(1+\zeta)
\ee
where $\zeta$ is the comoving curvature perturbation.

It follows using Eq.~(\ref{Eqdotn}) that the number density in the perturbed universe $\tilde n_i$ is related to
the one of a homogeneous cosmology $n_i$ thanks to
\be
\tilde n_i = n_i  -\frac{\eta\zeta}{3}\partial_\eta n_i = (1+\zeta)n_i
-\frac{\eta \zeta}{3} a {\cal J}_i\,,
\ee
where we used that for radiation domination $a \propto \eta$ and
$\partial_\eta a/a = 1/\eta$. At the end of BBN, when all reactions are
inefficient, all net creation rates ${\cal J}_i$ vanish and we find
that all number densities are simply rescaled by $1+\zeta$. Note that
this could have been anticipated because for super-Hubble modes, $\zeta$ is equal to
density perturbations on isocurvature surfaces~\citep{Vernizzi:2004nc}. Since
abundances are ratios of number densities, they remain unchanged by
the long mode perturbation. Hence, provided we can neglect the effect
of modes which are smaller than the Hubble radius during BBN, standard
adiabatic cosmological perturbations have no effect on BBN
predictions. Intuitively, an adiabatic perturbation enhances baryon
and photons number densities in the same proportions, leaving $\eta$
constant. Since this analysis is based on long modes only, one cannot
infer the consequences of fluctuations on scales smaller than the
Hubble radius during BBN from the homogeneous cosmology results. For these
small scales, only a complete treatment of all perturbed equations
dictating the evolution of species can lead to a meaningful result.

%TODO Cyril Maybe cite Iocco or Serpico I forgot because they have some work on gradient modes.

Finally, it is worth noting that long modes of entropy perturbation, that is perturbations which do not
modify the total energy density (and for which $\zeta=0$) but which modify the ratios $\rho_{\rm b}/\rho_\gamma$ have an effect on
abundances. Since during BBN the total energy density is dominated by
radiation, and long mode entropy perturbation can be rephrased as a
perturbation in $\rho_{\rm b}$ alone, that is a perturbation of $\eta$. As a consequence, its effect is directly evaluated by
 the sensitivity of BBN final abundances on $\eta$.

\subsection{Measurement of baryon abundance from BBN}\label{SecStatsBaryons}

The observed abundances are related to the underlying
one by an assumed normal distribution, that is for each isotope
observed, the likelihood is
\be\label{ApproxNormal2}
P(Y_i^{\rm obs} | Y_i) = {\cal N}\left[Y_i, \sigma^{\rm
  obs}_{i}\right](Y_i^{\rm obs})\,.
\ee
The observed values and their standard deviations are reported in
\S~\ref{s:obs}. 
Then we can consider
\be
P(Y_i^{\rm obs}|\omega_{\rm b})  \equiv \int \dd Y_i P(Y_i^{\rm obs} | Y_i) P(Y_i | \omega_{\rm b})\,, \nonumber
\ee
and we deduce using Eqs.~(\ref{ApproxNormal1}) and
(\ref{ApproxNormal2}) that it follows approximately a normal distribution
\be
P(Y_i^{\rm obs}|\omega_{\rm b}) \simeq {\cal N}\left[\bar{Y}_i\left(\omega_{\rm b}\right), \Sigma_i(\omega_{\rm b})\right](Y_i^{\rm obs})
\ee
with $\Sigma^2_i(\omega_{\rm b}) \equiv (\sigma^{\rm
  obs}_{i})^2+\left[\sigma^{\rm th}_i\left(\omega_{\rm
      b}\right)\right]^2$. In practice we do not use the
approximation~(\ref{ApproxNormal1}) and use instead the full result of our
Monte-Carlo  method to estimate the distribution of abundances due to reaction
rates uncertainties. However, the results obtained are extremely
similar given that (\ref{ApproxNormal1}) is a very good approximation.

If we then use an uniform prior on $\omega_{\rm b}$, that is if we do not use our knowledge from CMB observations, the
posterior distribution for $\omega_{\rm b}$ is immediately given from $P(\omega_{\rm b} |Y_i^{\rm
  obs})\propto P(Y_i^{\rm obs}|\omega_{\rm b})$. Otherwise we use the CMB prior to build the posterior distribution for $\omega_{\rm b}$ as
\bea
P(\omega_{\rm b} | Y_i^{\rm obs},{\rm CMB})  &\propto& P(Y_i^{\rm
  obs},\omega_{\rm b} | {\rm CMB}) \nonumber\\
&=& P(Y_i^{\rm  obs}|\omega_{\rm b}) P(\omega_{\rm b}|{\rm CMB}).
\eea
When including all observed abundances, we use that the total probability is the product of individual ones. In practice, only
${}^4{\rm He}$ and ${}^2{\rm H}$ are determined with enough precise experimental precision to be taken into account.

\begin{figure}[!htb]
\includegraphics[width=\mycolumnwidth]{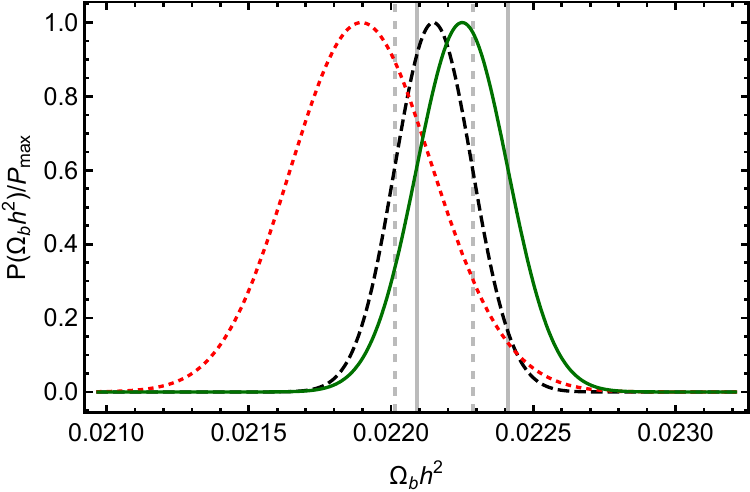}
\caption{$P(\Omega_{\rm b}h^2)$ normalized to a unit maximum. {\it Green continuous line
    :} CMB prior distribution. {\it Black dashed
    line} BBN posterior distribution (BBN+CMB). {\it Red dotted line :} baryon abundance
  distribution determined only from BBN. The vertical gray lines are
  the $\pm\sigma$ CMB (continuous) and CMB+BBN (dashed) bounds. \label{FigPosteriorEta}}
\end{figure}

The CMB prior distribution and the posterior distribution from the observations of BBN are plotted in Fig.~\ref{FigPosteriorEta}. The
posterior bounds from BBN and CMB are $\Omega_{\rm b} h^2 = 0.02215\pm
0.00014$, whereas from BBN alone on gets $\Omega_{\rm b} h^2 = 0.02190\pm
0.00025$. Note that if we had used a code which predicts ${\rm D}/{\rm
H}\simeq 2.61\times10^{-5}$ instead of ${\rm D}/{\rm H}\simeq
2.46\times10^{-5}$ for the Planck parameters, we would have obtained the constraint $\Omega_{\rm b} h^2 = 0.02271\pm
0.00025$ from BBN alone, having exactly the same central value than found in~\citet{Parthenope2}, using  {\tt PArthENoPE}.

\subsection{Neutrino chemical potential from BBN}\label{SecNeutrinoPhi}

If there is a neutrino asymmetry, that is a non-vanishing neutrino
chemical potential, then for each flavor we have necessarily 
\be\label{RelateChemicalNeutrinos}
\mu_\nu+\mu_{\bar \nu}=0
\ee 
because of processes like $\nu + \bar \nu
\leftrightarrow e^- + e^+ \leftrightarrow 2 \gamma$ and photons have a
vanishing chemical potential. Furthermore cosmological expansion affects only the particle momenta and $\xi_\nu \equiv
\mu_\nu/T_\nu$ is frozen at its initial value, whose consequences on BBN can be investigated~\citep{SerpicoRaffelt,IoccoReport,Simha:2008mt}.

The neutrino asymmetry (defined for each neutrino flavor) is defined by the excess
of neutrinos over antineutrinos as 
\bea\label{DefEtaNu}
\eta_{\nu}\equiv \frac{n_{\nu}-n_{\bar \nu}}{n_\gamma} &=&
\frac{I_+^{(1,1)}(0,\xi_\nu) -I_+^{(1,1)}(0,-\xi_\nu)
}{2 I_-^{(1,1)}}\nonumber\\
&\simeq& \frac{1}{12
  \zeta(3)}\left(\frac{T_\nu}{T}\right)^3\left(\pi^2 \xi_\nu + \xi_\nu^3\right).
\eea
The neutrino oscillations imply that the various flavors must reach an
equilibrium for which the chemical potentials are equal~\cite{Wong:2002fa,Dolgov:2002ab}.
Assuming accordingly that the asymmetry is the same for all flavors, the first
modification of the neutrino asymmetry is an excess of energy densities stored in neutrinos which can be
absorbed by a redefinition of the number of neutrino generations $N_\nu$ as
\bea
\tilde N_\nu &=& N_\nu\left(1+\frac{I_+^{(2,1)}(0,\xi_\nu) + I_+^{(2,1)}(0,-\xi_\nu)}{2 I_+^{(2,1)}}\right)\nonumber\\
&\simeq&N_\nu\left(1+\frac{30\xi_\nu^2}{7 \pi^2}+\frac{15\xi_\nu^4}{7 \pi^4}\right)\,.
\eea
This effect is very small for small values of $\xi_\nu$ since it is at least quadratic in $\xi_\nu$.

There is a second modification for the weak-interaction rates in which one must use instead of Eq.~(\ref{DefgnuANDg})
\be
g_\nu(E) \equiv \begin{cases} 
      g^+_\nu(E)\equiv\frac{1}{{\rm e}^{\left(E/T_\nu-\xi_\nu\right)}+1} & n \to p\,\, {\rm rates}\\
     g^-_\nu(E)\equiv \frac{1}{{\rm e}^{\left(E/T_\nu+\xi_\nu\right)}+1} & p\to n\,\,  {\rm rates}\,.\\
   \end{cases}
\ee
For instance in the definition (\ref{Defchipm}) of $\chi_\pm$, one
must use $g_\nu^\pm$ instead of $g_\nu$. To show this we used that the Pauli-blocking factor of the
antineutrino (resp. neutrino) is still related to the distribution of the
neutrino (resp. antineutrino) according to (\ref{PauliBlockingMagic})
thanks to the property (\ref{RelateChemicalNeutrinos}). This
modification is the most important one and modifies the thermodynamical equilibrium ratio between neutrons and protons
(\ref{DetailedBalanceCorrections}) because from Eq.~(\ref{muweak}) we
get $\mu_n-\mu_p = -\mu_\nu$. We get that at thermodynamical equilibrium
\be
\frac{n_n}{n_p} = \left.\frac{n_n}{n_p}\right|_{\xi_\nu=0} \times {\rm e}^{-\xi_\nu}\,,
\ee
and one expects a modification of the freeze-out abundance (for small $\xi_\nu$) of order
\be\label{EstimateDeltaYnphinu}
\frac{\Delta Y^F_n}{Y^F_n} \simeq \frac{\Delta \YP}{\YP} \simeq -(1-Y_n^F) \xi_\nu\,.
\ee

Finally there is a third modification when one considers the effect of incomplete
neutrino decoupling~\cite{Grohs:2016cuu} but this is negligible since
the effect without the asymmetry is already very small. For small $\xi_\nu$ we find the
abundance modifications
\beas\label{Sensitivityphinu}
\frac{\Delta \YP}{\YP} &\simeq&-0.96\,\xi_{\nu}\\
\frac{\Delta {\rm D}/{\rm H}}{{\rm D}/{\rm H}} &\simeq&-0.53\,\xi_{\nu}\\
\frac{\Delta {}^3{\rm He}/{\rm H}}{{}^3{\rm He}/{\rm H}}&\simeq&-0.18 \,\xi_{\nu}\\
\frac{\Delta {}^7{\rm Li}/{\rm H}}{{}^7{\rm Li}/{\rm H}} &\simeq&-0.62\,\xi_{\nu}\,.
\eeas 
The modification of $\YP$ is in very good agreement with the estimation (\ref{EstimateDeltaYnphinu}) since $(1-Y_n^F) \simeq
0.92$. Since BBN has no free parameter when assuming non-degenerate
neutrinos ($\xi_\nu=0$), the observational constraints on $\YP$ and
${\rm D}/{\rm H}$ can be used to obtain bounds on the degeneracy
parameter $\xi_\nu$ if we consider instead that it is
unknown. 

\begin{figure}[!htb]
\includegraphics[width=\mycolumnwidth]{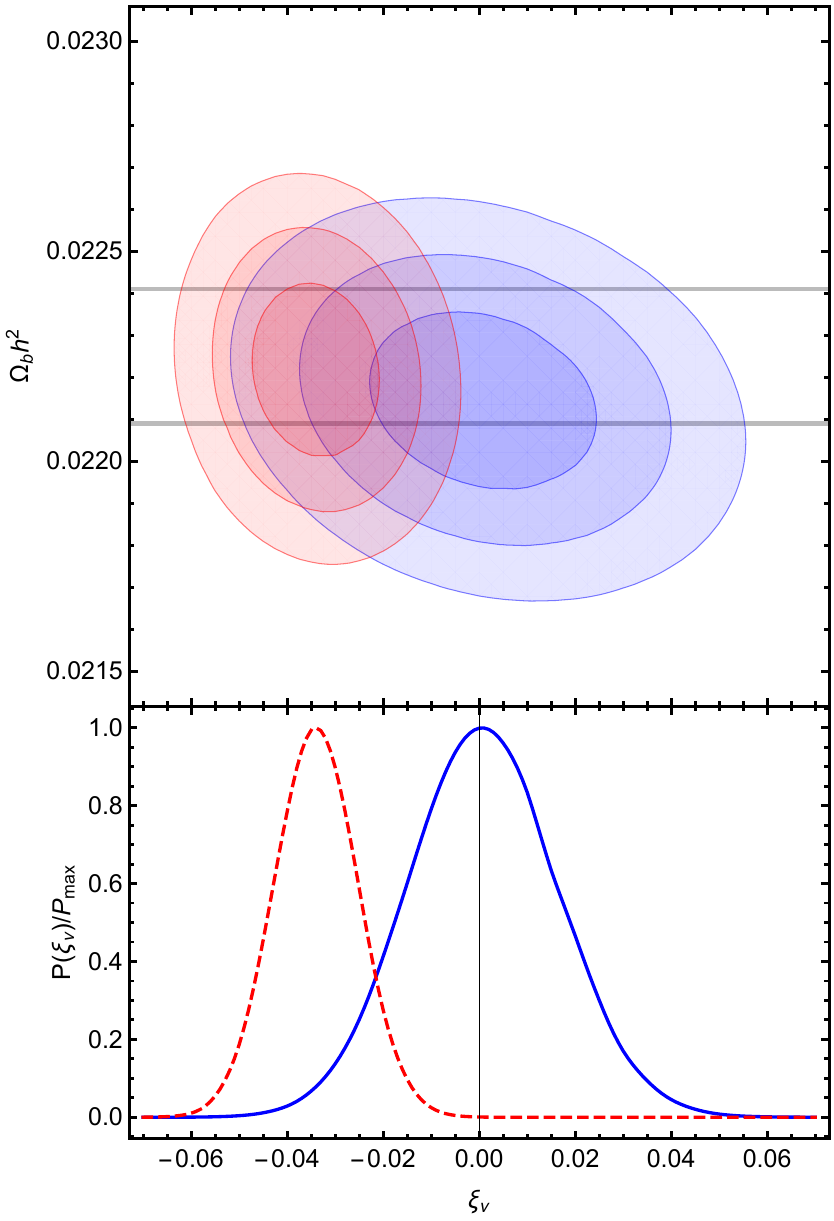}
\caption{{\it Top : } $P(\Omega_{\rm b}h^2,\xi_\nu)$ with $68.27\%$,
  $95.45\%$ and $99.73\%$ contours.  {\it Blue : } using the ${}^4{\rm He}$ bounds
  (\ref{YPAbundance}). {\it Red :} using the bounds $\YP=0.2551 \pm
  0.0022$ of \citet{Izo14}. The gray horizontal bars are the $\pm\sigma$
  CMB constraints on baryon abundance.  {\it Bottom : } marginalized distribution
  for $\xi_\nu$. Continuous line uses (\ref{YPAbundance}) whereas the
  dashed line uses  the bounds $\YP=0.2551 \pm  0.0022$ of \citet{Izo14}.  \label{Figphinu}}
\end{figure}

We repeat the analysis of \S~\ref{SecStatsBaryons}, using the pair of
cosmological parameters $(\omega_{\rm b},\xi_\nu)$ instead of
$\omega_{\rm b}$ alone. However, we still assume that the CMB prior
determines only $\omega_{\rm b}$.  Again, for each species the
uncertainty from nuclear rates $\sigma^{\rm th}_i(\omega_{\rm
  b},\xi_\nu)$ is obtained from a Monte-Carlo method. The posterior
distribution for $(\omega_{\rm b},\xi_\nu)$ from BBN combined with
CMB is plotted in Fig.~\ref{Figphinu}. Once marginalized over baryon abundance we get 
\be\label{Boundphinu}
\xi_\nu = 0.001 \pm 0.016\,,
\ee 
which is a much tighter
constraint than the constraints 
%from combined astrophysical probes~\citep[$-0.02 \pm 0.51$]{Nunes:2017xon}, 
from CMB alone~\citep[${\xi_\nu=-0.002}^{+0.053}_{-0.060}$]{Oldengott:2017tzj} or than earlier BBN constraints such as \citet[$\xi_\nu=0.037\pm0.026$]{Simha:2008mt}, thanks to
the recent improvement on both deuterium and ${}^4{\rm He}$ abundance
measurements. 

Even though the bounds (\ref{Boundphinu}) are tighter, they are also surprisingly still highly compatible with a vanishing
neutrino chemical potential. Since we find $|\xi_\nu| <
0.016\,(68\%\,{\rm CL})$, it corresponds to $(\tilde N_\nu
-N_\nu)/N_\nu < 0.011 \%$ and $\eta_\nu < 4.0\times 10^{-3}$. 

Note that if we had used {\tt Parthenope} \citep{Parthenope}, as is the case in \citet{Planck2016}, which predicts ${\rm D}/{\rm H} \simeq
2.61\times10^{-5}$ rather than ${\rm D}/{\rm H} \simeq 2.46\times10^{-5}$ for Planck parameters,  we would have found
rather $\xi_\nu = 0.021 \pm 0.016$ less compatible with a vanishing neutrino chemical potential. Indeed, in our case the observed
value for $\YP$ (resp. ${\rm D}/{\rm H}$) is lower (resp. higher) than the averaged value inferred from the CMB
measured baryon abundance, and given the linearized dependences
(\ref{Sensitivityphinu}) which have the same sign, this results in a
compensation and a central value very close to zero. However if the
prediction for deuterium from a BBN code is also above its observed value, then both $\YP$ and ${\rm D}/{\rm H}$ tend to favor a positive $\xi_\nu$.

\subsection{Number of neutrinos}\label{SecNumberNeutrinos}

%TODO Cyril. Write to say which date from Planck, which method etc... 
%TODO comparer aux courbes du Cyburt et Olive et aussi 
We repeat the analysis of \S~\ref{SecStatsBaryons} but for the pairs
of cosmological parameters $(\Omega_{\rm b}h^2,N_\nu)$ instead of the
baryon abundance alone. We use the CMB prior $P(\Omega_{\rm b}h^2,N_\nu | {\rm CMB})$ obtained from the
Monte-Carlo Markov chains {\tt  base\_nnu\_plikHM\_TTTEEE\_lowTEB} of
\citet{Planck2016} (TT+TE+EE+lowP analysis). The constraints from CMB are given directly in terms of $N_{\rm eff}$ (see \S~\ref{SecNeff}) since they are
obtained as the model-independent gravitational contribution of relativistic degrees of
freedom. However, during BBN the effective number of neutrinos evolves
as can be seen in Fig.~\ref{FigNeff}, and this evolution is
model-dependent. We assume that if there are extra relativistic degrees of freedom, they do
not share any of the energy which is brought to neutrinos by
incomplete decoupling. Hence for these extra neutrinos, or what is
described phenomenologically as extra neutrinos, $z_\nu=1$ throughout
BBN. Given these assumptions Eq.~(\ref{DefNeff}) which was derived for
$N_\nu=3$ becomes
\be\label{GeneralNeff}
N_{\rm eff} = 3 \left(\frac{z_\nu  z^{\rm stand}}{z}\right)^4+ (N_\nu-3) \left(\frac{z^{\rm stand}}{z}\right)^4\,,
\ee
from which we deduce in particular that $\dd N_{\rm eff}/\dd N_\nu =
\left(z^{\rm stand}/z\right)^4\simeq 1.0090$. In fact, this relation is
unchanged as long as the energy exchange between the plasma and the
neutrino sector remains identical to Eq.~(\ref{EqDefN}), that is if the function
${\cal N}(T)$ is unchanged. More generally the effective number of
neutrinos is insensitive to any type of neutrino spectral distortions
which leaves the total neutrino energy density unchanged. Hence, as
long as the energy transfer with the plasma is the same, we can also
assume that the new neutrino degrees of freedom share a part of the
energy transfer during the incomplete decoupling phase, and still use Eq.~(\ref{GeneralNeff}). This relation allows to convert the distribution $P(\Omega_{\rm b}h^2,N_{\rm eff} | {\rm
  CMB})$ into $P(\Omega_{\rm b}h^2,N_\nu | {\rm CMB})$. This approach is different from the one chosen by \citet{Cyb16} where
Eq.~(\ref{DefNeff}) is assumed to hold even for $N_\nu \neq
3$. However, given the rather large observational bounds on $N_\nu$
obtained, and the smallness of $z_\nu^4$, this difference in the
conversion of neutrino numbers is not crucial. 

\begin{figure}[!htb]
\includegraphics[width=\mycolumnwidth]{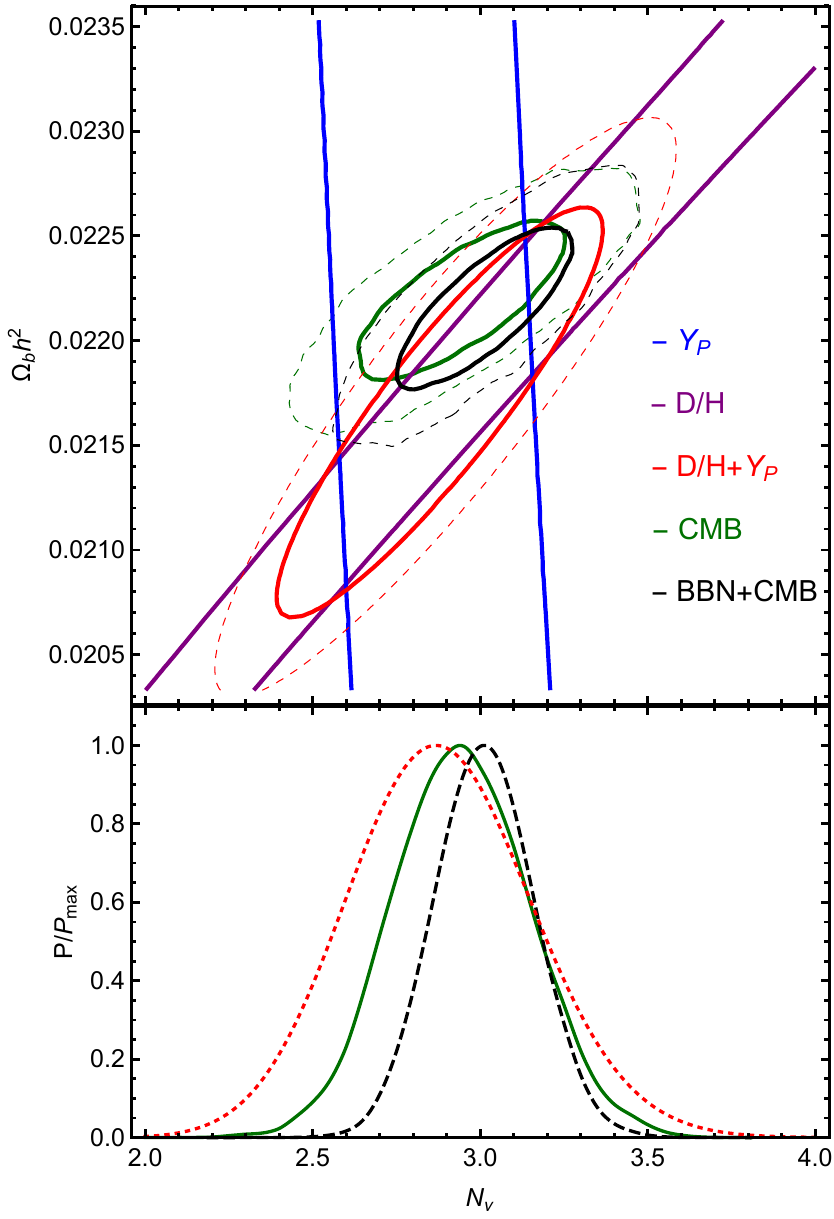}
\caption{Top : $P(\Omega_{\rm b} h^2,N_\nu)$. with $68.27\%$ and
  $95.45\%$ contours for different combinations of data. Bottom :
  $P(N_\nu)$ from marginalization. Continuous green is from CMB only,
  dotted red from BBN only, and dashed black is the combination of BBN
  and CMB. Note that the average value of $N_\nu$ for the combination of BBN and
  CMB is not between the corresponding averages obtained from CMB and
  BBN considered separately. There is no contradiction since the
  nearly elliptic preferred regions in the
  $(\Omega_{\rm b} h^2,N_\nu)$ space for BBN and CMB taken separately
  overlap away from the line defined by their respective average points.\label{FigNneutrinos}}
\end{figure}

The posteriors obtained are depicted in Fig.~\ref{FigNneutrinos}. They are to be
compared with the plots of \citet[Fig. 10]{Cyb16}. Furthermore, it presents a
significant improvement compared to \citet[Fig. 7]{Coo18}
thanks to the inclusion of the measured $\YP$ in the BBN
constraint. We used the full results of the chains for the CMB prior, but it can be
very well represented by a two-dimensional Gaussian distribution with $\Omega_{\rm
  b}h^2=0.022197\pm0.000245$, $N_\nu = 2.945\pm0.203$ and a
rather strong correlation coefficient $r=0.7699$, since this
approximation affects only marginally the posteriors obtained. Once marginalized over baryon abundance, the constraints on the
numbers of neutrinos are
\be
N_\nu = \begin{cases}   2.95 \pm 0.20& {\rm CMB} \\
2.88 \pm 0.27& {\rm BBN} \\
3.01 \pm 0.15& {\rm BBN+CMB.} \\
   \end{cases}
\ee
As for the marginalized baryon abundances we get
%TODO corriger cela.
\be
100\times\Omega_{\rm b}h^2 = \begin{cases}  2.220  \pm 0.025& {\rm CMB} \\
2.168 \pm 0.055& {\rm BBN} \\
2.216 \pm 0.022& {\rm BBN+CMB}. \\
   \end{cases}
\ee
Note that if we had used a code which predicts ${\rm D}/{\rm
H}\simeq 2.61\times10^{-5}$ instead of ${\rm D}/{\rm H}\simeq 2.46\times10^{-5}$ for the Planck
parameters, we would have obtained $N_\nu= 2.84 \pm 0.27$ from BBN
alone, corresponding to $N_{\rm eff} = 2.88 \pm 0.27$, whose central value is
similar to the one obtained in
\citet[Eq. 26]{Parthenope2}. Since we use consistently the CMB prior
$P(\Omega_{\rm b}h^2,N_\nu | {\rm CMB})$ when obtaining constraints from BBN and CMB, we cannot compare our
results with those of \citet{Parthenope2} in that case, since they use
only a CMB prior on $\Omega_{\rm b}h^2$.

\section*{Conclusion}

%TODO Check that agrees with numbers given
It is widely acknowledged that Cosmology has entered in the "precision era"; this should also apply to big bang nucleosynthesis. For both D and \qua\ isotopes the precision on
the abundances deduced from observations have reached the percent level.   
The precision on primordial D abundance prediction by BBN codes is now limited to a few 
percents because of the uncertainties on the \dpg, \ddn\ and \ddp\ thermonuclear reaction 
rates \citep{DiV14,Coc15}. Ongoing experiments  \cite{Gus17}  from LUNA at the Gran Sasso 
underground facility, supplemented by theoretical works \cite{Mar16} are expected to improve
the situation. Here, we concentrated on the prediction of \qua\ primordial abundance. 
Uncertainties on $\YP$ related to experimental data come from the neutron lifetime    
 $879.5 \,(\pm0.8)\,{\rm s}$ (but that may be affected by systematic uncertainties \citep{PDG17}),
 leading to a $\Delta\YPfour=1.7$ (Eq.~\ref{RuleOfThumb}) uncertainties and from 
 the \npg, \ddn\ and \ddp\ reaction rates ($\sim$1\% factor uncertainty \citep{AndoEtAl2006,Bayes17}) leading
 to $\Delta\YPfour\lesssim$0.5 in total (Eq.~\ref{q:dsigv}). These uncertainties are small, compared with
 the observational uncertainty of $\Delta\YPfour=40$  (Eq.~\ref {YPAbundance}). However, the
 predicted \qua\ primordial abundance includes corrections to the
 "bare" weak rates: zero--temperature radiative corrections, finite nucleon mass corrections, finite temperature radiative corrections,
weak-magnetism, QED plasma effects and incomplete neutrino decoupling that, in total, shift the abundance by $\delta\YPfour=44.7$ (Table~\ref{TableCorrections}), i.e. larger than the uncertainties.

It is thus of the utmost importance to precisely calculate all these corrections, in order
to limit theoretical uncertainties. Here, they are for the first time all included and calculated in a self consistent way allowing to take into account the correlations between them.
In addition, it was verified that all satisfy detailed balance, a crucial point since it
directly affects the neutron/proton number ratio, and hence the  \qua\ abundance.
Table~\ref{TableCorrections} details the contributions of these corrections to the
\hli\ primordial abundance.
We did not calculate the effect of incomplete neutrino decoupling, but use the 
results of   \citet{Parthenope}.  This amounts to ignoring the spectral  distortions,
but this affects $\YPfour$ by approximately less than one unit. 
However we find results for incomplete neutrino decoupling effects
which are very similar to \citet{Grohs:2015tfy}, and most notably we
find a coherence on the sign of light element abundances variations
which is different from \citet{Mangano2005}. Given the coupling we find
with finite nucleon mass effect, this is the very last correction which requires careful
evaluation to fully settle the weak rates corrections.

In this work, we have been using a network of $\approx$400 nuclear reactions (and their reverse),
from neutron to sodium to encompass the light element big bang nucleosynthesis up to the CNO 
isotopes \citep{Coc12a}. We used the most up to date reaction rates, and in particular those
involved in deuterium production \citep{Bayes16,Bayes17}, but even more important, the weak
rates with their carefully calculated corrections. Hence, we claim that our predicted abundances (Table~\ref{TableUncertainty})
are the most accurate to date, not only for D, \tro, \sep\ and CNO isotopes but also for \qua.
Our predicted deuterium and \qua\ primordial abundances are in agreement with observations, 
within error bars (Fig.~\ref{FigEta} and Table~\ref{t:hlix}), and in particular, there is no need for extra relativistic 
degrees of freedom (i.e. $N_\nu>3$). Our predicted lithium abundance remains a factor of $\approx$3
above the Spite plateau. The solution of this problem probably involves several mechanisms including
stellar depletion and possibly some physics beyond the standard model. Finally, our prediction of
CNO abundance (Table~\ref{TableUncertainty}) does not completely rule out the possibility that they may influence the evolution 
of some of the first, Population III, stars \cite{Eks08}.  

Last, but not least, we provide
at~\url{http://www2.iap.fr/users/pitrou/primat.htm}, a freely
available {\it Mathematica} code
that includes all the physics discussed in this work. We expect that it will be used and hopefully modified
in order to include new physics (see e.g. \citet{IoccoReport,Mat17a}), or just to update some reaction rates
\footnote{They can be provided in tabular form, in a format similar to the Starlib one \citep{Starlib} at~\url{https://starlib.github.io/Rate-Library/},
or as (possibly parameter dependent) analytical formulae.}.

\bigskip
%@@@@

%TODO Say that our results are super complete but do not include
%neutrinos non-thermal spectrum with neutrino oscillations, and that
%this is also a fraction for $\YPfour$. 

%\ifphysrep
%\else
%\bibliography{BiblioBBN}
%\fi

\begin{acknowledgments}

We are indebted to our collaborators on these topics:  Pierre Descouvemont, Brian Fields, St\'ephane Goriely,  Fa\"{\i}rouz Hammache, 
Christian Iliadis, Keith Olive, Patrick Petitjean.
C.P. thanks A. Sirlin and W. J. Marciano for discussions on radiative corrections and neutron lifetime, D. Seckel for comments on \cite{Seckel1993},
and G. Lavaux for discussions on Bayesian inference. We also thank
warmly Monique and Fran{\c c}ois Spite for their kind help regarding the
Lithium data, and Maxim Pospelov for discussions on radiative corrections.

\end{acknowledgments}

\appendix

\section{Thermodynamics}\label{AppThermo}

\subsection{Thermodynamical quantities}\label{AppThermoQuantities}

From the distribution function of a given species, it is
possible to define macroscopic quantities such as number density $n$,
energy density $\rho$, and pressure $P$. Due to the isotropy of the FL
spacetime, the distribution function depends only on the magnitude of
spatial momenta, that is it is of the form $f(t,p)$. Furthermore
isotropy implies that there is no anisotropic stress. Omitting the
time dependence to alleviate the notation, we define
\beas\label{ThermoFromf}
n&=& g \int f(p) \frac{4\pi p^2 \dd p}{(2\pi)^3} \\
\rho&=& g \int f(p) E\frac{4\pi p^2 \dd p}{(2\pi)^3}\\ 
P&=& g \int f(p)\frac{p^2}{3E} \frac{4\pi p^2 \dd p}{(2\pi)^3} \slabel{DefPressure1}
\eeas
with $E= \sqrt{p^2+m^2}$ and $g$ the number of spin degrees of
freedom of the species considered ($g=1$ for neutrinos and $g=2$ for
all other species). Using $E \dd E = p \dd p$, and given that
both Fermi-Dirac and Bose-Einstein distributions are given as
functions of $E$ and not $p$, it is often more convenient to rewrite
these expressions as
\begin{subeqnarray}\label{ThermoQuantities}
n&=& \frac{g}{2\pi^2} \int f(E) p E \dd E\\
\rho&=& \frac{g}{2\pi^2} \int f(E) p E^2 \dd E\\
P&=& \frac{g}{6\pi^2} \int f(E) p^3 \dd E \slabel{DefPressure2}
\end{subeqnarray}
with $p=\sqrt{E^2-m^2}$.

Bose-Einstein and Fermi-Dirac with statistical physics convention for
chemical potential are defined as
\be\label{FDBE}
g^\pm_{T,\mu}(E) \equiv \frac{1}{{\rm e}^{\frac{E-\mu}{T}} \pm 1}
\ee
with upper (lower) sign for fermions (bosons). 
For these distributions we find from Eqs.~(\ref{ThermoQuantities})
\beas
n(t) &=& \frac{g T^3}{2\pi^2} I_{\pm}^{(1,1)}(x,\xi)\\
\rho(t) &=& \frac{g T^4}{2\pi^2} I_{\pm}^{(2,1)}(x,\xi)\\
P(t) &=& \frac{g T^4}{6\pi^2} I_{\pm}^{(0,3)}(x,\xi)
\eeas
with $x\equiv m/T$ and $\xi \equiv \mu/T$ and the integrals
\beas\label{DefIpq1}
I_\pm^{(m,n)}(x,y) &=&\int_x^\infty \frac{u^m (u^2-x^2)^{n/2}}{{\rm
    e}^{u-y}\pm1}\dd u \,,\\
&=&\int_0^\infty \frac{(v^2+x^2)^{(m-1)/2} v^{n+1}}{{\rm e}^{\sqrt{v^2+x^2}-y}\pm1}\dd v .
\eeas
When the chemical potential can be neglected ($\xi \ll1$) we also use the notation
\be\label{DefIpq2}
I_\pm^{(m,n)}(x) \equiv I_\pm^{(m,n)}(x,0)\,.
\ee
Finally when the mass can also be neglected ($x\ll 1$) we define
\be\label{DefIpq3}
I_\pm^{(m,n)}\equiv I_\pm^{(m,n)}(0,0)\,,
\ee
whose most useful values are reported in table~\ref{FigImn}.
\begin{table}[!htb]
\caption{Integrals involved in the expressions of thermodynamical
  quantities when $x\ll 1$ and $\xi \ll 1$.\label{FigImn}}
\begin{tabularx}{\ifphysrep 0.4\columnwidth \else 0.75\columnwidth \fi}{AAA}
\toprule
   & Bosons ($-$) & Fermions ($+$)\\
 \midrule
  $I^{(0,1)}_\pm$&$\pi^2/6$ &$\pi^2/12$  \\
  $I^{(1,1)}_\pm$&$2 \zeta(3)$ &$3 \zeta(3)/2$  \\
  $I^{(2,1)}_\pm$& $\pi^4/15$&  $7 \pi^4 /120$\\
  $I^{(0,3)}_\pm$& $\pi^4/15$ &  $7 \pi^4/120 $\\
\bottomrule
\end{tabularx}
\end{table}
When the temperature is much smaller than the mass of the particles,
that is for $x=m/T \gg 1$, we find the approximate relations
\beas
n&\simeq& g T^3 \left(\frac{x}{2 \pi}\right)^{3/2} {\rm e}^{\xi-x} \slabel{nLowT}\\
\rho &\simeq& \left(m +\frac{3}{2} T\right) n\,.\slabel{rhoLowT}
\eeas

\subsection{Chemical potential of electrons}\label{AppChemicalPotential}

\begin{figure}[!htb]
     \includegraphics[width=\mycolumnwidth]{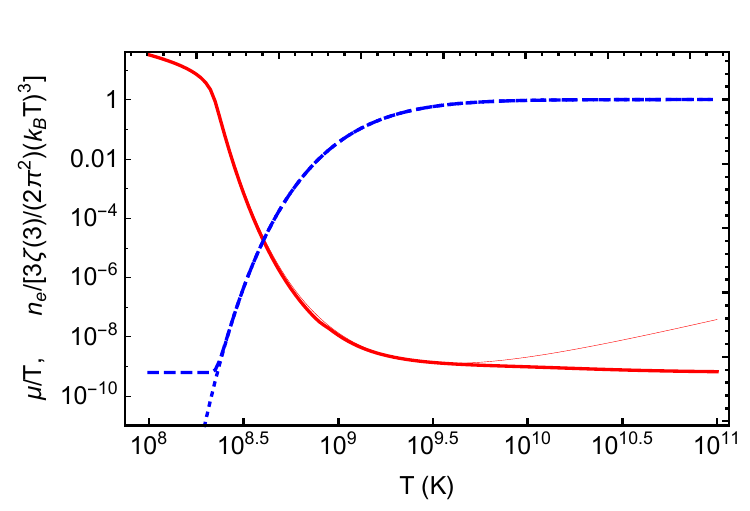}
     \caption{
{\it Continuous red lines} : Thick line is $\mu_e/T$ computed exactly from
Eq.~(\ref{TrueChemicalPotentialConstraint}), thin line is the
approximation (\ref{ApproxChemicalPotential}). 
{\it Dashed blue line} : $2\pi^2/[3 \zeta(3)]\,n_e/(k_B T)^3$ with the chemical potential taken into account.  
{\it Dotted blue line} : $2\pi^2/[3 \zeta(3)]\,n_e/(k_BT)^3$ with vanishing chemical  potential.}
\label{FigChemicalPotential}
\end{figure}

In chemical equilibrium $\mu_{e^-}=-\mu_{e^+}\equiv \mu_e$ because of
reactions $2\gamma \leftrightarrow e^++e^-$.  Hence the net negative charge density in electrons and positrons is
\be
n_{e^-}-n_{e^+} = \frac{2 T^3}{2\pi^2} \left[I_+^{(1,1)}(x,\xi)-I_+^{(1,1)}(x,-\xi)\right]
\ee
with $x\equiv m_e/T$ and $\xi \equiv \mu_e/T$. The chemical potential of electrons
and positrons must adapt to ensure the electric neutrality of the
Universe, hence it is constrained by
\be\label{TrueChemicalPotentialConstraint}
n_{e^-}-n_{e^+} = n_p\,.
\ee
Let us estimate the chemical potential at the end of BBN for
simplicity. Indeed, since $\xi$ increases with time, it is enough to check that its final
value at the end of BBN is small. We use therefore the final value  of baryon-to-photon ratio $\eta$ [defined in
Eq.~(\ref{DefEta})] and use that the baryonic matter is essentially in
protons and ${}^4{\rm He}$ nuclei. We find
\be
n_{e^-}-n_{e^+}  = (1-\YP/2) \eta_0\, \bar n_\gamma T^3\,. 
\ee
In that limit, the electric charge density of electrons and positrons is given by
\be 
n_{e^-}-n_{e^+} \simeq 4 T^3 \left(\frac{x}{2\pi}\right)^{3/2} {\rm e}^{-x} \sinh \xi\,.
\ee
Combining these two equations we obtain 
\be\label{ApproxChemicalPotential}
\xi \simeq {\rm Arcinsh}\left(\frac{\eta (1- \YP/2){\rm e}^x \sqrt{2} \zeta(3)}{\sqrt{\pi} x^{3/2}}\right)\,.
\ee
The evolution of the electrons chemical potential is depicted in
Fig.~\ref{FigChemicalPotential}. $\xi=\mu_e/T$ stays much below
$10^{-5}$ whenever $T>5\times 10^8 {\rm K}$. For lower temperatures, $\xi$
rises but it is only because electron-positron annihilations are
nearly complete and the chemical potential adapts so that the relic density of
electrons is equal to the number of protons. That is, whenever the
chemical potential of electrons rises at low temperature, the amount
of electrons is insignificant due to the low value of the baryon to
photon ratio. Hence for all practical purposes, from Eq.~(\ref{DotS}) we
deduce that whenever there is no heat exchange with the neutrinos, the
total entropy of the plasma is conserved, that is $\dot S_{\rm pl} =
\dd(s_{\rm pl}
a^3)/\dd t=0$.
\subsection{Nucleons at thermodynamical equilibrium}\label{SecMaxwellian}

The chemical potential of neutrons and protons are not negligible
individually, and they are constrained to give the number density
in the low temperature limit~(\ref{nLowT}). However the difference $\mu_n-\mu_p$ is negligible with
respect to $T$. Indeed we can neglect $\xi_e \equiv \mu_e/T$ (see \S~\ref{AppChemicalPotential}) and assuming it
is also the case for neutrinos (see \S \ref{SecNeutrinoPhi} for degenerate neutrinos), and given that at
equilibrium we have necessarily
\be\label{muweak}
\mu_n+\mu_\nu = \mu_p+\mu_e \quad \Rightarrow \quad
\mu_n-\mu_p=\mu_e-\mu_\nu\,,
\ee
then we deduce $\mu_n-\mu_p\ll 1$.

Hence from (\ref{nLowT}) we obtain that the neutron to proton ratio is
\be\label{DetailedBalanceCorrections}
\frac{n_n}{n_p} = \left(\frac{m_n}{m_p}\right)^{3/2}{\rm
  e}^{-\frac{\Gap}{T}}\simeq {\rm
  e}^{-\frac{\Gap}{T}} \left(1+\frac{3\Gap}{2 m_N}\right)\,.
\ee

It is possible to determine the chemical potential of neutrons and
protons from the expression (\ref{nLowT}) and the baryon-to-photon number ratio $\eta$. For instance for protons, we must have
\be
\bar n_p \equiv \frac{n_p}{T^3} = 2
\left(\frac{x_p}{2\pi}\right)^{3/2}{\rm e}^{\xi_p-x_p}=\eta
\frac{n_p}{n_{\rm b}} \bar n_\gamma = Y_p \eta \bar n_\gamma
\ee
with $x_p \equiv m_p/T$ and $\xi_p=\mu_p/T$, which can be used to solve for $\xi_p$. In particular the distribution
function of protons is approximately given by 
\be
f_p(p) \simeq  {\rm e}^{\frac{\mu_p-E}{T}} \simeq {\rm exp}\left(\xi_p-x_p -\frac{p^2}{2m_p T}\right)
\ee
and thus it is approximated by the Maxwellian distribution
\bea\label{Lowfprotons}
f_p(p) &\simeq& \frac{n_p}{2}\left(\frac{2\pi}{m_p T}\right)^{3/2} {\rm exp}\left(-\frac{p^2}{2m_p T}\right)\\
&\simeq& Y_p \frac{\eta \zeta(3)\sqrt{8}}{\sqrt{\pi}}\left(\frac{T}{m_p}\right)^{3/2}{\rm exp}\left(-\frac{p^2}{2
    m_p T}\right)\,.\nonumber
\eea
From this last equation we deduce that $f_p(p) \ll 1$, justifying that
we neglect Pauli-blocking effects in reactions rates for protons (and
similarly for neutrons).

The distribution function for an isotope $i$ is exactly similar with the
obvious replacement $Y_p \to Y_i$, $n_p \to n_i$ and $m_p \to m_i$. The distribution of velocities is defined as (omitting the $i$ index of the isotope considered)
\be
\phi_\mathrm{MB}(v)\dd v \equiv \frac{1}{n}\frac{\dd n}{\dd p}
\dd p=\frac{2}{n} f(p)\frac{4\pi p^2 \dd
  p}{(2\pi)^3}\,,
\ee
where $ m v = p$, and using the kinetic energy $E \equiv \tfrac{1}{2}
m v^2 \,\Rightarrow\, \dd E = m v \dd v$, it takes the form given in
Eq. (\ref{q:mb}).

\subsection{Abundances at nuclear statistical equilibrium}

\begin{figure}[!htb]
\centering
    \includegraphics[width=\mycolumnwidth]{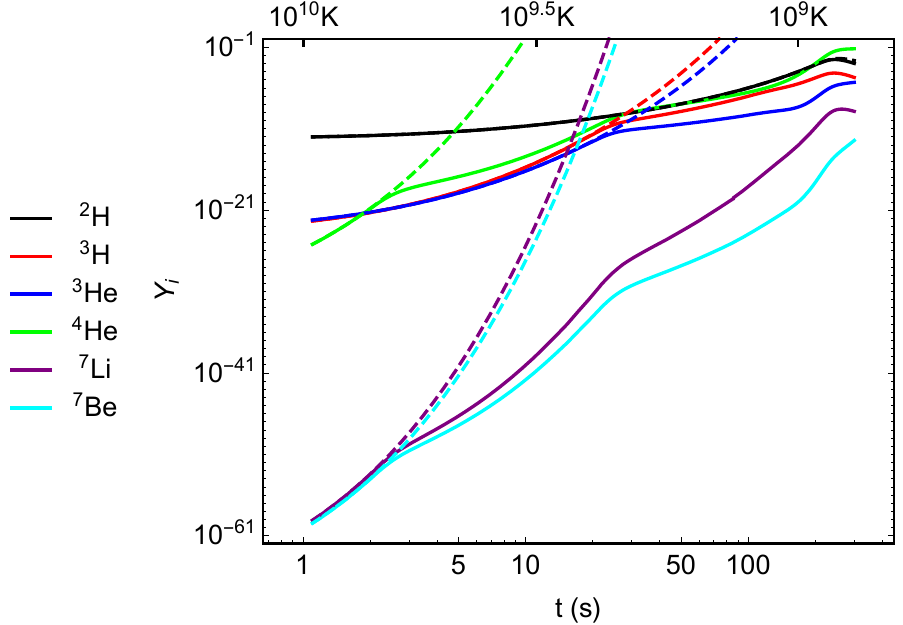}
\caption{Evolution of  the first elements abundances in solid lines, together with the
  nuclear statistical equilibrium values in dashed lines. The
  deuterium abundance stays very close to its NSE value until the
  time it is more efficiently destroyed than formed around $t \simeq 200 \,{\rm
  s}$.
\label{FigNSE}}
\end{figure}

At nuclear statistical equilibrium (NSE), a given isotope $i$ is at chemical
equilibrium with its nucleons, hence the relation between chemical potentials
\be\label{ChemicalPotentialSpecies}
\mu_i = Z_i \mu_p +(A_i-Z_i) \mu_n\,.
\ee
From Eq.~(\ref{nLowT}) 
\be\label{niequilibrium}
n_i= g_i \left(\frac{m_i T}{2\pi}\right)^{3/2}
{\rm e}^{\xi_i-x_i}\,,\quad g_i \equiv 2 s_i +1\,,
\ee
where $s_i$ is the spin of species $i$. Using Eq.~(\ref{ChemicalPotentialSpecies}) and (\ref{niequilibrium}), we get 
\ifphysrep
\be\label{niThermal}
n^{\rm NSE}_i = \frac{g_i
  m^{3/2}_i}{2^{A_i}}\left(\frac{n_p}{m^{3/2}_p}\right)^{Z_i}
\left(\frac{n_n}{m^{3/2}_n}\right)^{A_i-Z_i} \left(\frac{2
    \pi}{T}\right)^{\frac{3(A_i-1)}{2}}{\rm e}^{B_i/T}
\ee
\else
\bea\label{niThermal}
n^{\rm NSE}_i &=& \frac{g_i
  m^{3/2}_i}{2^{A_i}}\left(\frac{n_p}{m^{3/2}_p}\right)^{Z_i}
\left(\frac{n_n}{m^{3/2}_n}\right)^{A_i-Z_i} \nonumber\\
&&\times \left(\frac{2
    \pi}{T}\right)^{\frac{3(A_i-1)}{2}}\,{\rm e}^{B_i/T}
\eea
\fi
where we defined the binding energy
\be
B_i \equiv Z_i m_p + (A_i-Z_i) m_n - m_i\,.
\ee
Using (\ref{DefEta}), the abundances are then given at NSE by
\ifphysrep
\be\label{YiThermal}
Y^{\rm NSE}_i = g_i \zeta(3)^{A_i-1}2^{\frac{3A_i
      -5}{2}} \pi^{\frac{1-A_i}{2}}\left(\frac{m_i T^{A_i-1}}{m_p^{Z_i}
    m_n^{A_i-Z_i}}\right)^{3/2} \eta^{A_i-1} Y_p^{Z_i} Y_n^{A_i-Z_i}
{\rm e}^{B_i/T}\,.
\ee
\else
\bea\label{YiThermal}
Y^{\rm NSE}_i &=& g_i \zeta(3)^{A_i-1}2^{\frac{3A_i
      -5}{2}} \pi^{\frac{1-A_i}{2}}\left(\frac{m_i T^{A_i-1}}{m_p^{Z_i}
    m_n^{A_i-Z_i}}\right)^{3/2} \nonumber\\
&&\times \eta^{A_i-1} Y_p^{Z_i} Y_n^{A_i-Z_i}
{\rm e}^{B_i/T}\,.
\eea
\fi
We check in particular that for neutrons and protons $B_n=B_p=0$ and
$A_n=A_p=1$, so we get the tautological relation $Y_n^{\rm NSE}=Y_n$ and $Y_p^{\rm NSE}=Y_p$.
In order to obtain the NSE values from (\ref{YiThermal}), we use
\citet{Aud17} for the masses and spins of nuclear elements. In Fig. \ref{FigNSE}, we plot jointly the abundances of the main isotopes together with their NSE values to
check that they are equal at high temperatures.

\section{Weak reactions rates}\label{Apprates}

In this section, we gather all the technicalities required to obtain
the theoretical forms of the weak-rates including the most relevant
corrections.

\subsection{General expressions}\label{AppWeakGeneralExpressions}

Let us consider first the reactions with neutrons in the initial
state. Throughout this section we use the mostly minus metric
signature since it is the most common in particle physics. The general
expression for each of the three reactions involved is of the form~\cite{Lopez1997,Fidler:2017pkg}
\ifphysrep
\be\label{GeneralGamma}
n_n \Gamma = \int \Pi_{i} [\dd^3 \gr{p}_i] 
(2\pi)^4 \delta^4\left(\fourp{p}_n -\fourp{p}_p+\coeffa_\nu {\fourp{p}}_\nu +\coeffa_e \fourp{p}_e\right) \left| M\right|^2  f_n({E}_n) [1-f_p({E}_p)] f_\nu(\coeffa_\nu {E}_\nu)
f_e(\coeffa_e {E}_e)
\ee
\else
\bea\label{GeneralGamma}
&&n_n \Gamma = \int \Pi_{i} [\dd^3 \gr{p}_i] 
(2\pi)^4 \delta^4\left(\fourp{p}_n -\fourp{p}_p+\coeffa_\nu {\fourp{p}}_\nu +\coeffa_e \fourp{p}_e\right) \nonumber\\
&&\quad\times \left| M\right|^2  f_n({E}_n) [1-f_p({E}_p)] f_\nu(\coeffa_\nu {E}_\nu)
f_e(\coeffa_e {E}_e)
\eea
\fi
where the $\fourp{p}_i$ with $i=n,p,e,\nu$ are the four-momenta of
particles, and we used the compact notation for the relativistic volume element
\be
[\dd^3 \gr{p}]\equiv\frac{\dd^3 \gr{p}}{2 E (2\pi)^3} = \frac{4\pi p^2
\dd p}{2 E (2\pi^3)}\,.
\ee
The factor $\coeffa_\nu$ is $+1$ if the neutrino is in the initial
state and $-1$ otherwise with a similar definition for $\coeffa_e$ and
the electron or positron. In (\ref{GeneralGamma}), it is apparent that the Dirac function ensures energy and momentum
conservation. For a Fermi-Dirac distribution 
\be
g(-E) = 1-g(E)\,,
\ee
implying that the coefficients $\coeffa_e,\coeffa_\nu$ also ensure that
either it is the distribution function which appears if the lepton is
in the initial state, or the Pauli-blocking factor if it is in the
final state.

A first simplification consists in neglecting the Pauli-blocking
factor of the final state proton due to the very low value of the
baryon-to-photon ratio [see Eq.~(\ref{Lowfprotons})]. Having no
dependence on the distribution function of the final protons, we can simplify Eq.~(\ref{GeneralGamma})
further by performing the integral on spatial momenta of protons, using the spatial
part of the Dirac function. The reaction rates reduces to
\ifphysrep
\be\label{GeneralGamma2}
n_n \Gamma = \int \frac{ \dd^3 \gr{p}_n \dd^3 \gr{p}_e \dd^3
  \gr{p}_\nu }{2^4 (2\pi)^8}\delta\left(E_n-E_p+\coeffa_e E_e
  +\coeffa_\nu E_\nu\right) \frac{\left| M\right|^2}{E_n E_p E_e E_\nu}  f_n(E_n)
f_\nu(\coeffa_\nu {E}_\nu) f_e(\coeffa_e {E}_e) 
\ee
\else
\bea\label{GeneralGamma2}
n_n \Gamma &=& \int \frac{ \dd^3 \gr{p}_n \dd^3 \gr{p}_e \dd^3
  \gr{p}_\nu }{2^4 (2\pi)^8}\delta\left(E_n-E_p+\coeffa_e E_e
  +\coeffa_\nu E_\nu\right) \nonumber\\
&&\times \frac{\left| M\right|^2}{E_n E_p E_e E_\nu}  f_n(E_n)
f_\nu(\coeffa_\nu {E}_\nu) f_e(\coeffa_e {E}_e) 
\eea
\fi
where the proton momentum and energies are related by
\be
\gr{p}_p=\gr{p}_n+\coeffa_\nu\gr{p}_\nu+\coeffa_e\gr{p}_e\,,\quad E_p =
\sqrt{\gr{p}_p\cdot \gr{p}_p +m_p^2}\,.
\ee 
For a given reaction, $\left| M\right|^2$ is the corresponding
matrix-element of the weak interaction summed over all initial and
final states, computed from the interaction Hamiltonian. For the
reactions (\ref{AllWeakReactions}), the relevant part of the
interaction Hamiltonian is a coupling of weak currents in the form 
%(see e.g. \citet{Fidler:2017pkg,Seckel1993})
\be\label{DefHi}
{\cal H}_I = \frac{G_F}{\sqrt{2}}J^\mu_{e\nu} J_{pn,\,\mu}
\ee
where $G_F$ is the Fermi constant. The electron-neutron weak current
is purely left chiral, and of the form
\be
J^\mu_{e\nu} = \bar{\bm \nu}\gamma^\mu(1-\gamma^5) \gr{e}\,.
\ee
$\gr{e}$ and ${\bm \nu}$ are the fermionic quantum fields of electrons
and neutrinos, and the $\gamma^\mu$ are the matrices of the Clifford
algebra [we use the conventions of \cite{Fidler:2017pkg}].  The
proton-neutron current is of the form~\citep[Eqs.~1 and A2]{Ivanov:2012qe}
\be\label{HadronicCurrent}
J^{\mu}_{pn} = \cos \theta_{\rm C} \bar{\gr{p}} \left(\gamma^\mu(1-g_A
  \gamma^5)+\ii \frac{f_{\rm wm}}{m_N} 2\Sigma^{\mu\nu} q_\nu\right)\gr{n}
\ee
where $\gr{p}$ and ${\bm n}$ are the Fermionic quantum fields of
protons and neutrons and $2\Sigma^{\mu\nu} \equiv \ii/2
(\gamma^\mu\gamma^\nu-\gamma^\nu\gamma^\mu)$. 
$g_A$ is the axial current constant for nucleons (also sometimes
written $C_A/C_V$ or $\lambda$ in the literature), and $f_{\rm wm}$ is
the weak magnetism constant whose numerical value is given by~\cite{HorowitzLi,Horowitz:2001xf,Ivanov:2012qe}
\be
f_{\rm wm} =\frac{\mu_p-\mu_n}{2}\simeq \frac{1.793-(-1.913)}{2}\simeq 1.853\,.
\ee
$\theta_{\rm C}$ is the Cabbibo-Kobayashi-Maskawa (CKM) angle ($\cos
\theta_{\rm C}$ also noted $V_{ud}$ in the literature), $q^\mu$ is the
nucleon four-momentum transfer, that is the difference
between the final nucleon four-momentum and the initial one. The most recent numerical values for the constants in weak interactions are gathered in appendix~\ref{ParticleValues}.

It is possible to show that from
Eqs.~(\ref{DefHi})-(\ref{HadronicCurrent}) the matrix element is of
the form [see e.g. \cite{Fidler:2017pkg}]
\be\label{DefMLLMRRMLR2}
\frac{|M|^2}{2^7 G_F^2}= c_{LL }{\cal M}_{LL} +
c_{RR} {\cal M}_{RR} +c_{LR} {\cal M}_{LR}
\ee
with the coupling factors
\beas\label{DefgLLgRRgLR}
c_{LL} &\equiv& \frac{(1+g_A)^2}{4}\\
c_{RR} &\equiv& \frac{(1-g_A)^2}{4}\\
c_{LR} &\equiv &\frac{g_A^2-1}{4}\,.
\eeas
If weak interactions for nucleons were purely left-chiral, that is
with $g_A=1$, only the ${\cal M}_{LL}$ would contribute. ${\cal
  M}_{RR}$ contains the contribution of the right-chiral part of the interaction, and ${\cal M}_{LR}$ is an
interference between left and right chiral contributions. 

Let us ignore first the contribution from weak-magnetism that are
investigated further in \S~\ref{AppWM}. We then find
\beas\label{papbpcpd}
{\cal M}_{LL} &=& (\fourp{p}_n \cdot \fourp{p}_\nu)(\fourp{p}_p \cdot \fourp{p}_e)\\
{\cal M}_{RR} &=& (\fourp{p}_n \cdot \fourp{p}_e)(\fourp{p}_p \cdot \fourp{p}_\nu)\\
{\cal M}_{LR} &=& m_p m_n (\fourp{p}_\nu \cdot \fourp{p}_e)\,.
\eeas
It is worth noting that all weak reactions (\ref{AllWeakReactions})
have in fact the same matrix element. First the three reactions with a
neutron in the initial state are obtained by crossing
symmetry. Crossing symmetry for the electron amounts for instance to
the formal replacement $\fourp{p}_e \to - \fourp{p}_e$ with an overall minus
sign~\cite{Fidler:2017pkg} and it is straightforward to check that
this leaves the matrix element invariant. The same property arises
obviously for neutrinos. Second, the three reverse reactions (with an initial
proton and a final neutron) are obtained by time reversal, and since there is no CP
violation at this level, there is also a time-reversal symmetry so as
to ensure the CPT symmetry. Hence all reverse rates have exactly the
same matrix element. 

\subsection{Fokker-Planck expansion}\label{SecFokker}

The integral in (\ref{GeneralGamma2}) is $8$-dimensional when one
removes the Dirac function. Due to the isotropy of all distributions,
this can be reduced to a $5$-dimensional integral. This is the method
followed by \citet{Lopez1997}. Here we follow a much simpler route
by performing a Fokker-planck expansion in the energy transferred in
reactions. Even though our method is different from
\citet{Seckel1993}, it follows a method which is similar in
spirit. As we shall see, this results in one-dimensional integrals which are much faster to evaluate. 

At low temperature, it is enough to assume that nucleons follow an isotropic
Maxwellian distribution of velocities at the plasma temperature
$T$ given by Eq.~(\ref{Lowfprotons}). Hence the following integrals are obtained
\beas\label{IntMaxwellian}
2\int f_N(\gr{p}) \frac{\dd^3 \gr{p}}{(2\pi)^3} &=& n_N\,,\slabel{Nucleonrule1}\\
2\int f_N(\gr{p}) \frac{p^i}{m_N}\frac{\dd^3 \gr{p}}{(2\pi)^3} &=&0\,, \slabel{Nucleonrule2}\\ 
2\int f_N(\gr{p}) \frac{p^i p^j}{m_N^2}\frac{\dd^3 \gr{p}}{(2\pi)^3} &=& 
\frac{T}{m_N} \delta^{ij} n_N\,. \slabel{Nucleonrule3}
\eeas
In particular contracting with $\delta_{ij}$ we recover the expression for the pressure of nucleons
in the low temperature limit
\be
P_N = 2 \int f_N(\gr{p}) \frac{p^2}{3m_N}\frac{\dd^3 \gr{p}}{(2\pi)^3}
= T n_N\,.
\ee
For electron or neutrino distributions, since we have assumed
isotropy, we deduce the property
\be
\int g(E) p^\alpha E^\beta p^i p^j \frac{\dd^3 \gr{p}}{(2\pi)^3} = \frac{\delta^{ij}}{3}\int g(E) p^{\alpha+2} E^\beta \frac{\dd^3 \gr{p}}{(2\pi)^3}
\ee
where $\alpha$ and $\beta$ are some numbers. From isotropy we also find that
\be
\int g(E) p^\alpha E^\beta p^i \frac{\dd^3 \gr{p}}{(2\pi)^3} =0\,.
\ee
Hence for all practical purposes, we can perform the replacements 
\be\label{CutRule}
p^i p^j  \to  p^2 \delta^{ij}/3\,,\qquad p^i \to 0\,.
\ee 
on all species, resulting in great simplifications.

We assess the importance of corrections to the Born approximations as
explained in \S~\ref{SecFM}, that is in powers of $\epsilon\equiv T/m_N$. To evaluate the order of each term, we consider that the momentum or energies of neutrinos are of order $T
\sim \Gap$, that is factors of the type $E_e/m_N$ or $E_\nu/m_N$ are
of order $\epsilon$. Furthermore, from (\ref{IntMaxwellian}) a factor $\gr{p}_n/m_n$ is of order $\sqrt{T/M}
\sim \sqrt{\Gap/M}$ and thus $\sqrt{\epsilon}$. However since only
even powers of the spatial momentum of nucleons must appear [see
Eqs.~(\ref{IntMaxwellian})], we shall encounter terms of the type $|\gr{p}_p/m_n^2|$ which are of order $\epsilon$.

The Fokker-Planck expansion consists in expanding the energy difference between the
nucleons, $E_n-E_p$ around the lowest order value $\Gap=m_n-m_p$.
Keeping only the lowest corrections this expansion reads
\be\label{DiffEnEp}
E_n-E_p = \Gap+\delta Q_1 +\delta Q_2 +\delta Q_3
\ee
\beas
\delta Q_1&\equiv&-\frac{\gr{p}_n \cdot \gr{q}}{m_N}\\
\delta Q_2 &\equiv & - \frac{|\gr{q}|^2}{2 m_N}\\
\delta Q_3 &\equiv&
\frac{|\gr{p}_n|^2}{2}\left(\frac{1}{m_n}-\frac{1}{m_p}\right)\simeq 
-\frac{|\gr{p}_n|^2 \Gap}{2 m_N^2}\,.
\eeas
where $\gr{q}\equiv \gr{p}_p-\gr{p}_n = \coeffa_\nu
\gr{p}_\nu+\coeffa_e \gr{p}_e$ is the
spatial momentum transferred. The first term in (\ref{DiffEnEp}) is the
lowest order, or Born approximation, that is the only appearing when
considering the infinite nucleon mass approximation. The second term is an order $\sqrt{\epsilon}$ 
correction, and the third term is an order $\epsilon$ correction. Finally the last term is of order $T \Gap/m_N$ so it is an
order $\epsilon$ correction as well. It is the only corrective term
for which it is crucial to take into account the difference of mass between
neutrons and protons.  
Using Eq. (\ref{DiffEnEp}), we expand the Dirac delta function on
energies as
\ifphysrep
\be\label{DiracExpansion}
\delta\left(E_n-E_p+\coeffa_e E_e +\coeffa_\nu E_\nu\right)  \simeq \delta(\Sigma) + \delta'(\Sigma)\left(\sum_{i=1}^3\delta
  Q_i\right)+\frac{1}{2}\delta''(\Sigma) (\delta Q_1)^2\,,
\ee
\else
\bea\label{DiracExpansion}
&&\delta\left(E_n-E_p+\coeffa_e E_e
  +\coeffa_\nu E_\nu\right)  \simeq \\
&&\qquad \delta(\Sigma) + \delta'(\Sigma)\left(\sum_{i=1}^3\delta
  Q_i\right)+\frac{1}{2}\delta''(\Sigma) (\delta Q_1)^2\,,\nonumber
\eea
\fi
where $\Sigma \equiv \Gap +\coeffa_e E_e +\coeffa_\nu E_\nu$.

We must then expand the matrix element and the energies appearing in
Eq.~(\ref{GeneralGamma2}). It proves much easier to expand all these
contributions together. Furthermore, whenever a term is already of
order $\epsilon$, we know that it should multiply only the Born term
of the expansion (\ref{DiracExpansion}), so we can apply the
simplification rule (\ref{CutRule}). With this method we find
\beas\label{RulesMEEEE}
\frac{{\cal M}_{LL}}{\Pi_i E_i} &\to& 1 -\frac{\gr{p}_n}{m_N} \cdot
\left(\frac{\gr{p}_e}{E_e}+\frac{\gr{p}_\nu}{E_\nu}\right)
-\frac{\coeffa_\nu|\gr{p}_\nu|^2}{m_N E_\nu}\slabel{Expansion1}\\
\frac{{\cal M}_{RR}}{\Pi_i E_i} &\to& 1 -\frac{\gr{p}_n}{m_N} \cdot
\left(\frac{\gr{p}_e}{E_e}+\frac{\gr{p}_\nu}{E_\nu}\right)
-\frac{\coeffa_e|\gr{p}_e|^2}{m_N E_e}\slabel{Expansion2}\\
\frac{{\cal M}_{LR}}{\Pi_i E_i} &\to&
\left(1-\frac{|\gr{p}_n|^2}{m_N^2}\right)\left(1-\frac{\gr{p}_e \cdot \gr{p}_\nu}{E_e E_\nu}\right) \slabel{Expansion3}\,.
\eeas
The second term in Eqs.~(\ref{Expansion1}) and (\ref{Expansion2}) is
of order $\sqrt{\epsilon}$ and the last term in these equations is of
order $\epsilon$. Hence the second term needs to be coupled with the
order $\sqrt{\epsilon}$ term in the Dirac delta expansion
(\ref{DiracExpansion}) which is $\delta'(\Sigma) \delta Q_1$, and
simplified with the rules (\ref{CutRule}). 

There are four steps to complete this Fokker-Planck
expansion. 
\begin{enumerate}
\item First, using Eqs.~(\ref{RulesMEEEE}) and (\ref{DiracExpansion}) in
the reaction rates (\ref{GeneralGamma2}) we perform the integral on
the initial neutron momentum with the rules
(\ref{IntMaxwellian}). 
\item Second, we can replace the differential elements
for the integral on electron and neutrino momenta with $\dd^3 p \to
4\pi p^2 \dd p$ because we have already performed all angular
averages. 
\item We are left with a two dimensional integral
on the electron and neutrino momentum magnitudes $p_e=|\gr{p}_e|$ and
$p_\nu=|\gr{p}_\nu|$. Let us note $E_\nu=p_\nu$ in order to write the
result in a easily readable form. Third, we perform the integral on  $E_\nu$ using the Dirac delta and their derivatives. Whenever a
Dirac delta derivative appears, it means that we have to perform
integration by parts to convert it into a normal Dirac delta. This
will introduce derivatives with respect to the $E_\nu$ applied on the
neutrino distribution function or Pauli-blocking factor. Also for a
given reaction it might appear that the value of $E_\nu$ constrained
by the Dirac delta is not physical for that reaction if
$\coeffa_\nu=1$ and physical if $\coeffa_\nu=-1$, or vice-versa. This is the
reason why we consider the total reaction rate of the reactions (\ref{EqWeak1}) and (\ref{EqWeak2}). Once their rates are added, the
Dirac delta automatically selects either the neutrino in the initial
state, with the corresponding distribution function, or the neutrino
in the final state, with the associated Pauli-blocking
factor. Eventually once the rates (\ref{EqWeak1}) and (\ref{EqWeak2})
are added, we might forget about $\coeffa_\nu$, that is about the
position of the neutrino. We need only to compute two rates, one where
the electrons is in the initial state [reaction~(\ref{EqWeak3})], and
one where it is a positron which is in the final state [the sum of reactions~(\ref{EqWeak1}) and (\ref{EqWeak2})].
\item Finally, we need to determine the procedure to convert the rate with a neutron in the initial state into the reverse rate with a proton in the initial state. Even if the matrix
  element is the same for all reactions, as explained
  in~\S~\ref{AppWeakGeneralExpressions}, the method to perform a
  finite mass expansion is not symmetric under the interchange $p
  \leftrightarrow n$. Indeed we chose to expand the momentum of the
  final nucleon around the initial one, and we remove the integral on
  the final nucleon momenta. It is apparent on Eqs.~(\ref{papbpcpd})
  that the electron (resp. neutrino) momentum is contracted with the
  neutron (resp. proton) in the   $LL$ term but this is the opposite
  in the $RR$ term. Since the coupling factors of these terms are
  interchanged by the replacement $g_A \to -g_A$, we can deduce the
  rates with an initial proton from those with an initial neutron
  using the rule $g_A \to -g_A$. Obviously the argument of the Dirac
  delta contains now $E_p - E_n= -\Gap + \dots$ instead of $E_n-E_p=
  \Gap + \dots$ so we must also apply the rule $\Gap \to
  -\Gap$. Finally when considering a reverse reaction, the electron in
  the initial state turns into a positron in the final state so we
  must also apply the rule $E_e \to -E_e$, that is change the electron
  distribution function to a Pauli-blocking factor or
  vice-versa. These are the rules used extensively in \S~\ref{SecWeak} to deduce
  the reverse rates.
\end{enumerate}
Having sketched the details of the procedure, we are in position to
give the results. The Born approximation results are obvious under
this expansion. From Eqs.~(\ref{RulesMEEEE}) we obtain in the limit
\beas
\frac{{\cal M}_{LL}}{\Pi_i E_i} &\simeq& \frac{{\cal M}_{RR}}{\Pi_i
  E_i}  \simeq 1\,, \\
\frac{{\cal M}_{LR}}{\Pi_i E_i} &\simeq &\left(1-\frac{\gr{p}_e \cdot
    \gr{p}_\nu}{E_e E_\nu}\right)\,.\slabel{MLRBorn}
\eeas
Furthermore in the Dirac expansion (\ref{DiracExpansion}) we keep only
$\delta(\Sigma)$, and the average over electrons and neutrino spatial
momenta  removes the momentum dependent part of
Eq.~(\ref{MLRBorn}). The Born rates are gathered and commented in
\S~\ref{SecBornMainText}. 

The first corrections in this Fokker-Planck expansion, and which are
due to finite nucleon mass, are reported in the next section.

\subsection{Finite nucleon mass corrections}\label{AppFM}

The finite nucleon mass corrections take the form~(\ref{GammaFMnoRC}) [or
(\ref{GammaFMRC}) if radiative corrections are added]. The function $\chi^{\rm FM}_\pm$ is 
\begin{widetext}
\bea\label{chiFMFull}
\chi^{\rm FM}_\pm(E,g_A) &=& \tilde c_{LL}\frac{p^2}{m_N E}g^{(2,0)}_\nu(E_\nu^\mp) g(-E) -\tilde
c_{RR}\frac{E_\nu^\mp}{m_N} g^{(2,0)}_\nu(E_\nu^\mp) g(-E) \\
&&+\left(\tilde c_{LL}+\tilde c_{RR}\right)\frac{T}{m_N}\left(g^{(2,1)}_\nu(E_\nu^\mp)
  g(-E)\frac{p^2}{E}-g^{(3,1)}_\nu(E_\nu^\mp)
  g(-E)\right)\nonumber\\
&&+\left(\tilde c_{LL}+\tilde c_{RR}+\tilde
  c_{LR}\right)\left[\frac{T}{2 m_N}\left(g^{(4,2)}_\nu(E_\nu^\mp)
  g(-E)+g^{(2,2)}_\nu(E_\nu^\mp) g(-E) p^2\right) \right.\nonumber\\
&&\qquad\qquad\qquad \qquad\qquad\left.+ \frac{1}{2
  m_N}\left(g^{(4,1)}_\nu(E_\nu^\mp) g(-E)+g^{(2,1)}_\nu(E_\nu^\mp)
  g(-E)p^2\right)\right]\nonumber\\
&&-\left(\tilde c_{LL}+\tilde c_{RR}+\tilde
  c_{LR}\right)\frac{3 T}{2}\left[1-\left(\frac{m_n}{m_p}\right)^{\pm1}\right] g^{(2,1)}_\nu(E_\nu^\mp)
  g(-E)\nonumber\\
&&+\tilde
  c_{LR}\left[-\frac{3 T}{m_N} g^{(2,0)}_\nu(E_\nu^\mp)  g(-E)+\frac{p^2}{3 m_N
      E} g^{(3,1)}_\nu(E_\nu^\mp) g(-E)+\frac{p^2 T}{3 m_N E}g^{(3,2)}_\nu(E_\nu^\mp) g(-E)  \right]\nonumber
\eea
\end{widetext}
where $p=\sqrt{E^2-m_e^2}$, $E_\nu^\mp=E\mp\Gap$. We defined the reduced couplings
\beas\label{gtildeLLRRLR}
\tilde c_{LL} &\equiv&\frac{4}{1+3 g_A^2} c_{LL}\\
\tilde c_{RR} &\equiv&\frac{4}{1+3 g_A^2} c_{RR}\\
\tilde c_{LR} &\equiv&\frac{4}{1+3 g_A^2} c_{LR}\,,
\eeas
and the functions [with the notation (\ref{DefgnuANDg})]
\be
g_\nu^{(n,p)}(E_\nu) \equiv \frac{\partial^p [(E_\nu)^n
    g_\nu(E_\nu)]}{\partial E_\nu^p}\,.
\ee
For the neutron beta decay, the expression (\ref{chiFMFull}) for $T=0$
gives (with $E_\nu \equiv \Gap-E$ here)
\ifphysrep
\be\label{chiFMneutrondecay}
\left.\chi^{\rm FM}_+(E,g_A)\right|_{T=0} =\tilde c_{LL}\frac{p^2 E_\nu^2}{m_N E} +\tilde
c_{RR}\frac{E_\nu^3}{m_N} -\left(\tilde c_{LL}+\tilde c_{RR}+\tilde c_{LR}\right)
\frac{\left(2 E_\nu^3+E_\nu p^2\right)}{m_N}+\tilde  c_{LR}\frac{p^2 E_\nu^2 }{m_N E} 
\ee
\else
\bea\label{chiFMneutrondecay}
&&\left.\chi^{\rm FM}_+(E,g_A)\right|_{T=0} = \tilde c_{LL}\frac{p^2 E_\nu^2}{m_N E} +\tilde
c_{RR}\frac{E_\nu^3}{m_N} \\
&&-\left(\tilde c_{LL}+\tilde c_{RR}+\tilde c_{LR}\right)
\frac{\left(2 E_\nu^3+E_\nu p^2\right)}{m_N}+\tilde  c_{LR}\frac{p^2 E_\nu^2 }{m_N E}\,. \nonumber
\eea
\fi

\subsection{Weak-magnetism corrections}\label{AppWM}

Because the weak-magnetic term enters in the current
(\ref{HadronicCurrent}) with a factor $q_\nu/m_N$, with the momentum
transfer to nucleon being of the order of $T$, it is an order $\epsilon$
correction and thus vanishes at the Born approximation level. The matrix
element is corrected by the addition of
\begin{widetext}
\bea\label{EqMWM}
\frac{|M|^2}{2^7 G_F^2} &\supset& f_{\rm wm} g_A \frac{m_n + m_p}{2}
\left({\cal M}_{LL}-{\cal M}_{RR}\right)\\
&+&\frac{f_{\rm wm}}{2}\left[\frac{m_n + m_p}{2}\left({\cal M}_{LR}-{\cal M}_{LL}- {\cal M}_{RR}- (\fourp{p}_n \cdot  \fourp{p}_p)(\fourp{p}_\nu \cdot \fourp{p}_e)\right)+m_p  (\fourp{p}_n \cdot \fourp{p}_e)(\fourp{p}_n \cdot \fourp{p}_\nu)+m_n  (\fourp{p}_p \cdot \fourp{p}_e)(\fourp{p}_p \cdot \fourp{p}_\nu)\right]\nonumber
\eea
\end{widetext}
where we used the expressions (\ref{papbpcpd}). It is possible to
show, although rather tedious, that the contributions of the second
line vanish at first order in $\epsilon$. This remarkable
simplification has been noticed earlier by~\citet{Seckel1993}. Hence
the weak-magnetism contributes exactly as the axial vector current coupled to the vector current. It is taken into account
by changing the coupling constants (\ref{DefgLLgRRgLR}) to the values
(\ref{DefgLLgRRgLRwithWM}), modifying accordingly the constants~(\ref{gtildeLLRRLR}) appearing in the finite nucleon mass corrections.

Only the first two terms of Eq.~(\ref{chiFMFull}) can in principle
contribute to $\lambda_0$ since weak-magnetism does no affect the sum
$c_{LL}+c_{RR}$ nor $c_{LR}$. However, it is straightforward to show
that they result in a vanishing contribution since
%\bea
%\lambda_0^{\rm wm}&=&\frac{4 f_{\rm wm} g_A}{m_N(1+3
%  g_A^2)}\int_0^{\sqrt{\Gap^2-1}} \dd p p^2
%(E-\Gap)^2\left(\frac{p^2}{E}+(E-\Gap)\right)\nonumber\\
%&=&\frac{4 f_{\rm wm} g_A}{m_N(1+3 g_A^2)}\int_m^\Gap \dd
%E \partial_E\left[(E-\Gap)^3 (E^2-m^2)^{3/2}\right]\nonumber\\
%&=&0\,.
%\eea
\bea
\lambda_0^{\rm wm}&\propto&\int_0^{\sqrt{\Gap^2-1}} \dd p p^2
(E-\Gap)^2\left(\frac{p^2}{E}+(E-\Gap)\right)\nonumber\\
&=&\tfrac{1}{3}\int_m^\Gap \dd
E \partial_E\left[(E-\Gap)^3 (E^2-m^2)^{3/2}\right]\nonumber\\
&=&0\,.
\eea
Note that when the weak-magnetism contribution is coupled with radiative corrections,
this result is no longer valid and there is a small change in
$\lambda_0$ that we give in \S~\ref{SecMainWM}.

\subsection{Mandelstam variables}

Let us define the Mandelstam variables for $\nu+n \to e+p$. 
\beas
s &\equiv& (\fourp{p}_\nu + \fourp{p}_n)^2 = (\fourp{p}_e +\fourp{p}_p)^2\\
t &\equiv& (\fourp{p}_\nu - \fourp{p}_e)^2 = (\fourp{p}_n -\fourp{p}_p)^2\\
u &\equiv& (\fourp{p}_\nu - \fourp{p}_p)^2 = (\fourp{p}_n - \fourp{p}_e)^2\,.
\eeas
The total matrix element $|M|^2$ deduced from
Eqs.~(\ref{DefHi})-(\ref{HadronicCurrent}), is the sum of
Eqs.~(\ref{DefMLLMRRMLR2}) and (\ref{EqMWM}), where we use the
definitions (\ref{papbpcpd}). Once expressed with the Mandelstam
variables, we checked that it takes the form given in \citet[App. A]{Lopez1997} (published version and with $m_1=m_\nu$, $m_2=m_n$,
$m_3=m_e$ and $m_4=m_p$), except for a few differences.
First, we do not include their term $t_3$ because this is second order
  in finite nucleon mass effects. Hence, we kept only contributions which are
  linear in $f_{\rm wm}$. Second, we find that there is a typo in $t_5$ of \citet{Lopez1997}, since
  there should be a factor $g_A$ instead of $g_A^2$.

We also performed a comparison with \citet{Seckel1993} and found that
it matches our result and the one of \citet{Lopez1997} if we replace
its $f_2$ by $2 f_{\rm wm}$  (again ignoring terms which are quadratic
in $f_2$ and the parameter $f_{ps}$ in \citet{Seckel1993}).

\subsection{Radiative corrections and Sirlin's universal function}\label{SecSirlin}

When considering the effect of radiative corrections, the function ${\cal C}$ defined in Eq.~(\ref{DefRC}) is given by
\be\label{CSirlin}
{\cal C}(E,k_{\rm max}) = 4\ln \frac{m_Z}{m_p} +\ln \frac{m_p}{m_A}+2C + A_g+g(E,k_{\rm max})
\ee
with
\be
C\simeq 0.891\,,\qquad A_g \simeq -0.34\,,\qquad m_A \simeq 1.2\,{\rm GeV}\,.
\ee
$g(E,k_{\rm max})$ is Sirlin's universal function
\ifphysrep
\bea\label{SirlinFunction}
g(E,k_{\rm max}) &=& 3 \ln
\frac{m_p}{m_e}-\frac{3}{4}+\frac{4}{\beta}L\left(\frac{2\beta}{1+\beta}\right)\\
&&+4[R(\beta)-1]\left[\frac{k_{\rm
      max}}{3E}-\frac{3}{2}+\ln\left(2\frac{k_{\rm max}}{m_e}\right)\right]+R(\beta)\left[2(1+\beta^2)+\frac{k_{\rm max}^2}{6E^2}-4 \beta R(\beta)\right]\,,\nonumber
\eea
\else
\bea\label{SirlinFunction}
g(E,k_{\rm max}) &=& 3 \ln
\frac{m_p}{m_e}-\frac{3}{4}+\frac{4}{\beta}L\left(\frac{2\beta}{1+\beta}\right)
\\
&+&4[R(\beta)-1]\left[\frac{k_{\rm
      max}}{3E}-\frac{3}{2}+\ln\left(2\frac{k_{\rm max}}{m_e}\right)\right]\nonumber\\
&+&R(\beta)\left[2(1+\beta^2)+\frac{k_{\rm max}^2}{6E^2}-4 \beta R(\beta)\right]\,,\nonumber
\eea
\fi
with $\beta = p/E=\sqrt{E^2-m_e^2}/E$ and
\be
R(\beta) \equiv\frac{E}{p}\ln\left(\frac{p+E}{m}\right) = \frac{1}{2\beta} \ln\left(\frac{1+\beta}{1-\beta}\right) =\frac{\arctan(\beta)}{\beta} \nonumber
\ee
and where the Spence function is defined as
\be
L[x]=\int_0^x \frac{\ln(1-t)}{t}\dd t\,.
\ee
It is possible to use an expansion of the Spence function. We report
it here to correct typos in \citet{Dicus1982} and subsequently in \citet{LopezTurner1998,SmithFuller}
\ifphysrep
\be\frac{1}{\beta}L\left(\frac{2\beta}{1+\beta}\right) \simeq
-\frac{1}{(1+\beta)^6}\left(2+11\beta+\tfrac{224}{9}\beta^2+\tfrac{89}{3}\beta^3+\tfrac{1496}{75}\beta^4+\tfrac{596}{75}\beta^5+\tfrac{128}{49}\beta^6\right)\,.
\ee
\else
\bea
&&\frac{1}{\beta}L\left(\frac{2\beta}{1+\beta}\right) \simeq
-\frac{1}{(1+\beta)^6}\times\\
&&\left(2+11\beta+\tfrac{224}{9}\beta^2+\tfrac{89}{3}\beta^3+\tfrac{1496}{75}\beta^4+\tfrac{596}{75}\beta^5+\tfrac{128}{49}\beta^6\right)\nonumber\,.
\eea
\fi

Finally in order to obtain even more accurate results, it is also possible to refine the expression for the radiative
corrections by using \citet[Eq. 15]{Czarnecki2004} which is a
resummation of higher order corrections. Unless specified, this is the
radiative correction that we use so we report it here.
\ifphysrep
\bea\label{RadiativeMarcianoSirlin}
R(E,k_{\rm max}) &=&\left[1+\frac{\alpha_{\rm
      FS}}{2\pi}\left(g(E,k_{\rm max})-3\ln\frac{m_p}{2
      \Gap}\right)\right]\\
&&\times\left(L+\frac{\alpha_{\rm FS}}{\pi}
  C+\frac{\alpha_{\rm FS}}{2\pi} \delta\right)\left[S+\frac{\alpha_{\rm
      FS}(m_p)}{2\pi}\left(\ln\frac{m_p}{m_A}+A_g\right)+{\rm NLL}\right]\nonumber
\eea
\else
\bea\label{RadiativeMarcianoSirlin}
R(E,k_{\rm max}) &=&\left[1+\frac{\alpha_{\rm
      FS}}{2\pi}\left(g(E,k_{\rm max})-3\ln\frac{m_p}{2
      \Gap}\right)\right]\\
&&\times\left(L+\frac{\alpha_{\rm FS}}{\pi}
  C+\frac{\alpha_{\rm FS}}{2\pi} \delta\right)\nonumber\\
&&\times\left[S+\frac{\alpha_{\rm
      FS}(m_p)}{2\pi}\left(\ln\frac{m_p}{m_A}+A_g\right)+{\rm NLL}\right]\nonumber
\eea
\fi
\be
\alpha_{\rm FS}(m_p)\simeq \frac{1}{134}\,,\quad \frac{\alpha_{\rm
    FS}}{2\pi}\delta \simeq-0.00043\,,\quad{\rm NLL}\simeq-10^{-4}\nonumber
\ee
\be
L\simeq1.02094\,,\qquad S\simeq 1.02248
\ee

\subsection{Bremsstrahlung}\label{AppBS}

We review in the next section the bremsstrahlung correction to the
neutron decay, also called radiative neutron decay when the photon is detected~\citep{2010PhRvC..81c5503C}.
We then extend it to the other weak processes during BBN. This allows
to find the difference between the correct treatment of bremsstrahlung
and how it was partially included in the radiative correction factor
detailed above in~\S~\ref{SecSirlin}. We report subsequently the detailed
expression of this bremsstrahlung correction.

\subsubsection{Neutron decay}

Let us consider first the bremsstrahlung for the decay of neutron. We note the emitted photon momentum $\fourp{k}$. Its spatial part is
$\gr{k}$, that we decompose into energy $k$ and direction
$\hat{\gr{k}}$ as $\gr{k} = k \hat{\gr{k}}$. We note the electron
energy $E$ and its spatial momentum $p$ for simplicity.
In the infinite nucleon mass approximation, the bremsstrahlung
differential decay rate is of the form [see
e.g. \citet[Eq. B12]{Ivanov:2012qe},
\citet[Eq. 4]{2010PhRvC..81c5503C}, \citet[Eq. 6.6]{Dicus1968} or \citet[Eq. A40]{Ivanov:2017fra}]
\ifphysrep
\be
\frac{\dd \Gamma^{\rm BS}_{n \to}}{\dd E \dd k \dd \mu}=\frac{\alpha_{\rm FS}
    \myk}{2\pi} \frac{\beta
  E_\nu^2}{k}\left[\frac{\beta^2(1-\mu^2)}{(1-\beta
    \mu)^2}\left(1+\frac{k}{E}\right)+\frac{k^2}{E^2}\frac{1}{1-\beta \mu}\right]\nonumber
\ee
\else
\bea
\frac{\dd \Gamma^{\rm BS}_{n \to}}{\dd E \dd k \dd \mu}&=&\frac{\alpha_{\rm FS}
    \myk}{2\pi} \frac{\beta
  E_\nu^2}{k}\\
&\times&\left[\frac{\beta^2(1-\mu^2)}{(1-\beta
    \mu)^2}\left(1+\frac{k}{E}\right)+\frac{k^2}{E^2}\frac{1}{1-\beta \mu}\right]\nonumber
\eea
\fi
where
\be
\mu \equiv \gr{p}_e \cdot \hat{\gr{k}}
\ee
is the (cosinus) of the angle between the photon momentum and the
electron momentum. The neutrino energy is constrained from energy
conservation to be
\be
E_\nu = \Gap-E-k\,.
\ee
Performing the integral on $\mu$ leads to
\be\label{dGbetadecay}
\frac{\dd \Gamma^{\rm BS}_{n \to}}{\dd k \dd E}=\frac{\alpha_{\rm FS}
    \myk}{2\pi k} E_\nu^2
F_+(E,k)\,.
\ee
with
\bea
F_\pm(E,k) &\equiv& A(E,k) \pm k B(E,k)\\
A(E,k)&\equiv&(2 E^2 + k^2)\ln\left(\frac{E+p}{E-p}\right)-4 p E \nonumber\\
B(E,k)&\equiv&2E \ln\left(\frac{E+p}{E-p}\right)-4 p\,.\nonumber
\eea
The total decay rate cannot be obtained from a simple integration of
the form
\be
\Gamma^{\rm BS}_{n \to } = \int_{m_e}^{E_{\rm max}} \dd E \int_0^{\Gap-E} \dd k \frac{\dd
  \Gamma^{\rm BS}}{\dd k \dd E}\,.
\ee
since there is an infrared divergence. However this divergence cancels
a corresponding divergence in the pure radiative corrections. The
usual procedure consists in letting the photon having a mass and
taking the limit $m_\gamma \to 0$ [see e.g. \citet{Ivanov:2012qe} or \citet{Dicus1968}].

\subsubsection{Bremsstrahlung for other reactions}

The other rates of bremsstrahlung can be deduced from crossing
symmetry. Indeed, if one considers the process $n+e^+ \to p + \bar\nu
$ it is obtained from crossing symmetry of beta decay. A crossing
symmetry is performed by inverting the four-momentum of the
corresponding particle. Hence changing the position of the final
electron to an initial positron amounts to $E \to -E$ and $p \to
-p$. This leads to $F_+ \to F_-$ in (\ref{dGbetadecay}). In fact it is
simple to check that if the electron (or positron) and the photon are
on the same side, one should use $F_+$ and if they are on opposite
sides, we should use $F_-$. The contributions of bremsstrahlung to the
reaction rates is then straightforward, provided it is clear that we
should take the infrared regularized contribution, and they read
\ifphysrep
\beas\label{BSnpFormal1}
\Gamma_{n \to p\gamma} &=&\frac{\alpha_{\rm FS}
  \myk}{2\pi}\int_{m_e}^\infty \dd E \left[g(-E) \widetilde{\chi}_+^{+}(E)+g(E) \widetilde{\chi}_+^{-}(-E)\right]\\
\Gamma_{p\to n\gamma} &=&\left.\Gamma_{n\to
    p\gamma}\right|_{\widetilde{\chi}_+^{s}\to \widetilde{\chi}_-^{s}}
\eeas
\else
\beas\label{BSnpFormal1}
\Gamma_{n \to p\gamma} &=&\frac{\alpha_{\rm FS}
  \myk}{2\pi}\int_{m_e}^\infty \dd E \\
&\times&\left[g(-E) \widetilde{\chi}_+^{+}(E)+g(E) \widetilde{\chi}_+^{-}(-E)\right]\nonumber\\
\Gamma_{p\to n\gamma} &=&\left.\Gamma_{n\to
    p\gamma}\right|_{\widetilde{\chi}_+^{s}\to \widetilde{\chi}_-^{s}}
\eeas
\fi
with the definitions
\bea
\widetilde{\chi}_\pm^{s}(E) &\equiv& \int_0^\infty \frac{\dd k}{k}
F_{s}(E,k)\widetilde{\chi}_\pm(k+E)\\
\widetilde{\chi}_\pm(E) &\equiv& (E^\mp_\nu)^2 g^\mp_\nu(E_\nu) \,,\quad E^\mp_\nu = E\mp\Gap\,.
\eea

\subsubsection{Correction for bremsstrahlung}\label{SecCorrectionBS}

However, the bremsstrahlung radiative corrections
included in \S~\ref{SecCCRTnull}, that is in the factor~(\ref{DefRC}), are those of
neutron beta decay.  Indeed the radiative correction
function $R$ in Eqs.~(\ref{CCRn}) and (\ref{CCRp}) is the sum of a pure radiative correction and a
bremsstrahlung correction. It is decomposed as
\be 
R(E,k_{\rm max}) = R^{\rm pure}(E)+R^{\rm BS}(E,k_{\rm max})
\ee
with the BS part being
\be\label{RBSContains}
R^{\rm BS}(E, k_{\rm max}) \equiv \frac{\alpha_{\rm
    FS}}{2\pi}\int_0^{k_{\rm max}} \frac{\dd k}{k}\frac{(k_{\rm max}-k)^2}{k_{\rm max}^2}\frac{F_+(E,k)}{Ep}\,.
\ee
In Eqs.~(\ref{CCRn}) and (\ref{CCRp}) the choice was made to take
$k_{\rm max}$ as being the energy of the neutrino because this is the
case for neutron beta decay. However the maximum energy of the photon
emitted is not the neutrino energy in the other
reactions. Furthermore, the distribution function for the neutrino
appearing in Eqs.~(\ref{CCRn}) and (\ref{CCRp}) is incorrectly taken
in the limit in which the photon energy is so soft that it does not
affect the neutrino energy. These are the two shortcomings that we
need to correct for by adding the difference between the correct
correction [Eqs.~(\ref{BSnpFormal1})] and the
approximate contribution [it is formally Eqs.~(\ref{CCRn}) and (\ref{CCRp}) with $R
\to R^{\rm BS}$]. We write this bremsstrahlung correction in a form which contains explicitly no infrared divergence and this is
given by Eqs.~(\ref{BSnpFormal}) with the definitions
\begin{widetext}
\bea\label{BSCorrection}
\gamma^{\rm BS}_{n \to p+e} &=& \int_0^{|\Gap-E|}
  \frac{\dd k}{k}
  F_+(E,k)\left[\widetilde{\chi}_+(E+k)-(|\Gap-E|-k)^2g_\nu(E-\Gap)\right]+\int_{|\Gap-E|}^\infty
  \frac{\dd k}{k} F_+(E,k)\widetilde{\chi}_+(E+k)\nonumber\\
\gamma^{\rm BS}_{n+e \to p}&=& \int_0^{\Gap+E}
  \frac{\dd k}{k} \left[F_-(E,k)\widetilde{\chi}_+(k-E)-F_+(E,k)
    (|\Gap+E|-k)^2g_\nu(-E-\Gap)\right]\nonumber\\
&+&\int_{\Gap+E}^\infty
  \frac{\dd k}{k} F_-(E,k)\widetilde{\chi}_+(k-E)
\eea
\bea\label{BSCorrection2}
\gamma^{\rm BS}_{p \to n+e} &=& \int_0^{\Gap+E}
  \frac{\dd k}{k} F_+(E,k)\left[\widetilde{\chi}_-(E+k)-(|E+\Gap|-k)^2g_\nu(E+\Gap)\right]+\int_{\Gap+E}^\infty
  \frac{\dd k}{k} F_+(E,k)\widetilde{\chi}_-(E+k)\nonumber\\
\gamma^{\rm BS}_{p+e \to n}&=& \int_0^{|\Gap-E|}
  \frac{\dd k}{k} \left[F_-(E,k)\widetilde{\chi}_-(k-E)-F_+(E,k)
    (|E-\Gap|-k)^2g_\nu(-E+\Gap)\right]\nonumber\\
&+&\int_{|\Gap-E|}^\infty
  \frac{\dd k}{k} F_-(E,k)\widetilde{\chi}_-(k-E)\,.
\eea
\end{widetext}
With these corrections added to Eqs.~(\ref{CCRn}) and (\ref{CCRp}),
the total effect of bremsstrahlung is taken into account. However,
with this new contribution the rates no longer satisfy the detailed
balance relation  (\ref{MagicDetailedBalanceBorn}). Indeed the emission of a final photon needs to be
compensated by the absorption of photons from the thermal bath to lead to
a thermodynamical equilibrium. It is only when the finite temperature
radiative corrections are taken into account, with stimulated emission
and absorption, that the total rates can satisfy detailed
balance. This is shown explicitly in Fig. \ref{FigCorrectionCCRTRates}.

Note that by construction the bremsstrahlung corrections do
not modify the radiative corrections for $\lambda^{\rm RC0}_0$ because the
neutron beta decay in vacuum, is the only reaction for which bremsstrahlung is
fully taken into account already.

Finally note that there was some intuition in \citet{BrownSawyer}
about the incorrect treatment of real-photon processes. Indeed in this
 reference, the authors advocate that one should add the process
 $n+\bar \nu +e^+ \to p+\gamma$, that they call the {\it five-body process}, since it
 cannot be a correction to $n+\bar \nu +e^+ \to p$ which is forbidden
 energetically. This is indeed correct and it corresponds to the last
 term in $\delta \Gamma^{\rm BS}_{n+e \to p}$, where the photon emitted must have an energy larger than $\Gap+E$. What we find is
 that not only this process needs to be added, but {\it all other processes need
 to be corrected} except for the neutron beta decay.

\subsection{Finite temperature radiative corrections}\label{AppFullRCT}

The finite temperature radiative corrections are made of
Eqs.~(\ref{RCT1}), (\ref{RCT2}) and (\ref{RCT3}). These last two
contributions involve implicitly principal parts in the apparently
divergent part of the integrals. After rearrangement, they can be put in the form
\citep[Eq. (5.15)]{BrownSawyer}
\begin{widetext}
\beas
\Gamma^{\Delta E,T}_{n \to p}+\Gamma^{ep+ee,T}_{n \to p} &=&
\frac{\alpha_{\rm FS} \myk}{2\pi}\int_{m_e}^\infty \dd E
[\chi_+(E)+\chi_+(-E)] \left[-\frac{2 \pi^2 T^2 E}{3 p}+\int_{m_e}^\infty
  \dd E' F^T(E,E')\right]\\
\Gamma^{\Delta E,T}_{p \to n}+\Gamma^{ep+ee,T}_{p \to n}
&=&\left. \left(\Gamma^{\Delta E,T}_{n \to p}+\Gamma^{ep+ee,T}_{n \to p}
\right)\right|_{\chi_+ \to \chi_-}
\eeas
\beas
F^T(E,E') &\equiv& -\frac{1}{4} \ln^2 \left(\frac{p+p'}{p-p'}\right)^2
\left\{g'(E') \frac{p' E^2}{p E'}(E+E')+g(E')\frac{E^2}{p
  p'}\left[E' + \frac{m_e^2 E}{{E'}^2}\right]\right\}-g(E')\left[4E \frac{p'}{p}+2 E' L(E,E')\right]\nonumber\\
&&+\ln\left(\frac{p+p'}{p-p'}\right)^2
\left\{g'(E')\left[{p'}^2\frac{E}{E'}\left(\frac{m_e^2}{p^2}+2\right)-E^2\frac{p'}{p}L[E,E']\right]\nonumber\right.\\
&&\qquad \qquad\qquad \qquad\left.+g(E')\left[\frac{E m_e^2}{p^2{E'}^2}({E'}^2+2p^2+m_e^2)-\frac{E^2+{E'}^2}{E+E'}-\frac{E^2
  {E'}}{p p'}L(E,E')\right]\right\}\\
L(E,E')&\equiv&\ln \left(\frac{E E' + p p'+m_e^2}{E E' - p p'+m_e^2}\right)\,,
\eeas
\end{widetext}
where $g'(E) \equiv \partial_{E} g(E)$.

This form is simpler numerically if the double integration on $E$ and $E'$
is performed on the variables $E_\Sigma \equiv E+E'$ and $E_\Delta
\equiv E-E'$ using $2 \int \int \dd E \dd E' = \int \int \dd E_\Sigma
\dd E_{\Delta}$.

The first contribution (\ref{RCT1}) is made of real photon processes,
such as absorption and stimulated emission, see Fig.~\ref{FigFD2}, but not simple emission
which has been already accounted for as bremsstrahlung (see
appendix~\ref{AppBS}). To obtain it we start from the real photons
processes \citep[Eq.~B28]{BrownSawyer}
\ifphysrep
\beas\label{EqAllRealPhotons}
\Gamma^{\gamma}_{n \to p} &=&\frac{\alpha_{\rm FS}
  \myk}{2\pi}\int_{m_e}^\infty \dd E \left[g(-E) \widetilde{\chi}_+^{\gamma,+}(E)+g(E)
  \widetilde{\chi}_+^{\gamma,-}(-E)\right]\\
\Gamma^{\gamma}_{p \to n}&=&\left.\Gamma^{\gamma}_{n \to p}\right|_{\widetilde{\chi}_+^{\gamma,s}\to \widetilde{\chi}_-^{\gamma,s}}
\eeas
\else
\beas\label{EqAllRealPhotons}
\Gamma^{\gamma}_{n \to p} &=&\frac{\alpha_{\rm FS}
  \myk}{2\pi}\int_{m_e}^\infty \dd E \\
&\times&\left[g(-E) \widetilde{\chi}_+^{\gamma,+}(E)+g(E)
  \widetilde{\chi}_+^{\gamma,-}(-E)\right]\nonumber\\
\Gamma^{\gamma}_{p \to n}&=&\left.\Gamma^{\gamma}_{n \to p}\right|_{\widetilde{\chi}_+^{\gamma,s}\to \widetilde{\chi}_-^{\gamma,s}}
\eeas
\fi
where we defined
\ifphysrep
\be
\widetilde{\chi}_\pm^{\gamma,s}(E) \equiv \int_0^\infty \frac{\dd k}{k}
\left[(1+f(k))F_{s}(E,k)\widetilde{\chi}_\pm(E+k)\right.\left.+f(k) F_{-s}(E,k)\widetilde{\chi}_\pm(E-k)\right],
\ee
\else
\bea
\widetilde{\chi}_\pm^{\gamma,s}(E) &\equiv& \int_0^\infty \frac{\dd k}{k}
\left[(1+f(k))F_{s}(E,k)\widetilde{\chi}_\pm(E+k)\right.\nonumber\\
&&\qquad\quad\left.+f(k) F_{-s}(E,k)\widetilde{\chi}_\pm(E-k)\right],
\eea
\fi
and $f(k)\equiv g^-(k/T)=1/[{\rm exp}(k/T)-1]$. Note that from the
property
\be\label{BasicLawDetailedBalancek}
1+f(k)=-f(-k)={\rm e}^{\frac{k}{T}} f(k)
\ee
combined with the property (\ref{BasicLawDetailedBalance}), the
detailed balance relation (\ref{MagicDetailedBalanceBorn}) for the
real photon processes (\ref{EqAllRealPhotons}) is automatically
satisfied\footnote{At this stage this is only formal since this presents infrared divergences.}.

Then we replace the factor $[1+f(k)]$ by $f(k)$ so as to keep only
stimulated emission, since bremsstrahlung processes
(\ref{BSnpFormal1}) which are taken into account separately are
obtained by $f(k)\to 0$. A consequence is that the real photon processes without bremsstrahlung do no satisfy formally the detailed balance relation. It is only when all real photon processes are added that it
is recovered, as illustrated in Fig.  (\ref{FigCorrectionCCRTRates}).

Finally, in order to obtain the form (\ref{RCT1}) for which it is apparent that there is no infrared
divergence, it is necessary to add an infrared diverging contribution 
\citep[Eq. (B53)]{BrownSawyer} that can be considered as part of the
wave-function radiative correction (it is partially coming from the diagram~\ref{FDTRC1}) and which is
\ifphysrep
\beas\label{BrownB53}
\Gamma^{\gamma,Z}_{n \to p} &=&\frac{\alpha_{\rm FS}
  \myk}{2\pi}\int_{m_e}^\infty \dd E \left[g(-E) \widetilde{\chi}_+^{\gamma,Z}(E)+g(E)
  \widetilde{\chi}_+^{\gamma,Z}(-E)\right]\\
\Gamma^{\gamma,Z}_{p \to n}&=&\left.\Gamma^{\gamma,Z}_{n \to
  p}\right|_{\widetilde{\chi}_+^{\gamma,Z} \to \widetilde{\chi}_-^{\gamma,Z}}
\eeas
\else
\beas\label{BrownB53}
\Gamma^{\gamma,Z}_{n \to p} &=&\frac{\alpha_{\rm FS}
  \myk}{2\pi}\int_{m_e}^\infty \dd E \\
&\times&\left[g(-E) \widetilde{\chi}_+^{\gamma,Z}(E)+g(E)
  \widetilde{\chi}_+^{\gamma,Z}(-E)\right]\nonumber\\
\Gamma^{\gamma,Z}_{p \to n}&=&\left.\Gamma^{\gamma,Z}_{n \to
  p}\right|_{\widetilde{\chi}_+^{\gamma,Z} \to \widetilde{\chi}_-^{\gamma,Z}}
\eeas
\fi

\be
\widetilde{\chi}_\pm^{\gamma,Z}(E) \equiv -2\int_0^\infty \frac{\dd k}{k}
f(k)A(E,k)\widetilde{\chi}_\pm(E)\,.
\ee
Again on Eqs.~(\ref{BrownB53}) it is apparent using the property (\ref{BasicLawDetailedBalance}) that these extra terms
formally satisfy the detailed balance relation (\ref{MagicDetailedBalanceBorn}).

\section{Nuclear reactions} \label{AppReactionConventions}

\subsection{Conventions for nuclear reaction rates}

Let us consider for simplicity a two-body reaction of the type $k+l
\to i+j$ and its reverse reaction $i+j \to k+l$. From Eqs.~(\ref{CYi})
and (\ref{Gammatogamma2}) their contributions to the evolution of the abundance $Y_i$ take the form
\begin{equation} 
\dot Y_i \supset Y_kY_l \rho_nN_A\langle
\sigma v \rangle_{kl\rightarrow{ij}}- Y_iY_j \rho_nN_A\langle \sigma v \rangle_{ij\rightarrow{kl}}\,,
\label{q:nucl}
\end{equation} 
with
\begin{equation} 
\rho_{\rm n}\equiv\frac{n_{\rm b}}{N_A}\equiv\;n_{\rm b}{\cdot}u\;\; \mathrm{(g/cm^3)}.
\label{q:rhon}
\end{equation}  
We see that $\rho_{\rm n}$ has the dimension of a mass density since $u$ is the atomic mass unit, that, by definition is related to Avogadro's number by $N_A.u$=1g.
By convention, $N_A$ has been introduced in the definition of the reaction rate,
hence what is tabulated for a given reaction is $N_A \langle \sigma v \rangle$ and one must
multiply by $\rho_{\rm n}$ in order to build the rates (e.g. $\Gamma_{kl \to
  ij}$) which appear in the general form
Eq.~(\ref{GeneralYidot}). More generally for a reaction of the type
$i_1+\dots +i_p \to j_1+\dots j_q$, the contribution to the evolution
of the species $i_1$ is
\be
\dot Y_{i_1}\supset -Y_{i_1}\dots Y_{i_p} \rho_{\rm n}^{p-1} N_A^{p-1}\gamma_{i_1\dots i_p \to j_1\dots j_q}\,.
\ee
Conventionally what is tabulated is $N_A^{p-1}\gamma_{i_1\dots i_p \to
  j_1\dots j_q}$~\footnote{More precisely, from (\ref{GammaTogamma}) the factor is
  $N_A^{(N_{i_1}+\dots + N_{i_p})-1}$ if the stoichiometric
  coefficients are not unity.} and one must multiply by
$\rho_{\rm n}^{p-1}$ to build the reaction rate $\Gamma_{i_1\dots i_p
  \to j_1\dots j_q}$ appearing in Eq.~(\ref{GeneralYidot}). The method is
similar for a reaction of the type $j_1+\dots j_q \to i_1+\dots +i_p$
which creates species $i_1$. See also appendix E of \citet{Serpico:2004gx} on this topic.  

To summarize, in nucleosynthesis calculations (as in chemistry) we are concerned with number densities. However, it is convenient to normalize them
to Avogadro's number whose dimension can be considered\footnote{A subject of controversy, but irrelevant as soon as it is consistent with
the definition of the reaction rate units.}  to be the inverse of a mass. The form (\ref{q:nucl}) is purely
conventional and $\rho_{\rm n}$ is just
another manner to define a number density even though it has the
dimension of a mass density.

\subsection{Baryonic density and nucleonic density}

In practice, except for BBN, the nucleonic density of Eq.~(\ref{q:rhon}) is usually identified with the atomic matter density, and the nuclear energy source is calculated independently as nuclear flow $\times$ Q
%TODO Maybe write a better and correct footnote there.
%\footnote{In BBN, the energy release is transformed in kinetic energy, and for coupled species with the same temperature, all kinetic energies decrease with the same %scaling thanks to cosmological expansion, hence with no consequence in the Friedmann equation.}. 
This corresponds to the approximation 
\be
A{\cdot}u\;{\approx}\;Zm_p+ (A-Z)m_n-B(A,Z)+Zm_e\,,
\ee 
i.e. an error of  $\approx$1\%, completely negligible in stellar
modeling. For BBN where D/H observations reach the percent level of
uncertainty, it is worth considering the difference between
nucleonic density and baryonic density. 

The baryonic density  $\Omega_{\mathrm{b}}{\cdot}h^2$ deduced from CMB observations is the atomic density  i.e. taking into
account the $^4$He binding energy and the mass of the electrons. Given
that the fraction of baryons in the form of helium is $\YP \equiv 4
\,Y_{{}^4{\rm He}}$ and the rest is in the form of hydrogen, the average mass of baryons is \citep{Steigman2006}
\bea
m_{\rm b}  &\equiv& \myxi \ma \\
\myxi & \equiv & \frac{\YP}{4}
\frac{m_{^4\mathrm{He}}}{\ma}+(1-\YP)\frac{m_{^1\mathrm{H}}}{\ma}\label{Eqxivalue}\\
&=&\frac{m_{^1\mathrm{H}}}{\ma} \left(1-1.75891\times 10^{-3}\frac{\YP}{0.24709}\right)\nonumber
\eea
where $m_{^4\mathrm{He}}$ and $m_{^1\mathrm{H}}$ are the {\em atomic}
masses (see appendix~\ref{ParticleValues}). Using $\YP \simeq 0.24709$ for
the final BBN Helium abundance (it is the most relevant abundance for CMB since stellar formation has not
started during CMB formation) we get
\be
\myxi \simeq 1.006052 \,,\quad \myxi^{-1}\simeq 0.993984\,.
\ee
Hence, from Eqs.~(\ref{Defnb}) and (\ref{q:rhon}), one should use in Eq.~(\ref{q:nucl})
\be
\rho_{\rm n} =\myxi^{-1} \rho_{\rm b}\,,
\ee
where $\rho_{\rm b}$ is obtained from Eq.~(\ref{EqBaryonsScaling}). One can estimate the error introduced if
we ignore this subtlety and use $\myxi=1$ in nuclear reaction. Numerically we found
\beas
\Delta \YP&=&5.6\times 10^{-4}\\
\Delta\left(\frac{{\rm D}}{{\rm H}}\right)&=&-2.4\times 10^{-7}\\
\Delta\left(\frac{{{}^3{\rm He}}}{{\rm H}}\right)&=&3.7\times 10^{-8}\\
\Delta\left(\frac{{{}^7 {\rm Li}}}{{\rm H}}\right)&=&7.1\times 10^{-12}\,.
\eeas

Finally, since the abundances deduced from observations, other than
$^4$He, are expressed as number ratios, hence, e.g. the observed D/H can be directly compared to the BBN $Y_\mathrm{D}/Y_\mathrm{^1H}$  calculated ratio.  The $^4$He (pseudo-)mass fraction, $\YP$ deduced from spectroscopic observations is defined as~\cite{Pag92,Izo94}
%%%%%%%%%%%%%%%%%%%%%%%%%%%%%%%%%%%%%%%%%%%%%%%%%%%%%%%%%%%%%%%%%%%%%%%%%%%%%%%%%%%%%%%%
\begin{equation} 
\YP\equiv\frac{4y}{4y+1}
\end{equation}
%%%%%%%%%%%%%%%%%%%%%%%%%%%%%%%%%%%%%%%%%%%%%%%%%%%%%%%%%%%%%%%%%%%%%%%%%%%%%%%%%%%%%%%%
with
$y{\equiv}n_{^4\mathrm{He}}/n_\mathrm{^1H}{\equiv}Y_{^4\mathrm{He}}/Y_\mathrm{^1H}$
being the {\em number} ratio, resulting in a definition of $\YP$
identical to the BBN one.

\section{Numerical values}\label{ParticleValues}

%We should use \cite{Ser17} which gives $879.5(8)$.

The values of $\cos \theta_C =V_{\rm ud}$ and $g_A = C_A/C_V$ are taken from the Particle Data
Group (PDG) \citep{PDG17}. The neutron decay rate $\tau_n =
879.5(\pm0.8)\,{\rm s}$ is from \citet[Fig. 22]{Ser17} and includes only
experiments after 2000. This value is slightly lower and with less
experimental error than the previously admitted value $\tau_n =
880.2(1.1)\,{\rm s}$ of the PDG.
We report errors on numerical parameters only if they are meaningful
for BBN. Cosmological parameters are taken from \citet{Planck2016}.
\begin{table}[!h]
\centering
\caption{Numerical values used for the BBN code. \label{TableNumValues}}
\ifphysrep
\begin{tabular}{ll}
\else
\begin{tabularx}{0.8\columnwidth}{ll}
\fi
\toprule
 $u$ & $931.494061 \,{\rm MeV} $ \\
 $m_n$ & $939.565360\,{\rm MeV}$\\ 
 $m_p$& $938.272029\,{\rm MeV}$\\
$m_Z$ & $91.1876\,{\rm GeV}$\\
$m_W$ & $80.385\,{\rm GeV}$\\
$m_{^4\mathrm{He}}$ (\rm atomic) &$4.00260325413\,\ma$\\
$m_{^1\mathrm{H}}$ (\rm atomic) & $1.00782503223\,\ma$\\
\midrule
$g_A$& $1.2723(23)$ \\
$\cos \theta_C$ & $0.97420(20)$\\
$f_{\rm WM}$ & $1.853$\\
$G_F$ & $1.1663787\times 10^{-5} \,{\rm GeV}^{-2}$\\
$\tau_n$ & $879.5(8)\,{\rm s}$\\
$r_p$& $0.841\times 10^{-15} \,{\rm m}$\\
$\alpha_{\rm FS}$ & $1/137.03599911$\\\midrule
$T_0$ & $2.7255\,(\pm 0.0006)\,{\rm K}$\\
 $h=H/H_{100}$ & $0.6727\,(\pm0.0066)$ \\
$h^2 \Omega_b$ & $0.022250\,(\pm 0.00016)$\\
$h^2 \Omega_c$ & $0.1198\,(\pm0.0015)$\\
$\rho_{100}^{\rm crit} \equiv 3  H_{100}^2/(8\pi G)$&$1.87847\times 10^{-29}\,{\rm g}/{\rm cm}^3$\\
\bottomrule
\ifphysrep
\end{tabular}
\else
\end{tabularx}
\fi
\end{table}

%\ifphysrep
\bibliography{BiblioBBN}

\begin{thebibliography}{193}
\expandafter\ifx\csname natexlab\endcsname\relax\def\natexlab#1{#1}\fi
\expandafter\ifx\csname bibnamefont\endcsname\relax
  \def\bibnamefont#1{#1}\fi
\expandafter\ifx\csname bibfnamefont\endcsname\relax
  \def\bibfnamefont#1{#1}\fi
\expandafter\ifx\csname citenamefont\endcsname\relax
  \def\citenamefont#1{#1}\fi
\expandafter\ifx\csname url\endcsname\relax
  \def\url#1{\texttt{#1}}\fi
\expandafter\ifx\csname urlprefix\endcsname\relax\def\urlprefix{URL }\fi
\providecommand{\bibinfo}[2]{#2}
\providecommand{\eprint}[2][]{\url{#2}}

\bibitem[{\citenamefont{{Abers}} \emph{et~al.}(1968)\citenamefont{{Abers},
  {Dicus}, {Norton}, and {Quinn}}}]{Dicus1968}
\bibinfo{author}{\bibnamefont{{Abers}}, \bibfnamefont{E.~S.}},
  \bibinfo{author}{\bibfnamefont{D.~A.} \bibnamefont{{Dicus}}},
  \bibinfo{author}{\bibfnamefont{R.~E.} \bibnamefont{{Norton}}}, and
  \bibinfo{author}{\bibfnamefont{H.~R.} \bibnamefont{{Quinn}}},
  \bibinfo{year}{1968}, \bibinfo{journal}{Phys. Rev.}
  \textbf{\bibinfo{volume}{167}}, \bibinfo{pages}{1461}.

\bibitem[{Ade \emph{et~al.}(2016)\citenamefont{Ade} \emph{et~al.}}]{Planck2016}
\bibinfo{author}{\bibnamefont{Ade}, \bibfnamefont{P.~A.~R.}}, \emph{et~al.}
  (\bibinfo{collaboration}{Planck}), \bibinfo{year}{2016},
  \bibinfo{journal}{Astron. Astrophys.} \textbf{\bibinfo{volume}{594}},
  \bibinfo{pages}{A13}.

\bibitem[{\citenamefont{{Adelberger}}
  \emph{et~al.}(2011)\citenamefont{{Adelberger}, {Garc{\'{\i}}a}, {Robertson},
  {Snover}, {Balantekin}, {Heeger}, {Ramsey-Musolf}, {Bemmerer}, {Junghans},
  {Bertulani}, {Chen}, {Costantini}} \emph{et~al.}}]{Ade11}
\bibinfo{author}{\bibnamefont{{Adelberger}}, \bibfnamefont{E.~G.}},
  \bibinfo{author}{\bibfnamefont{A.}~\bibnamefont{{Garc{\'{\i}}a}}},
  \bibinfo{author}{\bibfnamefont{R.~G.~H.} \bibnamefont{{Robertson}}},
  \bibinfo{author}{\bibfnamefont{K.~A.} \bibnamefont{{Snover}}},
  \bibinfo{author}{\bibfnamefont{A.~B.} \bibnamefont{{Balantekin}}},
  \bibinfo{author}{\bibfnamefont{K.}~\bibnamefont{{Heeger}}},
  \bibinfo{author}{\bibfnamefont{M.~J.} \bibnamefont{{Ramsey-Musolf}}},
  \bibinfo{author}{\bibfnamefont{D.}~\bibnamefont{{Bemmerer}}},
  \bibinfo{author}{\bibfnamefont{A.}~\bibnamefont{{Junghans}}},
  \bibinfo{author}{\bibfnamefont{C.~A.} \bibnamefont{{Bertulani}}},
  \bibinfo{author}{\bibfnamefont{J.-W.} \bibnamefont{{Chen}}},
  \bibinfo{author}{\bibfnamefont{H.}~\bibnamefont{{Costantini}}},
  \emph{et~al.}, \bibinfo{year}{2011}, \bibinfo{journal}{Rev. Mod. Phys.}
  \textbf{\bibinfo{volume}{83}}, \bibinfo{pages}{195}.

\bibitem[{\citenamefont{{Aliotta}} \emph{et~al.}(2001)\citenamefont{{Aliotta},
  {Raiola}, {Gy{\"u}rky}, {Formicola}, {Bonetti}, {Broggini}, {Campajola},
  {Corvisiero}, {Costantini}, {D'Onofrio}, {F{\"u}l{\"o}p}, {Gervino}}
  \emph{et~al.}}]{Ali01}
\bibinfo{author}{\bibnamefont{{Aliotta}}, \bibfnamefont{M.}},
  \bibinfo{author}{\bibfnamefont{F.}~\bibnamefont{{Raiola}}},
  \bibinfo{author}{\bibfnamefont{G.}~\bibnamefont{{Gy{\"u}rky}}},
  \bibinfo{author}{\bibfnamefont{A.}~\bibnamefont{{Formicola}}},
  \bibinfo{author}{\bibfnamefont{R.}~\bibnamefont{{Bonetti}}},
  \bibinfo{author}{\bibfnamefont{C.}~\bibnamefont{{Broggini}}},
  \bibinfo{author}{\bibfnamefont{L.}~\bibnamefont{{Campajola}}},
  \bibinfo{author}{\bibfnamefont{P.}~\bibnamefont{{Corvisiero}}},
  \bibinfo{author}{\bibfnamefont{H.}~\bibnamefont{{Costantini}}},
  \bibinfo{author}{\bibfnamefont{A.}~\bibnamefont{{D'Onofrio}}},
  \bibinfo{author}{\bibfnamefont{Z.}~\bibnamefont{{F{\"u}l{\"o}p}}},
  \bibinfo{author}{\bibfnamefont{G.}~\bibnamefont{{Gervino}}}, \emph{et~al.},
  \bibinfo{year}{2001}, \bibinfo{journal}{Nuclear Physics A}
  \textbf{\bibinfo{volume}{690}}, \bibinfo{pages}{790}.

\bibitem[{\citenamefont{{Amsler}} \emph{et~al.}(2008)\citenamefont{{Amsler},
  {Doser}, {Antonelli}, {Asner}, {Babu}, {Baer}, {Band}, {Barnett}, {Bergren},
  {Beringer}, {Bernardi}, {Bertl}} \emph{et~al.}}]{PDG08}
\bibinfo{author}{\bibnamefont{{Amsler}}, \bibfnamefont{C.}},
  \bibinfo{author}{\bibfnamefont{M.}~\bibnamefont{{Doser}}},
  \bibinfo{author}{\bibfnamefont{M.}~\bibnamefont{{Antonelli}}},
  \bibinfo{author}{\bibfnamefont{D.~M.} \bibnamefont{{Asner}}},
  \bibinfo{author}{\bibfnamefont{K.~S.} \bibnamefont{{Babu}}},
  \bibinfo{author}{\bibfnamefont{H.}~\bibnamefont{{Baer}}},
  \bibinfo{author}{\bibfnamefont{H.~R.} \bibnamefont{{Band}}},
  \bibinfo{author}{\bibfnamefont{R.~M.} \bibnamefont{{Barnett}}},
  \bibinfo{author}{\bibfnamefont{E.}~\bibnamefont{{Bergren}}},
  \bibinfo{author}{\bibfnamefont{J.}~\bibnamefont{{Beringer}}},
  \bibinfo{author}{\bibfnamefont{G.}~\bibnamefont{{Bernardi}}},
  \bibinfo{author}{\bibfnamefont{W.}~\bibnamefont{{Bertl}}}, \emph{et~al.},
  \bibinfo{year}{2008}, \bibinfo{journal}{Physics Letters B}
  \textbf{\bibinfo{volume}{667}}, \bibinfo{pages}{1}.

\bibitem[{\citenamefont{{Anders}} \emph{et~al.}(2014)\citenamefont{{Anders},
  {Trezzi}, {Menegazzo}, {Aliotta}, {Bellini}, {Bemmerer}, {Broggini},
  {Caciolli}, {Corvisiero}, {Costantini}, {Davinson}, {Elekes}}
  \emph{et~al.}}]{And14}
\bibinfo{author}{\bibnamefont{{Anders}}, \bibfnamefont{M.}},
  \bibinfo{author}{\bibfnamefont{D.}~\bibnamefont{{Trezzi}}},
  \bibinfo{author}{\bibfnamefont{R.}~\bibnamefont{{Menegazzo}}},
  \bibinfo{author}{\bibfnamefont{M.}~\bibnamefont{{Aliotta}}},
  \bibinfo{author}{\bibfnamefont{A.}~\bibnamefont{{Bellini}}},
  \bibinfo{author}{\bibfnamefont{D.}~\bibnamefont{{Bemmerer}}},
  \bibinfo{author}{\bibfnamefont{C.}~\bibnamefont{{Broggini}}},
  \bibinfo{author}{\bibfnamefont{A.}~\bibnamefont{{Caciolli}}},
  \bibinfo{author}{\bibfnamefont{P.}~\bibnamefont{{Corvisiero}}},
  \bibinfo{author}{\bibfnamefont{H.}~\bibnamefont{{Costantini}}},
  \bibinfo{author}{\bibfnamefont{T.}~\bibnamefont{{Davinson}}},
  \bibinfo{author}{\bibfnamefont{Z.}~\bibnamefont{{Elekes}}}, \emph{et~al.},
  \bibinfo{year}{2014}, \bibinfo{journal}{Phys, Rev. Lett.}
  \textbf{\bibinfo{volume}{113}}, \bibinfo{eid}{042501}.

\bibitem[{\citenamefont{{Ando}} \emph{et~al.}(2006)\citenamefont{{Ando},
  {Cyburt}, {Hong}, and {Hyun}}}]{AndoEtAl2006}
\bibinfo{author}{\bibnamefont{{Ando}}, \bibfnamefont{S.}},
  \bibinfo{author}{\bibfnamefont{R.~H.} \bibnamefont{{Cyburt}}},
  \bibinfo{author}{\bibfnamefont{S.~W.} \bibnamefont{{Hong}}}, and
  \bibinfo{author}{\bibfnamefont{C.~H.} \bibnamefont{{Hyun}}},
  \bibinfo{year}{2006}, \bibinfo{journal}{Phys. Rev.}
  \textbf{\bibinfo{volume}{C74}}, \bibinfo{eid}{025809}.

\bibitem[{\citenamefont{{Angulo}} \emph{et~al.}(1999)\citenamefont{{Angulo},
  {Arnould}, {Rayet}, {Descouvemont}, {Baye}, {Leclercq-Willain}, {Coc},
  {Barhoumi}, {Aguer}, {Rolfs}, {Kunz}, {Hammer}} \emph{et~al.}}]{NACRE}
\bibinfo{author}{\bibnamefont{{Angulo}}, \bibfnamefont{C.}},
  \bibinfo{author}{\bibfnamefont{M.}~\bibnamefont{{Arnould}}},
  \bibinfo{author}{\bibfnamefont{M.}~\bibnamefont{{Rayet}}},
  \bibinfo{author}{\bibfnamefont{P.}~\bibnamefont{{Descouvemont}}},
  \bibinfo{author}{\bibfnamefont{D.}~\bibnamefont{{Baye}}},
  \bibinfo{author}{\bibfnamefont{C.}~\bibnamefont{{Leclercq-Willain}}},
  \bibinfo{author}{\bibfnamefont{A.}~\bibnamefont{{Coc}}},
  \bibinfo{author}{\bibfnamefont{S.}~\bibnamefont{{Barhoumi}}},
  \bibinfo{author}{\bibfnamefont{P.}~\bibnamefont{{Aguer}}},
  \bibinfo{author}{\bibfnamefont{C.}~\bibnamefont{{Rolfs}}},
  \bibinfo{author}{\bibfnamefont{R.}~\bibnamefont{{Kunz}}},
  \bibinfo{author}{\bibfnamefont{J.~W.} \bibnamefont{{Hammer}}}, \emph{et~al.},
  \bibinfo{year}{1999}, \bibinfo{journal}{Nuclear Physics A}
  \textbf{\bibinfo{volume}{656}}, \bibinfo{pages}{3}.

\bibitem[{\citenamefont{{Angulo}} \emph{et~al.}(2005)\citenamefont{{Angulo},
  {Casarejos}, {Couder}, {Demaret}, {Leleux}, {Vanderbist}, {Coc}, {Kiener},
  {Tatischeff}, {Davinson}, {Murphy}, {Achouri}} \emph{et~al.}}]{Ang05}
\bibinfo{author}{\bibnamefont{{Angulo}}, \bibfnamefont{C.}},
  \bibinfo{author}{\bibfnamefont{E.}~\bibnamefont{{Casarejos}}},
  \bibinfo{author}{\bibfnamefont{M.}~\bibnamefont{{Couder}}},
  \bibinfo{author}{\bibfnamefont{P.}~\bibnamefont{{Demaret}}},
  \bibinfo{author}{\bibfnamefont{P.}~\bibnamefont{{Leleux}}},
  \bibinfo{author}{\bibfnamefont{F.}~\bibnamefont{{Vanderbist}}},
  \bibinfo{author}{\bibfnamefont{A.}~\bibnamefont{{Coc}}},
  \bibinfo{author}{\bibfnamefont{J.}~\bibnamefont{{Kiener}}},
  \bibinfo{author}{\bibfnamefont{V.}~\bibnamefont{{Tatischeff}}},
  \bibinfo{author}{\bibfnamefont{T.}~\bibnamefont{{Davinson}}},
  \bibinfo{author}{\bibfnamefont{A.~S.} \bibnamefont{{Murphy}}},
  \bibinfo{author}{\bibfnamefont{N.~L.} \bibnamefont{{Achouri}}},
  \emph{et~al.}, \bibinfo{year}{2005}, \bibinfo{journal}{Astrophys.~J.~Lett.}
  \textbf{\bibinfo{volume}{630}}, \bibinfo{pages}{L105}.

\bibitem[{\citenamefont{{Aoki}} \emph{et~al.}(2009)\citenamefont{{Aoki},
  {Barklem}, {Beers}, {Christlieb}, {Inoue}, {Garc{\'{\i}}a P{\'e}rez},
  {Norris}, and {Carollo}}}]{Aok09}
\bibinfo{author}{\bibnamefont{{Aoki}}, \bibfnamefont{W.}},
  \bibinfo{author}{\bibfnamefont{P.~S.} \bibnamefont{{Barklem}}},
  \bibinfo{author}{\bibfnamefont{T.~C.} \bibnamefont{{Beers}}},
  \bibinfo{author}{\bibfnamefont{N.}~\bibnamefont{{Christlieb}}},
  \bibinfo{author}{\bibfnamefont{S.}~\bibnamefont{{Inoue}}},
  \bibinfo{author}{\bibfnamefont{A.~E.} \bibnamefont{{Garc{\'{\i}}a
  P{\'e}rez}}}, \bibinfo{author}{\bibfnamefont{J.~E.} \bibnamefont{{Norris}}},
  and \bibinfo{author}{\bibfnamefont{D.}~\bibnamefont{{Carollo}}},
  \bibinfo{year}{2009}, \bibinfo{journal}{Astrophys.~J.}
  \textbf{\bibinfo{volume}{698}}, \bibinfo{pages}{1803}.

\bibitem[{\citenamefont{{Arai}} \emph{et~al.}(2011)\citenamefont{{Arai},
  {Aoyama}, {Suzuki}, {Descouvemont}, and {Baye}}}]{Ara11}
\bibinfo{author}{\bibnamefont{{Arai}}, \bibfnamefont{K.}},
  \bibinfo{author}{\bibfnamefont{S.}~\bibnamefont{{Aoyama}}},
  \bibinfo{author}{\bibfnamefont{Y.}~\bibnamefont{{Suzuki}}},
  \bibinfo{author}{\bibfnamefont{P.}~\bibnamefont{{Descouvemont}}}, and
  \bibinfo{author}{\bibfnamefont{D.}~\bibnamefont{{Baye}}},
  \bibinfo{year}{2011}, \bibinfo{journal}{Phys. Rev. Lett.}
  \textbf{\bibinfo{volume}{107}}, \bibinfo{eid}{132502}.

\bibitem[{\citenamefont{Arbey}(2012)}]{AlterBBN}
\bibinfo{author}{\bibnamefont{Arbey}, \bibfnamefont{A.}}, \bibinfo{year}{2012},
  \bibinfo{journal}{Comput. Phys. Commun.} \textbf{\bibinfo{volume}{183}},
  \bibinfo{pages}{1822}.

\bibitem[{\citenamefont{{Asplund}} \emph{et~al.}(2006)\citenamefont{{Asplund},
  {Lambert}, {Nissen}, {Primas}, and {Smith}}}]{Asp06}
\bibinfo{author}{\bibnamefont{{Asplund}}, \bibfnamefont{M.}},
  \bibinfo{author}{\bibfnamefont{D.~L.} \bibnamefont{{Lambert}}},
  \bibinfo{author}{\bibfnamefont{P.~E.} \bibnamefont{{Nissen}}},
  \bibinfo{author}{\bibfnamefont{F.}~\bibnamefont{{Primas}}}, and
  \bibinfo{author}{\bibfnamefont{V.~V.} \bibnamefont{{Smith}}},
  \bibinfo{year}{2006}, \bibinfo{journal}{Astrophys. J.}
  \textbf{\bibinfo{volume}{644}}, \bibinfo{pages}{229}.

\bibitem[{\citenamefont{{Audi}} \emph{et~al.}(2017)\citenamefont{{Audi},
  {Kondev}, {Wang}, {Huang}, and {Naimi}}}]{Aud17}
\bibinfo{author}{\bibnamefont{{Audi}}, \bibfnamefont{G.}},
  \bibinfo{author}{\bibfnamefont{F.~G.} \bibnamefont{{Kondev}}},
  \bibinfo{author}{\bibfnamefont{M.}~\bibnamefont{{Wang}}},
  \bibinfo{author}{\bibfnamefont{W.~J.} \bibnamefont{{Huang}}}, and
  \bibinfo{author}{\bibfnamefont{S.}~\bibnamefont{{Naimi}}},
  \bibinfo{year}{2017}, \bibinfo{journal}{Chinese Physics C}
  \textbf{\bibinfo{volume}{41}}, \bibinfo{eid}{030001}.

\bibitem[{\citenamefont{{Aver}} \emph{et~al.}(2015)\citenamefont{{Aver},
  {Olive}, and {Skillman}}}]{Ave15}
\bibinfo{author}{\bibnamefont{{Aver}}, \bibfnamefont{E.}},
  \bibinfo{author}{\bibfnamefont{K.~A.} \bibnamefont{{Olive}}}, and
  \bibinfo{author}{\bibfnamefont{E.~D.} \bibnamefont{{Skillman}}},
  \bibinfo{year}{2015}, \bibinfo{journal}{"JCAP"} \textbf{\bibinfo{volume}{7}},
  \bibinfo{eid}{011}.

\bibitem[{\citenamefont{{Balashev}}
  \emph{et~al.}(2016)\citenamefont{{Balashev}, {Zavarygin}, {Ivanchik},
  {Telikova}, and {Varshalovich}}}]{Bal16}
\bibinfo{author}{\bibnamefont{{Balashev}}, \bibfnamefont{S.~A.}},
  \bibinfo{author}{\bibfnamefont{E.~O.} \bibnamefont{{Zavarygin}}},
  \bibinfo{author}{\bibfnamefont{A.~V.} \bibnamefont{{Ivanchik}}},
  \bibinfo{author}{\bibfnamefont{K.~N.} \bibnamefont{{Telikova}}}, and
  \bibinfo{author}{\bibfnamefont{D.~A.} \bibnamefont{{Varshalovich}}},
  \bibinfo{year}{2016}, \bibinfo{journal}{MNRAS}
  \textbf{\bibinfo{volume}{458}}, \bibinfo{pages}{2188}.

\bibitem[{\citenamefont{{Bania}} \emph{et~al.}(2002)\citenamefont{{Bania},
  {Rood}, and {Balser}}}]{Ban02}
\bibinfo{author}{\bibnamefont{{Bania}}, \bibfnamefont{T.~M.}},
  \bibinfo{author}{\bibfnamefont{R.~T.} \bibnamefont{{Rood}}}, and
  \bibinfo{author}{\bibfnamefont{D.~S.} \bibnamefont{{Balser}}},
  \bibinfo{year}{2002}, \bibinfo{journal}{Nature}
  \textbf{\bibinfo{volume}{415}}, \bibinfo{pages}{54}.

\bibitem[{\citenamefont{{Barbagallo}}
  \emph{et~al.}(2016)\citenamefont{{Barbagallo}, {Musumarra}, {Cosentino},
  {Maugeri}, {Heinitz}, {Mengoni}, {Dressler}, {Schumann}, {K{\"a}ppeler},
  {Colonna}, {Finocchiaro}, {Ayranov}} \emph{et~al.}}]{Bar16}
\bibinfo{author}{\bibnamefont{{Barbagallo}}, \bibfnamefont{M.}},
  \bibinfo{author}{\bibfnamefont{A.}~\bibnamefont{{Musumarra}}},
  \bibinfo{author}{\bibfnamefont{L.}~\bibnamefont{{Cosentino}}},
  \bibinfo{author}{\bibfnamefont{E.}~\bibnamefont{{Maugeri}}},
  \bibinfo{author}{\bibfnamefont{S.}~\bibnamefont{{Heinitz}}},
  \bibinfo{author}{\bibfnamefont{A.}~\bibnamefont{{Mengoni}}},
  \bibinfo{author}{\bibfnamefont{R.}~\bibnamefont{{Dressler}}},
  \bibinfo{author}{\bibfnamefont{D.}~\bibnamefont{{Schumann}}},
  \bibinfo{author}{\bibfnamefont{F.}~\bibnamefont{{K{\"a}ppeler}}},
  \bibinfo{author}{\bibfnamefont{N.}~\bibnamefont{{Colonna}}},
  \bibinfo{author}{\bibfnamefont{P.}~\bibnamefont{{Finocchiaro}}},
  \bibinfo{author}{\bibfnamefont{M.}~\bibnamefont{{Ayranov}}}, \emph{et~al.},
  \bibinfo{year}{2016}, \bibinfo{journal}{Phys. Rev. Lett.}
  \textbf{\bibinfo{volume}{117}}, \bibinfo{eid}{152701}.

\bibitem[{\citenamefont{{Becchetti}}
  \emph{et~al.}(1992)\citenamefont{{Becchetti}, {Brown}, {Liu}, {J{\"a}necke},
  {Roberts}, {Kolata}, {Smith}, {Lamkin}, {Morsad}, {Warner}, {Boyd}, and
  {Kalen}}}]{Bec92}
\bibinfo{author}{\bibnamefont{{Becchetti}}, \bibfnamefont{F.~D.}},
  \bibinfo{author}{\bibfnamefont{J.~A.} \bibnamefont{{Brown}}},
  \bibinfo{author}{\bibfnamefont{W.~Z.} \bibnamefont{{Liu}}},
  \bibinfo{author}{\bibfnamefont{J.~W.} \bibnamefont{{J{\"a}necke}}},
  \bibinfo{author}{\bibfnamefont{D.~A.} \bibnamefont{{Roberts}}},
  \bibinfo{author}{\bibfnamefont{J.~J.} \bibnamefont{{Kolata}}},
  \bibinfo{author}{\bibfnamefont{R.~J.} \bibnamefont{{Smith}}},
  \bibinfo{author}{\bibfnamefont{K.}~\bibnamefont{{Lamkin}}},
  \bibinfo{author}{\bibfnamefont{A.}~\bibnamefont{{Morsad}}},
  \bibinfo{author}{\bibfnamefont{R.~E.} \bibnamefont{{Warner}}},
  \bibinfo{author}{\bibfnamefont{R.~N.} \bibnamefont{{Boyd}}}, and
  \bibinfo{author}{\bibfnamefont{J.~D.} \bibnamefont{{Kalen}}},
  \bibinfo{year}{1992}, \bibinfo{journal}{Nucl.~Phys.~A}
  \textbf{\bibinfo{volume}{550}}, \bibinfo{pages}{507}.

\bibitem[{\citenamefont{Bernstein} \emph{et~al.}(1989)\citenamefont{Bernstein,
  Brown, and Feinberg}}]{Bernstein1989}
\bibinfo{author}{\bibnamefont{Bernstein}, \bibfnamefont{J.}},
  \bibinfo{author}{\bibfnamefont{L.~S.} \bibnamefont{Brown}}, and
  \bibinfo{author}{\bibfnamefont{G.}~\bibnamefont{Feinberg}},
  \bibinfo{year}{1989}, \bibinfo{journal}{Rev. Mod. Phys.}
  \textbf{\bibinfo{volume}{61}}, \bibinfo{pages}{25}.

\bibitem[{\citenamefont{Birrell} \emph{et~al.}(2014)\citenamefont{Birrell,
  Yang, and Rafelski}}]{Birrell:2014uka}
\bibinfo{author}{\bibnamefont{Birrell}, \bibfnamefont{J.}},
  \bibinfo{author}{\bibfnamefont{C.-T.} \bibnamefont{Yang}}, and
  \bibinfo{author}{\bibfnamefont{J.}~\bibnamefont{Rafelski}},
  \bibinfo{year}{2014}, \bibinfo{journal}{Nucl. Phys.}
  \textbf{\bibinfo{volume}{B890}}, \bibinfo{pages}{481}.

\bibitem[{\citenamefont{{Blas}} \emph{et~al.}(2011)\citenamefont{{Blas},
  {Lesgourgues}, and {Tram}}}]{CLASSPaper}
\bibinfo{author}{\bibnamefont{{Blas}}, \bibfnamefont{D.}},
  \bibinfo{author}{\bibfnamefont{J.}~\bibnamefont{{Lesgourgues}}}, and
  \bibinfo{author}{\bibfnamefont{T.}~\bibnamefont{{Tram}}},
  \bibinfo{year}{2011}, \bibinfo{journal}{JCAP} \textbf{\bibinfo{volume}{7}},
  \bibinfo{eid}{034}.

\bibitem[{\citenamefont{{Bonifacio}}
  \emph{et~al.}(2007)\citenamefont{{Bonifacio}, {Molaro}, {Sivarani}, {Cayrel},
  {Spite}, {Spite}, {Plez}, {Andersen}, {Barbuy}, {Beers}, {Depagne}, {Hill}}
  \emph{et~al.}}]{Bon07}
\bibinfo{author}{\bibnamefont{{Bonifacio}}, \bibfnamefont{P.}},
  \bibinfo{author}{\bibfnamefont{P.}~\bibnamefont{{Molaro}}},
  \bibinfo{author}{\bibfnamefont{T.}~\bibnamefont{{Sivarani}}},
  \bibinfo{author}{\bibfnamefont{R.}~\bibnamefont{{Cayrel}}},
  \bibinfo{author}{\bibfnamefont{M.}~\bibnamefont{{Spite}}},
  \bibinfo{author}{\bibfnamefont{F.}~\bibnamefont{{Spite}}},
  \bibinfo{author}{\bibfnamefont{B.}~\bibnamefont{{Plez}}},
  \bibinfo{author}{\bibfnamefont{J.}~\bibnamefont{{Andersen}}},
  \bibinfo{author}{\bibfnamefont{B.}~\bibnamefont{{Barbuy}}},
  \bibinfo{author}{\bibfnamefont{T.~C.} \bibnamefont{{Beers}}},
  \bibinfo{author}{\bibfnamefont{E.}~\bibnamefont{{Depagne}}},
  \bibinfo{author}{\bibfnamefont{V.}~\bibnamefont{{Hill}}}, \emph{et~al.},
  \bibinfo{year}{2007}, \bibinfo{journal}{Astron. Astrophys.}
  \textbf{\bibinfo{volume}{462}}, \bibinfo{pages}{851}.

\bibitem[{\citenamefont{{Broggini}}
  \emph{et~al.}(2012)\citenamefont{{Broggini}, {Canton}, {Fiorentini}, and
  {Villante}}}]{Bro12}
\bibinfo{author}{\bibnamefont{{Broggini}}, \bibfnamefont{C.}},
  \bibinfo{author}{\bibfnamefont{L.}~\bibnamefont{{Canton}}},
  \bibinfo{author}{\bibfnamefont{G.}~\bibnamefont{{Fiorentini}}}, and
  \bibinfo{author}{\bibfnamefont{F.~L.} \bibnamefont{{Villante}}},
  \bibinfo{year}{2012}, \bibinfo{journal}{JCAP} \textbf{\bibinfo{volume}{6}},
  \bibinfo{eid}{030}.

\bibitem[{\citenamefont{Brown and Sawyer}(2001)}]{BrownSawyer}
\bibinfo{author}{\bibnamefont{Brown}, \bibfnamefont{L.~S.}}, and
  \bibinfo{author}{\bibfnamefont{R.~F.} \bibnamefont{Sawyer}},
  \bibinfo{year}{2001}, \bibinfo{journal}{Phys. Rev.}
  \textbf{\bibinfo{volume}{D63}}, \bibinfo{pages}{083503}.

\bibitem[{\citenamefont{{Bystritsky}}
  \emph{et~al.}(2008)\citenamefont{{Bystritsky}, {Gerasimov}, {Krylov},
  {Parzhitskii}, {Dudkin}, {Kaminskii}, {Nechaev}, {Padalko}, {Petrov},
  {Mesyats}, {Filipowicz}, {Wozniak}} \emph{et~al.}}]{Bys08b}
\bibinfo{author}{\bibnamefont{{Bystritsky}}, \bibfnamefont{V.~M.}},
  \bibinfo{author}{\bibfnamefont{V.~V.} \bibnamefont{{Gerasimov}}},
  \bibinfo{author}{\bibfnamefont{A.~R.} \bibnamefont{{Krylov}}},
  \bibinfo{author}{\bibfnamefont{S.~S.} \bibnamefont{{Parzhitskii}}},
  \bibinfo{author}{\bibfnamefont{G.~N.} \bibnamefont{{Dudkin}}},
  \bibinfo{author}{\bibfnamefont{V.~L.} \bibnamefont{{Kaminskii}}},
  \bibinfo{author}{\bibfnamefont{B.~A.} \bibnamefont{{Nechaev}}},
  \bibinfo{author}{\bibfnamefont{V.~N.} \bibnamefont{{Padalko}}},
  \bibinfo{author}{\bibfnamefont{A.~V.} \bibnamefont{{Petrov}}},
  \bibinfo{author}{\bibfnamefont{G.~A.} \bibnamefont{{Mesyats}}},
  \bibinfo{author}{\bibfnamefont{M.}~\bibnamefont{{Filipowicz}}},
  \bibinfo{author}{\bibfnamefont{J.}~\bibnamefont{{Wozniak}}}, \emph{et~al.},
  \bibinfo{year}{2008}, \bibinfo{journal}{Nuclear Instruments and Methods in
  Physics Research A} \textbf{\bibinfo{volume}{595}}, \bibinfo{pages}{543}.

\bibitem[{\citenamefont{{Cambier}} \emph{et~al.}(1982)\citenamefont{{Cambier},
  {Primack}, and {Sher}}}]{Cambier1982}
\bibinfo{author}{\bibnamefont{{Cambier}}, \bibfnamefont{J.-L.}},
  \bibinfo{author}{\bibfnamefont{J.~R.} \bibnamefont{{Primack}}}, and
  \bibinfo{author}{\bibfnamefont{M.}~\bibnamefont{{Sher}}},
  \bibinfo{year}{1982}, \bibinfo{journal}{Nucl. Phys.}
  \textbf{\bibinfo{volume}{B209}}, \bibinfo{pages}{372}.

\bibitem[{\citenamefont{{Casella}} \emph{et~al.}(2002)\citenamefont{{Casella},
  {Costantini}, {Lemut}, {Limata}, {Bonetti}, {Broggini}, {Campajola},
  {Corvisiero}, {Cruz}, {D'Onofrio}, {Formicola}, {F{\"u}l{\"o}p}}
  \emph{et~al.}}]{Cas02}
\bibinfo{author}{\bibnamefont{{Casella}}, \bibfnamefont{C.}},
  \bibinfo{author}{\bibfnamefont{H.}~\bibnamefont{{Costantini}}},
  \bibinfo{author}{\bibfnamefont{A.}~\bibnamefont{{Lemut}}},
  \bibinfo{author}{\bibfnamefont{B.}~\bibnamefont{{Limata}}},
  \bibinfo{author}{\bibfnamefont{R.}~\bibnamefont{{Bonetti}}},
  \bibinfo{author}{\bibfnamefont{C.}~\bibnamefont{{Broggini}}},
  \bibinfo{author}{\bibfnamefont{L.}~\bibnamefont{{Campajola}}},
  \bibinfo{author}{\bibfnamefont{P.}~\bibnamefont{{Corvisiero}}},
  \bibinfo{author}{\bibfnamefont{J.}~\bibnamefont{{Cruz}}},
  \bibinfo{author}{\bibfnamefont{A.}~\bibnamefont{{D'Onofrio}}},
  \bibinfo{author}{\bibfnamefont{A.}~\bibnamefont{{Formicola}}},
  \bibinfo{author}{\bibfnamefont{Z.}~\bibnamefont{{F{\"u}l{\"o}p}}},
  \emph{et~al.}, \bibinfo{year}{2002}, \bibinfo{journal}{Nucl. Phys. A}
  \textbf{\bibinfo{volume}{706}}, \bibinfo{pages}{203}.

\bibitem[{\citenamefont{{Cassisi} and {Castellani}}(1993)}]{Cas93}
\bibinfo{author}{\bibnamefont{{Cassisi}}, \bibfnamefont{S.}}, and
  \bibinfo{author}{\bibfnamefont{V.}~\bibnamefont{{Castellani}}},
  \bibinfo{year}{1993}, \bibinfo{journal}{Astrophys.~J.~Supp.}
  \textbf{\bibinfo{volume}{88}}, \bibinfo{pages}{509}.

\bibitem[{\citenamefont{{Chakraborty}}
  \emph{et~al.}(2011)\citenamefont{{Chakraborty}, {Fields}, and
  {Olive}}}]{Cha11}
\bibinfo{author}{\bibnamefont{{Chakraborty}}, \bibfnamefont{N.}},
  \bibinfo{author}{\bibfnamefont{B.~D.} \bibnamefont{{Fields}}}, and
  \bibinfo{author}{\bibfnamefont{K.~A.} \bibnamefont{{Olive}}},
  \bibinfo{year}{2011}, \bibinfo{journal}{Phys. Rev.}
  \textbf{\bibinfo{volume}{D83}}, \bibinfo{eid}{063006}.

\bibitem[{\citenamefont{{Chapman}}(1997)}]{Chapman1997}
\bibinfo{author}{\bibnamefont{{Chapman}}, \bibfnamefont{I.~A.}},
  \bibinfo{year}{1997}, \bibinfo{journal}{Phys. Rev.}
  \textbf{\bibinfo{volume}{D55}}, \bibinfo{pages}{6287}.

\bibitem[{\citenamefont{{Charbonnel} and {Primas}}(2005)}]{Cha05}
\bibinfo{author}{\bibnamefont{{Charbonnel}}, \bibfnamefont{C.}}, and
  \bibinfo{author}{\bibfnamefont{F.}~\bibnamefont{{Primas}}},
  \bibinfo{year}{2005}, \bibinfo{journal}{Astron. Astrophys.}
  \textbf{\bibinfo{volume}{442}}, \bibinfo{pages}{961}.

\bibitem[{\citenamefont{{Clayton}}(1983)}]{Clayton}
\bibinfo{author}{\bibnamefont{{Clayton}}, \bibfnamefont{D.~D.}},
  \bibinfo{year}{1983}, \emph{\bibinfo{title}{{Principles of stellar evolution
  and nucleosynthesis}}} (\bibinfo{publisher}{Chicago: University of Chicago
  Press}).

\bibitem[{\citenamefont{{Coc}}(2013)}]{Zakopane}
\bibinfo{author}{\bibnamefont{{Coc}}, \bibfnamefont{A.}}, \bibinfo{year}{2013},
  \bibinfo{journal}{Acta Physica Polonica B} \textbf{\bibinfo{volume}{44}},
  \bibinfo{pages}{521}.

\bibitem[{\citenamefont{{Coc}} \emph{et~al.}(2012)\citenamefont{{Coc},
  {Goriely}, {Xu}, {Saimpert}, and {Vangioni}}}]{Coc12a}
\bibinfo{author}{\bibnamefont{{Coc}}, \bibfnamefont{A.}},
  \bibinfo{author}{\bibfnamefont{S.}~\bibnamefont{{Goriely}}},
  \bibinfo{author}{\bibfnamefont{Y.}~\bibnamefont{{Xu}}},
  \bibinfo{author}{\bibfnamefont{M.}~\bibnamefont{{Saimpert}}}, and
  \bibinfo{author}{\bibfnamefont{E.}~\bibnamefont{{Vangioni}}},
  \bibinfo{year}{2012}, \bibinfo{journal}{Astrophys.~J.}
  \textbf{\bibinfo{volume}{744}}, \bibinfo{eid}{158}.

\bibitem[{\citenamefont{{Coc}} \emph{et~al.}(2015)\citenamefont{{Coc},
  {Hammache}, and {Kiener}}}]{CHK}
\bibinfo{author}{\bibnamefont{{Coc}}, \bibfnamefont{A.}},
  \bibinfo{author}{\bibfnamefont{F.}~\bibnamefont{{Hammache}}}, and
  \bibinfo{author}{\bibfnamefont{J.}~\bibnamefont{{Kiener}}},
  \bibinfo{year}{2015}, \bibinfo{journal}{European Physical Journal A}
  \textbf{\bibinfo{volume}{51}}, \bibinfo{eid}{34}.

\bibitem[{\citenamefont{{Coc}} \emph{et~al.}(2009)\citenamefont{{Coc}, {Olive},
  {Uzan}, and {Vangioni}}}]{Coc09}
\bibinfo{author}{\bibnamefont{{Coc}}, \bibfnamefont{A.}},
  \bibinfo{author}{\bibfnamefont{K.~A.} \bibnamefont{{Olive}}},
  \bibinfo{author}{\bibfnamefont{J.-P.} \bibnamefont{{Uzan}}}, and
  \bibinfo{author}{\bibfnamefont{E.}~\bibnamefont{{Vangioni}}},
  \bibinfo{year}{2009}, \bibinfo{journal}{Phys. Rev.}
  \textbf{\bibinfo{volume}{D79}}, \bibinfo{eid}{103512}.

\bibitem[{\citenamefont{Coc} \emph{et~al.}(2015)\citenamefont{Coc, Petitjean,
  Uzan, Vangioni, Descouvemont, Iliadis, and Longland}}]{Coc15}
\bibinfo{author}{\bibnamefont{Coc}, \bibfnamefont{A.}},
  \bibinfo{author}{\bibfnamefont{P.}~\bibnamefont{Petitjean}},
  \bibinfo{author}{\bibfnamefont{J.-P.} \bibnamefont{Uzan}},
  \bibinfo{author}{\bibfnamefont{E.}~\bibnamefont{Vangioni}},
  \bibinfo{author}{\bibfnamefont{P.}~\bibnamefont{Descouvemont}},
  \bibinfo{author}{\bibfnamefont{C.}~\bibnamefont{Iliadis}}, and
  \bibinfo{author}{\bibfnamefont{R.}~\bibnamefont{Longland}},
  \bibinfo{year}{2015}, \bibinfo{journal}{Phys. Rev.}
  \textbf{\bibinfo{volume}{D92}}, \bibinfo{pages}{123526}.

\bibitem[{\citenamefont{Coc} \emph{et~al.}(2014)\citenamefont{Coc, Uzan, and
  Vangioni}}]{Coc14}
\bibinfo{author}{\bibnamefont{Coc}, \bibfnamefont{A.}},
  \bibinfo{author}{\bibfnamefont{J.-P.} \bibnamefont{Uzan}}, and
  \bibinfo{author}{\bibfnamefont{E.}~\bibnamefont{Vangioni}},
  \bibinfo{year}{2014}, \bibinfo{journal}{JCAP}
  \textbf{\bibinfo{volume}{1410}}, \bibinfo{pages}{050}.

\bibitem[{\citenamefont{{Coc} and {Vangioni}}(2010)}]{CV10}
\bibinfo{author}{\bibnamefont{{Coc}}, \bibfnamefont{A.}}, and
  \bibinfo{author}{\bibfnamefont{E.}~\bibnamefont{{Vangioni}}},
  \bibinfo{year}{2010}, in \emph{\bibinfo{booktitle}{{Big-Bang nucleosynthesis
  with updated nuclear data}}}, volume \bibinfo{volume}{202} of
  \emph{\bibinfo{series}{Journal of Physics Conference Series}}, p.
  \bibinfo{pages}{012001}.

\bibitem[{\citenamefont{{Coc} and {Vangioni}}(2014)}]{NIC2014}
\bibinfo{author}{\bibnamefont{{Coc}}, \bibfnamefont{A.}}, and
  \bibinfo{author}{\bibfnamefont{E.}~\bibnamefont{{Vangioni}}},
  \bibinfo{year}{2014}, in \emph{\bibinfo{booktitle}{Proceedings, XIII Nuclei
  in the Cosmos, Debrecen, Hungary, July 7-11, 2014, PoS (NIC XIII)}},
  p.~\bibinfo{pages}{22}.

\bibitem[{\citenamefont{{Coc} and {Vangioni}}(2017)}]{CV17}
\bibinfo{author}{\bibnamefont{{Coc}}, \bibfnamefont{A.}}, and
  \bibinfo{author}{\bibfnamefont{E.}~\bibnamefont{{Vangioni}}},
  \bibinfo{year}{2017}, \bibinfo{journal}{International Journal of Modern
  Physics E} \textbf{\bibinfo{volume}{26}}, \bibinfo{eid}{1741002}.

\bibitem[{\citenamefont{Coc} \emph{et~al.}(2002)\citenamefont{Coc,
  Vangioni-Flam, Cass\'e, and Rabiet}}]{Coc02}
\bibinfo{author}{\bibnamefont{Coc}, \bibfnamefont{A.}},
  \bibinfo{author}{\bibfnamefont{E.}~\bibnamefont{Vangioni-Flam}},
  \bibinfo{author}{\bibfnamefont{M.}~\bibnamefont{Cass\'e}}, and
  \bibinfo{author}{\bibfnamefont{M.}~\bibnamefont{Rabiet}},
  \bibinfo{year}{2002}, \bibinfo{journal}{Phys. Rev.}
  \textbf{\bibinfo{volume}{D65}}, \bibinfo{pages}{043510}.

\bibitem[{\citenamefont{{Coc}} \emph{et~al.}(2004)\citenamefont{{Coc},
  {Vangioni-Flam}, {Descouvemont}, {Adahchour}, and {Angulo}}}]{Coc04}
\bibinfo{author}{\bibnamefont{{Coc}}, \bibfnamefont{A.}},
  \bibinfo{author}{\bibfnamefont{E.}~\bibnamefont{{Vangioni-Flam}}},
  \bibinfo{author}{\bibfnamefont{P.}~\bibnamefont{{Descouvemont}}},
  \bibinfo{author}{\bibfnamefont{A.}~\bibnamefont{{Adahchour}}}, and
  \bibinfo{author}{\bibfnamefont{C.}~\bibnamefont{{Angulo}}},
  \bibinfo{year}{2004}, \bibinfo{journal}{Astrophys. J.}
  \textbf{\bibinfo{volume}{600}}, \bibinfo{pages}{544}.

\bibitem[{\citenamefont{Consiglio} \emph{et~al.}(2017)\citenamefont{Consiglio,
  de~Salas, Mangano, Miele, Pastor, and Pisanti}}]{Parthenope2}
\bibinfo{author}{\bibnamefont{Consiglio}, \bibfnamefont{R.}},
  \bibinfo{author}{\bibfnamefont{P.~F.} \bibnamefont{de~Salas}},
  \bibinfo{author}{\bibfnamefont{G.}~\bibnamefont{Mangano}},
  \bibinfo{author}{\bibfnamefont{G.}~\bibnamefont{Miele}},
  \bibinfo{author}{\bibfnamefont{S.}~\bibnamefont{Pastor}}, and
  \bibinfo{author}{\bibfnamefont{O.}~\bibnamefont{Pisanti}},
  \bibinfo{year}{2017}, \eprint{1712.04378}.

\bibitem[{\citenamefont{{Cooke}}(2015)}]{Coo15}
\bibinfo{author}{\bibnamefont{{Cooke}}, \bibfnamefont{R.~J.}},
  \bibinfo{year}{2015}, \bibinfo{journal}{Astrophys.~J.~Lett.}
  \textbf{\bibinfo{volume}{812}}, \bibinfo{eid}{L12}.

\bibitem[{\citenamefont{{Cooke}} \emph{et~al.}(2014)\citenamefont{{Cooke},
  {Pettini}, {Jorgenson}, {Murphy}, and {Steidel}}}]{Coo14}
\bibinfo{author}{\bibnamefont{{Cooke}}, \bibfnamefont{R.~J.}},
  \bibinfo{author}{\bibfnamefont{M.}~\bibnamefont{{Pettini}}},
  \bibinfo{author}{\bibfnamefont{R.~A.} \bibnamefont{{Jorgenson}}},
  \bibinfo{author}{\bibfnamefont{M.~T.} \bibnamefont{{Murphy}}}, and
  \bibinfo{author}{\bibfnamefont{C.~C.} \bibnamefont{{Steidel}}},
  \bibinfo{year}{2014}, \bibinfo{journal}{Astrophys. J.}
  \textbf{\bibinfo{volume}{781}}, \bibinfo{eid}{31}.

\bibitem[{\citenamefont{{Cooke}} \emph{et~al.}(2016)\citenamefont{{Cooke},
  {Pettini}, {Nollett}, and {Jorgenson}}}]{Coo16}
\bibinfo{author}{\bibnamefont{{Cooke}}, \bibfnamefont{R.~J.}},
  \bibinfo{author}{\bibfnamefont{M.}~\bibnamefont{{Pettini}}},
  \bibinfo{author}{\bibfnamefont{K.~M.} \bibnamefont{{Nollett}}}, and
  \bibinfo{author}{\bibfnamefont{R.}~\bibnamefont{{Jorgenson}}},
  \bibinfo{year}{2016}, \bibinfo{journal}{Astrophys. J.}
  \textbf{\bibinfo{volume}{830}}, \bibinfo{eid}{148}.

\bibitem[{\citenamefont{{Cooke}} \emph{et~al.}(2018)\citenamefont{{Cooke},
  {Pettini}, and {Steidel}}}]{Coo18}
\bibinfo{author}{\bibnamefont{{Cooke}}, \bibfnamefont{R.~J.}},
  \bibinfo{author}{\bibfnamefont{M.}~\bibnamefont{{Pettini}}}, and
  \bibinfo{author}{\bibfnamefont{C.~C.} \bibnamefont{{Steidel}}},
  \bibinfo{year}{2018}, \bibinfo{journal}{\apj} \textbf{\bibinfo{volume}{855}},
  \bibinfo{eid}{102}.

\bibitem[{\citenamefont{{Cooper}} \emph{et~al.}(2010)\citenamefont{{Cooper},
  {Chupp}, {Dewey}, {Gentile}, {Mumm}, {Nico}, {Thompson}, {Fisher}, {Kremsky},
  {Wietfeldt}, {Beise}, {Kiriluk}} \emph{et~al.}}]{2010PhRvC..81c5503C}
\bibinfo{author}{\bibnamefont{{Cooper}}, \bibfnamefont{R.~L.}},
  \bibinfo{author}{\bibfnamefont{T.~E.} \bibnamefont{{Chupp}}},
  \bibinfo{author}{\bibfnamefont{M.~S.} \bibnamefont{{Dewey}}},
  \bibinfo{author}{\bibfnamefont{T.~R.} \bibnamefont{{Gentile}}},
  \bibinfo{author}{\bibfnamefont{H.~P.} \bibnamefont{{Mumm}}},
  \bibinfo{author}{\bibfnamefont{J.~S.} \bibnamefont{{Nico}}},
  \bibinfo{author}{\bibfnamefont{A.~K.} \bibnamefont{{Thompson}}},
  \bibinfo{author}{\bibfnamefont{B.~M.} \bibnamefont{{Fisher}}},
  \bibinfo{author}{\bibfnamefont{I.}~\bibnamefont{{Kremsky}}},
  \bibinfo{author}{\bibfnamefont{F.~E.} \bibnamefont{{Wietfeldt}}},
  \bibinfo{author}{\bibfnamefont{E.~J.} \bibnamefont{{Beise}}},
  \bibinfo{author}{\bibfnamefont{K.~G.} \bibnamefont{{Kiriluk}}},
  \emph{et~al.}, \bibinfo{year}{2010}, \bibinfo{journal}{Phys. Rev.}
  \textbf{\bibinfo{volume}{C81}}, \bibinfo{eid}{035503}.

\bibitem[{\citenamefont{Creminelli}
  \emph{et~al.}(2011)\citenamefont{Creminelli, Pitrou, and
  Vernizzi}}]{Creminelli:2011sq}
\bibinfo{author}{\bibnamefont{Creminelli}, \bibfnamefont{P.}},
  \bibinfo{author}{\bibfnamefont{C.}~\bibnamefont{Pitrou}}, and
  \bibinfo{author}{\bibfnamefont{F.}~\bibnamefont{Vernizzi}},
  \bibinfo{year}{2011}, \bibinfo{journal}{JCAP}
  \textbf{\bibinfo{volume}{1111}}, \bibinfo{pages}{025}.

\bibitem[{\citenamefont{{Cyburt}}(2004)}]{Cyb04}
\bibinfo{author}{\bibnamefont{{Cyburt}}, \bibfnamefont{R.~H.}},
  \bibinfo{year}{2004}, \bibinfo{journal}{Phys. Rev.}
  \textbf{\bibinfo{volume}{D70}}, \bibinfo{eid}{023505}.

\bibitem[{\citenamefont{{Cyburt} and {Davids}}(2008)}]{Cyb08a}
\bibinfo{author}{\bibnamefont{{Cyburt}}, \bibfnamefont{R.~H.}}, and
  \bibinfo{author}{\bibfnamefont{B.}~\bibnamefont{{Davids}}},
  \bibinfo{year}{2008}, \bibinfo{journal}{Phys. Rev.}
  \textbf{\bibinfo{volume}{C78}}, \bibinfo{eid}{064614}.

\bibitem[{\citenamefont{{Cyburt}} \emph{et~al.}(2008)\citenamefont{{Cyburt},
  {Fields}, and {Olive}}}]{Cyb08b}
\bibinfo{author}{\bibnamefont{{Cyburt}}, \bibfnamefont{R.~H.}},
  \bibinfo{author}{\bibfnamefont{B.~D.} \bibnamefont{{Fields}}}, and
  \bibinfo{author}{\bibfnamefont{K.~A.} \bibnamefont{{Olive}}},
  \bibinfo{year}{2008}, \bibinfo{journal}{JCAP} \textbf{\bibinfo{volume}{11}},
  \bibinfo{eid}{012}.

\bibitem[{\citenamefont{{Cyburt}} \emph{et~al.}(2016)\citenamefont{{Cyburt},
  {Fields}, {Olive}, and {Yeh}}}]{Cyb16}
\bibinfo{author}{\bibnamefont{{Cyburt}}, \bibfnamefont{R.~H.}},
  \bibinfo{author}{\bibfnamefont{B.~D.} \bibnamefont{{Fields}}},
  \bibinfo{author}{\bibfnamefont{K.~A.} \bibnamefont{{Olive}}}, and
  \bibinfo{author}{\bibfnamefont{T.-H.} \bibnamefont{{Yeh}}},
  \bibinfo{year}{2016}, \bibinfo{journal}{Rev. Mod. Phys.}
  \textbf{\bibinfo{volume}{88}}, \bibinfo{eid}{015004}.

\bibitem[{\citenamefont{{Cyburt} and {Pospelov}}(2012)}]{Cyb12}
\bibinfo{author}{\bibnamefont{{Cyburt}}, \bibfnamefont{R.~H.}}, and
  \bibinfo{author}{\bibfnamefont{M.}~\bibnamefont{{Pospelov}}},
  \bibinfo{year}{2012}, \bibinfo{journal}{International Journal of Modern
  Physics E} \textbf{\bibinfo{volume}{21}},
  \bibinfo{eid}{1250004-1-1250004-13}.

\bibitem[{\citenamefont{Czarnecki} \emph{et~al.}(2004)\citenamefont{Czarnecki,
  Marciano, and Sirlin}}]{Czarnecki2004}
\bibinfo{author}{\bibnamefont{Czarnecki}, \bibfnamefont{A.}},
  \bibinfo{author}{\bibfnamefont{W.~J.} \bibnamefont{Marciano}}, and
  \bibinfo{author}{\bibfnamefont{A.}~\bibnamefont{Sirlin}},
  \bibinfo{year}{2004}, \bibinfo{journal}{Phys. Rev.}
  \textbf{\bibinfo{volume}{D70}}, \bibinfo{pages}{093006}.

\bibitem[{\citenamefont{Czarnecki} \emph{et~al.}(2018)\citenamefont{Czarnecki,
  Marciano, and Sirlin}}]{Cza18}
\bibinfo{author}{\bibnamefont{Czarnecki}, \bibfnamefont{A.}},
  \bibinfo{author}{\bibfnamefont{W.~J.} \bibnamefont{Marciano}}, and
  \bibinfo{author}{\bibfnamefont{A.}~\bibnamefont{Sirlin}},
  \bibinfo{year}{2018}, \bibinfo{journal}{Phys. Rev. Lett.}
  \textbf{\bibinfo{volume}{120}}, \bibinfo{pages}{202002}.

\bibitem[{\citenamefont{{deBoer}} \emph{et~al.}(2014)\citenamefont{{deBoer},
  {G{\"o}rres}, {Smith}, {Uberseder}, {Wiescher}, {Kontos}, {Imbriani}, {Di
  Leva}, and {Strieder}}}]{deB14}
\bibinfo{author}{\bibnamefont{{deBoer}}, \bibfnamefont{R.~J.}},
  \bibinfo{author}{\bibfnamefont{J.}~\bibnamefont{{G{\"o}rres}}},
  \bibinfo{author}{\bibfnamefont{K.}~\bibnamefont{{Smith}}},
  \bibinfo{author}{\bibfnamefont{E.}~\bibnamefont{{Uberseder}}},
  \bibinfo{author}{\bibfnamefont{M.}~\bibnamefont{{Wiescher}}},
  \bibinfo{author}{\bibfnamefont{A.}~\bibnamefont{{Kontos}}},
  \bibinfo{author}{\bibfnamefont{G.}~\bibnamefont{{Imbriani}}},
  \bibinfo{author}{\bibfnamefont{A.}~\bibnamefont{{Di Leva}}}, and
  \bibinfo{author}{\bibfnamefont{F.}~\bibnamefont{{Strieder}}},
  \bibinfo{year}{2014}, \bibinfo{journal}{Phys. Rev.}
  \textbf{\bibinfo{volume}{C90}}, \bibinfo{eid}{035804}.

\bibitem[{\citenamefont{{Descouvemont}}
  \emph{et~al.}(2004)\citenamefont{{Descouvemont}, {Adahchour}, {Angulo},
  {Coc}, and {Vangioni-Flam}}}]{Des04}
\bibinfo{author}{\bibnamefont{{Descouvemont}}, \bibfnamefont{P.}},
  \bibinfo{author}{\bibfnamefont{A.}~\bibnamefont{{Adahchour}}},
  \bibinfo{author}{\bibfnamefont{C.}~\bibnamefont{{Angulo}}},
  \bibinfo{author}{\bibfnamefont{A.}~\bibnamefont{{Coc}}}, and
  \bibinfo{author}{\bibfnamefont{E.}~\bibnamefont{{Vangioni-Flam}}},
  \bibinfo{year}{2004}, \bibinfo{journal}{Atomic Data and Nuclear Data Tables}
  \textbf{\bibinfo{volume}{88}}, \bibinfo{pages}{203}.

\bibitem[{\citenamefont{{Di Valentino}} \emph{et~al.}(2014)\citenamefont{{Di
  Valentino}, {Gustavino}, {Lesgourgues}, {Mangano}, {Melchiorri}, {Miele}, and
  {Pisanti}}}]{DiV14}
\bibinfo{author}{\bibnamefont{{Di Valentino}}, \bibfnamefont{E.}},
  \bibinfo{author}{\bibfnamefont{C.}~\bibnamefont{{Gustavino}}},
  \bibinfo{author}{\bibfnamefont{J.}~\bibnamefont{{Lesgourgues}}},
  \bibinfo{author}{\bibfnamefont{G.}~\bibnamefont{{Mangano}}},
  \bibinfo{author}{\bibfnamefont{A.}~\bibnamefont{{Melchiorri}}},
  \bibinfo{author}{\bibfnamefont{G.}~\bibnamefont{{Miele}}}, and
  \bibinfo{author}{\bibfnamefont{O.}~\bibnamefont{{Pisanti}}},
  \bibinfo{year}{2014}, \bibinfo{journal}{Phys. Rev.}
  \textbf{\bibinfo{volume}{D90}}(\bibinfo{number}{2}), \bibinfo{eid}{023543}.

\bibitem[{\citenamefont{{Dicus}} \emph{et~al.}(1982)\citenamefont{{Dicus},
  {Kolb}, {Gleeson}, {Sudarshan}, {Teplitz}, and {Turner}}}]{Dicus1982}
\bibinfo{author}{\bibnamefont{{Dicus}}, \bibfnamefont{D.~A.}},
  \bibinfo{author}{\bibfnamefont{E.~W.} \bibnamefont{{Kolb}}},
  \bibinfo{author}{\bibfnamefont{A.~M.} \bibnamefont{{Gleeson}}},
  \bibinfo{author}{\bibfnamefont{E.~C.~G.} \bibnamefont{{Sudarshan}}},
  \bibinfo{author}{\bibfnamefont{V.~L.} \bibnamefont{{Teplitz}}}, and
  \bibinfo{author}{\bibfnamefont{M.~S.} \bibnamefont{{Turner}}},
  \bibinfo{year}{1982}, \bibinfo{journal}{Phys. Rev.}
  \textbf{\bibinfo{volume}{D26}}, \bibinfo{pages}{2694}.

\bibitem[{\citenamefont{{Dodelson} and {Turner}}(1992)}]{1992PhRvD..46.3372D}
\bibinfo{author}{\bibnamefont{{Dodelson}}, \bibfnamefont{S.}}, and
  \bibinfo{author}{\bibfnamefont{M.~S.} \bibnamefont{{Turner}}},
  \bibinfo{year}{1992}, \bibinfo{journal}{PHys. Rev.}
  \textbf{\bibinfo{volume}{D46}}, \bibinfo{pages}{3372}.

\bibitem[{\citenamefont{Dolgov} \emph{et~al.}(2002)\citenamefont{Dolgov,
  Hansen, Pastor, Petcov, Raffelt, and Semikoz}}]{Dolgov:2002ab}
\bibinfo{author}{\bibnamefont{Dolgov}, \bibfnamefont{A.~D.}},
  \bibinfo{author}{\bibfnamefont{S.~H.} \bibnamefont{Hansen}},
  \bibinfo{author}{\bibfnamefont{S.}~\bibnamefont{Pastor}},
  \bibinfo{author}{\bibfnamefont{S.~T.} \bibnamefont{Petcov}},
  \bibinfo{author}{\bibfnamefont{G.~G.} \bibnamefont{Raffelt}}, and
  \bibinfo{author}{\bibfnamefont{D.~V.} \bibnamefont{Semikoz}},
  \bibinfo{year}{2002}, \bibinfo{journal}{Nucl. Phys.}
  \textbf{\bibinfo{volume}{B632}}, \bibinfo{pages}{363}.

\bibitem[{\citenamefont{Dolgov} \emph{et~al.}(1997)\citenamefont{Dolgov,
  Hansen, and Semikoz}}]{Dolgov1997}
\bibinfo{author}{\bibnamefont{Dolgov}, \bibfnamefont{A.~D.}},
  \bibinfo{author}{\bibfnamefont{S.~H.} \bibnamefont{Hansen}}, and
  \bibinfo{author}{\bibfnamefont{D.~V.} \bibnamefont{Semikoz}},
  \bibinfo{year}{1997}, \bibinfo{journal}{Nucl. Phys.}
  \textbf{\bibinfo{volume}{B503}}, \bibinfo{pages}{426}.

\bibitem[{\citenamefont{Dolgov} \emph{et~al.}(1999)\citenamefont{Dolgov,
  Hansen, and Semikoz}}]{Dolgov1998}
\bibinfo{author}{\bibnamefont{Dolgov}, \bibfnamefont{A.~D.}},
  \bibinfo{author}{\bibfnamefont{S.~H.} \bibnamefont{Hansen}}, and
  \bibinfo{author}{\bibfnamefont{D.~V.} \bibnamefont{Semikoz}},
  \bibinfo{year}{1999}, \bibinfo{journal}{Nucl. Phys.}
  \textbf{\bibinfo{volume}{B543}}, \bibinfo{pages}{269}.

\bibitem[{\citenamefont{{Dvorkin}} \emph{et~al.}(2016)\citenamefont{{Dvorkin},
  {Vangioni}, {Silk}, {Petitjean}, and {Olive}}}]{Dvo16}
\bibinfo{author}{\bibnamefont{{Dvorkin}}, \bibfnamefont{I.}},
  \bibinfo{author}{\bibfnamefont{E.}~\bibnamefont{{Vangioni}}},
  \bibinfo{author}{\bibfnamefont{J.}~\bibnamefont{{Silk}}},
  \bibinfo{author}{\bibfnamefont{P.}~\bibnamefont{{Petitjean}}}, and
  \bibinfo{author}{\bibfnamefont{K.~A.} \bibnamefont{{Olive}}},
  \bibinfo{year}{2016}, \bibinfo{journal}{MNRAS}
  \textbf{\bibinfo{volume}{458}}, \bibinfo{pages}{L104}.

\bibitem[{\citenamefont{{Ekstr{\"o}m}}
  \emph{et~al.}(2008)\citenamefont{{Ekstr{\"o}m}, {Meynet}, {Chiappini},
  {Hirschi}, and {Maeder}}}]{Eks08}
\bibinfo{author}{\bibnamefont{{Ekstr{\"o}m}}, \bibfnamefont{S.}},
  \bibinfo{author}{\bibfnamefont{G.}~\bibnamefont{{Meynet}}},
  \bibinfo{author}{\bibfnamefont{C.}~\bibnamefont{{Chiappini}}},
  \bibinfo{author}{\bibfnamefont{R.}~\bibnamefont{{Hirschi}}}, and
  \bibinfo{author}{\bibfnamefont{A.}~\bibnamefont{{Maeder}}},
  \bibinfo{year}{2008}, \bibinfo{journal}{Astron.~Astrophys.}
  \textbf{\bibinfo{volume}{489}}, \bibinfo{pages}{685}.

\bibitem[{\citenamefont{Esposito} \emph{et~al.}(1999)\citenamefont{Esposito,
  Mangano, Miele, and Pisanti}}]{Esposito1998}
\bibinfo{author}{\bibnamefont{Esposito}, \bibfnamefont{S.}},
  \bibinfo{author}{\bibfnamefont{G.}~\bibnamefont{Mangano}},
  \bibinfo{author}{\bibfnamefont{G.}~\bibnamefont{Miele}}, and
  \bibinfo{author}{\bibfnamefont{O.}~\bibnamefont{Pisanti}},
  \bibinfo{year}{1999}, \bibinfo{journal}{Nucl. Phys.}
  \textbf{\bibinfo{volume}{B540}}, \bibinfo{pages}{3}.

\bibitem[{\citenamefont{Esposito}
  \emph{et~al.}(2000{\natexlab{a}})\citenamefont{Esposito, Mangano, Miele, and
  Pisanti}}]{Esposito1999}
\bibinfo{author}{\bibnamefont{Esposito}, \bibfnamefont{S.}},
  \bibinfo{author}{\bibfnamefont{G.}~\bibnamefont{Mangano}},
  \bibinfo{author}{\bibfnamefont{G.}~\bibnamefont{Miele}}, and
  \bibinfo{author}{\bibfnamefont{O.}~\bibnamefont{Pisanti}},
  \bibinfo{year}{2000}{\natexlab{a}}, \bibinfo{journal}{Nucl. Phys.}
  \textbf{\bibinfo{volume}{B568}}, \bibinfo{pages}{421}.

\bibitem[{\citenamefont{Esposito}
  \emph{et~al.}(2000{\natexlab{b}})\citenamefont{Esposito, Miele, Pastor,
  Peloso, and Pisanti}}]{Esposito:2000hi}
\bibinfo{author}{\bibnamefont{Esposito}, \bibfnamefont{S.}},
  \bibinfo{author}{\bibfnamefont{G.}~\bibnamefont{Miele}},
  \bibinfo{author}{\bibfnamefont{S.}~\bibnamefont{Pastor}},
  \bibinfo{author}{\bibfnamefont{M.}~\bibnamefont{Peloso}}, and
  \bibinfo{author}{\bibfnamefont{O.}~\bibnamefont{Pisanti}},
  \bibinfo{year}{2000}{\natexlab{b}}, \bibinfo{journal}{Nucl. Phys.}
  \textbf{\bibinfo{volume}{B590}}, \bibinfo{pages}{539}.

\bibitem[{\citenamefont{{Famiano}} \emph{et~al.}(2016)\citenamefont{{Famiano},
  {Balantekin}, and {Kajino}}}]{Fam16}
\bibinfo{author}{\bibnamefont{{Famiano}}, \bibfnamefont{M.~A.}},
  \bibinfo{author}{\bibfnamefont{A.~B.} \bibnamefont{{Balantekin}}}, and
  \bibinfo{author}{\bibfnamefont{T.}~\bibnamefont{{Kajino}}},
  \bibinfo{year}{2016}, \bibinfo{journal}{Phys. Rev.}
  \textbf{\bibinfo{volume}{C93}}, \bibinfo{eid}{045804}.

\bibitem[{\citenamefont{Fidler and Pitrou}(2017)}]{Fidler:2017pkg}
\bibinfo{author}{\bibnamefont{Fidler}, \bibfnamefont{C.}}, and
  \bibinfo{author}{\bibfnamefont{C.}~\bibnamefont{Pitrou}},
  \bibinfo{year}{2017}, \bibinfo{journal}{JCAP}
  \textbf{\bibinfo{volume}{1706}}(\bibinfo{number}{06}), \bibinfo{pages}{013}.

\bibitem[{\citenamefont{{Fields}}(2011)}]{Fie11}
\bibinfo{author}{\bibnamefont{{Fields}}, \bibfnamefont{B.~D.}},
  \bibinfo{year}{2011}, \bibinfo{journal}{Annual Review of Nuclear and Particle
  Science} \textbf{\bibinfo{volume}{61}}, \bibinfo{pages}{47}.

\bibitem[{\citenamefont{Fields} \emph{et~al.}(1993)\citenamefont{Fields,
  Dodelson, and Turner}}]{Fields:1992zb}
\bibinfo{author}{\bibnamefont{Fields}, \bibfnamefont{B.~D.}},
  \bibinfo{author}{\bibfnamefont{S.}~\bibnamefont{Dodelson}}, and
  \bibinfo{author}{\bibfnamefont{M.~S.} \bibnamefont{Turner}},
  \bibinfo{year}{1993}, \bibinfo{journal}{Phys. Rev.}
  \textbf{\bibinfo{volume}{D47}}, \bibinfo{pages}{4309}.

\bibitem[{\citenamefont{{Fornal} and {Grinstein}}(2018)}]{For18}
\bibinfo{author}{\bibnamefont{{Fornal}}, \bibfnamefont{B.}}, and
  \bibinfo{author}{\bibfnamefont{B.}~\bibnamefont{{Grinstein}}},
  \bibinfo{year}{2018}, \bibinfo{journal}{ArXiv e-prints} \eprint{1801.01124}.

\bibitem[{\citenamefont{Fornengo} \emph{et~al.}(1997)\citenamefont{Fornengo,
  Kim, and Song}}]{Fornengo:1997wa}
\bibinfo{author}{\bibnamefont{Fornengo}, \bibfnamefont{N.}},
  \bibinfo{author}{\bibfnamefont{C.~W.} \bibnamefont{Kim}}, and
  \bibinfo{author}{\bibfnamefont{J.}~\bibnamefont{Song}}, \bibinfo{year}{1997},
  \bibinfo{journal}{Phys. Rev.} \textbf{\bibinfo{volume}{D56}},
  \bibinfo{pages}{5123}.

\bibitem[{\citenamefont{{Fowler}} \emph{et~al.}(1967)\citenamefont{{Fowler},
  {Caughlan}, and {Zimmerman}}}]{Fow67}
\bibinfo{author}{\bibnamefont{{Fowler}}, \bibfnamefont{W.~A.}},
  \bibinfo{author}{\bibfnamefont{G.~R.} \bibnamefont{{Caughlan}}}, and
  \bibinfo{author}{\bibfnamefont{B.~A.} \bibnamefont{{Zimmerman}}},
  \bibinfo{year}{1967}, \bibinfo{journal}{Ann. Rev. of Astron. and~Astrophys.}
  \textbf{\bibinfo{volume}{5}}, \bibinfo{pages}{525}.

\bibitem[{\citenamefont{{Fr{\"o}berg}}(1955)}]{Fro55}
\bibinfo{author}{\bibnamefont{{Fr{\"o}berg}}, \bibfnamefont{C.-E.}},
  \bibinfo{year}{1955}, \bibinfo{journal}{Reviews of Modern Physics}
  \textbf{\bibinfo{volume}{27}}, \bibinfo{pages}{399}.

\bibitem[{\citenamefont{{Fu}} \emph{et~al.}(2015)\citenamefont{{Fu}, {Bressan},
  {Molaro}, and {Marigo}}}]{Fu15}
\bibinfo{author}{\bibnamefont{{Fu}}, \bibfnamefont{X.}},
  \bibinfo{author}{\bibfnamefont{A.}~\bibnamefont{{Bressan}}},
  \bibinfo{author}{\bibfnamefont{P.}~\bibnamefont{{Molaro}}}, and
  \bibinfo{author}{\bibfnamefont{P.}~\bibnamefont{{Marigo}}},
  \bibinfo{year}{2015}, \bibinfo{journal}{MNRAS}
  \textbf{\bibinfo{volume}{452}}, \bibinfo{pages}{3256}.

\bibitem[{\citenamefont{{Fu}} \emph{et~al.}(2018)\citenamefont{{Fu}, {Romano},
  {Bragaglia}, {Mucciarelli}, {Lind}, {Delgado Mena}, {Sousa}, {Randich},
  {Bressan}, {Sbordone}, {Martell}, {Korn}} \emph{et~al.}}]{Fu18}
\bibinfo{author}{\bibnamefont{{Fu}}, \bibfnamefont{X.}},
  \bibinfo{author}{\bibfnamefont{D.}~\bibnamefont{{Romano}}},
  \bibinfo{author}{\bibfnamefont{A.}~\bibnamefont{{Bragaglia}}},
  \bibinfo{author}{\bibfnamefont{A.}~\bibnamefont{{Mucciarelli}}},
  \bibinfo{author}{\bibfnamefont{K.}~\bibnamefont{{Lind}}},
  \bibinfo{author}{\bibfnamefont{E.}~\bibnamefont{{Delgado Mena}}},
  \bibinfo{author}{\bibfnamefont{S.~G.} \bibnamefont{{Sousa}}},
  \bibinfo{author}{\bibfnamefont{S.}~\bibnamefont{{Randich}}},
  \bibinfo{author}{\bibfnamefont{A.}~\bibnamefont{{Bressan}}},
  \bibinfo{author}{\bibfnamefont{L.}~\bibnamefont{{Sbordone}}},
  \bibinfo{author}{\bibfnamefont{S.}~\bibnamefont{{Martell}}},
  \bibinfo{author}{\bibfnamefont{A.~J.} \bibnamefont{{Korn}}}, \emph{et~al.},
  \bibinfo{year}{2018}, \bibinfo{journal}{Astron. Astrophys.}
  \textbf{\bibinfo{volume}{610}}, \bibinfo{eid}{A38}.

\bibitem[{\citenamefont{Gnedin and Gnedin}(1998)}]{Gnedin1997}
\bibinfo{author}{\bibnamefont{Gnedin}, \bibfnamefont{N.~Y.}}, and
  \bibinfo{author}{\bibfnamefont{O.~Y.} \bibnamefont{Gnedin}},
  \bibinfo{year}{1998}, \bibinfo{journal}{Astrophys. J.}
  \textbf{\bibinfo{volume}{509}}, \bibinfo{pages}{11}.

\bibitem[{\citenamefont{{G{\'o}mez I{\~n}esta}}
  \emph{et~al.}(2017)\citenamefont{{G{\'o}mez I{\~n}esta}, {Iliadis}, and
  {Coc}}}]{Bayes17}
\bibinfo{author}{\bibnamefont{{G{\'o}mez I{\~n}esta}}, \bibfnamefont{{\'A}.}},
  \bibinfo{author}{\bibfnamefont{C.}~\bibnamefont{{Iliadis}}}, and
  \bibinfo{author}{\bibfnamefont{A.}~\bibnamefont{{Coc}}},
  \bibinfo{year}{2017}, \bibinfo{journal}{ArXiv e-prints} \eprint{1710.01647}.

\bibitem[{\citenamefont{{Goriely}} \emph{et~al.}(2008)\citenamefont{{Goriely},
  {Hilaire}, and {Koning}}}]{TALYS}
\bibinfo{author}{\bibnamefont{{Goriely}}, \bibfnamefont{S.}},
  \bibinfo{author}{\bibfnamefont{S.}~\bibnamefont{{Hilaire}}}, and
  \bibinfo{author}{\bibfnamefont{A.~J.} \bibnamefont{{Koning}}},
  \bibinfo{year}{2008}, \bibinfo{journal}{Astron.~Astrophys.}
  \textbf{\bibinfo{volume}{487}}, \bibinfo{pages}{767}.

\bibitem[{\citenamefont{{Goudelis}}
  \emph{et~al.}(2016)\citenamefont{{Goudelis}, {Pospelov}, and
  {Pradler}}}]{Gou16}
\bibinfo{author}{\bibnamefont{{Goudelis}}, \bibfnamefont{A.}},
  \bibinfo{author}{\bibfnamefont{M.}~\bibnamefont{{Pospelov}}}, and
  \bibinfo{author}{\bibfnamefont{J.}~\bibnamefont{{Pradler}}},
  \bibinfo{year}{2016}, \bibinfo{journal}{Phys. Rev. Lett.}
  \textbf{\bibinfo{volume}{116}}, \bibinfo{eid}{211303}.

\bibitem[{\citenamefont{{Greife}} \emph{et~al.}(1995)\citenamefont{{Greife},
  {Gorris}, {Junker}, {Rolfs}, and {Zahnow}}}]{Gre95}
\bibinfo{author}{\bibnamefont{{Greife}}, \bibfnamefont{U.}},
  \bibinfo{author}{\bibfnamefont{F.}~\bibnamefont{{Gorris}}},
  \bibinfo{author}{\bibfnamefont{M.}~\bibnamefont{{Junker}}},
  \bibinfo{author}{\bibfnamefont{C.}~\bibnamefont{{Rolfs}}}, and
  \bibinfo{author}{\bibfnamefont{D.}~\bibnamefont{{Zahnow}}},
  \bibinfo{year}{1995}, \bibinfo{journal}{Zeitschrift fur Physik A Hadrons and
  Nuclei} \textbf{\bibinfo{volume}{351}}, \bibinfo{pages}{107}.

\bibitem[{\citenamefont{Grohs} \emph{et~al.}(2017)\citenamefont{Grohs, Fuller,
  Kishimoto, and Paris}}]{Grohs:2016cuu}
\bibinfo{author}{\bibnamefont{Grohs}, \bibfnamefont{E.}},
  \bibinfo{author}{\bibfnamefont{G.~M.} \bibnamefont{Fuller}},
  \bibinfo{author}{\bibfnamefont{C.~T.} \bibnamefont{Kishimoto}}, and
  \bibinfo{author}{\bibfnamefont{M.~W.} \bibnamefont{Paris}},
  \bibinfo{year}{2017}, \bibinfo{journal}{Phys. Rev.}
  \textbf{\bibinfo{volume}{D95}}, \bibinfo{pages}{063503}.

\bibitem[{\citenamefont{Grohs} \emph{et~al.}(2016)\citenamefont{Grohs, Fuller,
  Kishimoto, Paris, and Vlasenko}}]{Grohs:2015tfy}
\bibinfo{author}{\bibnamefont{Grohs}, \bibfnamefont{E.}},
  \bibinfo{author}{\bibfnamefont{G.~M.} \bibnamefont{Fuller}},
  \bibinfo{author}{\bibfnamefont{C.~T.} \bibnamefont{Kishimoto}},
  \bibinfo{author}{\bibfnamefont{M.~W.} \bibnamefont{Paris}}, and
  \bibinfo{author}{\bibfnamefont{A.}~\bibnamefont{Vlasenko}},
  \bibinfo{year}{2016}, \bibinfo{journal}{Phys. Rev.}
  \textbf{\bibinfo{volume}{D93}}, \bibinfo{pages}{083522}.

\bibitem[{\citenamefont{{Gruyters}}
  \emph{et~al.}(2016)\citenamefont{{Gruyters}, {Lind}, {Richard}, {Grundahl},
  {Asplund}, {Casagrande}, {Charbonnel}, {Milone}, {Primas}, and
  {Korn}}}]{Gru16}
\bibinfo{author}{\bibnamefont{{Gruyters}}, \bibfnamefont{P.}},
  \bibinfo{author}{\bibfnamefont{K.}~\bibnamefont{{Lind}}},
  \bibinfo{author}{\bibfnamefont{O.}~\bibnamefont{{Richard}}},
  \bibinfo{author}{\bibfnamefont{F.}~\bibnamefont{{Grundahl}}},
  \bibinfo{author}{\bibfnamefont{M.}~\bibnamefont{{Asplund}}},
  \bibinfo{author}{\bibfnamefont{L.}~\bibnamefont{{Casagrande}}},
  \bibinfo{author}{\bibfnamefont{C.}~\bibnamefont{{Charbonnel}}},
  \bibinfo{author}{\bibfnamefont{A.}~\bibnamefont{{Milone}}},
  \bibinfo{author}{\bibfnamefont{F.}~\bibnamefont{{Primas}}}, and
  \bibinfo{author}{\bibfnamefont{A.~J.} \bibnamefont{{Korn}}},
  \bibinfo{year}{2016}, \bibinfo{journal}{Astron. Astrophys.}
  \textbf{\bibinfo{volume}{589}}, \bibinfo{eid}{A61}.

\bibitem[{\citenamefont{{Gustavino}}(2017)}]{Gus17}
\bibinfo{author}{\bibnamefont{{Gustavino}}, \bibfnamefont{C.}},
  \bibinfo{year}{2017}, in \emph{\bibinfo{booktitle}{European Physical Journal
  Web of Conferences}}, volume \bibinfo{volume}{136}, p.
  \bibinfo{pages}{01009}.

\bibitem[{\citenamefont{{Hammache}}
  \emph{et~al.}(2013)\citenamefont{{Hammache}, {Coc}, {de S{\'e}r{\'e}ville},
  {Stefan}, {Roussel}, {Ancelin}, {Assi{\'e}}, {Audouin}, {Beaumel},
  {Franchoo}, {Fernandez-Dominguez}, {Fox}} \emph{et~al.}}]{Ham13}
\bibinfo{author}{\bibnamefont{{Hammache}}, \bibfnamefont{F.}},
  \bibinfo{author}{\bibfnamefont{A.}~\bibnamefont{{Coc}}},
  \bibinfo{author}{\bibfnamefont{N.}~\bibnamefont{{de S{\'e}r{\'e}ville}}},
  \bibinfo{author}{\bibfnamefont{I.}~\bibnamefont{{Stefan}}},
  \bibinfo{author}{\bibfnamefont{P.}~\bibnamefont{{Roussel}}},
  \bibinfo{author}{\bibfnamefont{S.}~\bibnamefont{{Ancelin}}},
  \bibinfo{author}{\bibfnamefont{M.}~\bibnamefont{{Assi{\'e}}}},
  \bibinfo{author}{\bibfnamefont{L.}~\bibnamefont{{Audouin}}},
  \bibinfo{author}{\bibfnamefont{D.}~\bibnamefont{{Beaumel}}},
  \bibinfo{author}{\bibfnamefont{S.}~\bibnamefont{{Franchoo}}},
  \bibinfo{author}{\bibfnamefont{B.}~\bibnamefont{{Fernandez-Dominguez}}},
  \bibinfo{author}{\bibfnamefont{S.}~\bibnamefont{{Fox}}}, \emph{et~al.},
  \bibinfo{year}{2013}, \bibinfo{journal}{Phys. Rev.}
  \textbf{\bibinfo{volume}{C88}}, \bibinfo{eid}{062802}.

\bibitem[{\citenamefont{{Hammache}}
  \emph{et~al.}(2010)\citenamefont{{Hammache}, {Heil}, {Typel}, {Galaviz},
  {S{\"u}mmerer}, {Coc}, {Uhlig}, {Attallah}, {Caamano}, {Cortina}, {Geissel},
  {Hellstr{\"o}m}} \emph{et~al.}}]{Ham10}
\bibinfo{author}{\bibnamefont{{Hammache}}, \bibfnamefont{F.}},
  \bibinfo{author}{\bibfnamefont{M.}~\bibnamefont{{Heil}}},
  \bibinfo{author}{\bibfnamefont{S.}~\bibnamefont{{Typel}}},
  \bibinfo{author}{\bibfnamefont{D.}~\bibnamefont{{Galaviz}}},
  \bibinfo{author}{\bibfnamefont{K.}~\bibnamefont{{S{\"u}mmerer}}},
  \bibinfo{author}{\bibfnamefont{A.}~\bibnamefont{{Coc}}},
  \bibinfo{author}{\bibfnamefont{F.}~\bibnamefont{{Uhlig}}},
  \bibinfo{author}{\bibfnamefont{F.}~\bibnamefont{{Attallah}}},
  \bibinfo{author}{\bibfnamefont{M.}~\bibnamefont{{Caamano}}},
  \bibinfo{author}{\bibfnamefont{D.}~\bibnamefont{{Cortina}}},
  \bibinfo{author}{\bibfnamefont{H.}~\bibnamefont{{Geissel}}},
  \bibinfo{author}{\bibfnamefont{M.}~\bibnamefont{{Hellstr{\"o}m}}},
  \emph{et~al.}, \bibinfo{year}{2010}, \bibinfo{journal}{Phys. Rev.}
  \textbf{\bibinfo{volume}{C82}}, \bibinfo{eid}{065803}.

\bibitem[{\citenamefont{Hannestad}(2002)}]{Hannestad2001}
\bibinfo{author}{\bibnamefont{Hannestad}, \bibfnamefont{S.}},
  \bibinfo{year}{2002}, \bibinfo{journal}{Phys. Rev.}
  \textbf{\bibinfo{volume}{D65}}, \bibinfo{pages}{083006}.

\bibitem[{\citenamefont{Hannestad and Madsen}(1995)}]{Hannestad1995}
\bibinfo{author}{\bibnamefont{Hannestad}, \bibfnamefont{S.}}, and
  \bibinfo{author}{\bibfnamefont{J.}~\bibnamefont{Madsen}},
  \bibinfo{year}{1995}, \bibinfo{journal}{Phys. Rev.}
  \textbf{\bibinfo{volume}{D52}}, \bibinfo{pages}{1764}.

\bibitem[{\citenamefont{{Heckler}}(1994)}]{Heckler1994}
\bibinfo{author}{\bibnamefont{{Heckler}}, \bibfnamefont{A.~F.}},
  \bibinfo{year}{1994}, \bibinfo{journal}{Phys. Rev.}
  \textbf{\bibinfo{volume}{D49}}, \bibinfo{pages}{611}.

\bibitem[{\citenamefont{{Hernanz}} \emph{et~al.}(1996)\citenamefont{{Hernanz},
  {Jose}, {Coc}, and {Isern}}}]{HJCI96}
\bibinfo{author}{\bibnamefont{{Hernanz}}, \bibfnamefont{M.}},
  \bibinfo{author}{\bibfnamefont{J.}~\bibnamefont{{Jose}}},
  \bibinfo{author}{\bibfnamefont{A.}~\bibnamefont{{Coc}}}, and
  \bibinfo{author}{\bibfnamefont{J.}~\bibnamefont{{Isern}}},
  \bibinfo{year}{1996}, \bibinfo{journal}{Astrophys.~J.~Lett.}
  \textbf{\bibinfo{volume}{465}}, \bibinfo{pages}{L27}.

\bibitem[{\citenamefont{Horowitz}(2002)}]{Horowitz:2001xf}
\bibinfo{author}{\bibnamefont{Horowitz}, \bibfnamefont{C.~J.}},
  \bibinfo{year}{2002}, \bibinfo{journal}{Phys. Rev.}
  \textbf{\bibinfo{volume}{D65}}, \bibinfo{pages}{043001}.

\bibitem[{\citenamefont{Horowitz and Li}(2000)}]{HorowitzLi}
\bibinfo{author}{\bibnamefont{Horowitz}, \bibfnamefont{C.~J.}}, and
  \bibinfo{author}{\bibfnamefont{G.}~\bibnamefont{Li}}, \bibinfo{year}{2000},
  \bibinfo{journal}{Phys. Rev.} \textbf{\bibinfo{volume}{D61}},
  \bibinfo{pages}{063002}.

\bibitem[{\citenamefont{{Hosford}} \emph{et~al.}(2009)\citenamefont{{Hosford},
  {Ryan}, {Garc{\'{\i}}a P{\'e}rez}, {Norris}, and {Olive}}}]{Hos09}
\bibinfo{author}{\bibnamefont{{Hosford}}, \bibfnamefont{A.}},
  \bibinfo{author}{\bibfnamefont{S.~G.} \bibnamefont{{Ryan}}},
  \bibinfo{author}{\bibfnamefont{A.~E.} \bibnamefont{{Garc{\'{\i}}a
  P{\'e}rez}}}, \bibinfo{author}{\bibfnamefont{J.~E.} \bibnamefont{{Norris}}},
  and \bibinfo{author}{\bibfnamefont{K.~A.} \bibnamefont{{Olive}}},
  \bibinfo{year}{2009}, \bibinfo{journal}{Astron. Astrophys.}
  \textbf{\bibinfo{volume}{493}}, \bibinfo{pages}{601}.

\bibitem[{\citenamefont{{Hou}} \emph{et~al.}(2015)\citenamefont{{Hou}, {He},
  {Kubono}, and {Chen}}}]{Hou15}
\bibinfo{author}{\bibnamefont{{Hou}}, \bibfnamefont{S.~Q.}},
  \bibinfo{author}{\bibfnamefont{J.~J.} \bibnamefont{{He}}},
  \bibinfo{author}{\bibfnamefont{S.}~\bibnamefont{{Kubono}}}, and
  \bibinfo{author}{\bibfnamefont{Y.~S.} \bibnamefont{{Chen}}},
  \bibinfo{year}{2015}, \bibinfo{journal}{Phys. Rev.}
  \textbf{\bibinfo{volume}{C91}}, \bibinfo{eid}{055802}.

\bibitem[{\citenamefont{{Howk}} \emph{et~al.}(2012)\citenamefont{{Howk},
  {Lehner}, {Fields}, and {Mathews}}}]{How12}
\bibinfo{author}{\bibnamefont{{Howk}}, \bibfnamefont{J.~C.}},
  \bibinfo{author}{\bibfnamefont{N.}~\bibnamefont{{Lehner}}},
  \bibinfo{author}{\bibfnamefont{B.~D.} \bibnamefont{{Fields}}}, and
  \bibinfo{author}{\bibfnamefont{G.~J.} \bibnamefont{{Mathews}}},
  \bibinfo{year}{2012}, \bibinfo{journal}{Nature}
  \textbf{\bibinfo{volume}{489}}, \bibinfo{pages}{121}.

\bibitem[{\citenamefont{{Iliadis}}(2007)}]{chbook}
\bibinfo{author}{\bibnamefont{{Iliadis}}, \bibfnamefont{C.}},
  \bibinfo{year}{2007}, \emph{\bibinfo{title}{{Nuclear Physics of Stars}}}
  (\bibinfo{publisher}{Wiley-VCH Verlag}).

\bibitem[{\citenamefont{{Iliadis}} \emph{et~al.}(2016)\citenamefont{{Iliadis},
  {Anderson}, {Coc}, {Timmes}, and {Starrfield}}}]{Bayes16}
\bibinfo{author}{\bibnamefont{{Iliadis}}, \bibfnamefont{C.}},
  \bibinfo{author}{\bibfnamefont{K.~S.} \bibnamefont{{Anderson}}},
  \bibinfo{author}{\bibfnamefont{A.}~\bibnamefont{{Coc}}},
  \bibinfo{author}{\bibfnamefont{F.~X.} \bibnamefont{{Timmes}}}, and
  \bibinfo{author}{\bibfnamefont{S.}~\bibnamefont{{Starrfield}}},
  \bibinfo{year}{2016}, \bibinfo{journal}{Astrophys.~J.}
  \textbf{\bibinfo{volume}{831}}, \bibinfo{eid}{107}.

\bibitem[{\citenamefont{{Iliadis}} \emph{et~al.}(2010)\citenamefont{{Iliadis},
  {Longland}, {Champagne}, {Coc}, and {Fitzgerald}}}]{Eval2}
\bibinfo{author}{\bibnamefont{{Iliadis}}, \bibfnamefont{C.}},
  \bibinfo{author}{\bibfnamefont{R.}~\bibnamefont{{Longland}}},
  \bibinfo{author}{\bibfnamefont{A.~E.} \bibnamefont{{Champagne}}},
  \bibinfo{author}{\bibfnamefont{A.}~\bibnamefont{{Coc}}}, and
  \bibinfo{author}{\bibfnamefont{R.}~\bibnamefont{{Fitzgerald}}},
  \bibinfo{year}{2010}, \bibinfo{journal}{Nucl. Phys.}
  \textbf{\bibinfo{volume}{A841}}, \bibinfo{pages}{31}.

\bibitem[{\citenamefont{{Iocco}} \emph{et~al.}(2007)\citenamefont{{Iocco},
  {Mangano}, {Miele}, {Pisanti}, and {Serpico}}}]{Ioc07}
\bibinfo{author}{\bibnamefont{{Iocco}}, \bibfnamefont{F.}},
  \bibinfo{author}{\bibfnamefont{G.}~\bibnamefont{{Mangano}}},
  \bibinfo{author}{\bibfnamefont{G.}~\bibnamefont{{Miele}}},
  \bibinfo{author}{\bibfnamefont{O.}~\bibnamefont{{Pisanti}}}, and
  \bibinfo{author}{\bibfnamefont{P.~D.} \bibnamefont{{Serpico}}},
  \bibinfo{year}{2007}, \bibinfo{journal}{Phys. Rev.}
  \textbf{\bibinfo{volume}{D75}}, \bibinfo{eid}{087304}.

\bibitem[{\citenamefont{{Iocco}} \emph{et~al.}(2009)\citenamefont{{Iocco},
  {Mangano}, {Miele}, {Pisanti}, and {Serpico}}}]{IoccoReport}
\bibinfo{author}{\bibnamefont{{Iocco}}, \bibfnamefont{F.}},
  \bibinfo{author}{\bibfnamefont{G.}~\bibnamefont{{Mangano}}},
  \bibinfo{author}{\bibfnamefont{G.}~\bibnamefont{{Miele}}},
  \bibinfo{author}{\bibfnamefont{O.}~\bibnamefont{{Pisanti}}}, and
  \bibinfo{author}{\bibfnamefont{P.~D.} \bibnamefont{{Serpico}}},
  \bibinfo{year}{2009}, \bibinfo{journal}{Phys. Rep.}
  \textbf{\bibinfo{volume}{472}}, \bibinfo{pages}{1}.

\bibitem[{\citenamefont{Ivanov} \emph{et~al.}(2017)\citenamefont{Ivanov,
  Höllwieser, Troitskaya, Wellenzohn, and Berdnikov}}]{Ivanov:2017fra}
\bibinfo{author}{\bibnamefont{Ivanov}, \bibfnamefont{A.~N.}},
  \bibinfo{author}{\bibfnamefont{R.}~\bibnamefont{Höllwieser}},
  \bibinfo{author}{\bibfnamefont{N.~I.} \bibnamefont{Troitskaya}},
  \bibinfo{author}{\bibfnamefont{M.}~\bibnamefont{Wellenzohn}}, and
  \bibinfo{author}{\bibfnamefont{{\relax Ya}.~A.} \bibnamefont{Berdnikov}},
  \bibinfo{year}{2017}, \bibinfo{journal}{Phys. Rev.}
  \textbf{\bibinfo{volume}{D95}}, \bibinfo{pages}{033007}.

\bibitem[{\citenamefont{Ivanov} \emph{et~al.}(2013)\citenamefont{Ivanov,
  Pitschmann, and Troitskaya}}]{Ivanov:2012qe}
\bibinfo{author}{\bibnamefont{Ivanov}, \bibfnamefont{A.~N.}},
  \bibinfo{author}{\bibfnamefont{M.}~\bibnamefont{Pitschmann}}, and
  \bibinfo{author}{\bibfnamefont{N.~I.} \bibnamefont{Troitskaya}},
  \bibinfo{year}{2013}, \bibinfo{journal}{Phys. Rev.}
  \textbf{\bibinfo{volume}{D88}}, \bibinfo{pages}{073002}.

\bibitem[{\citenamefont{{Izotov}} \emph{et~al.}(2014)\citenamefont{{Izotov},
  {Thuan}, and {Guseva}}}]{Izo14}
\bibinfo{author}{\bibnamefont{{Izotov}}, \bibfnamefont{Y.~I.}},
  \bibinfo{author}{\bibfnamefont{T.~X.} \bibnamefont{{Thuan}}}, and
  \bibinfo{author}{\bibfnamefont{N.~G.} \bibnamefont{{Guseva}}},
  \bibinfo{year}{2014}, \bibinfo{journal}{MNRAS}
  \textbf{\bibinfo{volume}{445}}, \bibinfo{pages}{778}.

\bibitem[{\citenamefont{{Izotov}} \emph{et~al.}(1994)\citenamefont{{Izotov},
  {Thuan}, and {Lipovetsky}}}]{Izo94}
\bibinfo{author}{\bibnamefont{{Izotov}}, \bibfnamefont{Y.~I.}},
  \bibinfo{author}{\bibfnamefont{T.~X.} \bibnamefont{{Thuan}}}, and
  \bibinfo{author}{\bibfnamefont{V.~A.} \bibnamefont{{Lipovetsky}}},
  \bibinfo{year}{1994}, \bibinfo{journal}{Astrophys.~J.}
  \textbf{\bibinfo{volume}{435}}, \bibinfo{pages}{647}.

\bibitem[{\citenamefont{{Izzo}} \emph{et~al.}(2015)\citenamefont{{Izzo}, {Della
  Valle}, {Mason}, {Matteucci}, {Romano}, {Pasquini}, {Vanzi}, {Jordan},
  {Fernandez}, {Bluhm}, {Brahm}, {Espinoza}} \emph{et~al.}}]{Izz15}
\bibinfo{author}{\bibnamefont{{Izzo}}, \bibfnamefont{L.}},
  \bibinfo{author}{\bibfnamefont{M.}~\bibnamefont{{Della Valle}}},
  \bibinfo{author}{\bibfnamefont{E.}~\bibnamefont{{Mason}}},
  \bibinfo{author}{\bibfnamefont{F.}~\bibnamefont{{Matteucci}}},
  \bibinfo{author}{\bibfnamefont{D.}~\bibnamefont{{Romano}}},
  \bibinfo{author}{\bibfnamefont{L.}~\bibnamefont{{Pasquini}}},
  \bibinfo{author}{\bibfnamefont{L.}~\bibnamefont{{Vanzi}}},
  \bibinfo{author}{\bibfnamefont{A.}~\bibnamefont{{Jordan}}},
  \bibinfo{author}{\bibfnamefont{J.~M.} \bibnamefont{{Fernandez}}},
  \bibinfo{author}{\bibfnamefont{P.}~\bibnamefont{{Bluhm}}},
  \bibinfo{author}{\bibfnamefont{R.}~\bibnamefont{{Brahm}}},
  \bibinfo{author}{\bibfnamefont{N.}~\bibnamefont{{Espinoza}}}, \emph{et~al.},
  \bibinfo{year}{2015}, \bibinfo{journal}{Astrophys.~J.~Lett.}
  \textbf{\bibinfo{volume}{808}}, \bibinfo{eid}{L14}.

\bibitem[{\citenamefont{{Kawabata}}
  \emph{et~al.}(2017)\citenamefont{{Kawabata}, {Fujikawa}, {Furuno}, {Goto},
  {Hashimoto}, {Ichikawa}, {Itoh}, {Iwasa}, {Kanada-En'yo}, {Koshikawa},
  {Kubono}, {Miyawaki}} \emph{et~al.}}]{Kaw17}
\bibinfo{author}{\bibnamefont{{Kawabata}}, \bibfnamefont{T.}},
  \bibinfo{author}{\bibfnamefont{Y.}~\bibnamefont{{Fujikawa}}},
  \bibinfo{author}{\bibfnamefont{T.}~\bibnamefont{{Furuno}}},
  \bibinfo{author}{\bibfnamefont{T.}~\bibnamefont{{Goto}}},
  \bibinfo{author}{\bibfnamefont{T.}~\bibnamefont{{Hashimoto}}},
  \bibinfo{author}{\bibfnamefont{M.}~\bibnamefont{{Ichikawa}}},
  \bibinfo{author}{\bibfnamefont{M.}~\bibnamefont{{Itoh}}},
  \bibinfo{author}{\bibfnamefont{N.}~\bibnamefont{{Iwasa}}},
  \bibinfo{author}{\bibfnamefont{Y.}~\bibnamefont{{Kanada-En'yo}}},
  \bibinfo{author}{\bibfnamefont{A.}~\bibnamefont{{Koshikawa}}},
  \bibinfo{author}{\bibfnamefont{S.}~\bibnamefont{{Kubono}}},
  \bibinfo{author}{\bibfnamefont{E.}~\bibnamefont{{Miyawaki}}}, \emph{et~al.},
  \bibinfo{year}{2017}, \bibinfo{journal}{Phys. Rev. Lett.}
  \textbf{\bibinfo{volume}{118}}, \bibinfo{eid}{052701}.

\bibitem[{\citenamefont{Kawano}(1992)}]{Kawano:1992ua}
\bibinfo{author}{\bibnamefont{Kawano}, \bibfnamefont{L.}},
  \bibinfo{year}{1992}, \bibinfo{title}{{Let's go: Early universe. 2.
  Primordial nucleosynthesis: The Computer way}}.

\bibitem[{\citenamefont{{Kernan}}(1993)}]{Kernan}
\bibinfo{author}{\bibnamefont{{Kernan}}, \bibfnamefont{P.~J.}},
  \bibinfo{year}{1993}, \emph{\bibinfo{title}{{Two astroparticle physics
  problems: Solar neutrinos and primordial He-4}}}, Ph.D. thesis,
  \bibinfo{school}{Ohio State University}.

\bibitem[{\citenamefont{{Kirsebom} and {Davids}}(2011)}]{Kir11}
\bibinfo{author}{\bibnamefont{{Kirsebom}}, \bibfnamefont{O.~S.}}, and
  \bibinfo{author}{\bibfnamefont{B.}~\bibnamefont{{Davids}}},
  \bibinfo{year}{2011}, \bibinfo{journal}{Phys. Rev.}
  \textbf{\bibinfo{volume}{C84}}, \bibinfo{eid}{058801}.

\bibitem[{\citenamefont{{Korn}} \emph{et~al.}(2006)\citenamefont{{Korn},
  {Grundahl}, {Richard}, {Barklem}, {Mashonkina}, {Collet}, {Piskunov}, and
  {Gustafsson}}}]{Kor06}
\bibinfo{author}{\bibnamefont{{Korn}}, \bibfnamefont{A.~J.}},
  \bibinfo{author}{\bibfnamefont{F.}~\bibnamefont{{Grundahl}}},
  \bibinfo{author}{\bibfnamefont{O.}~\bibnamefont{{Richard}}},
  \bibinfo{author}{\bibfnamefont{P.~S.} \bibnamefont{{Barklem}}},
  \bibinfo{author}{\bibfnamefont{L.}~\bibnamefont{{Mashonkina}}},
  \bibinfo{author}{\bibfnamefont{R.}~\bibnamefont{{Collet}}},
  \bibinfo{author}{\bibfnamefont{N.}~\bibnamefont{{Piskunov}}}, and
  \bibinfo{author}{\bibfnamefont{B.}~\bibnamefont{{Gustafsson}}},
  \bibinfo{year}{2006}, \bibinfo{journal}{Nature}
  \textbf{\bibinfo{volume}{442}}, \bibinfo{pages}{657}.

\bibitem[{\citenamefont{{Kusakabe}}
  \emph{et~al.}(2014)\citenamefont{{Kusakabe}, {Cheoun}, and {Kim}}}]{Kus14}
\bibinfo{author}{\bibnamefont{{Kusakabe}}, \bibfnamefont{M.}},
  \bibinfo{author}{\bibfnamefont{M.-K.} \bibnamefont{{Cheoun}}}, and
  \bibinfo{author}{\bibfnamefont{K.~S.} \bibnamefont{{Kim}}},
  \bibinfo{year}{2014}, \bibinfo{journal}{Phys. Rev.}
  \textbf{\bibinfo{volume}{D90}}, \bibinfo{eid}{045009}.

\bibitem[{\citenamefont{Lesgourgues}(2011)}]{CLASS}
\bibinfo{author}{\bibnamefont{Lesgourgues}, \bibfnamefont{J.}},
  \bibinfo{year}{2011}, \bibinfo{title}{\mbox{CLASS}},
  \bibinfo{howpublished}{\url{http://class-code.net/}}.

\bibitem[{\citenamefont{Lewis and Challinor}(1999)}]{CAMB}
\bibinfo{author}{\bibnamefont{Lewis}, \bibfnamefont{A.}}, and
  \bibinfo{author}{\bibfnamefont{A.}~\bibnamefont{Challinor}},
  \bibinfo{year}{1999}, \bibinfo{title}{\mbox{CAMB}},
  \bibinfo{howpublished}{\url{http://camb.info}}.

\bibitem[{\citenamefont{Lewis} \emph{et~al.}(2000)\citenamefont{Lewis,
  Challinor, and Lasenby}}]{Lewis:1999bs}
\bibinfo{author}{\bibnamefont{Lewis}, \bibfnamefont{A.}},
  \bibinfo{author}{\bibfnamefont{A.}~\bibnamefont{Challinor}}, and
  \bibinfo{author}{\bibfnamefont{A.}~\bibnamefont{Lasenby}},
  \bibinfo{year}{2000}, \bibinfo{journal}{Astrophys. J.}
  \textbf{\bibinfo{volume}{538}}, \bibinfo{pages}{473}.

\bibitem[{\citenamefont{{Lind}} \emph{et~al.}(2013)\citenamefont{{Lind},
  {Melendez}, {Asplund}, {Collet}, and {Magic}}}]{Lin13}
\bibinfo{author}{\bibnamefont{{Lind}}, \bibfnamefont{K.}},
  \bibinfo{author}{\bibfnamefont{J.}~\bibnamefont{{Melendez}}},
  \bibinfo{author}{\bibfnamefont{M.}~\bibnamefont{{Asplund}}},
  \bibinfo{author}{\bibfnamefont{R.}~\bibnamefont{{Collet}}}, and
  \bibinfo{author}{\bibfnamefont{Z.}~\bibnamefont{{Magic}}},
  \bibinfo{year}{2013}, \bibinfo{journal}{Astron.~Astrophys.}
  \textbf{\bibinfo{volume}{554}}, \bibinfo{eid}{A96}.

\bibitem[{\citenamefont{{Longland}}
  \emph{et~al.}(2010)\citenamefont{{Longland}, {Iliadis}, {Champagne},
  {Newton}, {Ugalde}, {Coc}, and {Fitzgerald}}}]{Eval1}
\bibinfo{author}{\bibnamefont{{Longland}}, \bibfnamefont{R.}},
  \bibinfo{author}{\bibfnamefont{C.}~\bibnamefont{{Iliadis}}},
  \bibinfo{author}{\bibfnamefont{A.~E.} \bibnamefont{{Champagne}}},
  \bibinfo{author}{\bibfnamefont{J.~R.} \bibnamefont{{Newton}}},
  \bibinfo{author}{\bibfnamefont{C.}~\bibnamefont{{Ugalde}}},
  \bibinfo{author}{\bibfnamefont{A.}~\bibnamefont{{Coc}}}, and
  \bibinfo{author}{\bibfnamefont{R.}~\bibnamefont{{Fitzgerald}}},
  \bibinfo{year}{2010}, \bibinfo{journal}{Nucl. Phys.}
  \textbf{\bibinfo{volume}{A841}}, \bibinfo{pages}{1}.

\bibitem[{\citenamefont{{Lopez} and {Turner}}(1999)}]{LopezTurner1998}
\bibinfo{author}{\bibnamefont{{Lopez}}, \bibfnamefont{R.~E.}}, and
  \bibinfo{author}{\bibfnamefont{M.~S.} \bibnamefont{{Turner}}},
  \bibinfo{year}{1999}, \bibinfo{journal}{Phys. Rev.}
  \textbf{\bibinfo{volume}{D59}}, \bibinfo{eid}{103502}.

\bibitem[{\citenamefont{Lopez} \emph{et~al.}(1997)\citenamefont{Lopez, Turner,
  and Gyuk}}]{Lopez1997}
\bibinfo{author}{\bibnamefont{Lopez}, \bibfnamefont{R.~E.}},
  \bibinfo{author}{\bibfnamefont{M.~S.} \bibnamefont{Turner}}, and
  \bibinfo{author}{\bibfnamefont{G.}~\bibnamefont{Gyuk}}, \bibinfo{year}{1997},
  \bibinfo{journal}{Phys. Rev.} \textbf{\bibinfo{volume}{D56}},
  \bibinfo{pages}{3191}.

\bibitem[{\citenamefont{{Ma}} \emph{et~al.}(1997)\citenamefont{{Ma},
  {Karwowski}, {Brune}, {Ayer}, {Black}, {Blackmon}, {Ludwig}, {Viviani},
  {Kievsky}, and {Schiavilla}}}]{Ma97}
\bibinfo{author}{\bibnamefont{{Ma}}, \bibfnamefont{L.}},
  \bibinfo{author}{\bibfnamefont{H.~J.} \bibnamefont{{Karwowski}}},
  \bibinfo{author}{\bibfnamefont{C.~R.} \bibnamefont{{Brune}}},
  \bibinfo{author}{\bibfnamefont{Z.}~\bibnamefont{{Ayer}}},
  \bibinfo{author}{\bibfnamefont{T.~C.} \bibnamefont{{Black}}},
  \bibinfo{author}{\bibfnamefont{J.~C.} \bibnamefont{{Blackmon}}},
  \bibinfo{author}{\bibfnamefont{E.~J.} \bibnamefont{{Ludwig}}},
  \bibinfo{author}{\bibfnamefont{M.}~\bibnamefont{{Viviani}}},
  \bibinfo{author}{\bibfnamefont{A.}~\bibnamefont{{Kievsky}}}, and
  \bibinfo{author}{\bibfnamefont{R.}~\bibnamefont{{Schiavilla}}},
  \bibinfo{year}{1997}, \bibinfo{journal}{Phys. Rev.}
  \textbf{\bibinfo{volume}{C55}}, \bibinfo{pages}{588}.

\bibitem[{\citenamefont{Mangano} \emph{et~al.}(2002)\citenamefont{Mangano,
  Miele, Pastor, and Peloso}}]{Mangano2001}
\bibinfo{author}{\bibnamefont{Mangano}, \bibfnamefont{G.}},
  \bibinfo{author}{\bibfnamefont{G.}~\bibnamefont{Miele}},
  \bibinfo{author}{\bibfnamefont{S.}~\bibnamefont{Pastor}}, and
  \bibinfo{author}{\bibfnamefont{M.}~\bibnamefont{Peloso}},
  \bibinfo{year}{2002}, \bibinfo{journal}{Phys. Lett.}
  \textbf{\bibinfo{volume}{B534}}, \bibinfo{pages}{8}.

\bibitem[{\citenamefont{Mangano} \emph{et~al.}(2005)\citenamefont{Mangano,
  Miele, Pastor, Pinto, Pisanti, and Serpico}}]{Mangano2005}
\bibinfo{author}{\bibnamefont{Mangano}, \bibfnamefont{G.}},
  \bibinfo{author}{\bibfnamefont{G.}~\bibnamefont{Miele}},
  \bibinfo{author}{\bibfnamefont{S.}~\bibnamefont{Pastor}},
  \bibinfo{author}{\bibfnamefont{T.}~\bibnamefont{Pinto}},
  \bibinfo{author}{\bibfnamefont{O.}~\bibnamefont{Pisanti}}, and
  \bibinfo{author}{\bibfnamefont{P.~D.} \bibnamefont{Serpico}},
  \bibinfo{year}{2005}, \bibinfo{journal}{Nucl. Phys.}
  \textbf{\bibinfo{volume}{B729}}, \bibinfo{pages}{221}.

\bibitem[{\citenamefont{Mangano} \emph{et~al.}(2006)\citenamefont{Mangano,
  Miele, Pastor, Pinto, Pisanti, and Serpico}}]{Mangano2006}
\bibinfo{author}{\bibnamefont{Mangano}, \bibfnamefont{G.}},
  \bibinfo{author}{\bibfnamefont{G.}~\bibnamefont{Miele}},
  \bibinfo{author}{\bibfnamefont{S.}~\bibnamefont{Pastor}},
  \bibinfo{author}{\bibfnamefont{T.}~\bibnamefont{Pinto}},
  \bibinfo{author}{\bibfnamefont{O.}~\bibnamefont{Pisanti}}, and
  \bibinfo{author}{\bibfnamefont{P.~D.} \bibnamefont{Serpico}},
  \bibinfo{year}{2006}, \bibinfo{journal}{Nucl. Phys.}
  \textbf{\bibinfo{volume}{B756}}, \bibinfo{pages}{100}.

\bibitem[{\citenamefont{Marciano and Sirlin}(2006)}]{Marciano:2005ec}
\bibinfo{author}{\bibnamefont{Marciano}, \bibfnamefont{W.~J.}}, and
  \bibinfo{author}{\bibfnamefont{A.}~\bibnamefont{Sirlin}},
  \bibinfo{year}{2006}, \bibinfo{journal}{Phys. Rev. Lett.}
  \textbf{\bibinfo{volume}{96}}, \bibinfo{pages}{032002}.

\bibitem[{\citenamefont{{Marcucci}}
  \emph{et~al.}(2016)\citenamefont{{Marcucci}, {Mangano}, {Kievsky}, and
  {Viviani}}}]{Mar16}
\bibinfo{author}{\bibnamefont{{Marcucci}}, \bibfnamefont{L.~E.}},
  \bibinfo{author}{\bibfnamefont{G.}~\bibnamefont{{Mangano}}},
  \bibinfo{author}{\bibfnamefont{A.}~\bibnamefont{{Kievsky}}}, and
  \bibinfo{author}{\bibfnamefont{M.}~\bibnamefont{{Viviani}}},
  \bibinfo{year}{2016}, \bibinfo{journal}{Phys. Rev. Lett.}
  \textbf{\bibinfo{volume}{116}}, \bibinfo{eid}{102501}.

\bibitem[{\citenamefont{{Marcucci}}
  \emph{et~al.}(2005)\citenamefont{{Marcucci}, {Viviani}, {Schiavilla},
  {Kievsky}, and {Rosati}}}]{Mar05}
\bibinfo{author}{\bibnamefont{{Marcucci}}, \bibfnamefont{L.~E.}},
  \bibinfo{author}{\bibfnamefont{M.}~\bibnamefont{{Viviani}}},
  \bibinfo{author}{\bibfnamefont{R.}~\bibnamefont{{Schiavilla}}},
  \bibinfo{author}{\bibfnamefont{A.}~\bibnamefont{{Kievsky}}}, and
  \bibinfo{author}{\bibfnamefont{S.}~\bibnamefont{{Rosati}}},
  \bibinfo{year}{2005}, \bibinfo{journal}{Phys. Rev.}
  \textbf{\bibinfo{volume}{C72}}, \bibinfo{eid}{014001}.

\bibitem[{\citenamefont{{Mathews} and {Kusakabe}}(2017)}]{Mat17a}
\bibinfo{author}{\bibnamefont{{Mathews}}, \bibfnamefont{G.~J.}}, and
  \bibinfo{author}{\bibfnamefont{M.}~\bibnamefont{{Kusakabe}}},
  \bibinfo{year}{2017}, \bibinfo{journal}{International Journal of Modern
  Physics E} \textbf{\bibinfo{volume}{26}}, \bibinfo{eid}{1702006}.

\bibitem[{\citenamefont{{Mathews}} \emph{et~al.}(2017)\citenamefont{{Mathews},
  {Kusakabe}, and {Kajino}}}]{MKK17}
\bibinfo{author}{\bibnamefont{{Mathews}}, \bibfnamefont{G.~J.}},
  \bibinfo{author}{\bibfnamefont{M.}~\bibnamefont{{Kusakabe}}}, and
  \bibinfo{author}{\bibfnamefont{T.}~\bibnamefont{{Kajino}}},
  \bibinfo{year}{2017}, \bibinfo{journal}{International Journal of Modern
  Physics E} \textbf{\bibinfo{volume}{26}}, \bibinfo{eid}{1741001}.

\bibitem[{\citenamefont{{Mel{\'e}ndez}}
  \emph{et~al.}(2010)\citenamefont{{Mel{\'e}ndez}, {Casagrande},
  {Ram{\'{\i}}rez}, {Asplund}, and {Schuster}}}]{Mel10}
\bibinfo{author}{\bibnamefont{{Mel{\'e}ndez}}, \bibfnamefont{J.}},
  \bibinfo{author}{\bibfnamefont{L.}~\bibnamefont{{Casagrande}}},
  \bibinfo{author}{\bibfnamefont{I.}~\bibnamefont{{Ram{\'{\i}}rez}}},
  \bibinfo{author}{\bibfnamefont{M.}~\bibnamefont{{Asplund}}}, and
  \bibinfo{author}{\bibfnamefont{W.~J.} \bibnamefont{{Schuster}}},
  \bibinfo{year}{2010}, \bibinfo{journal}{Astron. Astrophys.}
  \textbf{\bibinfo{volume}{515}}, \bibinfo{eid}{L3}.

\bibitem[{\citenamefont{{Mendes}} \emph{et~al.}(2012)\citenamefont{{Mendes},
  {L{\'e}pine-Szily}, {Descouvemont}, {Lichtenth{\"a}ler}, {Guimar{\~a}es}, {de
  Faria}, {Barioni}, {Pires}, {Morcelle}, {Pampa Condori}, {Morais},
  {Leistenschneider}} \emph{et~al.}}]{Men12}
\bibinfo{author}{\bibnamefont{{Mendes}}, \bibfnamefont{D.~R., Jr.}},
  \bibinfo{author}{\bibfnamefont{A.}~\bibnamefont{{L{\'e}pine-Szily}}},
  \bibinfo{author}{\bibfnamefont{P.}~\bibnamefont{{Descouvemont}}},
  \bibinfo{author}{\bibfnamefont{R.}~\bibnamefont{{Lichtenth{\"a}ler}}},
  \bibinfo{author}{\bibfnamefont{V.}~\bibnamefont{{Guimar{\~a}es}}},
  \bibinfo{author}{\bibfnamefont{P.~N.} \bibnamefont{{de Faria}}},
  \bibinfo{author}{\bibfnamefont{A.}~\bibnamefont{{Barioni}}},
  \bibinfo{author}{\bibfnamefont{K.~C.~C.} \bibnamefont{{Pires}}},
  \bibinfo{author}{\bibfnamefont{V.}~\bibnamefont{{Morcelle}}},
  \bibinfo{author}{\bibfnamefont{R.}~\bibnamefont{{Pampa Condori}}},
  \bibinfo{author}{\bibfnamefont{M.~C.} \bibnamefont{{Morais}}},
  \bibinfo{author}{\bibfnamefont{E.}~\bibnamefont{{Leistenschneider}}},
  \emph{et~al.}, \bibinfo{year}{2012}, \bibinfo{journal}{Phys. Rev.}
  \textbf{\bibinfo{volume}{C86}}, \bibinfo{eid}{064321}.

\bibitem[{\citenamefont{{Michaud}} \emph{et~al.}(1984)\citenamefont{{Michaud},
  {Fontaine}, and {Beaudet}}}]{Mic84}
\bibinfo{author}{\bibnamefont{{Michaud}}, \bibfnamefont{G.}},
  \bibinfo{author}{\bibfnamefont{G.}~\bibnamefont{{Fontaine}}}, and
  \bibinfo{author}{\bibfnamefont{G.}~\bibnamefont{{Beaudet}}},
  \bibinfo{year}{1984}, \bibinfo{journal}{Astrophys.~J.}
  \textbf{\bibinfo{volume}{282}}, \bibinfo{pages}{206}.

\bibitem[{\citenamefont{Mirbabayi and Zaldarriaga}(2015)}]{Mirbabayi:2014hda}
\bibinfo{author}{\bibnamefont{Mirbabayi}, \bibfnamefont{M.}}, and
  \bibinfo{author}{\bibfnamefont{M.}~\bibnamefont{Zaldarriaga}},
  \bibinfo{year}{2015}, \bibinfo{journal}{JCAP}
  \textbf{\bibinfo{volume}{1503}}(\bibinfo{number}{03}), \bibinfo{pages}{056}.

\bibitem[{\citenamefont{{Mukhamedzhanov}}
  \emph{et~al.}(2016)\citenamefont{{Mukhamedzhanov}, {Shubhchintak}, and
  {Bertulani}}}]{Muk16}
\bibinfo{author}{\bibnamefont{{Mukhamedzhanov}}, \bibfnamefont{A.~M.}},
  \bibinfo{author}{\bibnamefont{{Shubhchintak}}}, and
  \bibinfo{author}{\bibfnamefont{C.~A.} \bibnamefont{{Bertulani}}},
  \bibinfo{year}{2016}, \bibinfo{journal}{Phys. Rev.}
  \textbf{\bibinfo{volume}{C93}}, \bibinfo{eid}{045805}.

\bibitem[{\citenamefont{Nakamura} \emph{et~al.}(2017)\citenamefont{Nakamura,
  Hasahimoto, Ichimasa, and Arai}}]{Nakamura2017}
\bibinfo{author}{\bibnamefont{Nakamura}, \bibfnamefont{R.}},
  \bibinfo{author}{\bibfnamefont{M.-a.} \bibnamefont{Hasahimoto}},
  \bibinfo{author}{\bibfnamefont{R.}~\bibnamefont{Ichimasa}}, and
  \bibinfo{author}{\bibfnamefont{K.}~\bibnamefont{Arai}}, \bibinfo{year}{2017},
  \bibinfo{journal}{Int. J. Mod. Phys.}
  \textbf{\bibinfo{volume}{E26}}(\bibinfo{number}{08}),
  \bibinfo{pages}{1741003}.

\bibitem[{\citenamefont{{Neff}}(2011)}]{Nef11}
\bibinfo{author}{\bibnamefont{{Neff}}, \bibfnamefont{T.}},
  \bibinfo{year}{2011}, \bibinfo{journal}{Phys. Rev. Lett.}
  \textbf{\bibinfo{volume}{106}}, \bibinfo{eid}{042502}.

\bibitem[{\citenamefont{{Nollett} and {Burles}}(2000)}]{Nol00}
\bibinfo{author}{\bibnamefont{{Nollett}}, \bibfnamefont{K.~M.}}, and
  \bibinfo{author}{\bibfnamefont{S.}~\bibnamefont{{Burles}}},
  \bibinfo{year}{2000}, \bibinfo{journal}{PHys. Rev.}
  \textbf{\bibinfo{volume}{D61}}, \bibinfo{eid}{123505}.

\bibitem[{\citenamefont{Oldengott and Schwarz}(2017)}]{Oldengott:2017tzj}
\bibinfo{author}{\bibnamefont{Oldengott}, \bibfnamefont{I.~M.}}, and
  \bibinfo{author}{\bibfnamefont{D.~J.} \bibnamefont{Schwarz}},
  \bibinfo{year}{2017}, \bibinfo{journal}{Europhys. Lett.}
  \textbf{\bibinfo{volume}{119}}(\bibinfo{number}{2}), \bibinfo{pages}{29001}.

\bibitem[{\citenamefont{{Olive}}(2010)}]{Oli10}
\bibinfo{author}{\bibnamefont{{Olive}}, \bibfnamefont{K.~A.}},
  \bibinfo{year}{2010}, \bibinfo{journal}{ArXiv e-prints} \eprint{1005.3955}.

\bibitem[{\citenamefont{{Olive}} \emph{et~al.}(2012)\citenamefont{{Olive},
  {Petitjean}, {Vangioni}, and {Silk}}}]{Oli12}
\bibinfo{author}{\bibnamefont{{Olive}}, \bibfnamefont{K.~A.}},
  \bibinfo{author}{\bibfnamefont{P.}~\bibnamefont{{Petitjean}}},
  \bibinfo{author}{\bibfnamefont{E.}~\bibnamefont{{Vangioni}}}, and
  \bibinfo{author}{\bibfnamefont{J.}~\bibnamefont{{Silk}}},
  \bibinfo{year}{2012}, \bibinfo{journal}{MNRAS}
  \textbf{\bibinfo{volume}{426}}, \bibinfo{pages}{1427}.

\bibitem[{\citenamefont{{O'Malley}}
  \emph{et~al.}(2011)\citenamefont{{O'Malley}, {Bardayan}, {Adekola}, {Ahn},
  {Chae}, {Cizewski}, {Graves}, {Howard}, {Jones}, {Kozub}, {Lindhardt},
  {Matos}} \emph{et~al.}}]{OMa11}
\bibinfo{author}{\bibnamefont{{O'Malley}}, \bibfnamefont{P.~D.}},
  \bibinfo{author}{\bibfnamefont{D.~W.} \bibnamefont{{Bardayan}}},
  \bibinfo{author}{\bibfnamefont{A.~S.} \bibnamefont{{Adekola}}},
  \bibinfo{author}{\bibfnamefont{S.}~\bibnamefont{{Ahn}}},
  \bibinfo{author}{\bibfnamefont{K.~Y.} \bibnamefont{{Chae}}},
  \bibinfo{author}{\bibfnamefont{J.~A.} \bibnamefont{{Cizewski}}},
  \bibinfo{author}{\bibfnamefont{S.}~\bibnamefont{{Graves}}},
  \bibinfo{author}{\bibfnamefont{M.~E.} \bibnamefont{{Howard}}},
  \bibinfo{author}{\bibfnamefont{K.~L.} \bibnamefont{{Jones}}},
  \bibinfo{author}{\bibfnamefont{R.~L.} \bibnamefont{{Kozub}}},
  \bibinfo{author}{\bibfnamefont{L.}~\bibnamefont{{Lindhardt}}},
  \bibinfo{author}{\bibfnamefont{M.}~\bibnamefont{{Matos}}}, \emph{et~al.},
  \bibinfo{year}{2011}, \bibinfo{journal}{Phys. Rev.}
  \textbf{\bibinfo{volume}{C84}}, \bibinfo{eid}{042801}.

\bibitem[{\citenamefont{{Pagel}} \emph{et~al.}(1992)\citenamefont{{Pagel},
  {Simonson}, {Terlevich}, and {Edmunds}}}]{Pag92}
\bibinfo{author}{\bibnamefont{{Pagel}}, \bibfnamefont{B.~E.~J.}},
  \bibinfo{author}{\bibfnamefont{E.~A.} \bibnamefont{{Simonson}}},
  \bibinfo{author}{\bibfnamefont{R.~J.} \bibnamefont{{Terlevich}}}, and
  \bibinfo{author}{\bibfnamefont{M.~G.} \bibnamefont{{Edmunds}}},
  \bibinfo{year}{1992}, \bibinfo{journal}{Mon. Not. R. Astron. Soc.}
  \textbf{\bibinfo{volume}{255}}, \bibinfo{pages}{325}.

\bibitem[{\citenamefont{{Patrignani} and {Particle Data Group}}(2016 and 2017
  update)}]{PDG17}
\bibinfo{author}{\bibnamefont{{Patrignani}}, \bibfnamefont{C.}}, and
  \bibinfo{author}{\bibnamefont{{Particle Data Group}}}, \bibinfo{year}{2016
  and 2017 update}, \bibinfo{journal}{Chinese Physics C}
  \textbf{\bibinfo{volume}{40}}, \bibinfo{eid}{100001}.

\bibitem[{\citenamefont{{Pettini} and {Cooke}}(2012)}]{Pet12}
\bibinfo{author}{\bibnamefont{{Pettini}}, \bibfnamefont{M.}}, and
  \bibinfo{author}{\bibfnamefont{R.}~\bibnamefont{{Cooke}}},
  \bibinfo{year}{2012}, \bibinfo{journal}{MNRAS}
  \textbf{\bibinfo{volume}{425}}, \bibinfo{pages}{2477}.

\bibitem[{\citenamefont{Pisanti} \emph{et~al.}(2008)\citenamefont{Pisanti,
  Cirillo, Esposito, Iocco, Mangano, Miele, and Serpico}}]{Parthenope}
\bibinfo{author}{\bibnamefont{Pisanti}, \bibfnamefont{O.}},
  \bibinfo{author}{\bibfnamefont{A.}~\bibnamefont{Cirillo}},
  \bibinfo{author}{\bibfnamefont{S.}~\bibnamefont{Esposito}},
  \bibinfo{author}{\bibfnamefont{F.}~\bibnamefont{Iocco}},
  \bibinfo{author}{\bibfnamefont{G.}~\bibnamefont{Mangano}},
  \bibinfo{author}{\bibfnamefont{G.}~\bibnamefont{Miele}}, and
  \bibinfo{author}{\bibfnamefont{P.~D.} \bibnamefont{Serpico}},
  \bibinfo{year}{2008}, \bibinfo{journal}{Comput. Phys. Commun.}
  \textbf{\bibinfo{volume}{178}}, \bibinfo{pages}{956}.

\bibitem[{\citenamefont{Pitrou and Stebbins}(2014)}]{PitrouStebbins}
\bibinfo{author}{\bibnamefont{Pitrou}, \bibfnamefont{C.}}, and
  \bibinfo{author}{\bibfnamefont{A.}~\bibnamefont{Stebbins}},
  \bibinfo{year}{2014}, \bibinfo{journal}{Gen. Rel. Grav.}
  \textbf{\bibinfo{volume}{46}}(\bibinfo{number}{11}), \bibinfo{pages}{1806}.

\bibitem[{\citenamefont{{Pospelov} and {Pradler}}(2010)}]{Pos10}
\bibinfo{author}{\bibnamefont{{Pospelov}}, \bibfnamefont{M.}}, and
  \bibinfo{author}{\bibfnamefont{J.}~\bibnamefont{{Pradler}}},
  \bibinfo{year}{2010}, \bibinfo{journal}{Annual Review of Nuclear and Particle
  Science} \textbf{\bibinfo{volume}{60}}, \bibinfo{pages}{539}.

\bibitem[{\citenamefont{{Reggiani}}
  \emph{et~al.}(2017)\citenamefont{{Reggiani}, {Mel{\'e}ndez}, {Kobayashi},
  {Karakas}, and {Placco}}}]{Reg17}
\bibinfo{author}{\bibnamefont{{Reggiani}}, \bibfnamefont{H.}},
  \bibinfo{author}{\bibfnamefont{J.}~\bibnamefont{{Mel{\'e}ndez}}},
  \bibinfo{author}{\bibfnamefont{C.}~\bibnamefont{{Kobayashi}}},
  \bibinfo{author}{\bibfnamefont{A.}~\bibnamefont{{Karakas}}}, and
  \bibinfo{author}{\bibfnamefont{V.}~\bibnamefont{{Placco}}},
  \bibinfo{year}{2017}, \bibinfo{journal}{Astron.~Astrophys.}
  \textbf{\bibinfo{volume}{608}}, \bibinfo{eid}{A46}.

\bibitem[{\citenamefont{{Richard}} \emph{et~al.}(2005)\citenamefont{{Richard},
  {Michaud}, and {Richer}}}]{Ric05}
\bibinfo{author}{\bibnamefont{{Richard}}, \bibfnamefont{O.}},
  \bibinfo{author}{\bibfnamefont{G.}~\bibnamefont{{Michaud}}}, and
  \bibinfo{author}{\bibfnamefont{J.}~\bibnamefont{{Richer}}},
  \bibinfo{year}{2005}, \bibinfo{journal}{Astrophys. J.}
  \textbf{\bibinfo{volume}{619}}, \bibinfo{pages}{538}.

\bibitem[{\citenamefont{{Riemer-S{\o}rensen}}
  \emph{et~al.}(2017)\citenamefont{{Riemer-S{\o}rensen}, {Kotu{\v s}}, {Webb},
  {Ali}, {Dumont}, {Murphy}, and {Carswell}}}]{Rie17}
\bibinfo{author}{\bibnamefont{{Riemer-S{\o}rensen}}, \bibfnamefont{S.}},
  \bibinfo{author}{\bibfnamefont{S.}~\bibnamefont{{Kotu{\v s}}}},
  \bibinfo{author}{\bibfnamefont{J.~K.} \bibnamefont{{Webb}}},
  \bibinfo{author}{\bibfnamefont{K.}~\bibnamefont{{Ali}}},
  \bibinfo{author}{\bibfnamefont{V.}~\bibnamefont{{Dumont}}},
  \bibinfo{author}{\bibfnamefont{M.~T.} \bibnamefont{{Murphy}}}, and
  \bibinfo{author}{\bibfnamefont{R.~F.} \bibnamefont{{Carswell}}},
  \bibinfo{year}{2017}, \bibinfo{journal}{MNRAS}
  \textbf{\bibinfo{volume}{468}}, \bibinfo{pages}{3239}.

\bibitem[{\citenamefont{de~Salas and Pastor}(2016)}]{deSalas:2016ztq}
\bibinfo{author}{\bibnamefont{de~Salas}, \bibfnamefont{P.~F.}}, and
  \bibinfo{author}{\bibfnamefont{S.}~\bibnamefont{Pastor}},
  \bibinfo{year}{2016}, \bibinfo{journal}{JCAP}
  \textbf{\bibinfo{volume}{1607}}(\bibinfo{number}{07}), \bibinfo{pages}{051}.

\bibitem[{\citenamefont{{Sallaska}}
  \emph{et~al.}(2013)\citenamefont{{Sallaska}, {Iliadis}, {Champange},
  {Goriely}, {Starrfield}, and {Timmes}}}]{Starlib}
\bibinfo{author}{\bibnamefont{{Sallaska}}, \bibfnamefont{A.~L.}},
  \bibinfo{author}{\bibfnamefont{C.}~\bibnamefont{{Iliadis}}},
  \bibinfo{author}{\bibfnamefont{A.~E.} \bibnamefont{{Champange}}},
  \bibinfo{author}{\bibfnamefont{S.}~\bibnamefont{{Goriely}}},
  \bibinfo{author}{\bibfnamefont{S.}~\bibnamefont{{Starrfield}}}, and
  \bibinfo{author}{\bibfnamefont{F.~X.} \bibnamefont{{Timmes}}},
  \bibinfo{year}{2013}, \bibinfo{journal}{Astrophys. J.}
  \textbf{\bibinfo{volume}{207}}, \bibinfo{eid}{18}.

\bibitem[{\citenamefont{{Sawyer}}(1996)}]{Sawyer1996}
\bibinfo{author}{\bibnamefont{{Sawyer}}, \bibfnamefont{R.~F.}},
  \bibinfo{year}{1996}, \bibinfo{journal}{Phys. Rev.}
  \textbf{\bibinfo{volume}{D53}}, \bibinfo{pages}{4232}.

\bibitem[{\citenamefont{{Sbordone}}
  \emph{et~al.}(2010)\citenamefont{{Sbordone}, {Bonifacio}, {Caffau}, {Ludwig},
  {Behara}, {Gonz{\'a}lez Hern{\'a}ndez}, {Steffen}, {Cayrel}, {Freytag},
  {van't Veer}, {Molaro}, {Plez}} \emph{et~al.}}]{Sbo10}
\bibinfo{author}{\bibnamefont{{Sbordone}}, \bibfnamefont{L.}},
  \bibinfo{author}{\bibfnamefont{P.}~\bibnamefont{{Bonifacio}}},
  \bibinfo{author}{\bibfnamefont{E.}~\bibnamefont{{Caffau}}},
  \bibinfo{author}{\bibfnamefont{H.-G.} \bibnamefont{{Ludwig}}},
  \bibinfo{author}{\bibfnamefont{N.~T.} \bibnamefont{{Behara}}},
  \bibinfo{author}{\bibfnamefont{J.~I.} \bibnamefont{{Gonz{\'a}lez
  Hern{\'a}ndez}}},
  \bibinfo{author}{\bibfnamefont{M.}~\bibnamefont{{Steffen}}},
  \bibinfo{author}{\bibfnamefont{R.}~\bibnamefont{{Cayrel}}},
  \bibinfo{author}{\bibfnamefont{B.}~\bibnamefont{{Freytag}}},
  \bibinfo{author}{\bibfnamefont{C.}~\bibnamefont{{van't Veer}}},
  \bibinfo{author}{\bibfnamefont{P.}~\bibnamefont{{Molaro}}},
  \bibinfo{author}{\bibfnamefont{B.}~\bibnamefont{{Plez}}}, \emph{et~al.},
  \bibinfo{year}{2010}, \bibinfo{journal}{Astron. Astrophys.}
  \textbf{\bibinfo{volume}{522}}, \bibinfo{eid}{A26}.

\bibitem[{\citenamefont{{Schaeuble} and {King}}(2012)}]{Sch12}
\bibinfo{author}{\bibnamefont{{Schaeuble}}, \bibfnamefont{M.}}, and
  \bibinfo{author}{\bibfnamefont{J.~R.} \bibnamefont{{King}}},
  \bibinfo{year}{2012}, \bibinfo{journal}{PASP} \textbf{\bibinfo{volume}{124}},
  \bibinfo{pages}{164}.

\bibitem[{\citenamefont{{Schmid}} \emph{et~al.}(1997)\citenamefont{{Schmid},
  {Rice}, {Chasteler}, {Godwin}, {Kiang}, {Kiang}, {Laymon}, {Prior}, {Tilley},
  and {Weller}}}]{Sch97}
\bibinfo{author}{\bibnamefont{{Schmid}}, \bibfnamefont{G.~J.}},
  \bibinfo{author}{\bibfnamefont{B.~J.} \bibnamefont{{Rice}}},
  \bibinfo{author}{\bibfnamefont{R.~M.} \bibnamefont{{Chasteler}}},
  \bibinfo{author}{\bibfnamefont{M.~A.} \bibnamefont{{Godwin}}},
  \bibinfo{author}{\bibfnamefont{G.~C.} \bibnamefont{{Kiang}}},
  \bibinfo{author}{\bibfnamefont{L.~L.} \bibnamefont{{Kiang}}},
  \bibinfo{author}{\bibfnamefont{C.~M.} \bibnamefont{{Laymon}}},
  \bibinfo{author}{\bibfnamefont{R.~M.} \bibnamefont{{Prior}}},
  \bibinfo{author}{\bibfnamefont{D.~R.} \bibnamefont{{Tilley}}}, and
  \bibinfo{author}{\bibfnamefont{H.~R.} \bibnamefont{{Weller}}},
  \bibinfo{year}{1997}, \bibinfo{journal}{Phys. Rev.}
  \textbf{\bibinfo{volume}{C56}}, \bibinfo{pages}{2565}.

\bibitem[{\citenamefont{{Scholl}} \emph{et~al.}(2011)\citenamefont{{Scholl},
  {Fujita}, {Adachi}, {von Brentano}, {Fujita}, {G{\'o}rska}, {Hashimoto},
  {Hatanaka}, {Matsubara}, {Nakanishi}, {Ohta}, {Sakemi}}
  \emph{et~al.}}]{Sch11}
\bibinfo{author}{\bibnamefont{{Scholl}}, \bibfnamefont{C.}},
  \bibinfo{author}{\bibfnamefont{Y.}~\bibnamefont{{Fujita}}},
  \bibinfo{author}{\bibfnamefont{T.}~\bibnamefont{{Adachi}}},
  \bibinfo{author}{\bibfnamefont{P.}~\bibnamefont{{von Brentano}}},
  \bibinfo{author}{\bibfnamefont{H.}~\bibnamefont{{Fujita}}},
  \bibinfo{author}{\bibfnamefont{M.}~\bibnamefont{{G{\'o}rska}}},
  \bibinfo{author}{\bibfnamefont{H.}~\bibnamefont{{Hashimoto}}},
  \bibinfo{author}{\bibfnamefont{K.}~\bibnamefont{{Hatanaka}}},
  \bibinfo{author}{\bibfnamefont{H.}~\bibnamefont{{Matsubara}}},
  \bibinfo{author}{\bibfnamefont{K.}~\bibnamefont{{Nakanishi}}},
  \bibinfo{author}{\bibfnamefont{T.}~\bibnamefont{{Ohta}}},
  \bibinfo{author}{\bibfnamefont{Y.}~\bibnamefont{{Sakemi}}}, \emph{et~al.},
  \bibinfo{year}{2011}, \bibinfo{journal}{Phys. Rev.}
  \textbf{\bibinfo{volume}{C84}}, \bibinfo{eid}{014308}.

\bibitem[{\citenamefont{{Scholz}} \emph{et~al.}(2015)\citenamefont{{Scholz},
  {Heber}, {Heuser}, {Ziegerer}, {Geier}, and {Niederhofer}}}]{Sch15}
\bibinfo{author}{\bibnamefont{{Scholz}}, \bibfnamefont{R.-D.}},
  \bibinfo{author}{\bibfnamefont{U.}~\bibnamefont{{Heber}}},
  \bibinfo{author}{\bibfnamefont{C.}~\bibnamefont{{Heuser}}},
  \bibinfo{author}{\bibfnamefont{E.}~\bibnamefont{{Ziegerer}}},
  \bibinfo{author}{\bibfnamefont{S.}~\bibnamefont{{Geier}}}, and
  \bibinfo{author}{\bibfnamefont{F.}~\bibnamefont{{Niederhofer}}},
  \bibinfo{year}{2015}, \bibinfo{journal}{Astron. Astrophys.}
  \textbf{\bibinfo{volume}{574}}, \bibinfo{eid}{A96}.

\bibitem[{\citenamefont{Seckel}(1993)}]{Seckel1993}
\bibinfo{author}{\bibnamefont{Seckel}, \bibfnamefont{D.}},
  \bibinfo{year}{1993}, \eprint{hep-ph/9305311}.

\bibitem[{Serebrov \emph{et~al.}(2017)\citenamefont{Serebrov}
  \emph{et~al.}}]{Ser17}
\bibinfo{author}{\bibnamefont{Serebrov}, \bibfnamefont{A.~P.}}, \emph{et~al.},
  \bibinfo{year}{2017}, \eprint{1712.05663}.

\bibitem[{\citenamefont{Serpico} \emph{et~al.}(2004)\citenamefont{Serpico,
  Esposito, Iocco, Mangano, Miele, and Pisanti}}]{Serpico:2004gx}
\bibinfo{author}{\bibnamefont{Serpico}, \bibfnamefont{P.~D.}},
  \bibinfo{author}{\bibfnamefont{S.}~\bibnamefont{Esposito}},
  \bibinfo{author}{\bibfnamefont{F.}~\bibnamefont{Iocco}},
  \bibinfo{author}{\bibfnamefont{G.}~\bibnamefont{Mangano}},
  \bibinfo{author}{\bibfnamefont{G.}~\bibnamefont{Miele}}, and
  \bibinfo{author}{\bibfnamefont{O.}~\bibnamefont{Pisanti}},
  \bibinfo{year}{2004}, \bibinfo{journal}{JCAP}
  \textbf{\bibinfo{volume}{0412}}, \bibinfo{pages}{010}.

\bibitem[{\citenamefont{Serpico and Raffelt}(2005)}]{SerpicoRaffelt}
\bibinfo{author}{\bibnamefont{Serpico}, \bibfnamefont{P.~D.}}, and
  \bibinfo{author}{\bibfnamefont{G.~G.} \bibnamefont{Raffelt}},
  \bibinfo{year}{2005}, \bibinfo{journal}{Phys. Rev.}
  \textbf{\bibinfo{volume}{D71}}, \bibinfo{pages}{127301}.

\bibitem[{\citenamefont{Simha and Steigman}(2008)}]{Simha:2008mt}
\bibinfo{author}{\bibnamefont{Simha}, \bibfnamefont{V.}}, and
  \bibinfo{author}{\bibfnamefont{G.}~\bibnamefont{Steigman}},
  \bibinfo{year}{2008}, \bibinfo{journal}{JCAP}
  \textbf{\bibinfo{volume}{0808}}, \bibinfo{pages}{011}.

\bibitem[{\citenamefont{{Simonucci}}
  \emph{et~al.}(2013)\citenamefont{{Simonucci}, {Taioli}, {Palmerini}, and
  {Busso}}}]{Sim13}
\bibinfo{author}{\bibnamefont{{Simonucci}}, \bibfnamefont{S.}},
  \bibinfo{author}{\bibfnamefont{S.}~\bibnamefont{{Taioli}}},
  \bibinfo{author}{\bibfnamefont{S.}~\bibnamefont{{Palmerini}}}, and
  \bibinfo{author}{\bibfnamefont{M.}~\bibnamefont{{Busso}}},
  \bibinfo{year}{2013}, \bibinfo{journal}{Astrophys. J.}
  \textbf{\bibinfo{volume}{764}}, \bibinfo{eid}{118}.

\bibitem[{\citenamefont{{Sirlin}}(1967)}]{Sirlin1967}
\bibinfo{author}{\bibnamefont{{Sirlin}}, \bibfnamefont{A.}},
  \bibinfo{year}{1967}, \bibinfo{journal}{Phys. Rev.}
  \textbf{\bibinfo{volume}{164}}, \bibinfo{pages}{1767}.

\bibitem[{\citenamefont{{Smith} and {Fuller}}(2010)}]{SmithFuller}
\bibinfo{author}{\bibnamefont{{Smith}}, \bibfnamefont{C.~J.}}, and
  \bibinfo{author}{\bibfnamefont{G.~M.} \bibnamefont{{Fuller}}},
  \bibinfo{year}{2010}, \bibinfo{journal}{Phys. Rev.}
  \textbf{\bibinfo{volume}{D81}}, \bibinfo{eid}{065027}.

\bibitem[{\citenamefont{{Spite} and {Spite}}(1982)}]{Spi82}
\bibinfo{author}{\bibnamefont{{Spite}}, \bibfnamefont{F.}}, and
  \bibinfo{author}{\bibfnamefont{M.}~\bibnamefont{{Spite}}},
  \bibinfo{year}{1982}, \bibinfo{journal}{Astron. Astrophys.}
  \textbf{\bibinfo{volume}{115}}, \bibinfo{pages}{357}.

\bibitem[{\citenamefont{{Spite}} \emph{et~al.}(2012)\citenamefont{{Spite},
  {Spite}, and {Bonifacio}}}]{Spi12}
\bibinfo{author}{\bibnamefont{{Spite}}, \bibfnamefont{M.}},
  \bibinfo{author}{\bibfnamefont{F.}~\bibnamefont{{Spite}}}, and
  \bibinfo{author}{\bibfnamefont{P.}~\bibnamefont{{Bonifacio}}},
  \bibinfo{year}{2012}, \bibinfo{journal}{Memorie della Societa Astronomica
  Italiana Supplementi} \textbf{\bibinfo{volume}{22}}, \bibinfo{pages}{9}.

\bibitem[{\citenamefont{{Spite}} \emph{et~al.}(2015)\citenamefont{{Spite},
  {Spite}, {Caffau}, and {Bonifacio}}}]{Spi15}
\bibinfo{author}{\bibnamefont{{Spite}}, \bibfnamefont{M.}},
  \bibinfo{author}{\bibfnamefont{F.}~\bibnamefont{{Spite}}},
  \bibinfo{author}{\bibfnamefont{E.}~\bibnamefont{{Caffau}}}, and
  \bibinfo{author}{\bibfnamefont{P.}~\bibnamefont{{Bonifacio}}},
  \bibinfo{year}{2015}, \bibinfo{journal}{Astron. Astrophys.}
  \textbf{\bibinfo{volume}{582}}, \bibinfo{eid}{A74}.

\bibitem[{\citenamefont{Steigman}(2006)}]{Steigman2006}
\bibinfo{author}{\bibnamefont{Steigman}, \bibfnamefont{G.}},
  \bibinfo{year}{2006}, \bibinfo{journal}{JCAP}
  \textbf{\bibinfo{volume}{0610}}, \bibinfo{pages}{016}.

\bibitem[{\citenamefont{{Steigman}}(2007)}]{Ste07}
\bibinfo{author}{\bibnamefont{{Steigman}}, \bibfnamefont{G.}},
  \bibinfo{year}{2007}, \bibinfo{journal}{Annual Review of Nuclear and Particle
  Science} \textbf{\bibinfo{volume}{57}}, \bibinfo{pages}{463}.

\bibitem[{\citenamefont{{Tajitsu}} \emph{et~al.}(2016)\citenamefont{{Tajitsu},
  {Sadakane}, {Naito}, {Arai}, {Kawakita}, and {Aoki}}}]{Taj16}
\bibinfo{author}{\bibnamefont{{Tajitsu}}, \bibfnamefont{A.}},
  \bibinfo{author}{\bibfnamefont{K.}~\bibnamefont{{Sadakane}}},
  \bibinfo{author}{\bibfnamefont{H.}~\bibnamefont{{Naito}}},
  \bibinfo{author}{\bibfnamefont{A.}~\bibnamefont{{Arai}}},
  \bibinfo{author}{\bibfnamefont{H.}~\bibnamefont{{Kawakita}}}, and
  \bibinfo{author}{\bibfnamefont{W.}~\bibnamefont{{Aoki}}},
  \bibinfo{year}{2016}, \bibinfo{journal}{Astrophys.~J.}
  \textbf{\bibinfo{volume}{818}}, \bibinfo{eid}{191}.

\bibitem[{\citenamefont{Vangioni-Flam}
  \emph{et~al.}(2000)\citenamefont{Vangioni-Flam, Coc, and Cass\'e}}]{Van00}
\bibinfo{author}{\bibnamefont{Vangioni-Flam}, \bibfnamefont{E.}},
  \bibinfo{author}{\bibfnamefont{A.}~\bibnamefont{Coc}}, and
  \bibinfo{author}{\bibfnamefont{M.}~\bibnamefont{Cass\'e}},
  \bibinfo{year}{2000}, \bibinfo{journal}{Astron. Astrophys.}
  \textbf{\bibinfo{volume}{360}}, \bibinfo{pages}{15}.

\bibitem[{\citenamefont{{Vangioni-Flam}}
  \emph{et~al.}(2003)\citenamefont{{Vangioni-Flam}, {Olive}, {Fields}, and
  {Cass{\'e}}}}]{Van03}
\bibinfo{author}{\bibnamefont{{Vangioni-Flam}}, \bibfnamefont{E.}},
  \bibinfo{author}{\bibfnamefont{K.~A.} \bibnamefont{{Olive}}},
  \bibinfo{author}{\bibfnamefont{B.~D.} \bibnamefont{{Fields}}}, and
  \bibinfo{author}{\bibfnamefont{M.}~\bibnamefont{{Cass{\'e}}}},
  \bibinfo{year}{2003}, \bibinfo{journal}{Astrophys.~J.}
  \textbf{\bibinfo{volume}{585}}, \bibinfo{pages}{611}.

\bibitem[{\citenamefont{Vernizzi}(2005)}]{Vernizzi:2004nc}
\bibinfo{author}{\bibnamefont{Vernizzi}, \bibfnamefont{F.}},
  \bibinfo{year}{2005}, \bibinfo{journal}{Phys. Rev.}
  \textbf{\bibinfo{volume}{D71}}, \bibinfo{pages}{061301}.

\bibitem[{\citenamefont{{Wagoner}}(1969)}]{Wag69}
\bibinfo{author}{\bibnamefont{{Wagoner}}, \bibfnamefont{R.~V.}},
  \bibinfo{year}{1969}, \bibinfo{journal}{Astrophys.~J.~Supp.}
  \textbf{\bibinfo{volume}{18}}, \bibinfo{pages}{247}.

\bibitem[{\citenamefont{{Wagoner}}(1973)}]{Wagoner1973}
\bibinfo{author}{\bibnamefont{{Wagoner}}, \bibfnamefont{R.~V.}},
  \bibinfo{year}{1973}, \bibinfo{journal}{Astrophys.~J.}
  \textbf{\bibinfo{volume}{179}}, \bibinfo{pages}{343}.

\bibitem[{\citenamefont{{Wagoner}} \emph{et~al.}(1967)\citenamefont{{Wagoner},
  {Fowler}, and {Hoyle}}}]{Wagoner1967}
\bibinfo{author}{\bibnamefont{{Wagoner}}, \bibfnamefont{R.~V.}},
  \bibinfo{author}{\bibfnamefont{W.~A.} \bibnamefont{{Fowler}}}, and
  \bibinfo{author}{\bibfnamefont{F.}~\bibnamefont{{Hoyle}}},
  \bibinfo{year}{1967}, \bibinfo{journal}{Astrophys.~J.~Supp.}
  \textbf{\bibinfo{volume}{148}}, \bibinfo{pages}{3}.

\bibitem[{\citenamefont{{Wang}} \emph{et~al.}(2011)\citenamefont{{Wang},
  {Bertulani}, and {Balantekin}}}]{Wan11}
\bibinfo{author}{\bibnamefont{{Wang}}, \bibfnamefont{B.}},
  \bibinfo{author}{\bibfnamefont{C.~A.} \bibnamefont{{Bertulani}}}, and
  \bibinfo{author}{\bibfnamefont{A.~B.} \bibnamefont{{Balantekin}}},
  \bibinfo{year}{2011}, \bibinfo{journal}{Phys. Rev.}
  \textbf{\bibinfo{volume}{C83}}, \bibinfo{eid}{018801}.

\bibitem[{\citenamefont{{Weinberg}}(1972)}]{Weinberg1972}
\bibinfo{author}{\bibnamefont{{Weinberg}}, \bibfnamefont{S.}},
  \bibinfo{year}{1972}, \emph{\bibinfo{title}{{Gravitation and Cosmology:
  Principles and Applications of the General Theory of Relativity}}}
  (\bibinfo{publisher}{Wiley}).

\bibitem[{\citenamefont{Weinberg}(2003)}]{Weinberg:2003sw}
\bibinfo{author}{\bibnamefont{Weinberg}, \bibfnamefont{S.}},
  \bibinfo{year}{2003}, \bibinfo{journal}{Phys. Rev.}
  \textbf{\bibinfo{volume}{D67}}, \bibinfo{pages}{123504}.

\bibitem[{\citenamefont{Wietfeldt and Greene}(2011)}]{Wie11}
\bibinfo{author}{\bibnamefont{Wietfeldt}, \bibfnamefont{F.~E.}}, and
  \bibinfo{author}{\bibfnamefont{G.~L.} \bibnamefont{Greene}},
  \bibinfo{year}{2011}, \bibinfo{journal}{Rev. Mod. Phys.}
  \textbf{\bibinfo{volume}{83}}, \bibinfo{pages}{1173}.

\bibitem[{\citenamefont{{Wilkinson}}(1982)}]{Wilkinson1982}
\bibinfo{author}{\bibnamefont{{Wilkinson}}, \bibfnamefont{D.~H.}},
  \bibinfo{year}{1982}, \bibinfo{journal}{Nucl. Phys.}
  \textbf{\bibinfo{volume}{A377}}, \bibinfo{pages}{474}.

\bibitem[{\citenamefont{Wong}(2002)}]{Wong:2002fa}
\bibinfo{author}{\bibnamefont{Wong}, \bibfnamefont{Y.~Y.~Y.}},
  \bibinfo{year}{2002}, \bibinfo{journal}{Phys. Rev.}
  \textbf{\bibinfo{volume}{D66}}, \bibinfo{pages}{025015}.

\bibitem[{\citenamefont{{Xu}} \emph{et~al.}(2013)\citenamefont{{Xu},
  {Takahashi}, {Goriely}, {Arnould}, {Ohta}, and {Utsunomiya}}}]{NACRE2}
\bibinfo{author}{\bibnamefont{{Xu}}, \bibfnamefont{Y.}},
  \bibinfo{author}{\bibfnamefont{K.}~\bibnamefont{{Takahashi}}},
  \bibinfo{author}{\bibfnamefont{S.}~\bibnamefont{{Goriely}}},
  \bibinfo{author}{\bibfnamefont{M.}~\bibnamefont{{Arnould}}},
  \bibinfo{author}{\bibfnamefont{M.}~\bibnamefont{{Ohta}}}, and
  \bibinfo{author}{\bibfnamefont{H.}~\bibnamefont{{Utsunomiya}}},
  \bibinfo{year}{2013}, \bibinfo{journal}{Nucl. Phys.}
  \textbf{\bibinfo{volume}{A918}}, \bibinfo{pages}{61}.

\bibitem[{\citenamefont{{Yamazaki}}
  \emph{et~al.}(2014)\citenamefont{{Yamazaki}, {Kusakabe}, {Kajino}, {Mathews},
  and {Cheoun}}}]{Yam14}
\bibinfo{author}{\bibnamefont{{Yamazaki}}, \bibfnamefont{D.~G.}},
  \bibinfo{author}{\bibfnamefont{M.}~\bibnamefont{{Kusakabe}}},
  \bibinfo{author}{\bibfnamefont{T.}~\bibnamefont{{Kajino}}},
  \bibinfo{author}{\bibfnamefont{G.~J.} \bibnamefont{{Mathews}}}, and
  \bibinfo{author}{\bibfnamefont{M.-K.} \bibnamefont{{Cheoun}}},
  \bibinfo{year}{2014}, \bibinfo{journal}{Phys. Rev.}
  \textbf{\bibinfo{volume}{D90}}, \bibinfo{eid}{023001}.

\bibitem[{\citenamefont{{Young}} \emph{et~al.}(2014)\citenamefont{{Young},
  {Clayton}, {Filippone}, {Geltenbort}, {Ito}, {Liu}, {Makela}, {Morris},
  {Plaster}, {Saunders}, {Seestrom}, and {Vogelaar}}}]{You14}
\bibinfo{author}{\bibnamefont{{Young}}, \bibfnamefont{A.~R.}},
  \bibinfo{author}{\bibfnamefont{S.}~\bibnamefont{{Clayton}}},
  \bibinfo{author}{\bibfnamefont{B.~W.} \bibnamefont{{Filippone}}},
  \bibinfo{author}{\bibfnamefont{P.}~\bibnamefont{{Geltenbort}}},
  \bibinfo{author}{\bibfnamefont{T.~M.} \bibnamefont{{Ito}}},
  \bibinfo{author}{\bibfnamefont{C.-Y.} \bibnamefont{{Liu}}},
  \bibinfo{author}{\bibfnamefont{M.}~\bibnamefont{{Makela}}},
  \bibinfo{author}{\bibfnamefont{C.~L.} \bibnamefont{{Morris}}},
  \bibinfo{author}{\bibfnamefont{B.}~\bibnamefont{{Plaster}}},
  \bibinfo{author}{\bibfnamefont{A.}~\bibnamefont{{Saunders}}},
  \bibinfo{author}{\bibfnamefont{S.~J.} \bibnamefont{{Seestrom}}}, and
  \bibinfo{author}{\bibfnamefont{R.~B.} \bibnamefont{{Vogelaar}}},
  \bibinfo{year}{2014}, \bibinfo{journal}{Journal of Physics G Nuclear Physics}
  \textbf{\bibinfo{volume}{41}}, \bibinfo{eid}{114007}.

\bibitem[{\citenamefont{{Zavarygin}}
  \emph{et~al.}(2017)\citenamefont{{Zavarygin}, {Webb}, {Dumont}, and
  {Riemer-S{\o}rensen}}}]{Zav17}
\bibinfo{author}{\bibnamefont{{Zavarygin}}, \bibfnamefont{E.~O.}},
  \bibinfo{author}{\bibfnamefont{J.~K.} \bibnamefont{{Webb}}},
  \bibinfo{author}{\bibfnamefont{V.}~\bibnamefont{{Dumont}}}, and
  \bibinfo{author}{\bibfnamefont{S.}~\bibnamefont{{Riemer-S{\o}rensen}}},
  \bibinfo{year}{2017}, \bibinfo{journal}{ArXiv e-prints} \eprint{1706.09512}.

\bibitem[{\citenamefont{{Zavarygin}}
  \emph{et~al.}(2018)\citenamefont{{Zavarygin}, {Webb}, {Riemer-S{\o}rensen},
  and {Dumont}}}]{Zav18}
\bibinfo{author}{\bibnamefont{{Zavarygin}}, \bibfnamefont{E.~O.}},
  \bibinfo{author}{\bibfnamefont{J.~K.} \bibnamefont{{Webb}}},
  \bibinfo{author}{\bibfnamefont{S.}~\bibnamefont{{Riemer-S{\o}rensen}}}, and
  \bibinfo{author}{\bibfnamefont{V.}~\bibnamefont{{Dumont}}},
  \bibinfo{year}{2018}, \bibinfo{journal}{ArXiv e-prints} \eprint{1801.04704}.

\end{thebibliography}
%\else
%\fi

%%%%%%%%%%%%%%%%%%%%%
\clearpage 
\end{document}